TOR VERGATA
UNIVERSITÀ DEGLI STUDI DI ROMA

Ph.D. program in
Computer Science, Control and GeoInformation
XXXVII Cycle

# Bringing Order Amidst Chaos: On the Role of Artificial Intelligence in Secure Software Engineering

**Matteo Esposito**

TOR VERGATA
UNIVERSITÀ DEGLI STUDI DI ROMA

Ph.D. program in
Computer Science, Control and GeoInformation
XXXVII Cycle

**Matteo Esposito**

# Bringing Order Amidst Chaos: On the Role of Artificial Intelligence in Secure Software Engineering

| **Thesis Committee** | **Reviewers** |
|---|---|
| Prof. Davide Falessi (Advisor) | Prof. Maya Daneva |
| Prof. Francesco Quaglia | Prof. Tomi Männistö |

A.Y. 2023/2024

Author's address:
**Matteo Esposito**
**Dipartimento di Ingegneria Cive e Ingegneria Informatica (DICII)**
**Università degli Studi di Roma "Tor Vergata"**
**Via del Politecnico 1, 00132 Roma, Italy**
**e-mail:** m.esposito@ing.uniroma2.it

**www:** sere.ing.uniroma2.it/matteo-esposito

# Contents

# List of Figures

# List of Tables

# List of Algorithms

# Dedication

*To my **family**, for their unconditional support, the cornerstone of my life. While this chapter ends, a new one begins and the real journey is just starting!*

*To my advisor, **Prof. Davide Falessi**, for his expert guidance in introducing me to research and trust in my abilities. As all PhD have their own story, ours was a scientifically challenging one, but we came through it!*

*To **Prof. Valentina Lenarduzzi** and **Prof. Davide Taibi**, whose expertise and example marked a turning point in my journey, offering support, hope, and inspiration when I needed it the most. I will always treasure it. I hope to be as inspiring as you one day!*

*To my **fellow PhD colleagues** and companions in this adventure of growth, discovery, and shared challenges. Our curiosity and ideas may one day shape our fields!*

*To all the **faculty**, **staff**, and **friends**, for making this journey richer, funny, and unforgettable. It may have been a rollercoaster, but it was a cool one!*

# Abstract

***Context:*** *Developing secure and reliable software is an enduring challenge in software engineering (SE). The current evolving landscape of technology brings myriad opportunities and threats, creating a dynamic environment where chaos and order vie for dominance. Secure software engineering (SSE) faces the continuous challenge of addressing vulnerabilities that threaten the security of software systems and have broader socio-economic implications, as they can endanger critical national infrastructure and cause significant financial losses. Researchers and practitioners investigated methodologies such as Static Application Security Testing Tools (SASTTs) and artificial intelligence (AI) such as machine learning (ML) and large language models (LLM) to identify and mitigate these vulnerabilities, each possessing unique advantages and limitations.*

***Aim:*** *In this thesis, we aim to bring order to the chaos caused by the haphazard usage of AI in SSE contexts without considering the differences that specific domain holds and can impact the accuracy of AI.*

***Methodology:*** *Our Methodology features a mix of empirical strategies to evaluate effort-aware metrics, analysis of SASTTs, method-level analysis, and evidence-based strategies, such as systematic dataset review, to characterize vulnerability prediction datasets.*

***Results**: Our main results include insights into the limitations of current static analysis tools in identifying software vulnerabilities effectively, such as the identification of gaps in the coverage of SASTTs regarding vulnerability types, the scarce relationship among vulnerability severity scores, an increase in defect prediction accuracy by leveraging just-in-time modeling, and the threats of untouched methods.*

***Conclusions**: In conclusion, this thesis highlights the complexity of SSE and the potential of in-depth context knowledge in enhancing the accuracy of AI in vulnerability and defect prediction methodologies. Our comprehensive analysis contributes to the adoption and research on the effectiveness of prediction models benefiting practitioners and researchers.*

# Chapter 1

# Introduction

*This chapter introduces the context and the rationale of our thesis.*

## 1.1 Context

---

In software engineering, **defect prediction** is essential for improving software reliability and quality, guiding efforts to address potential issues preemptively. Researchers have developed various models for defect prediction, using metrics such as product attributes [35, 155], process details [270], historical defect data [286], change-inducing fixes [200], and even leveraging deep learning for feature engineering directly from source code [400]. However, despite extensive work on predicting defects in commits and classes [165], there has been limited exploration of the potential benefits of integrating these predictive models. We investigate how defect predictions for methods and classes may be improved by incorporating **just-in-time (JIT) defect prediction**, which leverages commit-defectiveness data.

To enhance classifier accuracy in defect prediction, we focus on **effort-aware metrics (EAMs)**, which evaluate how effectively a classifier ranks defect-prone entities [259]. For example, *PofBx* measures the percentage of bugs a developer can identify by inspecting the top $x$ percent of the code. Our work addresses a critical gap by normalizing EAMs to better account for entities' size and importance. This leads to an optimized ranking process and a higher defect detection rate with minimal inspection efforts.

In today's world, software underpins everything from household devices to critical systems in transportation and healthcare [426, 196]. However, as systems grow in complexity, so does the security threat landscape. Vulnerabilities in software represent a costly risk to businesses and a significant threat to national infrastructure [378]. Effective identification of these vulnerabilities relies on **Static Application Security Testing Tools (SASTTs)**, which analyze code for potential weaknesses without executing it [82]. These tools are often benchmarked against the **Common Weakness Enumeration (CWE)** framework [263] and the **Juliet Test Suite** [186], both of which classify and document known software vulnerabilities. Yet, SASTTs frequently generate false positives, driving researchers to explore machine learning as a complementary solution.

Vulnerability prediction has emerged as a proactive measure to prevent potential exploits, helping to secure systems by identifying software sections that may harbor vulnerabilities [355]. Since machine learning models rely on large, validated datasets, **dataset availability and quality** are crucial for reproducible research in vulnerability prediction studies [151]. Addressing this, we provide a systematic review of existing datasets alongside a tool, **VALIDATE**, to streamline dataset discovery and enhance research replicability in vulnerability prediction.

Moreover, the increasing accumulation of unresolved vulnerabilities or **security debt** has underscored the need for prioritizing remediation efforts. Like technical debt, unresolved vulnerabilities can grow exponentially if not addressed, posing significant risks to organizations' security [348]. By introducing a framework for vulnerability prioritization, our work aids organizations in focusing resources on high-risk vulnerabilities to reduce security debt and maintain a strong security posture [358].

In mission-critical environments, the rapid identification of potential risks is essential. Preliminary Security Risk Analysis (PSRA) is used in sectors like healthcare, aerospace, and finance to evaluate whether a scenario might pose a security risk without identifying specific vulnerabilities. **Large Language Models (LLMs)** offer promising potential for PSRA by rapidly processing and synthesizing information to assist in risk assessment. In this work, we present a case study on applying LLMs for PSRA, evaluating a fine-tuned model's proficiency against seven human experts. This study demonstrates LLMs' capability to augment traditional risk analysis and shows that AI may provide valuable support for cybersecurity risk management.

Through these contributions, we address critical areas in defect prediction, vulnerability detection, and risk assessment, advancing both the effectiveness of predictive models and the accessibility of tools for research reproducibility. Our approach strengthens traditional software engineering practices and opens new avenues for AI-driven security solutions and hybrid classical-quantum computing in software reliability.

## 1.2 Goal and Research Questions

---

Developing secure and reliable software is an enduring challenge in software engineering (SE). The current evolving landscape of technology brings myriad opportunities and threats, creating a dynamic environment where chaos and order vie for dominance. Amidst the SE tasks, the role of artificial intelligence (AI) emerges as a beacon of hope, offering promising avenues for enhancing the security and resilience of software systems.

We aim to bring order in the chaos caused by the usage of AI in different contexts without considering the details of the specific domain in which secure software engineering exploits AI. Hence, we focus on two specific domains: **defect** prediction and **vulnerability** detection and prioritization.

Figure 1.2.1 presents the thesis structure, Research Questions (*RQs*), and the published works that helped answer the questions. We seek two answers to two main *RQs*:

$RQ_1$ What are metrics and methods to improve **defect** prediction?

$RQ_2$ Which tools and techniques can support **vulnerability** detection and prediction?

Each main *RQ* has sub-research questions that refine and narrow down the focus of the main ones. Finally, alongside the main *RQs*, we contribute two papers to the emerging field of Quantum Software Engineering (QSE) to democratize access to QC resources and lower the entry barrier to QC computation and programming.

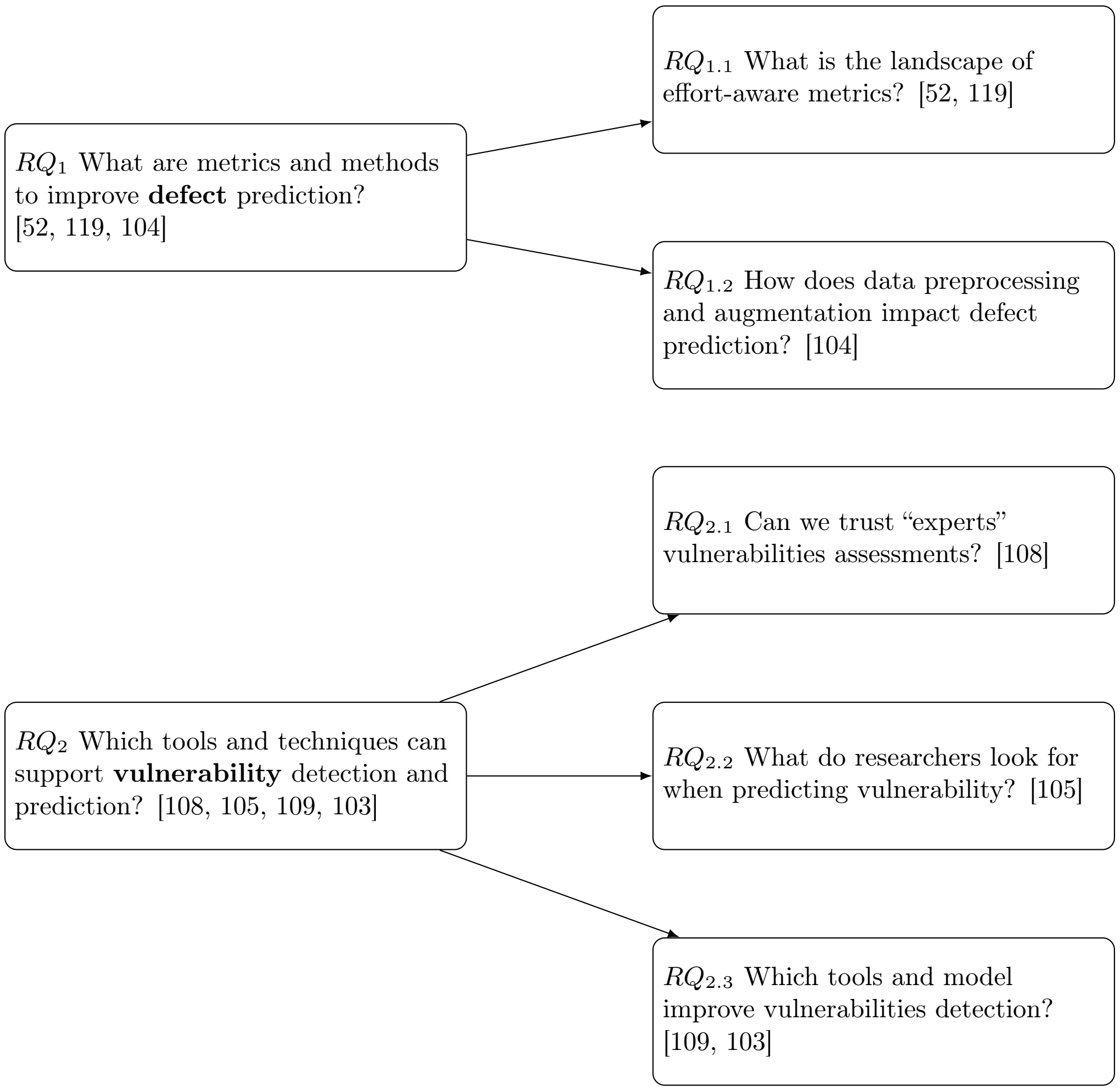

Figure 1.2.1: Research Questions and Published Works

As mentioned before, we aim to contribute, via our work, to the overall AI adoption in SE. To better describe our contributions from the point of view of the AI pipeline, we leverage one broad definition of the AIOps pipeline [324] and map each *RQs* contribution to a specific stage of the pipeline.

AIOps (Artificial Intelligence for IT Operations) uses AI and machine learning to streamline and enhance IT workflows. It starts with extracting and validating data from IT systems, then preprocessing to prepare it for analysis. Models are then trained, tuned, validated, and deployed [324].

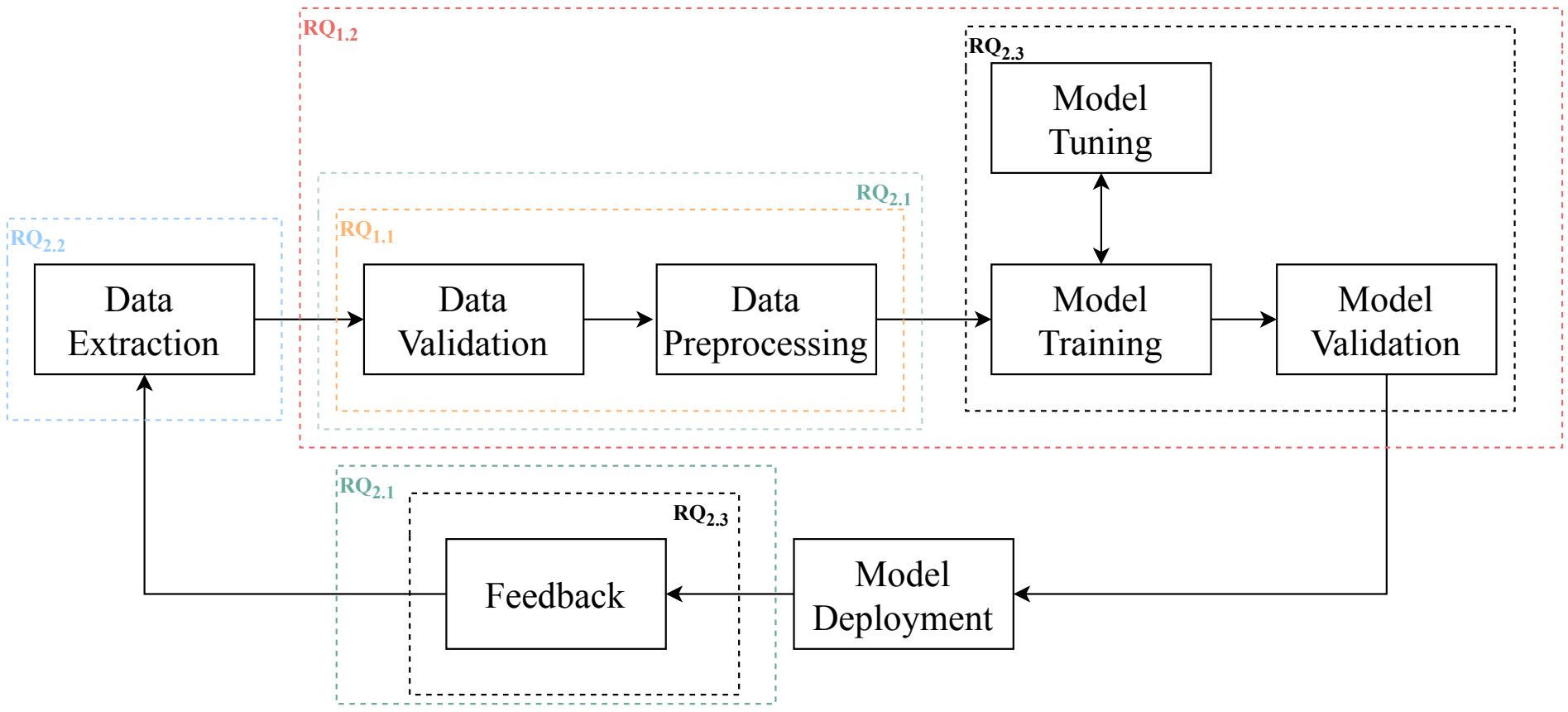

Figure 1.2.2: Thesis Contributions to the AIOps pipeline

Figure 1.2.2 presents the map between sub-RQs and the main AIOps stages. According to Figure 1.2.2, our findings contribute to specific stages covering most of the AIOps pipeline. On the one hand, $RQ_1$ is mainly focused on contributions that validate data and propose novel preprocessing techniques as well as model training, tuning, and validation. On the other hand, $RQ_2$ also covers contributions in the data extraction and the feedback stages.

## 1.3 Contributions

---

Our work advances defect prediction and vulnerability detection techniques through several targeted contributions. First, we conduct a systematic mapping study on **effort-aware metrics (EAMs)**, uncovering trends and revealing issues in the inconsistent use of these metrics. To enhance accuracy and comparability, we propose a **normalization approach** for EAMs, which improves the ranking of classifiers and demonstrates the need for multiple EAMs to evaluate classifier performance fully. We also introduce a computational tool that supports reproducibility and generalizability in defect prediction studies by simplifying EAM computation and ensuring consistent terminology. Similarly, to advance **JIT defect prediction**, we evaluate CDP versus within-project MDP approaches, showing how leveraging JIT information can increase prediction accuracy in both scenarios ($RQ_{1.1}$).

Furthermore, we explore **hidden risks in untouched code** by analyzing the frequency and defectiveness of UM and TM. This analysis clarifies how data segmentation impacts defect prediction and offers models that improve bugginess predictions based on isolated UM and TM data ($RQ_{1.2}$).

In vulnerability detection, we examine the **relationship between CVE, CWE, and SQ rules**, highlighting ambiguities that may mislead practitioners. Our correlation analysis between **NVD severity** and default severity shows how severity ratings influence vulnerability management practices ($RQ_{2.1}$). We also provide the first comprehensive review of **vulnerability prediction datasets** and introduce **VALIDATE**, an online tool that facilitates dataset discovery based on specific features ($RQ_{2.2}$).

Our systematic evaluation of **static analysis tools (SASTTs)** for Java considers the most extensive set of tools and CWEs to date, offering insights into uncovered CWEs and test cases, method-level effectiveness, and effort-aware analysis in vulnerability detection. This comprehensive landscape gives researchers and practitioners an in-depth understanding of SASTTs' utility. We also explore **risk analysis for mission-critical systems** by pioneering the use of large language models (LLMs) in predictive safety risk analysis (PSRA). Our case study demonstrates how LLMs, when fine-tuned, can perform comparably to human experts, revealing the potential of AI in enhancing risk assessment ($RQ_{2.3}$).

Finally, as side-contributions to the emerging new field of QSE, to democratize quantum computing access for classical developers, we introduce $Classi|Q\rangle$, a framework using **ASTs** and **QPLPs** to bridge classical and quantum computing. Complementing this, we design **QCSHQD**, a service-oriented tool that integrates quantum computing into traditional workflows, providing an accessible path for classical developers to harness quantum resources.

Our contributions collectively enhance defect prediction and vulnerability detection by introducing new tools, systematic analyses, and methodologies. Our work also paves new avenues in AI-driven risk assessment and quantum software engineering, addressing traditional software engineering challenges and emerging needs in advanced computation.

# Chapter 2

# State of the Art

*This Chapter presents the background literature and related work to our thesis. For each area of research, we compiled the state of the art and compared the previous methodologies and results with the ones obtained in our contributions to the SSE field.*

## 2.1 Effort Aware Metrics

---

### 2.1.1 Accuracy metrics

Accuracy metrics evaluate the ability of a classifier to provide correct classifications. Examples of accuracy metrics include the following:

- True Positive (TP): The class is actually defective and is predicted to be defective.
- False Negative (FN): The class is actually defective and is predicted to be non-defective.
- True Negative (TN): The class is actually non-defective and is predicted to be non-defective.
- False Positive (FP): The class is actually non-defective and is predicted to be defective.
- **Precision**: $\frac{TP}{TP+FP}$.
- **Recall**: $\frac{TP}{TP+FN}$.
- **F1-score**: $\frac{2*Precision*Recall}{Precision+Recall}$.
- **AUC** (Area Under the Receiving Operating Characteristic Curve) [313] is the area under the curve, of true positive rate versus false positive rate, that is defined by setting multiple thresholds. AUC has the advantage of being threshold-independent.

- **MCC** (Matthews Correlation Coefficient) is commonly used in assessing the performance of classifiers dealing with unbalanced data [249], and is defined as: $\frac{TP*TN-FP*FN}{\sqrt{(TP+FP)(TP+FN)(TN+FP)(TN+FN)}}$. Its interpretation is similar to correlation measures, i.e., $MCC < 0.2$ is considered to be low, $0.2 \leq MCC < 0.4$—fair, $0.4 \leq MCC < 0.6$—moderate, $0.6 \leq MCC < 0.8$—strong, and $MCC \geq 0.8$—very strong.

- **Gmeasure**: $\frac{2*Recall*(1-pf)}{Recall+(1-pf)}$ is the harmonic mean between recall and probability of false alarm (pf ), which denotes the ratio of the number of non-defective modules that are wrongly classified as defective to the total number of non-defective modules as $\frac{FP}{FP+TN}$.[63]

A drawback of the metrics above is that they somehow assume that the costs associated with testing activities are the same for each entity, which is not reasonable in practice. For example, costs for unit testing and code reviews are roughly proportional to the size of the entity under test.

### 2.1.2 Effort-aware metrics

The rationale behind EAM is that they focus on effort reduction gained by using classifiers [259].

In general, there are two types of EAM: normalized by size or not normalized by size. The most known not-normalized EAM is called *PofB* [64, 399, 417, 385] which is defined as the proportion of defective entities identified by analyzing the first x% of the code base as ranked according to their probabilities, as provided by the prediction model, to be defective. The better the ranking, the higher the PofB, the higher the support provided during testing. For instance, a method with a PofB10 of 30% means that 30% of defective entities have been found by analyzing 10% of the codebase using the ranking provided by the method.

Since the PofBX of a perfect ranking is still costly, it is interesting to compare the ranking provided by a prediction model with a perfect ranking; this helps understanding how the prediction model performed compared to a perfect model. Therefore, Mende et al [259], as inspired by Arisholm et al. [25], proposed *Popt* which measures the ranking accuracy provided by a prediction model by taking into account how it is worse than a perfect ranking and how it is better than a random ranking. Popt is defined as the area $\Delta opt$ between the optimal model and the prediction model. In the optimal model, all instances are ordered by decreasing fault density, and in the predicted model, all instances are ordered by decreasing predicted defectiveness. The equation of computing Popt is shown below, where a larger Popt value means a smaller difference between the optimal and predicted model: Popt $=1 - \Delta opt$ [428].

Popt and PofB are two different metrics describing two different aspects of the accuracy of a model. Popt and PofB rank entities in two different ways: Popt according to bug density (i.e., bug probability divided by entity size), PofB according to bug probability. Therefore, the ranking of classifiers provided by Popt and PofB might differ. Finally, Popt is more realistic than PofB as the ranking is based on density rather than probability. However, Popt is harder to interpret than PofB as a classifier with the double of Popt does not provide the double of benefits to its user. Thus, in our thesis, we try to bring the best of PofB and Popt by proposing a new EAM metric that ranks entities similarly to both Popt and PofB.

In the following, we describe additional EAMS.

- *Norm(Popt)*: is introduced by [130] and coincides with Popt20
- *PCI@20%* and *PMI@20%*: have been introduced b [173] and [67] respec-

tively and they represent the Proportion of Changes Inspected and Proportion of Modules Inspected, respectively, when 20% LOC are inspected. Note that these metrics are about the ranking of modules in general rather than about the ranking of defective modules. The idea behind these two similar metrics is that context switches shall be minimized to support effective testing. Specifically, a larger PMI@20% indicates that developers need to inspect more files under the same volume of LOC to inspect. Thus bug prediction models should strive to reduce PMI@20% while trying to increase Popt [315] simultaneously.

- *PFI@20%*: has been introduced by [315] and it coincides with PMI@20 [67] when the module is a file.

- *IFA*: "returns the number of initial false alarms encountered before the first real defective module is found" [67]. This effort-aware performance metric has been considerably influenced by previous work on automatic software fault location[205]. When IFA is high then there are many false positives before detecting the first defective module. [67].

- *Peffort*: has been introduced by [87] and it is similar to our proposed NPofB. Peffort uses the LOC metric as a proxy for inspection effort. Peffort evaluates a ranking of entities based on the number of predicted defects divided by size. In contrast, our NPofB evaluates a ranking of entities based on the predicted defectiveness divided by size.

### 2.1.3 Evaluations

As EAMs drive and impact the results of prediction models evaluations, it is important to discuss studies about how to evaluate prediction models. The evaluation of prediction models performed in studies has been largely discussed.

Many papers explicitly criticized specific empirical evaluations. For instance, [160] criticized the use of the ScottKnottESD test in [372].

[352] found that the choice of classifier has less impact on results than the researcher group. Thus, they suggest conducting blind analysis, improve reporting protocols, and conduct more intergroup studies. [374] replied for a possible explanation for the results aside from the researcher's bias; however, after a few months, [354] concluded that the problem of the researcher's bias remains.

[439] criticized [261] because their results are not satisfactory for practical use due to the small percentage of defective modules. [439] suggest using accuracy metrics, such as Recall and Precision, instead of pd or pf. [260] replied that it is often required to lower precision to achieve higher recall and that there are many domains where low precision is useful. [260], in contrast to [439], advised researchers to avoid the use of precision metric; they suggest the use of more stable metrics (i.e., recall (pd) and false alarm rates) for datasets with a large proportion of negative (i.e. not defective) instances.

[117] reports on the importance of preserving the order of data between the training and testing set. Afterward, the same issue was deeply discussed in [132] Thus, results are unrealistic if the underlying evaluation does not preserve the order of data.

[115] show that dormant defects impact classifiers' accuracy and hence its evaluation. Specifically, an entity, such as a class or method used in the training/testing set, can be labeled in the ground-truth as defective only after the contained defect is fixed. Since defects can sleep for months or years [66, 5] then the entity erroneously seems to be not defective until the defect it contains is fixed. Thus, [5] suggest to ignore the most recent releases to avoid that dormant

defects impact classifiers' accuracy.

[353] commented on the low extent to which published analyses based on the NASA defect datasets are meaningful and comparable.

Very recently [267] proposed a new approach and a new performance metric (the Ratio of Relevant Areas) for assessing a defect proneness model by considering only parts of a ROC curve. They also show the differences and how their metric is more reliable and less misleading compared to the existing ones.

## 2.2 Defect Prediction

In the context of class-level bugginess prediction, various papers were published suggesting methodologies, tools, and metrics to increase the prediction's accuracy. In the context of metrics, two main papers describe a catalog of metrics. Code metrics were first introduced by Menzies et al. [261] showing that despite the dispute of "McCabes versus Halstead versus lines of code counts," the relevant thing was how the attributes were used to build predictors. In this context, the prediction was targeted to more useful prioritization of resource-bound code exploration. Process metrics, on the other hand, were first introduced by Weyuker et al. [405]. They have shown that considering change information about the files, like adding developer information as a prediction factor, such as the number of developers that made changes to each file both in the prior release and cumulatively raised the accuracy of correctly identified bugs, from a 20% to 85%.

Zimmermann et al. [450] Investigated an important paramount question: Where do bugs come from? Their work was centered around three specific releases of the Eclipse project. They showed that the combination of complexity metrics can predict bugs, suggesting that the more complex the code is, the more defects it has, so practically attributing to code complexity the root of software bugs.

Few studies have recently focused on predicting methods' bugginess rather than classes. Giger et al. [146] and Hata et al. [158] first introduced the concept of process and code metrics, also used by Pascarella et al. [298].

McIntosh and Kamei [254] presented bug prediction models at the meth-ods level. The models were based on change metrics and source code metrics typically used in bug prediction. Their results indicate that change metrics significantly outperform source code metrics in general., Still, their study did not consider, like we do, the differences between a touched and an untouched method.

Many recent bug-predicting studies are focusing on predicting at the level of the commit [317, 173, 207, 190, 254, 124] or even of the line [312] into a commit.

In the context of predicting Just-in-time a buggy commit, Kamei and Shihab [190] showed that rapid changes in the properties of fix-inducing changes could impact the performance and interpretation of JIT models. Indeed, fluctuation in the relevance of some metrics at a specific point of the development process, combined with a shift in the development object and the involved entities, may severely drop the model's performance.

Fan et al. [124], as Kamei and Shihab [190], remarked that the labeling part indeed covers the most crucial role and that the model's performance is not influenced only by the amount of data. Inadequate labeling induces great confusion in the model, i.e., JIT models, in which performance can severely drop.

Chen et al. [66] firstly introduced the concept of *dormant defects* (though called dormant bugs). They showed that dormant defects are fixed faster and by more experienced developers. Similarly, Rodriguez-Perez et al. [332] and Costa et al. [80] show that the time to fix a defect, i.e., the sleeping phenomenon, is on average about one year. Thus, we conclude that dataset creation will miss most defects on releases less than a year old. In a recent paper [5] we reported that many defects are dormant and many classes snore. Specifically, on average among projects, most of the defects in a project are dormant for more than 20% of the existing releases, and 2) in most projects the missing rate is more than 25% even if we remove the last 50% of releases. Concerning previous work on dormant defects, including ours, in our thesis, we quantify, for the first time, the effect of dormant defects on defect prediction accuracy and their evaluation, and we provide and evaluate a countermeasure.

Many papers have suggested the use of feature selection to improve the performance of classifiers. For instance, Shivaji et al. [357] used a feature selection technique for bug-prediction using classifiers like Naive Bayes and Support Vector Machine (SVM), reporting that a reduction of 4.1% to 12.52% of the original feature's set yielded optimal classification result increasing both speed and accuracy.

### 2.2.1 Dataset creation from mining version control systems

Regarding the dataset creation, similar to what was done by Giger et al. [146], we used the link between Jira Tickets and Bug-fixing commit to get the original labels for the bugginess. Still, as discovered by Ahluwalia et al. [5], dormant bugs do exist, so following their same workflow, we proceeded to label as buggy the same methods in all their dormant states. A problem of paramount importance when creating defect prediction models is represented by noise in the underlying datasets. In this context, different works study the noise sources and foresee countermeasures for that. Kim et al. [201] measured the impact of noise on defect prediction models and provided guidelines for acceptable noise levels. They also propose a noise detection and elimination algorithm to address this problem. However, the noise studied and removed is supposed to be random.

Tantithamthavorn et al. [373] found that (1) issue report mislabelling is not random; (2) precision is rarely impacted by mislabeled issue reports, suggesting that practitioners can rely on the accuracy of modules labeled as defective by models that are trained using noisy data; (3) however, models trained on noisy data typically achieve about 60% of the recall of models trained on clean data. Herzig et al. [166] reported that 39%

### 2.2.2 Combining heterogeneous predictions

While countless studies investigated how to predict the defectiveness of commits [165, 254, 296, 161, 208, 173, 124, 385, 333, 297, 146], or classes [188, 372, 188, 39, 375, 371, 162, 170, 418, 236, 70, 184, 93, 291, 364, 435, 213, 428, 300, 164, 316, 354, 163, 22, 32, 40, 207, 267, 268, 380, 436, 180, 63, 88] in a separate fashion, to the best of our knowledge, no study other than Pascarella et al. [296], investigated how heterogeneous predictions can benefit one another.

Another family of studies that combines heterogeneous information is the ensemble model, which has been used in the context of defect prediction as a way to combine the prediction of several classifiers [210, 301, 382, 420].

### 2.2.3 Method Defectiveness Prediction

The first proposing of lowering the granularity of defective prediction have been Menzies et al. [261] and Tosun et al. [381]. Giger et al. [146] were the first to perform an MDP study. Specifically, Giger et al. [146] defined a set of product and process features and found that both product and process features support MDP (i.e., F-Measure=86%).

Our paper has been highly inspired by Pascarella et al. [297]. Specifically,

Pascarella et al. [297] provide negative results regarding the performance of MDP. In other words, using the same design of Giger et al. [146], they show that MDP is as accurate as a random approach, i.e., the obtained AUC is about 0.51. We share with them several design decisions, including:

- The use of process metrics as features for MDP. Specifically, "The addition of alternative features based on textual, code smells, and developer-related factors improve the performance of the existing models only marginally, if at all." [297]
- The use of a realistic validation procedure. However, they performed a walk-forward procedure, whereas we performed a simple split by preserving. However, both procedures are realistic since they preserve the order of data [117].

The differences in design include:

- We use an advanced SZZ implementation (RA-SZZ) whereas they use Relink [415].
- The use of a different definition of a defective entity. In our research, an entity is defective from when the defect was injected until the last release before the defect has been fixed.
- We use a different set of classifiers.
- We use effort-aware metrics such as PofB.
- We selected a different set of projects from which we derived the datasets. The change was since we needed the same dataset to produce commit, method, and class data.

The differences in results include that the accuracy achieved by MDP, even without leveraging JIT, is much better than a random approach. Specifically, According to Figure 3.1.19, the median AUC across classifiers and datasets is 0.81 without leveraging JIT and 0.96 when leveraging JIT. Moreover, the proportion of defective methods is lower by about an order of magnitude in our datasets than in their datasets. Those differences are due to the set of changes in the design, and we prefer not to speculate on which specific change caused the difference in results.

To our knowledge, no study is investigating MDP other than Pascarella et al. [297] and Giger et al. [146].

Defect prediction can focus on finer-grained software entities other than methods, such as commits (JIT) and statements [312]. However, these types of entities (commits and statements) seem more helpful when ranked at the moment of a commit instead of during the testing phase (our target phase in our thesis).

## 2.3 Static Analysis Security Testing Tools

---

In this section, we present the background and the related work using SASTTs. Undiscovered vulnerabilities can lead to costly impacts on software firms [377]. Cavusoglu et al. [54] show that a fast fix is essential to avoid loss connected to the vulnerability and its public disclosure. Louridas [240] present the considerable cost associated with inadequate software testing infrastructure; thus, a vulnerability discovery tool is needed [94]. SASTTs are a practical choice to actively discover software vulnerabilities during development [51, 351, 45, 441]. The National Vulnerabilities Database NVD [281] collects the Common Vulnerabilities and Exposures (**CVE**) cataloged via the CWEs [246]. SASTTs are developed to target CWEs for vulnerability discovery. While multiple SASTTs can identify the same vulnerability, only a few can detect more specific and challenging CWEs [288]. NIST provides the SARD Test Suites [44] targetting the most common Programming Languages for benchmarking SASTTs. SASTT analysis can support code review. Tufano et al. [387] very recently conducted an exhaustive examination of the state of the art in code review. The authors evaluate ChatGPT's effectiveness and found that it falls short of outperforming human expertise in code review. Therefore, human intervention remains necessary, regardless of the extent of automation implemented.

The wide spreading of software vulnerability increases over time, and new vulnerability discovery is becoming a matter of every day [272]. To aid developers in reducing the vulnerability present in the latest releases of software, various SASTTs exist [28, 288, 277, 394] targeting different and specific vulnerabilities hence, showing diverse capabilities among them [59]. The most important and

used SASTTs [288, 220, 59, 152, 28, 16, 277, 214, 222, 111] include the following:

FindSecBugs: Security plugin for SpotBugs analyzing Java code for vulnerabilities [138]. Infer Facebook's scalable static analysis tool for Java, C, and C++ detecting errors and vulnerabilities [114]. JLint: Open-source Java code analyzer for identifying errors and security issues [27]. PMD: Java static analysis tool providing detailed reports on issues like unused code [308]. SonarQube: Open-source code quality management tool supporting multiple languages [363]. Snyk Code: Identifies code issues in Java and Python projects, offering detailed reports [362]. Spotbugs: Free Java bytecode analyzer for problems and vulnerabilities with bytecode analysis [366]. VCG: Security review tool for various languages, including Java, with customizable checks [393].

Yang et al. [419], discusses the challenges associated with SASTTs in software development due to difficulties handling large numbers of reported vulnerabilities, a high rate of false warnings, and a lack of guidance in fixing reported vulnerabilities. To address these challenges, the authors propose a set of approaches called Priv. The authors show that it effectively identifies and prioritizes vulnerability warnings, reduces the rate of false positives, and provides complete and correct fix suggestions for a significant percentage of evaluated warnings.

### 2.3.1 Evaluations on test suites

Table 2.3.1 presents the critical aspect of related work on evaluating SASTTs on publicly available benchmarks such as JTS. Oyetoyan et al. [288] discuss the inclusion of tools to aid agile development pipeline in focusing on security aspects; the authors choose to target the 112 CWE provided by the JTS (v. 1.2) testing a total of six different tools, focusing on the accuracy metrics. Li et al.

Table 2.3.1: Related work on the usage of SATTs on JTS.

| | **Oyetoyan et al. [288]** | **Li et al. [220]** | **Charest et al. [59]** | **Goseva-Popstojanova and Perhinschi [152]** | **Arusoaie et al. [28]** | **Alqaradaghi et al. [16]** | **Nguyen-Duc et al. [277]** | **Our Work** |
|---|---|---|---|---|---|---|---|---|
| **Granularity** | File | File * | Method | Functions / Methods | File | Method | File | Method |
| **PL** | Java | Java | Java | Java, C++ | C++ | Java | Java | Java |
| **CWE** | 112 | OWASP [287] | 4 | 22+19 | 51 | MITRE [264] | 112 | 112 |
| **Aspects** | Accuracy, Interview | Accuracy, Usability | Accuracy | Accuracy | Accuracy | Accuracy | Accuracy, Interview | Accuracy |
| **Dataset** | Juliet v1.2 | Juliet | Juliet v1.2 | Juliet, OSS | Toyota [383] | Juliet v1.3 | Juliet v1.3 | Juliet v1.3 |
| **#Tool** | 5+1 | 5 | 4 | 3 | 12 | 4 | 7 | 8 |
| **Raw Results** | No | No | No | No | No | No | **Yes** | **Yes** |
| **Scripts** | **Yes** | No | No | No | **Yes** | No | No | **Yes** |
| **Datasets** | **Yes** | No | No | No | **Yes** | No | **Yes** | **Yes** |

[220] presents the challenge of using IDE-integrated plugins for vulnerability discovery; the authors target 29 CWE from the OWASP 2017 TOP 10, focusing on accuracy and usability [361] metrics, testing a total of five different tools. Li et al. [220] highlighted the differences in what the SASTTs claim to be able to discover and what they can; in the same vain, our work discusses this issue in Section 3.1.2.2. Charest et al. [59] discuss the importance of the SASTTs and their actual discovery capabilities; the authors compared four open-source tools on four specific CWEs from the JTC, focusing on java methods rather than classes. Goseva-Popstojanova and Perhinschi [152] compared SASTTs targetting Java and C++ programming languages resulting, in some cases, in performance worse than the random flip of a coin; the authors tested three SASTTs on 41 CWEs belonging to JTC and Open-Source Software (OSS), targeting functions an methods and focusing on accuracy metrics. Arusoaie et al. [28] investigate on the accuracy of custom developed SASTTs for detecting C++ vulnerability; the authors tested 12 SASTTs targetting 51 CWE from the Toyota ITC Test

Suite [356] focusing on accuracy metrics such as detection rate and false positive rate. Alqaradaghi et al. [16] conduct a comparison experiment with open source SASTTs targeting specific CWE highliting differences in SASTTs ability to discover certain CWEs; the authors targeted six of the official "Weaknesses in the 2020 CWE Top 25 Most Dangerous Software Weaknesses" [264], testing four open source SASTTs focusing on accuracy metrics. Nguyen-Duc et al. [277] investigate the combination of SASTTs results to improve the final result; the authors targeted the 112 CWE from the JTT, testing seven SASTTs, focusing on files and delivering a positive result if and only if all SASTTs agreed on positive.

Li et al. [220] perform a comprehensive study on using five IDE plugins to report vulnerability issues. The authors identify a disparity between the advertised detection capabilities and the actual detection performance of the five plugins, resulting in a high false positive rate. Our work focuses on discovering vulnerable methods [152, 16] targeting 112 CWEs from the JTT for Java, testing six open source SASTTs and two commercial ones.

## 2.4 Vulnerability Prediction

This section introduces ML for Software Engineering (**SE**) and discusses related works in VPS, emphasising reproducibility.

ML can support SE in many tasks [434]. Zhang and Tsai [433] defined the field of SE as a "fertile ground where many software development and maintenance tasks could be formulated as learning problems and approached in terms of learning algorithms". ML can tackle problems that humans usually have a mere grasp or no knowledge at all [433] or optimise solutions to otherwise well-known problems but with inefficient solutions[346, 339]. Researchers focus on systematically reviewing the tasks of SE that benefit from the ML "unparalleled capabilities" [421, 247, 434, 147]. For instance, Lyu et al. [242] discuss the development process that benefited from emerging AIOps models. Kapur and Sodhi [193] discuss the effort estimation based on metrics such as software features similarity and developer activity. Durelli et al. [100] conduct a mapping study on ML applications for software testing.

The activity of mining software repositories (**MSR**) led to ideas and challenges for empirical studies in SE [29, 396, 248]. Hassan et al. [157] point out the effectiveness of MSR in empirically validating new ideas and techniques. MSR activity supports the creation of datasets containing useful information for predictive models [390, 1, 266, 200, 239, 104]. Several studies describe tools, techniques, and advantages in automatic mining [273, 228, 409]. Bavota [36] presents issues in mining software repositories, including the lack of meaningful content in commit messages [167], misclassification of data [24, 373] and the missing link between ticket and commit [30]. Zhou et al. [447] investigate on how automatically identify security patches through commit-related data; Zou et al. [451] uses ML to protect the return pointer. Finally, Vandehei et al. [391] and Falessi et al. [115] highlight the importance, in the data preparation stage, of correctly labelling the data according to the ticketing system and the information from the version control system (**VCS**) platform.

Finally, Ibrahim et al. [175] discuss the significant threat posed by software vulnerabilities in opensource projects, which can compromise system integrity, availability, and confidentiality. Their analysis of the top 100 PHP opensource projects reveals that 27% of them exhibit security vulnerabilities. Increasing cyber-warfare [15] and data breaches [197] threaten critical infrastructure and user privacy [378, 178, 223]. Furthermore, our recent study [108] reveals that vulnerability default severity might be inaccurate.

### 2.4.1 Machine Learning for Vulnerability Prediction Studies

ML for VPS led to several studies that approached challenges and novel ideas in the field [368, 350, 224, 414, 285, 320, 368]. Croft et al. [83] conduct an SLR in the data preparation phase of a VPS study. They show that data is the crucial component of any data-driven application; nevertheless, the preparation phase of a dataset is still full of challenges. Our study relies on Croft et al. [83] study selection as all datasets require the data preparation phase. We focus on the entire dataset rather than only on the data preparation phase. Thus, our selection of studies extends from Croft et al. [83]. We conduct a new SLR adopting Croft et al. [83] search strings and inclusion and exclusion criteria. We expand the search scope by adding two specific keywords, "dataset" and "repository", and by broadening the time range, including studies published until January 2023. Finally, it is essential to emphasise that although Croft et al. [83] focuses on the data preparation phase of VPS, we focus on characterising existing VPS datasets across 9 dimensions.

Jabeen et al. [177] experimented on the effectiveness of different ML and statistical techniques for software vulnerability prediction. The authors use goodness-of-fit and criteria on prediction capabilities to assess the performances. Jabeen et al. [177] show that ML techniques are more effective than statistical ones.

Zheng et al. [445] investigate the factors that can affect the vulnerability detection capabilities of ML models. Zheng et al. [445] use the CountVectorizer[1] to extract features from text, to improve the performance of conventional ML models. The authors show that deep learning models can perform better than traditional ML models. Zhu et al. [448] highlight a "perception gap" between deep learning and human experts in understanding code semantics. In real-world scenarios, deep learning-based methods underperform by over 50% compared to controlled experiments, prompting a deep dive into this phenomenon and exploring current solutions to narrow the gap.

Partenza et al. [295] assess the capabilities of a periodic neural network called ASTNN. To experiment with ASTNN the authors used the Juliet test case suite outperforming the authors' previous research on project Achilles. However, Partenza et al. [295] show that the same neural network performance dropped when tested against OWASP real-world vulnerabilities. The author's research highlights that ad hoc datasets, like Juliet, are unsuitable for ML training due to their asymmetries in the complexity of vulnerable and non-vulnerable codes

[1] `https://scikit-learn.org/stable/modules/generated/sklearn.feature_extraction.text.CountVectorizer.html`

and unconfirmed cases.

Yu et al. [430] discusses the application of active learning for vulnerability identification, introducing HARMLESS, an incremental support vector machine that achieves high recall by inspecting a small portion of source code files. Despite known challenges such as the high computational and annotation costs for new instances, which may reduce the anticipated advantages in human-effort cost reduction, HARMLESS demonstrates effective vulnerability detection.

Jabeen et al. [177] and Zheng et al. [445] and Partenza et al. [295] highlight the impacts of ML techniques on datasets confirming the fundamental idea of VALIDATE, i.e., researchers should focus on ML or statistical techniques aimed at improving previous results rather than reinventing the wheel (i.e., creating datasets). More specifically, Partenza et al. [295] reports problems using ad hoc datasets such as Juliet to train specific ML algorithms. Our SDR reveals that working in the VPS field frequently employs ad-hoc datasets for their ML training and validation phases. Researchers may use VALIDATE to address this problem to identify datasets that align with their innovative ideas and research goals.

### 2.4.2 Replicability and Reproducibility

Giray [147] analyses state of the art and challenges in the engineering of ML systems. They show that the random nature of ML-based systems hinders SE tasks in engineering; moreover, they discuss the lack of tools and a well-proven methodology for engineering processes. The replicability and reproducibility (**R&R**), are among the challenges in engineering ML systems. On a similar topic, Liu et al. [235] focus on R&R in Deep Learning for SE. They show that 10% of the studies have at least one research question about R&R; 62% of the studies do not provide a good source code or original data. The PROMISE repository [344] is the first repository of freely available datasets for SE. As with VALIDATE, with the PROMISE repository, the authors want to support researchers in new ideas rather than spending effort "reinventing the wheel", i.e., creating a new dataset that might already exist. PROMISE has been online since 2005 and supports all fields of SE. Based on the original PROMISE idea and the need for reproducibility highlighted in Liu et al. [235], we developed VALIDATE to serve a specific SE domain, i.e., VPS. It should be noted that PROMISE exists in three distinct versions[2,3,4]. At time of writing, there is only one operative version of PROMISE[4], although the oldest among the three. The advantages of VALIDATE compared to PROMISE are:

- VALIDATE dataset repository includes studies related to each dataset: Cheikhi and Abran [61] comprehensively overview the PROMISE and ISBSG dataset repositories. They reveal a lack of studies associated with specific PROMISE datasets; they note that 70 of 84 PROMISE datasets need more practical usage information.
- VALIDATE allows users to search for datasets based on attributes to improve PROMISE on this aspect: Cheikhi and Abran [61] show that only 37 of 84 PROMISE datasets comprehensively describe data and attributes of the dataset. Of course, this search is possible since VALIDATE is specific to a subdomain of SE.
- Public donated dataset peer review: VALIDATE allows the community to publicly donate VPS datasets by opening an issue report[5] and filling in the necessary filter values. While PROMISE allows users to donate their dataset by email, using GitHub issues allows a public peer review of the candidate dataset.

[2] `http://promise.site.uottawa.ca/SERepository/`
[3] `http://code.google.com/p/promisedata/`
[4] `http://openscience.us/repo`
[5] `https://go.validatetool.org/community`

- Community contributions: VALIDATE users can submit an issue requesting a change or fix a bug in VALIDATE.

PROMISE has the advantage, compared to VALIDATE, of providing the only means to share the repository about all SE subdomains.

### 2.4.3 The impact of datasets and feature selection on the accuracy of ML

To our knowledge, no previous study has investigated optimal dataset selection. Dataset selection is essential in machine learning-related tasks [445]. The dataset characteristics and the representativeness and the relevance of the data can impact the feature selection (FS) [13, 284], the model accuracy and the generalisation capabilities [83, 280]. In this sub-section, we present related work showcasing the importance of the dataset characteristics and selection process in the overall ML/DL application.

Alelyani et al. [13] show how dataset characteristics impact the stability of FS algorithms. The authors extensively analysed the inner properties of different datasets to assess how those affect the consistency of the FS algorithm. Furthermore, Oreski et al. [284] focused on seven characterisation methodologies for datasets and five FS techniques. The author reveals how specific characteristics of a dataset deeply impact the accuracy and the time complexity of FS techniques. Thus, the characteristics of datasets are essential in dimensionality reduction, hence critically influencing the accuracy and generalisation capabilities of predictive models.

Zheng et al. [445] investigate four factors influencing machine learning-based vulnerability detection, including data quality and classification models. As a result, selecting appropriate datasets impacts the performance of a classification or neural network. Likewise, [445, 280, 83] focusing on the data preparation phase of vulnerability prediction studies, emphasise the importance of carefully selecting appropriate datasets for specific research ideas. Nong et al. [280] address the shortage of systematic research on open science practices within software engineering, mainly focusing on deep learning-based software vulnerability detection. The authors performed a comprehensive literature review on 55 original studies, underscoring the importance of meticulous dataset selection. Their investigation reveals that the utilisation of imbalanced or artificially generated datasets resulted in overly optimistic performance assessments, thereby compromising the replicability of most techniques.

Finally, regarding the importance of FS algorithm, Zhang et al. [440] conducted the first extensive empirical study on the correlation between various application features and vulnerability proliferation. Seven FS techinques were applied to nine feature subsets selected from 34 collected features, this allowed the authors to discover that application complexity alone is not the sole determinant of vulnerability discovery. More specifically, human-related factors also significantly explain the proliferation of vulnerabilities.

## 2.5 Vulerability Prioritisation

---

Managing software vulnerabilities is essential for keeping users and data secure. The NVD uses CVEs to disclose a specific vulnerability in proprietary or open-source software publicly. CVEs standardise how security vulnerabilities are identified and tracked. CVEs aid organisations in managing and prioritising their effort [33, 427]. CVE specifies the vulnerabilities via CWE, a taxonomy that identifies and categorises common software weaknesses and vulnerabilities. Specialised tools can detect and prioritise vulnerabilities during software development by examining the source code without running it (i.e., static analysis).

SQ is an open-source platform that offers static code analysis to assist developers in identifying and resolving software vulnerabilities, bugs, and code smells[1]. We define the severity SQ assigns to each rule as "**default severity**". Vulnerability prioritisation is critical in software engineering as it helps identify and address security vulnerabilities promptly and effectively. Given the escalating complexity of software systems and the proliferation of potential threats, it is crucial to prioritize vulnerabilities based on their possible impact on the system [432, 232].

Many recent studies tackled the challenges posed by vulnerability prioritisation. Previous studies focused on creating an analysis model for automatic vulnerability prioritisation and enhancing base metrics and data provided by security advisors or specialised tools.

Huang et al. [172] formulate an analysis model using the fuzzy analytic hierarchy process. Their model aims to analyse the degree to which a specific vulnerability can affect the system, also considering human subjectivity in the final decision-making action in the prioritisation. Our work aims at unveiling the fundamental challenge of having a universal severity rating for each vulnerability. Hence multiple analysis models are required for each distinct analysis context.

[1] `https://sonarsource.com/`

Farris et al. [126] present a new software and network vulnerability management strategy called VULCON centred on time-to-vulnerability remediation and total vulnerability exposure. VULCON considers actual vulnerability scan reports, metadata about the identified vulnerabilities, asset criticality, available personnel, and custom performance metrics to prioritise vulnerabilities automatically. We address the issue of automatic vulnerability prioritisation as represented by SQ rules to test whether SQ rule default severity (i.e., an automatic approach) is a viable approximation of human expert prioritization (i.e., NVD scores).

Moreover, with the growing number of hardware and software vulnerabilities being discovered, manual classification of vulnerability types and prioritisation becomes increasingly tricky, justifying the need for automated machine learning classification [427]. Okutan et al. [283] developed an approach to curate vulnerability reports in real-time and map them to structured vulnerability attribute data using NLP and ML, automating the process and saving time compared to manual methods. Furthermore, Gonzalez et al. [150] highlights the importance of accurate and complete information in CVE reports to prevent software system vulnerability exploits. Hence, our work highlights the need for precise vulnerability prioritization to aid practitioners and researchers in accurately tackling the most relevant threat. Finally, Falessi and Voegele [121] showed that the negative consequences of a quality rule violation can change depending on the specific rule or context. Hence, in the same vein of our work, developers cannot effectively prioritise vulnerabilities unless they conduct context-specific validation and customisation of quality rules accordingly. Our work focuses on the correlation of NVD and SQ Rules vulnerabilities.

# Chapter 3

# Contributions

*This Chapter introduces all of our contributions to the Secure Software Engineering field. It is divided into three chapters: 1)* ***Defects****, in which we present our contributions to the area of defect measurement and predictions; 2)* ***Vulnerabilities****, in which we present our contributions to the area of vulnerability discovery, prediction, severity estimation, and risk analysis and management; 3)* ***Other Areas of Research****, in which we present our main contribution to the emerging field of Quantum Software Engineering.*

## 3.1 Metrics and Methods to Improve Defect Prediction

---

### 3.1.1 On Effort-aware Metrics for Defect Prediction

[Context] Advances in defect prediction models, aka classifiers, have been validated via accuracy metrics. Effort-aware metrics (EAMs) relate to benefits provided by a classifier in accurately ranking defective entities such as classes or methods. PofB is an EAM that relates to a user that follows a ranking of the probability that an entity is defective, provided by the classifier. Despite the importance of EAMs, there is no study investigating EAMs trends and validity. [AIM] The aim of this paper is twofold: 1) we reveal issues in EAMs usage, and 2) we propose and evaluate a normalization of PofBs (aka NPofBs), which is based on ranking defective entities by predicted defect density. [METHOD] We perform a systematic mapping study featuring 152 primary studies in major journals and an empirical study featuring 10 EAMs, 10 classifiers, two industrial, and 12 open-source projects. [RESULTS] Our systematic mapping study reveals that most studies using EAMs use only a single EAM (e.g., PofB20) and that some studies mismatched EAMs names. The main result of our empirical study is that NPofBs are statistically and by orders of magnitude higher than PofBs. [CONCLUSIONS] In conclusion, the proposed normalization of PofBs: (i) increases the realism of results as it relates to a better use of classifiers, and (ii) promotes the practical adoption of prediction models in industry as it shows higher benefits. Finally, we provide a tool to compute EAMs to support researchers in avoiding past issues in using EAMs.

#### 3.1.1.1 Study Design

In this work, we investigate the following research questions:

- **RQ1: Which EAMs are used in software engineering journal papers?** In this research question we investigate the trends in EAMs usage, i.e., which and how many EAMs are used in past studies. We are also interested in understanding if the same study uses multiple EAMs and if the EAMs are consistently defined and computed across different studies.

- **RQ2: Does the normalization improve PofBs?** In this research question we investigate if the normalization of PofBs brings higher accuracy. Higher accuracy means that if we analyze a percent of lines of code of the possibly defective entities, we cover a high number of defective entities following a ranking that is based on both the entities likelihood (to be defective) and its size rather than a ranking that is based only on the entities likelihood. If the normalization of PofBs brings higher accuracy, then studies reporting EAMs, unlike our normalized EAMs, underestimate the benefits of using a classifier for ranking defective classes. Moreover, the normalized EAMs shall be considered more realistic than EAMs since they relate to better classifiers.

- **RQ3: Does the ranking of classifiers change by normalizing PofBs?** In this research question we are interested in understanding if the best classifier of a PofB is also the best classifier of NPofB; i.e. if a classifier results as best in PofB10 then it might not be the best in NPofB10. Suppose the normalization changes the ranking of classifiers. In that case, past studies using PofB are misleading, i.e., past studies might not identify the classifier providing the highest benefit to the user in ranking defective classes.

- **RQ4: Does the ranking of classifiers change across normalized PofBs?** In this research question we are interested in understanding if multiple NPofBs are needed to support a comprehensive understanding of classifier accuracy. In other words, we want to know if different NPofBs rank classifiers in the same way. If different NPofBs rank classifiers differently, then results related to a single NPofB cannot be generalized to the overall ranking effectiveness provided by classifiers; i.e. if a classifier resulted as best in NPofB20 then it might not be the best in NPofB10.

**RQ1: Which EAMs are used in software engineering journal papers?**

To investigate the trends in EAMs usage, we carried out a mapping study (MS) in the first semester of 2021 by following the Kitchenham and Charters guidelines [204].

We performed the MS by applying the following query in the tile:

$$(bug\, OR\, defect)\; AND\; (prediction\, OR\, estimation)$$

To make the MS feasible to our effort constraints we focused on the top five journals in the software engineering areas: IEEE Transactions on Software Engineering, ACM Transactions on Software Engineering and Methodology, Empirical Software Engineering and Measurement, Journal of Systems and Software, and Information and Software Technology.

We excluded conferences since they pose space constraints. Specifically, we wanted to be sure that a limited use of EAM was a deliberate design decision of the authors rather than a decision to meet the (conference) space constraints.

We limited our search in the last ten years, i.e., [January 2010 - March 2021] given effort constraints. This search provided us a set of about 179 papers. Then we applied the following exclusion criteria:

- Comments and answer to comments kind of papers.
- Systematic and mapping study kind of papers.
- Practitioners' opinions kind of papers.
- Studies about models predicting things other than defectiveness such as ticket resolution time.

After applying the exclusion criteria, we focused the remainder of the MS on 152 primary studies.

Once we applied the above-mentioned exclusion criteria for each paper, we checked the name of the EAMs used and their definition (i.e., how it was computed). Thus, we started from an empty list of EAMs and we improved the list as we analyzed the papers. Both authors have independently performed the data extracting and synthesizing all papers after training on a small set of papers. The results of the authors perfectly coincided.

**RQ2: Does the normalization improve PofBs?**

In general, EAMs try to measure the ranking effectiveness of prediction models. The rationale behind EAMs is to measure the effort testers require to find a specific percent of defects by following a ranked set of entities possibly containing defects. Since the testing effort varies according to the size of the entities under test, we had the intuition that the ranking of entities, is more effective if it takes into consideration both the likelihood of the entity to be defective and also its size. Therefore, in this paper we propose and validate a new EAM that measures the ranking effectiveness of prediction models when the size of the ranked entities normalizes the ranking; i.e., it measures the effectiveness of an effort-aware ranking.

To investigate if the normalization increases PofBs we perform an empirical study based on within-project across-release class-level defect prediction. Specifically, we observe if the PofB of the same classifier on the same dataset increases after the normalization. As datasets we use the same two industry projects and 12 open-source projects we successfully used in a recent study [117]. We refer to the recent study for details about the size and characteristics of the projects.

**Independent variable** The independent variable of this research question is the presence or absence of normalization in computing PofBs. In this study, we use the term, feature, to refer to the input (e.g., CHURN) of a classifier. Our independent variable is the normalization of the ranking by size as this is what we conjecture influences the ranking effectiveness. We note that in some studies, that are different from the present one, the features are the independent variables.

**Dependent variables** The dependent variable of this research question is the score of PofB with and without the normalization. As PofBs we considered the spectrum from 10 to 90 with a step of 10. We neglected PofB0 since this is always zero and PofB100 since this is always 100. We also considered the AveragePofB as computed as the average between the PofBs from 0 to 100 with a step of 10. Thus we considered ten different PofBs.

In addition to comparing the two scores, with versus without the normalization, in this paper we observe the relative gain provided by the normalization as defined as $\frac{(NPofB-PofB)}{PofB}$ where NPofB represents the normalized score of PofB.

**Measurement procedure** For each project, we:

1. Perform preprocessing:

   - Normalization: we normalize the data with log10 as performed in a related study [179, 375].
   - Feature Selection: we filter the independent variables described above by using the correlation-based feature subset selection [156, 145, 207]. The approach evaluates the worth of a subset of features by considering the individual predictive capability of each feature, as well as the degree of redundancy between different features. The approach searches the space of feature subsets by a greedy hill-climbing augmented with a backtracking facility. The approach starts with an empty set of features and performs a forward search by considering all possible single-feature additions and deletions at a given point.
   - Balancing: we apply SMOTE [60, 3] so that each dataset is perfectly balanced.

2. Create the Train and Test datasets by adopting the above walk-forward validation technique. Specifically, our context is the within-project across-release class-level defect prediction. As a measurement procedure, we adopt the walk-forward validation technique suggested in a recent study [117]. In this technique, the project is first organized in releases. Afterwards, there is a loop *for n =2, n++, up to n= max releases* where the data of the initial *n-1* releases is used as training set, and the data of the last *n* release is used as testing set. This technique has the advantage of preserving the order of data and hence avoiding that data from the future is used to predict data in the past. Moreover, the technique is fully replicable as there is no random mechanism. The disadvantage is that it requires the project to have at least two releases. The random aspects in our classifiers, if any, are controlled by seeds that are used as a parameter, i.e., input, of the classifiers. Therefore, our classifiers are deterministic rather than stochastic, i.e., our results coincide over multiple runs on the same train-test pair. Thus, there is no need to perform a sensitivity analysis of our results. Our set of 14 projects, analyzed via a walk-forward technique, leads to a total of 71 datasets (i.e., 71 specific combinations of training and testing sets). For instance, since KeymindA consists of five releases, then walk-forward on KeymindA leads to 4 datasets. Again, we forward the reader to the previous study for further details about the datasets [117].

3. Compute predicted probability of defectiveness of each class by using each of the ten classifiers.

4. Compute PofBs and NPofBs.

As classifiers we used the ones used in a previous study [117]:

- Decision Table: Two major parts: schema, the set of features included in the table, and a body, labeled instances defined by features in the schema. Given an unlabeled instance, try matching instance to record in the table. [206]

- IBk: Also known as the k-nearest neighbor's algorithm (k-NN), which is a non-parametric method. The classification is based on the majority vote of its neighbors, with the object being assigned to the class most common among its k nearest neighbors [18]. K-nearest neighbors classifier run with k = 1 [4].

- J48: Generates a pruned C4.5 decision tree [319].
- KStar: Instance-based classifier using some similarity function. Uses an entropy-based distance function [74].
- Naive Bayes: Classifies records using estimator classes and applying Bayes theorem [185] i.e., it assumes that the contribution of an individual feature towards deciding the probability of a particular class is independent of other features in that project instance[253].
- SMO: John Platt's sequential minimal optimization algorithm for training a support vector classifier [307]
- Random Forest: Ensemble learning creating a collection of decision trees. Random trees correct for overfitting [48].
- Logistic Regression: It estimates the probabilities of the different possible outcomes of a categorically distributed dependent variable, given a set of independent variables. The estimation is performed through the logistic distribution function [212].
- BayesNet:Bayesian networks (BNs), also known as belief networks (or Bayes nets for short), belong to the family of probabilistic graphical models (GMs). These graphical structures are used to represent knowledge about an uncertain domain. In particular, each node in the graph represents a random variable, while the edges between the nodes represent probabilistic dependencies among the corresponding random variables. These conditional dependencies in the graph are often estimated by using known statistical and computational methods. Hence, BNs combine principles from graph theory, probability theory, computer science, and statistics [38].

- Bagging: Probably the most well-known sampling approach. Given a training set, bagging generates multiple bootstrapped training sets and calls the base model learning algorithm with each of them to yield a set of base models [209].

**Analysis Procedure** We compare the value of PofBs of the same classifier on the same dataset, with versus without the normalization.

Since our data strongly deviate from normality, the hypotheses of this research question are tested using the Wilcoxon signed-rank test [408]. The test is paired since the compared distributions, with versus without normalization, are related to the identical objects (i.e., the score of the same ten classifiers, ten classifiers, on the same 71 datasets). We also use the Cliff's delta (paired) to analyze the effect size [153]. In order to interpret the Cliff's delta (paired) effect size, we used the following standard interpretation [392]:

| **Cliff's delta value** | **Interpretation** |
|---|---|
| $>= 0.11$ | Small |
| $>= 0.28$ | Medium |
| $>= 0.43$ | Large |

Table 3.1.1: Interpretation of Cliff's delta effect size.

**RQ3: Does the ranking of classifiers change by normalizing PofBs?**

Since the normalization of PofBs results in higher accuracy (RQ2), it is interesting to understand the validity of past studies since they do not normalize

PofBs. Suppose the normalization changes the ranking of prediction models. In that case, past studies using PofBs are misleading, i.e., past studies might not identify the prediction model providing the highest benefit to the user.

The dependent variable is the rank of classifiers. The independent variable is the presence or absence of the PofB normalization.

In this research question, we leverage RQ2 results, i.e., the accuracy of 10 classifiers over 72 datasets grouped in 14 projects. To compare the rankings, we use the Spearman's rank correlation (Spearman, 1904) between the ranking of classifiers provided by the same PofB, with versus without the normalization, in each of the 72 datasets. To compare the rankings we use the Spearman's rank correlation [50] between the ranking of classifier provided by the same PofB, with versus without the normalization. In order to interpret Spearman's values, we used the following standard interpretation [11]:

| $\rho$ **Spearman's value** | **Interpretation** |
|---|---|
| $< 0.6$ | fair |
| $< 0.8$ | moderate |
| $< 0.9$ | very strong |
| $= 1$ | perfect |

Table 3.1.2: Interpretation of Rho Spearman's values.

We also compare, for each dataset and PofB, if the best classifier coincides after the normalization.

**RQ4: Does the ranking of classifiers change across normalized PofBs?**

Since past studies used a very limited set of EAMs (RQ1), it is interesting to understand if it is a valid design decision to use a limited set of NPofBs. Suppose different NPofBs rank classifiers differently. In that case, the results related to a single NPofB cannot be generalized to the overall ranking effectiveness provided by prediction models; i.e. if a prediction model resulted as best in NPofB20 then it might not be the best in NPofB10.

Table 3.1.3: EAM used in past studies.

| Popt20 | Popt30 | Popt40 | AveragePopt | PCI20 | PofB20 | Peffort | IFA | Norm(Popt) | PMI@20% | PFI@20% |
|---|---|---|---|---|---|---|---|---|---|---|
| 7 | 1 | 1 | 8 | 1 | 7 | 1 | 4 | 1 | 1 | 1 |

Table 3.1.4: Number of EAM used in past studies.

| # EAM | 0 | 1 | 2 | 3 | 4 | $\geq 5$ |
|---|---|---|---|---|---|---|
| # PS | 132 | 12 | 5 | 2 | 1 | 0 |

To compare the rankings we use the Spearman's rank correlation [50] between the ranking of classifiers provided by each pair of NPofBx, with x in the range [10, 90]. As in RQ3, in this research question we leverage RQ2 results. Specifically, each classifier, in each of the 72 datasets, has a ranking in the range [1,10] (as we used 10 classifiers) with a specific PofBx. We compute the Spearman's values across each combination of NPofBx, with x in the range [10, 90]. We also compare, for each dataset, the proportion of ten NPofBs sharing the same classifier as best.

#### 3.1.1.2 Study Results

**RQ1: Which EAMs are used in software engineering journal papers?**

Table 3.1.3 reports the EAMs used in software engineering journal papers. According to Table 3.1.3 the most used EAM is AveragePopt and PofB20.

Table 3.1.4 reports the number of EAMs used in past studies. According to Table 3.1.4 the majority of the studies used no EAM and hence ignored to validate the model according to their impact on effort. Moreover, the majority of studies using EAMs used a single EAM (i.e., 12 out of 20).

Table 3.1.5 reports the number of studies correctly or incorrectly naming

Table 3.1.5: Number of studies correctly or incorrectly naming EAM.

| | |
|---|---|
| # PS Correctly naming EAM | 13 |
| # PS Incorrectly naming EAM | 7 |

EAMs according to their original definitions [259, 64]. According to Table 3.1.5 seven out of 20 studies incorrectly named EAM.

**RQ2: Does the normalization improve PofBs?**

Figure 3.1.1 reports ten PofBs, and their normalization, of 10 classifiers over the 72 datasets grouped in 14 projects. Figure 3.1.2 reports the gain achieved by normalizing a specific PofB metric. According to Figure 3.1.2 the normalization increases the performance of the median classifier of all PofBs in all 14 projects.

Table 3.1.6 reports the average gain, across datasets and classifiers, in normalizing a specific EAM. According to Table 3.1.6 the relative gain doubles when decreasing the PofB metric; e.g., the relative gain in PofB10 is double than PofB20, which is the double of PofB30.

Table 3.1.6 reports the statistical test results comparing a PofB before and after the normalization. According to Table 3.1.6, the normalization significantly improves all ten PofB metrics. Therefore, we can reject H10 for all ten EAMs. Moreover, the effect size resulted as large for all ten EAMs.

**RQ3: Does the ranking of classifiers change by normalizing PofBs?**

Table 3.1.7 reports the correlation between the same PofB before and after the normalization. According to Table 3.1.7 the correlation is only fair between the same PofB before and after the normalization in nine out of 10 PofBs.

Figure 3.1.3 reports the proportion of times the same PofB metrics, with and without normalization, identifies the same classifier as best. According to

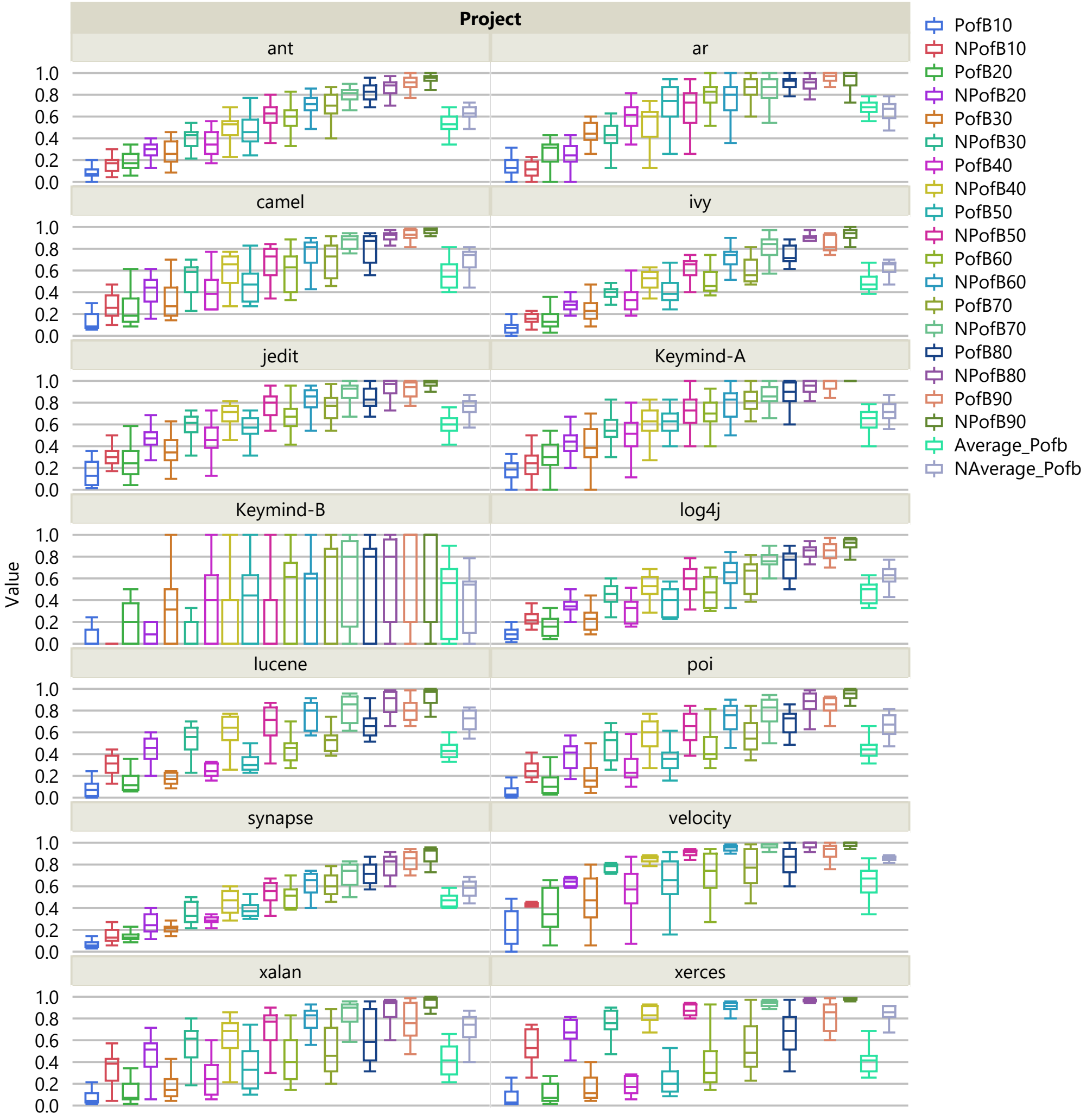

Figure 3.1.1: PofBs, and their normalization (NPofBs), of 10 classifiers over the 14 projects.

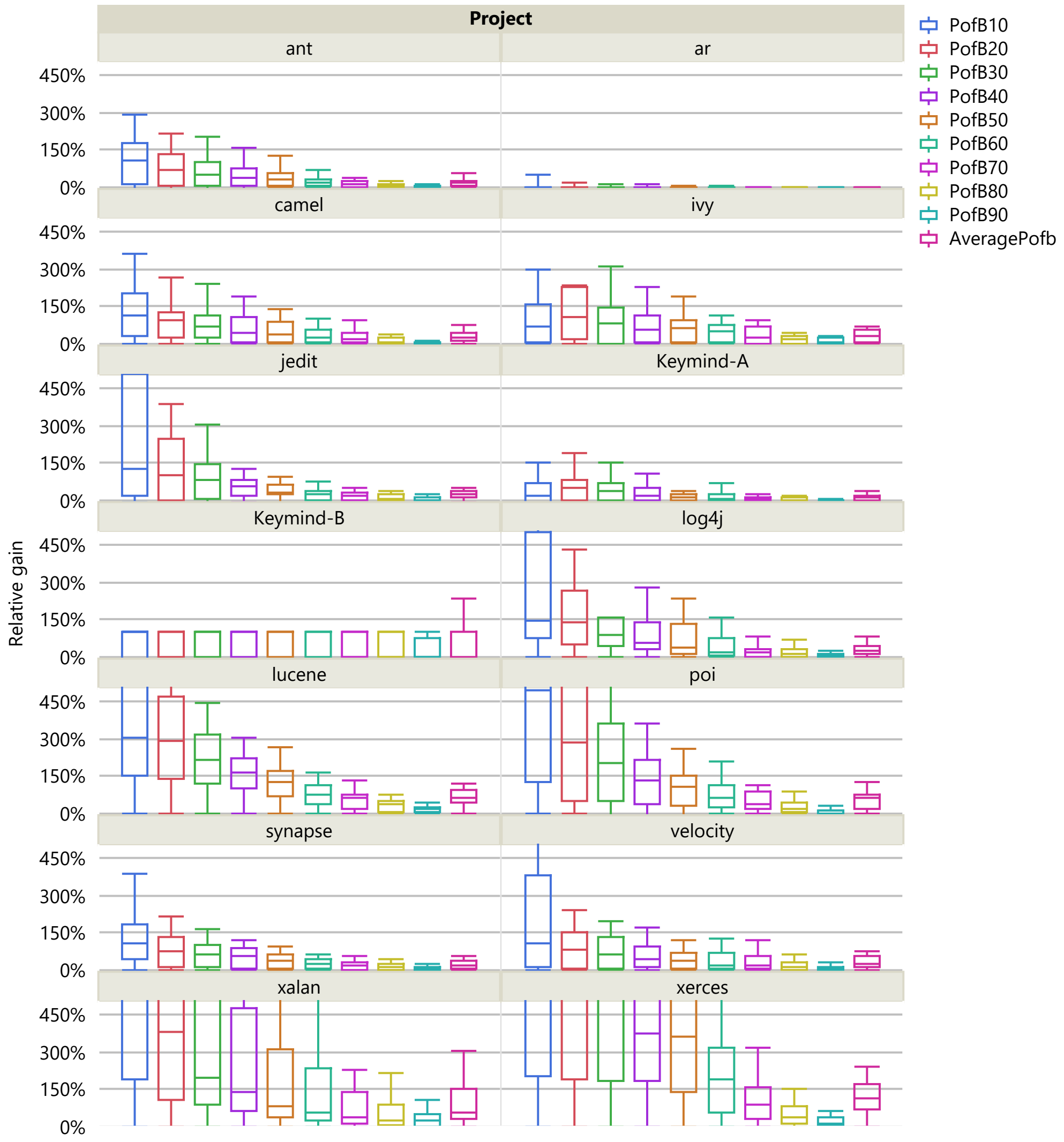

Figure 3.1.2: Average by project of the relative gain in normalizing a specific PofB.

Table 3.1.6: Relative gain, statistical test and Cliff's delta results comparing a PofB metric with its normalization.

| **Compared Metrics** | **Relative Gain** | **Test Statistic S** | **Pvalue** | **Cliff's Delta** |
|---|---|---|---|---|
| PofB10 vs. NPofB10 | 414% | 31267 | <0.0001 | 0.5804 |
| PofB20 vs. NPofB20 | 224% | 32089 | <0.0001 | 0.5726 |
| PofB30 vs. NPofB30 | 147% | 33804 | <0.0001 | 0.5588 |
| PofB40 vs. NPofB40 | 107% | 34122 | <0.0001 | 0.5497 |
| PofB50 vs. NPofB50 | 78% | 34119 | <0.0001 | 0.5217 |
| PofB60 vs. NPofB60 | 51% | 33533 | <0.0001 | 0.4965 |
| PofB70 vs. NPofB70 | 35% | 33700 | <0.0001 | 0.4695 |
| PofB80 vs. NPofB80 | 23% | 32107 | <0.0001 | 0.4318 |
| PofB90 vs. NPofB90 | 11% | 31094 | <0.0001 | 0.3540 |
| Average Pofb vs. NAverage Pofb | 38% | 34446 | <0.0001 | 0.5515 |

Table 3.1.7: Correlation between the same PofB before and after the normalization.

| **Compared metrics** | **Rho** | **Pvalue** | **Intepretation** |
|---|---|---|---|
| PofB10 vs. NPofB10 | 0.121 | <0.0001 | Fair |
| PofB20 vs. NPofB20 | 0.112 | <0.0001 | Fair |
| PofB30 vs. NPofB30 | 0.219 | <0.0001 | Fair |
| PofB40 vs. NPofB40 | 0.246 | <0.0001 | Fair |
| PofB50 vs. NPofB50 | 0.316 | <0.0001 | Fair |
| PofB60 vs. NPofB60 | 0.349 | <0.0001 | Fair |
| PofB70 vs. NPofB70 | 0.410 | <0.0001 | Fair |
| PofB80 vs. NPofB80 | 0.440 | <0.0001 | Fair |
| PofB90 vs. NPofB90 | 0.607 | <0.0001 | Moderate |
| Average_PofB vs. NAverage_PofB | 0.245 | <0.0001 | Fair |

Figure 3.1.3, the proportion of times the same PofB metrics, with and without normalization, identifies the same classifier as best changes across projects and PofBs. Specifically, in half of the datasets, the best classifier changes in all PofBs after the normalization. Table 3.1.8 summarizes Figure 3.1.3 by reporting in average per specific PofBs, the proportion of times the same PofB metrics, with and without normalization, identifies the same classifier as best. According to Table 3.1.8, in all ten PofBs the proportion of times that the same PofB metrics, with and without normalization, identifies the same classifier as best is less than half. Thus, the best classifier is more likely to change than to coincide when considering PofB after the normalization.

**RQ4: Does the ranking of classifiers change across normalized PofBs?**

Table 3.1.10 in the appendix reports the correlation among each couple of normalized PofBs. To better summarize Table 3.1.10, Figure 3.1.4 reports the distribution of the correlations among each couple of NPofBs. Table 3.1.9 reports the frequency of interpretations of correlation values among couples of NPofBs.

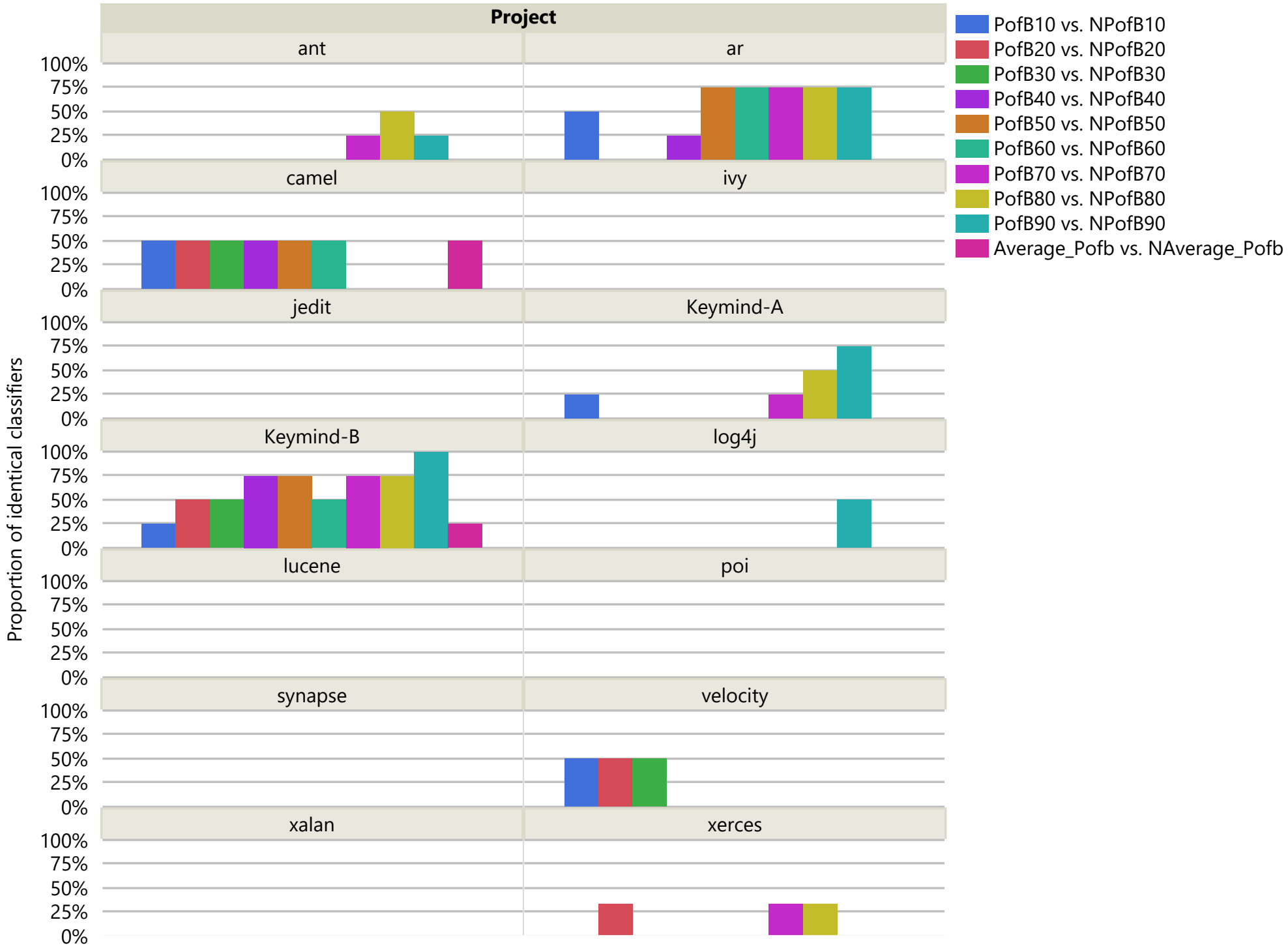

Figure 3.1.3: Proportion of times the same PofB metric, with and without normalization, identifies the same classifier as best.

Table 3.1.8: Proportion of times the same PofB metrics, with and without normalization, identifies the same classifier as best.

| **Compared metrics** | **Proportion of equivalent best classifiers** |
|---|---|
| PofB10 vs. NPofB10 | 15% |
| PofB20 vs. NPofB20 | 12% |
| PofB30 vs. NPofB30 | 10% |
| PofB40 vs. NPofB40 | 12% |
| PofB50 vs. NPofB50 | 17% |
| PofB60 vs. NPofB60 | 15% |
| PofB70 vs. NPofB70 | 22% |
| PofB80 vs. NPofB80 | 27% |
| PofB90 vs. NPofB90 | 29% |
| Average_PofB vs. NAverage_PofB | 5% |

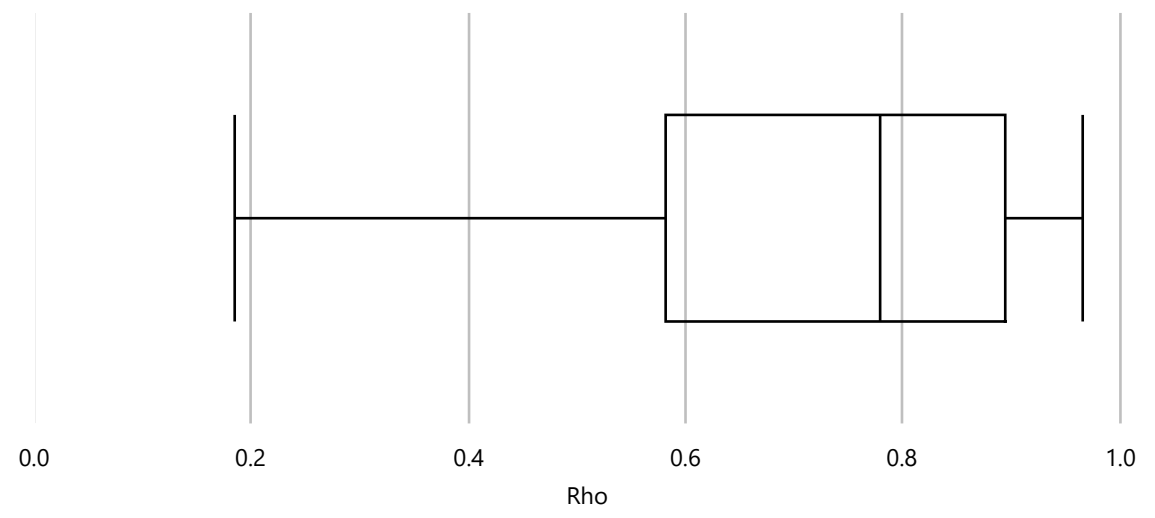

Figure 3.1.4: Distribution of correlations among each couple of NPofBs.

Table 3.1.9: Frequency of interpretations of correlations among couples of NPofBs.

| **Interpretation** | # |
|---|---|
| Inverse | 0 |
| Fair | 7 |
| Moderate | 8 |
| Very strong | 30 |
| Perfect | 0 |

According to Table 3.1.9 no couple of NPofBs is perfectly correlated. Moreover, only 30 out of 45 couples of NPofBs are very strongly correlated. In conclusion, Table 3.1.9 shows that the rankings of classifiers are far to be identical across different NPofBs.

Figure 3.1.5 reports the distribution of the proportion of times the best classifier for a dataset coincides across NPofBs. According to Figure 3.1.5, in only five out of 41 datasets the best classifier for a dataset coincides across the ten PofBs. Thus, in about 88% of the cases, the best classifier varies across NPofBs.

### 3.1.1.3 Discussion

This section discusses our main results, the possible explanations for the results, implications, and guidelines for practitioners and researchers.

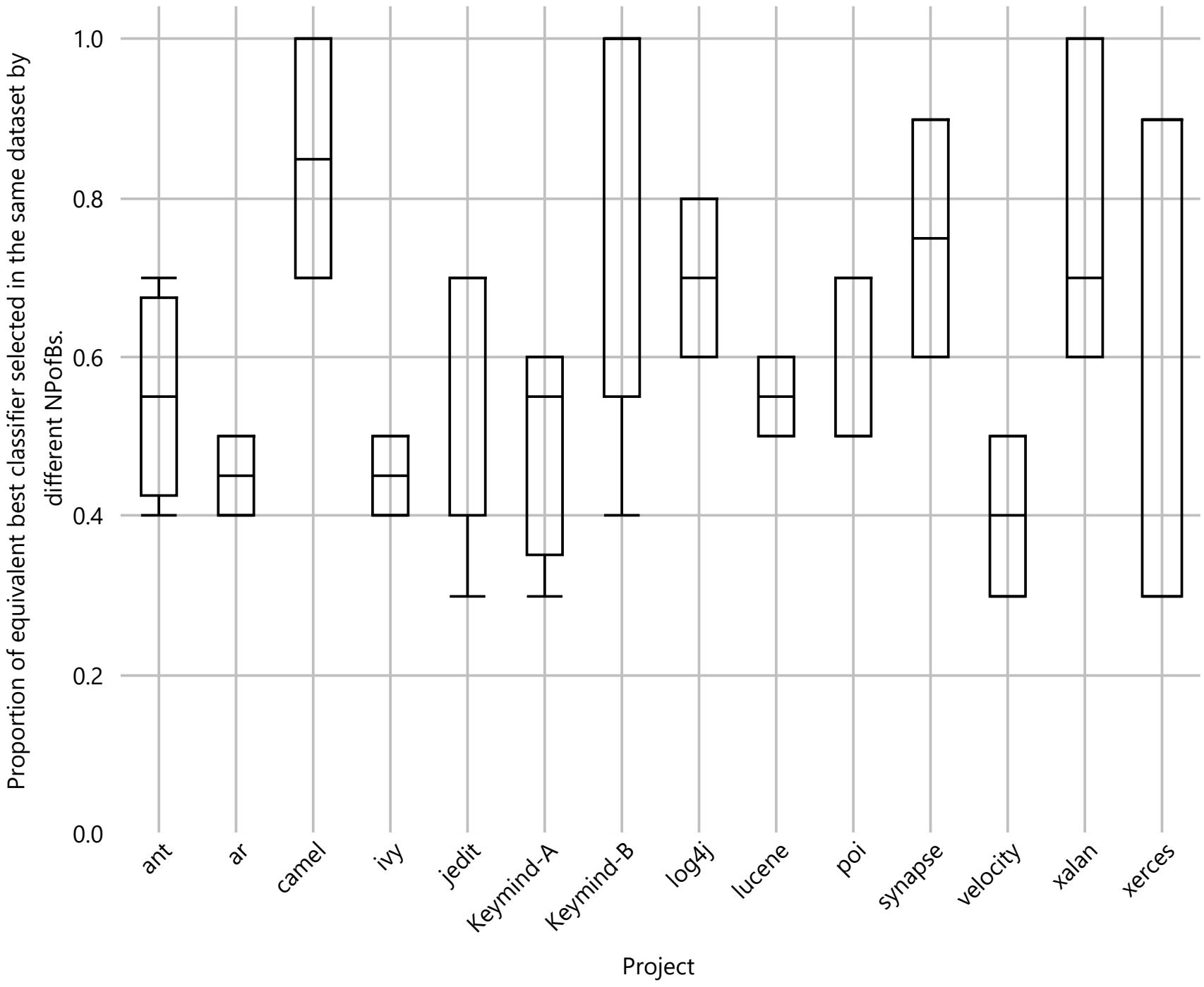

Figure 3.1.5: Distribution of the proportion of datasets where the best classifier coincides across NPofBs.

### RQ1: Which EAMs are used in software engineering journal papers?

The main result of RQ1 is that EAMs are used in a minority of defect prediction studies, i.e., 20 out of 152 software engineering journal papers. One possible reason is that EAMs do not make much sense in Just-in-time prediction studies, i.e., in studies predicting the defectiveness of commits. As a matter of fact, in the JIT context, the user is envisioned to consider the defectiveness prediction just after each single commit, and hence a JIT classifier cannot help the user in ranking the possibly defective entities (i.e., commits). However, we observed many JIT studies using EAMs and many non-JIT studies, i.e., studies predicting a class's defectiveness or method, not using EAMs. One possible reason for the low EAMs usage in non-JIT studies is the absence of a tool for EAMs computation. A further possible reason for the low EAMs adoption could be the lack of awareness about the importance of EAMs to evaluate the realistic benefits of using prediction models.

Another important result of RQ1 is that some EAMs are misnamed and that only one study used more than one EAM. Again, one possible reason for this result is the absence of a tool for EAMs computation.

The main implication of RQ1 is that a tool to automate EAMs computation would have supported a broader and more correct use of EAMs.

### RQ2: Does the normalization improve PofBs?

The main result of RQ2 is that the proposed normalization increases statistically and of orders of magnitude the PofBs. While the improvement is statistically significant on all 10 PofBs, we can see that the normalization increased the different PofBs differently. Specifically, the relative gain provided by the normalization resulted perfectly inversely correlated with the percent of analyzed

LOC related in the specific PofB; the highest gain was observed in PofB10. This result can be explained by the fact that the ranking quality looses benefit while the number of analyzed entities are high. In our case, it is obvious that when considering PofB90 many rankings can result equally beneficial to the user as long as the 10% of the not analyzed LOC are shared across such rankings. Thus, a better ranking is more visible in PofB10 than in PofB90.

We also note that the relative gain was higher in some datasets, i..e., xerces, than others, i.e., ar. The most probable reason is that the range of the relative gain is large when the accuracy without the normalization is low. Specifically, the PofBs in ar are much higher than in xerces. In other words, the normalization in ar has less chance of improving the PofBs since it is already high.

The main implication of RQ2 is that we need to use the normalized PofBs (aka NPofBs) rather than PofBs. The NPofBs are more realistic than PofBs as they relate to better use of classifiers. Moreover, studies reporting PofBs, rather than its normalization, underestimate the benefits of using classifiers for ranking defective classes and hence might have hindered the practical adoption of defect classifiers.

**RQ3: Does the ranking of classifiers change by normalizing PofBs?**

The main result of RQ3 is that the normalization changes the ranking of classifiers. Specifically, the correlation is only fair between the same PofB before and after the normalization in nine out of 10 PofBs. Moreover, in more than half of cases a classifier resulting as best with a PofB is not best with its normalization. The main implication of RQ2 is that past studies using the not normalized version of PofB likely highlighted a classifier as best despite another classifier brings the highest benefit in ranking defective classes to the user. Hence, RQ3 results call for replication of past studies using the not normalized version of PofBs.

**RQ4: Does the ranking of classifiers change across normalized PofBs?**

The main result of RQ4 is that no couple of NPofBs is perfectly correlated. Moreover, in 88% of the cases, the best classifier varies across the ten considered NPofBs. Thus, the main implication of RQ4 is that a single EAM does not exhaustively capture classifiers' ability to rank defective candidate classes. Thus, researchers shall use multiple EAMs to support a comprehensive understanding of classifiers accuracy. Past studies that validated classifiers via a single EAM shall be replicated to increase their results generalizability.

#### 3.1.1.4 The ACUME Tool

In this paper we provide a new tool called ACUME (ACcUracy MEtrics) which can compute PofB, the new proposed NPofB, Popt, IFA, and the performance metrics reported in Section 2.

To run, ACUME requires to:

- Download project files from GitHub;
- Place the csv files of the whole dataset in the data folder, and if needed in the test folder place your test files;
- Update configs.py file to your needs by following the instructions of ReadMe.md file.

To run the code, you need Python installed; more details are presented in the Readme file.

ACUME has been developed with clean code principles, kiss principle, and functional/OOP4 programming. There is linear complexity and minimal repetition of calculation to achieve a fast script with minimal memory usage by sacrificing readability for efficiency. Within approximately 1000 lines of code, there are two data classes: class DataEntity (no function) and ProcessedDataEntity (10 functions/methods), six stand-alone models - not class-dependent as well as many helper functions in each class.

Figure 3.1.6 reports the steps of the process on which ACUME works:

1. Step 1: The research teams create or reuse datasets. A name identifies the different datasets, e.g., KeymindA.csv, ant.csv.

2. Step 2: The research team provides the datatsets to a ML engine, such as Weka. The ML engine applies one or more defect prediction models that vary in classifiers, feature selection, balancing, etc.

3. Step 3: The ML engine outputs the predicted file. The predicted file is designed to be as simple as possible; for each predicted entity (e.g., a class), the predicted file reports the ID, the size, the predicted probability to be defective, and the actual defectiveness. The different predicted files are identified simply with a name, e.g., KeymindA_RandomForest_ withFeatureSelection.csv, ant_RandomForest_withFeatureSelection.csv, etc.

4. Step 4: The research team provides the predicted files to the ACUME tool.

5. Step 5: The ACUME tool outputs a single CSV file reporting, for each row, the performances of a predicted file in terms of accuracy metrics and EAMs.

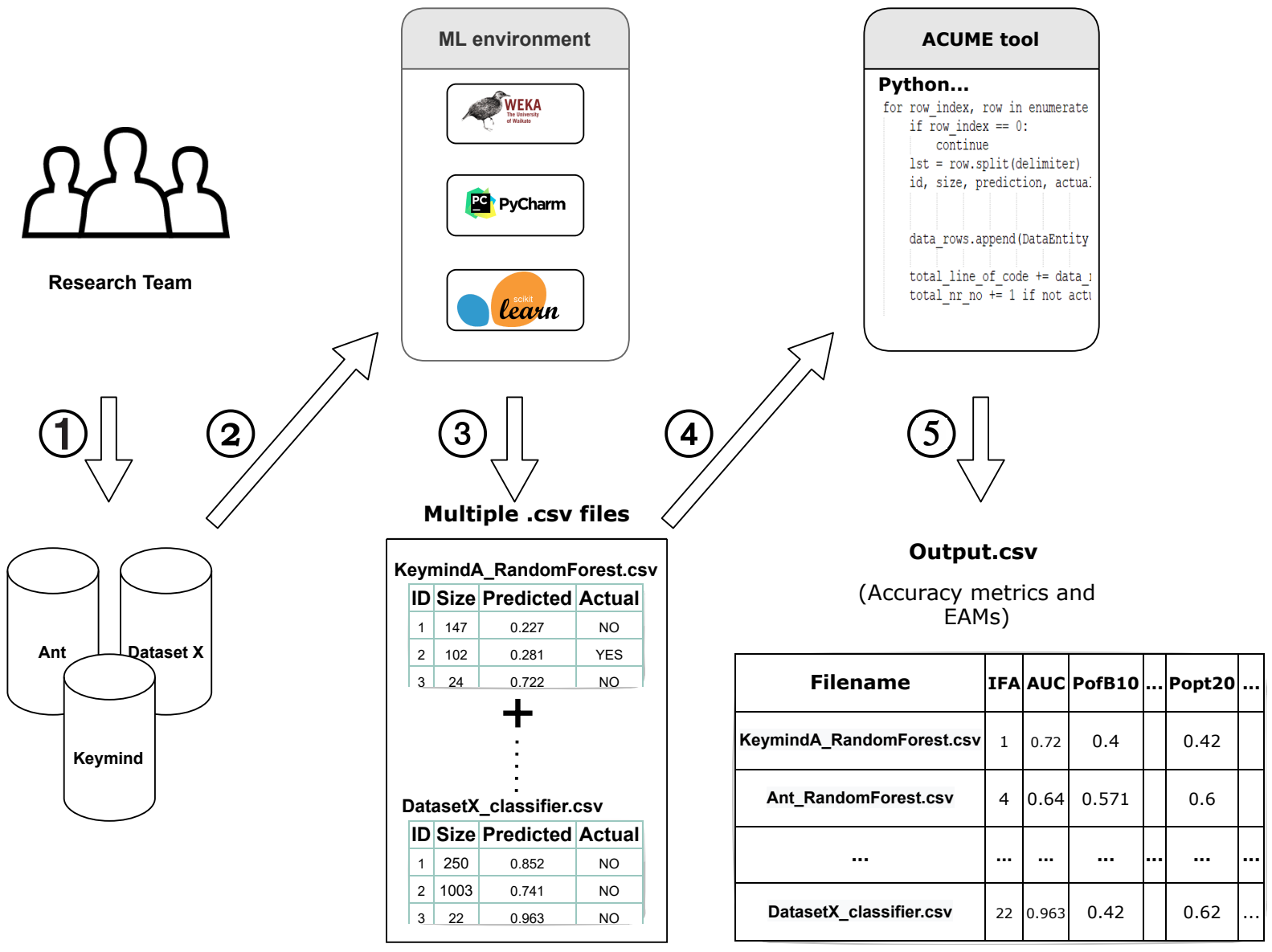

Figure 3.1.6: Distribution of the correlation among each couple of EAMs

We provide online material about ACUME[1].

ACUME has been used and validated by several researchers and students at the University of Rome Tor Vergata. We tested ACUME using a set of unit tests hence comparing the accuracy metrics computed by ACUME with expected values. To compute the expected values, we used a mixed approach according to metrics under test. Specifically, for metrics available in WEKA, such as AUC, we computed the expected values via WEKA on the project breast-cancer, as natively provided in WEKA. For metrics not available in WEKA, such as EAMs, we computed the expected values via Excels formulas on the ten projects used to address the research questions in this paper. The validation folder in the replication package reports the validation artifacts.

Finally, since its code is open, we welcome bug reports and feature requests.

[1] https://github.com/jonidacarka/ACUME.git

#### 3.1.1.5 Threats To Validity

In this section, we report the threats to validity of our study. The section is organized by threat type, i.e., Conclusion, Internal, Construct, and External.

**Conclusion**

Conclusion validity concerns issues that affect the ability to draw accurate conclusions regarding the observed relationships between the independent and dependent variables [413].

We tested all hypotheses with a non-parametric test (e.g., Wilcoxon Signed Rank) [408] which is prone to type-2 error, i.e., not rejecting a false hypothesis. We have rejected the hypotheses in all cases; thus, the likelihood of a type-2 error is null. Moreover, the alternative would have been using parametric tests (e.g., ANOVA) that are prone to type-1 error, i.e., rejecting a true hypothesis, which is less desirable than type-2 error in our context.

**Internal**

Internal validity concerns the influences that can affect the independent variables concerning causality [413]. A threat to internal validity is the lack of ground truth for class defectiveness, which could have been underestimated in our measurements. To avoid this threat, we used a set of projects already successfully used in our recent study [117]. As many other studies on defect prediction [375, 117, 23, 372, 140, 115, 391, 115], we do not differentiate among severity levels of the (predicted) defects.

### Construct

Construct validity is concerned with the degree to which our measurements indeed reflect what we claim to measure [413].

In order to make our empirical investigation reliable, we used the walk-forward technique as suggested in our recent study [117]. It could be that our results are impacted by our specific design choices, including classifiers, features, and accuracy metrics. In order to face this threat, we based our choice on past studies.

Despite many studies suggest the tuning of hyperparameters [139, 376], we used default hyperparameters due to resource constraints and to the static time-ordering design of our evaluation. Moreover, tuning might be relevant in studies aiming at improving the accuracy of classifiers whereas in this study tuning might be irrelevant as we aim at measuring the accuracy of classifiers, regardless of their tuning status. Finally, as our paper suggests, we plan to validate hyperparameters tuning via multiple and normalized PofBs.

Finally, the labeling of entities as defective or not has been subject to significant effort, and we still do not know how to perfectly label entities [391]. To avoid this type of noise in the data, we used data coming from the literature and largely used in the past [117].

### External

External validity is concerned with the extent to which the research elements (subjects, artifacts, etc.) are representative of actual elements [413].

This study used a large set of datasets and hence could be deemed of high generalization compared to similar studies. Moreover, in this study, we used both open-source and industry-type projects.

Finally, to promote reproducible research, all datasets, results, and scripts for this paper are available online[1].

#### 3.1.1.6 Conclusion

Despite the importance of EAMs, there is no study investigating EAMs usage trends and validity. Therefore, in this paper, we analyze trends in EAMs usage in the software engineering literature. Our systematic mapping study found 152 primary studies in major software engineering journals, and it shows that most studies using EAMs use only a single EAM (e.g., Popt or PofB20) and that some studies mismatched EAMs names.

To improve the internal validity of results provided by PofBs, we proposed normalization of PofBs. The normalization is based on ranking the defective candidates by the probability of the candidate to be defective divided by its size. We validated the normalization of PofBs by analyzing 10 PofBs, 10 classifiers, two industrial projects and 12 open-source projects. Our results show that the normalization increases statistically and of orders of magnitude the PofBs. Thus, past studies reporting PofBs underestimate the benefits of using classifiers for ranking defective classes and might have hindered the practical adoption of prediction models in the industry. The proposed normalization increases the realism of PofBs values as it relates to better use of classifiers and promotes the practical adoption of prediction models in the industry.

We showed that when considering the same dataset, in most of the cases, the best classifier for a PofB changes when considering the normalized version of that PofB. Thus, past studies that used the non-normalized version of PofBs likely identified the wrong best classifier, i.e., past studies likely identified as best a classifier not providing the highest benefit to the user in ranking defective classes.

In this paper, we also showed that multiple PofBs are needed to support a comprehensive understanding of classifiers accuracy. Thus, we provide a tool to compute EAMs automatically; this aims at supporting researchers in: 1) avoiding extra effort in EAMs computation as there is no available tool to compute EAMs, 2) increasing results reproducibility as the way to compute EAMs is shared across researchers, 3) increasing results validity by avoiding the observed EAMs misnaming and, 4) increase results generalizability by avoiding single EAMs usage.

In the future, we plan to replicate the validation of past defect prediction studies that considered a single EAM by including multiple EAMs. Hence, we want to check if the observed best classifier varies when considering multiple EAMs. We also plan to validate past studies by using the proposed normalization of PofBs rather than the original PofBs. Thus, we want to check if the observed best classifier varies when considering the proposed normalization. We also plan to investigate the opinions of developers on EAMs and specifically on the validated new NPofB metric. Finally, we will try to propose and evaluate EAMs, or new normalization, that more realistically measure the benefits provided by classifiers when ranking candidate defective entities.

#### 3.1.1.7 Appendix

Table 3.1.10: Correlations among specific couples of normalized PofBs.

| **Compared variables** | **Rho** | **Pvalue** | **Intepretation** |
|---|---|---|---|
| NAverage_PofB vs. NPofB10 | 0.783856 | <0.0001 | Very strong |

Continued on next page

| Continuation of Table 3.1.10 | | | |
|---|---|---|---|
| **Compared variables** | **Rho** | **Pvalue** | **Intepretation** |
| NAverage_PofB vs. NPofB20 | 0.85454 | <0.0001 | Very strong |
| NAverage_PofB vs. NPofB30 | 0.946433 | <0.0001 | Very strong |
| NAverage_PofB vs. NPofB40 | 0.964284 | <0.0001 | Very strong |
| NAverage_PofB vs. NPofB50 | 0.958544 | <0.0001 | Very strong |
| NAverage_PofB vs. NPofB60 | 0.922743 | <0.0001 | Very strong |
| NAverage_PofB vs. NPofB70 | 0.855365 | <0.0001 | Very strong |
| NAverage_PofB vs. NPofB80 | 0.809668 | <0.0001 | Very strong |
| NAverage_PofB vs. NPofB90 | 0.537388 | <0.0001 | Moderate |
| NPofB20 vs. NPofB10 | 0.932106 | <0.0001 | Very strong |
| NPofB30 vs. NPofB10 | 0.779776 | <0.0001 | Very strong |
| NPofB30 vs. NPofB20 | 0.846695 | <0.0001 | Very strong |
| NPofB40 vs. NPofB10 | 0.731393 | <0.0001 | Very strong |
| NPofB40 vs. NPofB20 | 0.800647 | <0.0001 | Very strong |
| NPofB40 vs. NPofB30 | 0.966495 | <0.0001 | Very strong |
| NPofB50 vs. NPofB10 | 0.651501 | <0.0001 | Moderate |
| NPofB50 vs. NPofB20 | 0.727009 | <0.0001 | Very strong |
| NPofB50 vs. NPofB30 | 0.909338 | <0.0001 | Very strong |
| NPofB50 vs. NPofB40 | 0.95835 | <0.0001 | Very strong |
| NPofB60 vs. NPofB10 | 0.578208 | <0.0001 | Moderate |
| NPofB60 vs. NPofB20 | 0.650734 | <0.0001 | Moderate |
| NPofB60 vs. NPofB30 | 0.823908 | <0.0001 | Very strong |
| NPofB60 vs. NPofB40 | 0.880341 | <0.0001 | Very strong |

Continued on next page

| Continuation of Table 3.1.10 | | | |
|---|---|---|---|
| **Compared variables** | **Rho** | **Pvalue** | **Intepretation** |
| NPofB60 vs. NPofB50 | 0.939176 | <0.0001 | Very strong |
| NPofB70 vs. NPofB10 | 0.486744 | <0.0001 | Fair |
| NPofB70 vs. NPofB20 | 0.562888 | <0.0001 | Moderate |
| NPofB70 vs. NPofB30 | 0.717087 | <0.0001 | Very strong |
| NPofB70 vs. NPofB40 | 0.781574 | <0.0001 | Very strong |
| NPofB70 vs. NPofB50 | 0.850415 | <0.0001 | Very strong |
| NPofB70 vs. NPofB60 | 0.930698 | <0.0001 | Very strong |
| NPofB80 vs. NPofB10 | 0.474829 | <0.0001 | Fair |
| NPofB80 vs. NPofB20 | 0.561707 | <0.0001 | Moderate |
| NPofB80 vs. NPofB30 | 0.684472 | <0.0001 | Moderate |
| NPofB80 vs. NPofB40 | 0.714814 | <0.0001 | Very strong |
| NPofB80 vs. NPofB50 | 0.777562 | <0.0001 | Very strong |
| NPofB80 vs. NPofB60 | 0.844337 | <0.0001 | Very strong |
| NPofB80 vs. NPofB70 | 0.916603 | <0.0001 | Very strong |
| NPofB90 vs. NPofB10 | 0.185601 | <0.0001 | Fair |
| NPofB90 vs. NPofB20 | 0.313019 | <0.0001 | Fair |
| NPofB90 vs. NPofB30 | 0.3962 | <0.0001 | Fair |
| NPofB90 vs. NPofB40 | 0.458927 | <0.0001 | Fair |
| NPofB90 vs. NPofB50 | 0.497088 | <0.0001 | Fair |
| NPofB90 vs. NPofB60 | 0.5854 | <0.0001 | Moderate |
| NPofB90 vs. NPofB70 | 0.71 | <0.0001 | Very strong |
| NPofB90 vs. NPofB80 | 0.739416 | <0.0001 | Very strong |

### 3.1.2 Uncovering the Hidden Risks: The Importance of Predicting Bugginess in Untouched Methods

Bugs in untouched code can be a ticking time bomb, but are they worth predicting? Our study dives deep into the importance of predicting bugginess in untouched methods and its impact on bug prediction accuracy. Our analysis of six open-source projects reveals the hidden risks of dormant bugs and the benefits of predicting in isolation untouched methods. Our findings significantly increase prediction accuracy, decrease train dataset sizes, and prove that untouched methods are an overlooked yet vital aspect of bug prediction.

#### 3.1.2.1 Empirical Design

In this work, we aim to address three research questions related to the bug proneness of software methods in the context of six open-source projects from the Apache Software Foundation ecosystem[2,3,4,5,6,7]. Table 3.1.11 reports the details of the used projects in terms of number of commits, number of buggy files, number of methods, number of buggy methods, number of releases, total number of bugs, number of bugs with affected version available in Jira, number of development days.

[2] https://github.com/apache/ant-ivy
[3] https://github.com/apache/bookkeeper
[4] https://github.com/apache/deltaspike
[5] https://github.com/apache/karaf
[6] https://github.com/apache/openwebbeans
[7] https://github.com/apache/qpid-jms

Table 3.1.11: Characteristics of the studied systems.

| Project Name | Commits | Buggy Files | Methods | Buggy Methods | Releases | Bugs | Bugs With AV | Days |
|---|---|---|---|---|---|---|---|---|
| ant-ivy | 2048 | 1915 | 34898 | 12117 | 9 | 629 | 567 | 5764 |
| bookkeeper | 1982 | 5171 | 61389 | 12427 | 14 | 435 | 170 | 3513 |
| deltaspike | 853 | 197 | 90973 | 5002 | 17 | 361 | 308 | 3536 |
| karaf | 6564 | 1345 | 171942 | 2861 | 42 | 2224 | 1620 | 4086 |
| openwebbeans | 2341 | 115 | 66052 | 6711 | 18 | 734 | 566 | 4621 |
| qpid-jms | 709 | 476 | 92568 | 9201 | 14 | 176 | 168 | 3022 |

### RQ1: Are methods more touched or untouched?

The first research question investigates the proportion of TMs over UMs and whether UMs deserve attention in bug prediction. The dependent variable is the number of UM and TM, while the independent variable is the type of method, i.e., touched or untouched. Our null hypothesis is $H_{01}$: there is no difference in the number of TM and UM. To measure the number of TM and UM, we labelled methods as touched or untouched based on their value of LOC touched. A method is labelled as untouched if LOC touched in the release is zero; otherwise, it is labelled as touched. To test the difference in the number of TM versus UM, we use the Wilcoxon signed-rank test[408].

### RQ2: Are buggy methods more touched or untouched?

The second research question investigates the distribution of untouched buggy methods (**UBM**) and whether they deserve to be predicted for bugginess. We hypothesise that UBM is more than touched buggy methods (**TBM**) since dormant bugs cause UM to be buggy, but no study measures how many UM are buggy. The dependent variable is the number of buggy methods, while the independent variable is the type of buggy method, i.e., touched or untouched. Our null hypothesis is $H_{02}$: there is no difference in the number of TBM and UBM.

To label methods as buggy or not, we use the "realistic approach" [424] and rely on the affected version reported in Jira. We selected the Apache projects with the highest proportion of Jira bug tickets having the affected version. We handled Jira bug tickets without the affected version as successfully proposed in a recent study [391]. To test the difference in the number of TBM versus UBM, we use the Wilcoxon signed-rank test.

**RQ3: Does predicting in isolation increase accuracy?**

The third research question aims to optimize the bugginess prediction of TM and UM. We hypothesize that some features might have biased the prediction of TM and UM, e.g., lines of code touched [438, 146] have a relevant predictive power for TM but not for UM. The dependent variable is the accuracy in bugginess prediction, while the independent variable is the type of prediction, i.e., baseline or in isolation. Previous approaches use all data in the training set to predict the bugginess of methods. In contrast, isolation prediction uses only UM in the training set to predict the bugginess of UM and only TM in training set to predict the bugginess of TM. To optimise the prediction of TM and UM, we investigate the effect of feature selection and dimensionality reduction techniques. Our null hypothesis: $H_{03}$: there is no difference in the accuracy baseline versus isolation prediction of buggy methods. To test the accuracy difference between standard and isolation predictions, we use the Wilcoxon signed-rank test. We employed 10 classifiers used in previous related papers [52].

We use the Wilcoxon test for the three null hypothesis because in each hypothesis, the pair of variables are not normally distributed; hence we can't use the paired t-test, and the Wilcoxon test offers a robust alternative in this case.

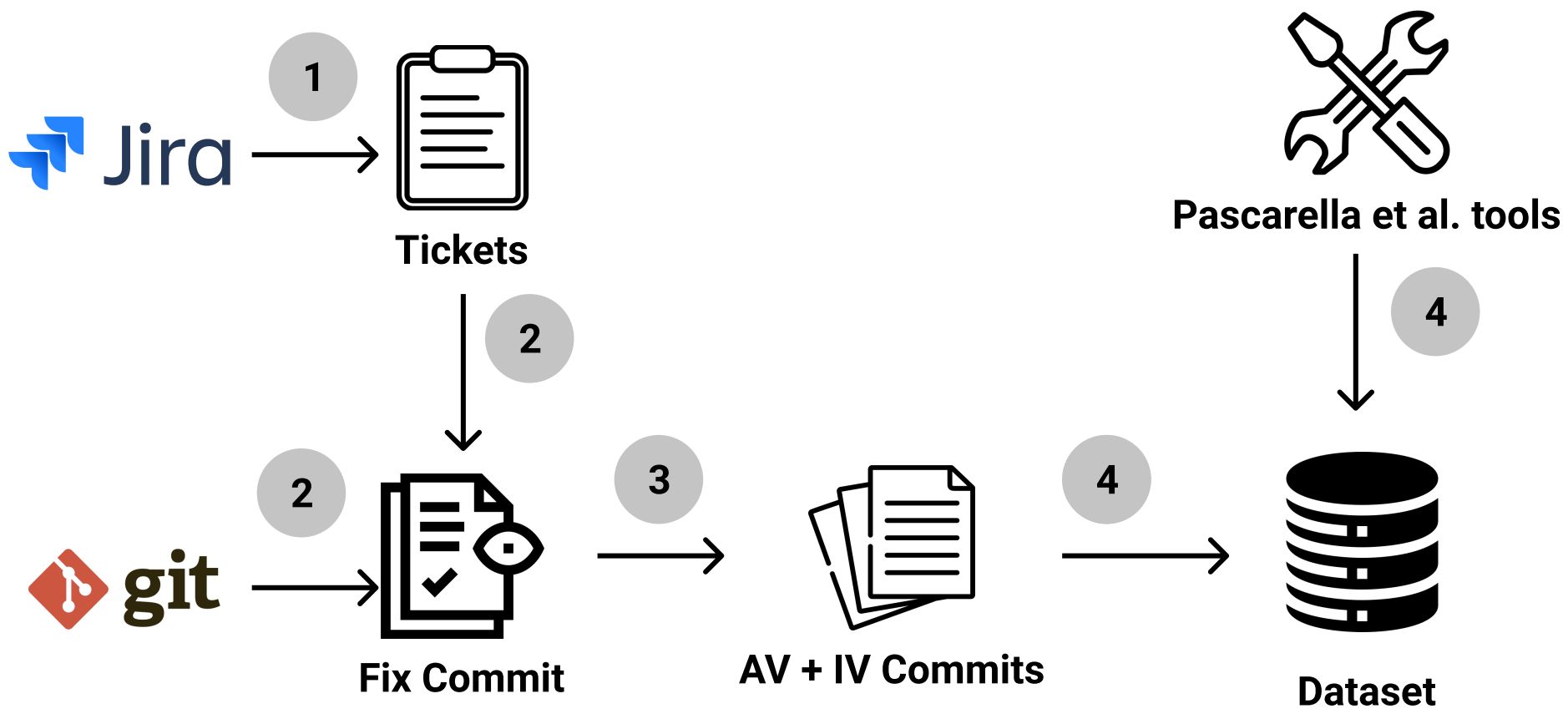

Figure 3.1.7: Dataset Creation.

**Overview of the dataset creation process.**

Figure 3.1.7 reports the steps of the *dataset creation* macro-step. **Step 1:** Retrieve from Apache Software Foundation's Jira instance[8] the list of all bugs-releated tickets **Step 2:** Link with the git history of the bug-fix and its commit **Step 3:** For each bug, we computed $\mathrm{P}_{\mathrm{MovingWindow}}$ as the average P among the last 1% of bugs as previously done by Vandehei et al. [391] and Falessi et al. [115]. We ordered the bugs by their fix date. Furthermore, we utilized the $\mathrm{P}_{\mathrm{ColdStart}}$ for $\mathrm{P}_{\mathrm{MovingWindow}}$ values that had an average of less than 1% of bugs. We computed the $IV$ for each bug as $(FV - OV) * P_{\mathrm{MovingWindow}}$. Moreover, if $FV$ equals $OV$, then $IV$ equals $FV$. We excluded bugs that were not post-release, and thus, we set $FV - OV$ equal to 1 to ensure that $IV$ is not equivalent to $FV$. For each bug, we label each version preceding the $IV$ as not affected, while versions from the $IV$ to the $FV$ as affected. Finally, the $FV$ was labelled as not affected. Please refer to Vandehei et al. [391] for the definitions of IV, OV, FV, and P. **Step 4:** We use Pascarella et al. tools [297] to get all method metrics and combine them with the information from Step 3. As features for predictions, we used both process and product metrics. Figure 3.1.8 reports the steps of the *split and filtering* macro-step. **Step 5:** We take only the first halves of the releases according to Ahluwalia et al. [5] **Step 6:** We partitioned the dataset, maintaining the order of the data, into a training set consisting of approximately 66% of the releases and a testing set comprising approximately 33% of the releases [117]. We selected this particular split proportion as it aligns with the recommended approach in ML [410], and it also bears similarities to the split proportion utilized in the bootstrap approach [117]. Finally, since we performed the split at the release level, and different releases may have varying numbers of commits, classes, and methods, the split proportion may vary across datasets.

[8] `https://issues.apache.org/jira/`

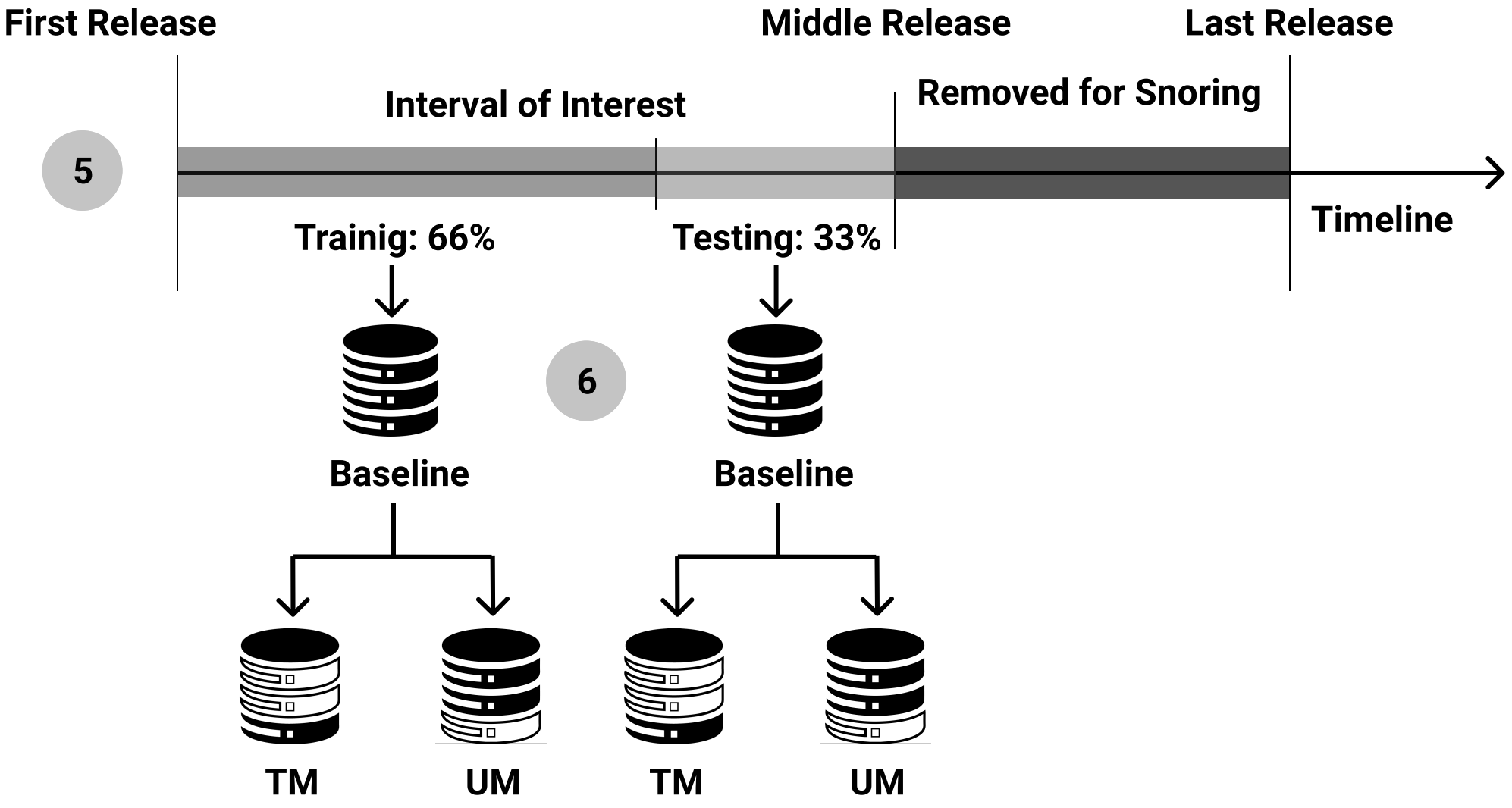

Figure 3.1.8: Overview of the split and filtering procedures.

Table 3.1.12: Wilcoxon test on the difference in the number of TM and UM, per each specific project.

| **Project Name** | bookkeeper | deltaspike | karaf | openwebbeans | qpid-jms |
|---|---|---|---|---|---|
| **P-Value** | 0.0001 | 0.0001 | 0.0001 | 0.0001 | 0.0001 |

#### 3.1.2.2 Results

**RQ1: Are methods more touched or untouched?**

The total number of methods resulted in 1,462,737 untouched and 73,282 touched. Figure 3.1.9 reports the distribution of TMs and UMs. Figure 3.1.10 reports the proportion of TMs and UMs across releases of the same project.

According to Figure 3.1.9 and Figure 3.1.10 we can observe that, on average per release and project, the majority of methods are untouched (i.e., 95%). Most releases have more UMs than TMs in all the studied projects. Furthermore, many releases across all projects (higher than 92%) contain mostly UM. However, in the case of bookkeeper, the proportion of UMs is only 20% in a release. These findings highlight the importance of considering UMs when predicting buggy methods and suggest that previous approaches based on process metrics may not be effective for most buggy methods.

Table 3.1.12 reports the result of applying the Wilcoxon test on the difference in the number of TMs and UMs. According to Table 3.1.12, the p-value is lower than alpha in all six projects. Therefore, we can reject $H_{01}$ and claim that the number of methods is significantly more untouched than touched.

**RQ2: Are buggy methods more touched or untouched?**

Figure 3.1.11 reports the distribution of TM and UM. To facilitate the interpretation, Figure 3.1.12 reports the proportion across releases of the same project

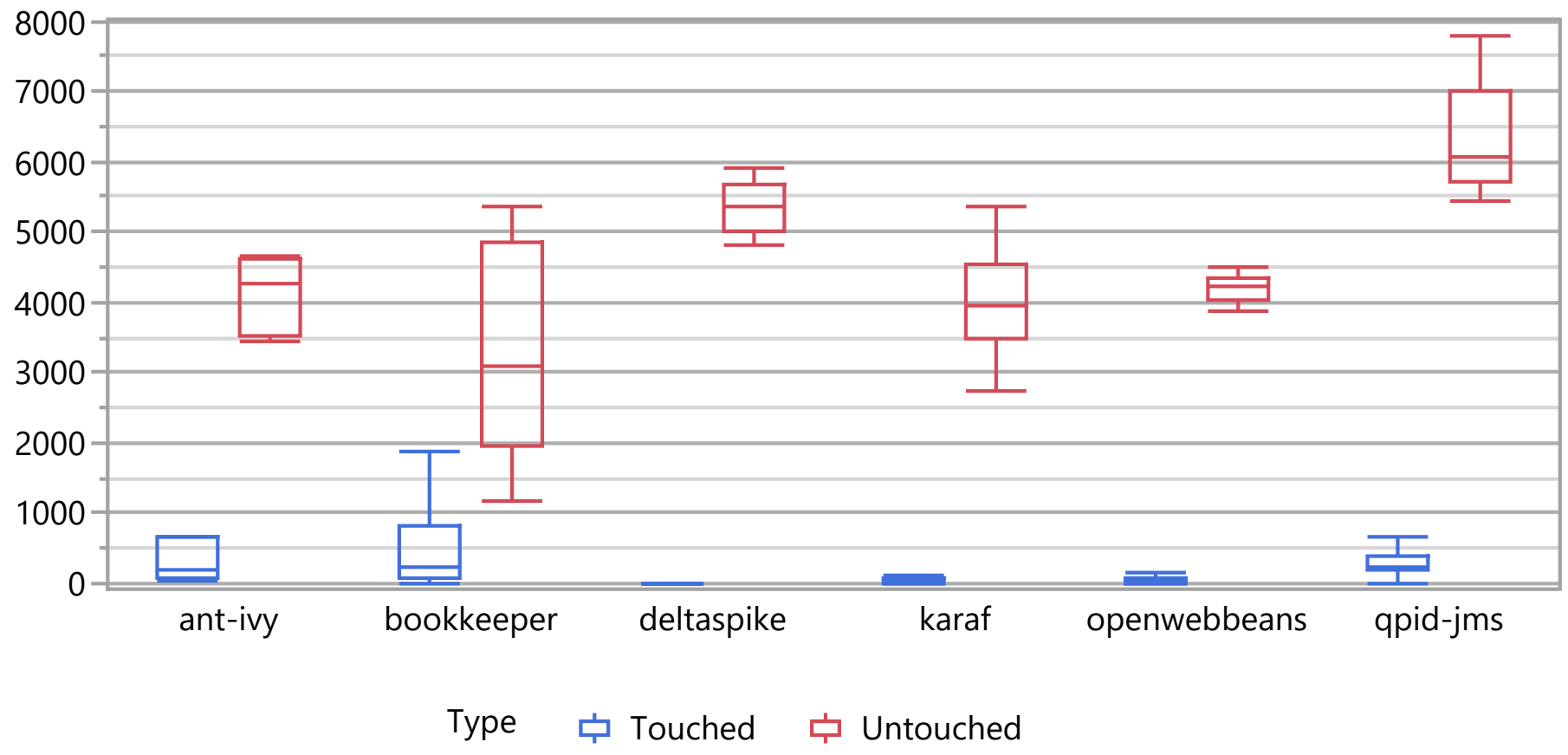

Figure 3.1.9: Distribution of TMs and UMs.

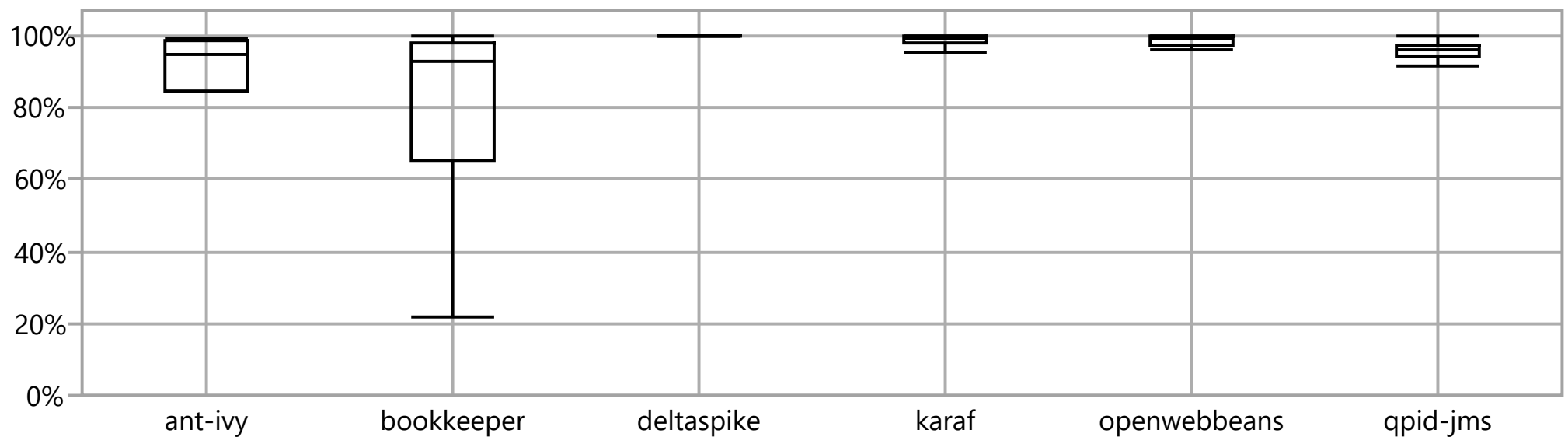

Figure 3.1.10: Proportion of UMs in different projects.

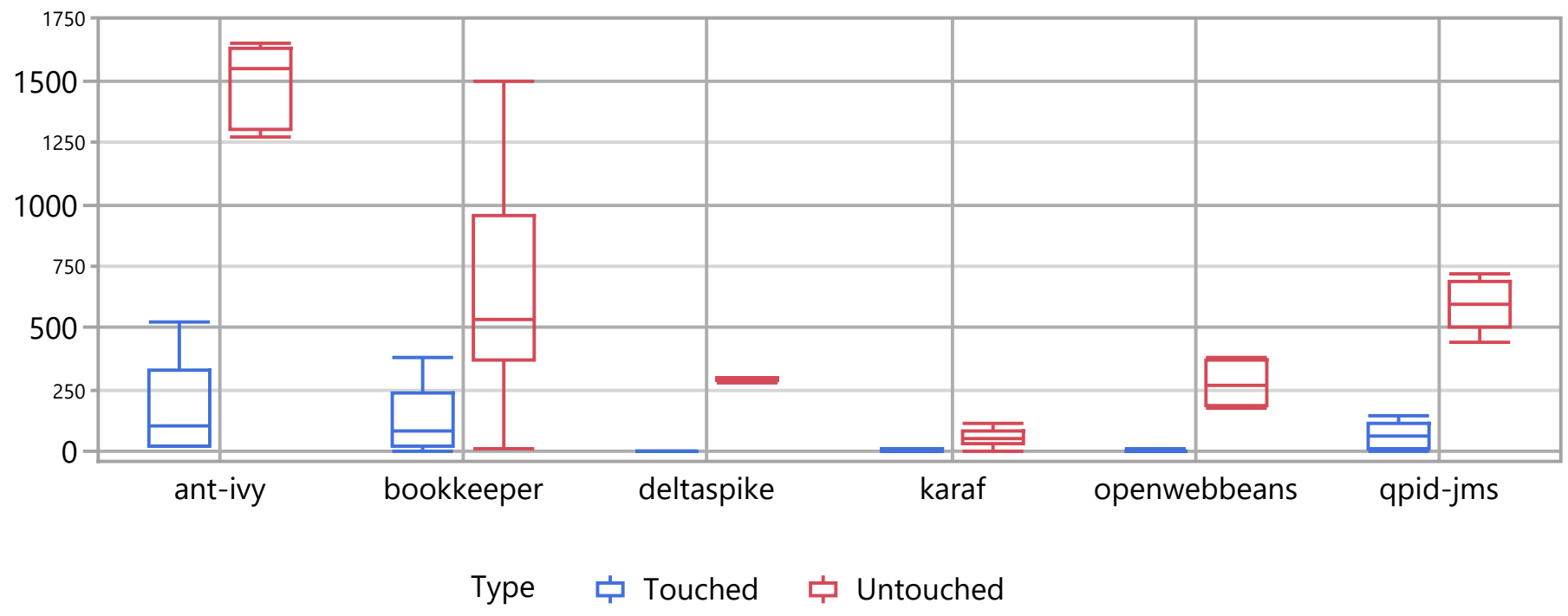

Figure 3.1.11: Distribution of TBMs and UBMs by project.

Table 3.1.13: Wilcoxon test on the difference in the proportion of buggy UM.

| Project Name | ant-ivy | book-keeper | deltaspike | openwebbeans | qpid-jms |
|---|---|---|---|---|---|
| **Method** | 0.0086 | 0.0038 | 0.0001 | 0.0001 | 0.0001 |

of the number of the buggy methods that are TM or UM.

Figure 3.1.11 and Figure 3.1.12 show that across all six projects and releases, on average 88% of buggy methods are untouched. Moreover, there are more buggy UM than buggy TM in all projects. Thirdly, in most releases of all projects, at least 83% of the buggy methods are UM. Lastly, the proportion of UM in buggy methods varies greatly between projects, with Bookeeper having the lowest proportion of UM in buggy methods (16%) compared to the other projects. These findings suggest that process metrics proposed in the past [243] for defect prediction may not be effective in identifying buggy methods, as most buggy methods are untouched and have no correlation with any process metric.

Table 3.1.13 reports the result of applying the Wilcoxon test on the difference in the number of buggy TM versus buggy UM. According to Table 3.1.13 the p-value is lower than alpha in all six projects for methods. Therefore, we can reject $H_{20}$ and claim that buggy UM are significantly more than buggy TM.

Table 3.1.14: Average gain provided by isolation in specific projects and type of methods.

| Project | F1 | | MAP | | AUC | | G_Measure | | MCC | | Popt20 | |
|---|---|---|---|---|---|---|---|---|---|---|---|---|
| | TM | UM | TM | UM | TM | UM | TM | UM | TM | UM | TM | UM |
| **ant-ivy** | 12% | 296% | 10% | 35% | 3% | 7% | 16% | 273% | 480% | 213% | 20% | 8% |
| **bookkeeper** | 158% | 60% | 13% | 7% | 20% | -3% | 158% | 83% | 69% | 24% | 9% | -1% |
| **deltaspike** | 55% | 323% | 928% | 1740% | 91% | 6% | 49% | 293% | -137% | 171% | 36% | 12% |
| **karaf** | 18% | 1129% | 6% | 6453% | 15% | 26% | 30% | 1209% | 14% | -58% | -7% | 9% |
| **openwebbeans** | -10% | 22% | 73% | 0% | 22% | -1% | -16% | 22% | -95% | 11% | -11% | 5% |
| **qpid-jms** | 61% | 85% | 25% | 307% | 9% | 4% | 88% | 83% | 102% | 61% | 32% | 15% |
| **Weighted Average by Project** | 85% | 489% | 19% | 2595% | 14% | 11% | 93% | 512% | 150% | 39% | 14% | 9% |
| **Weighted Average by Project and Methods** | 465% | | 2469% | | 10% | | 487% | | 37% | | 8% | |

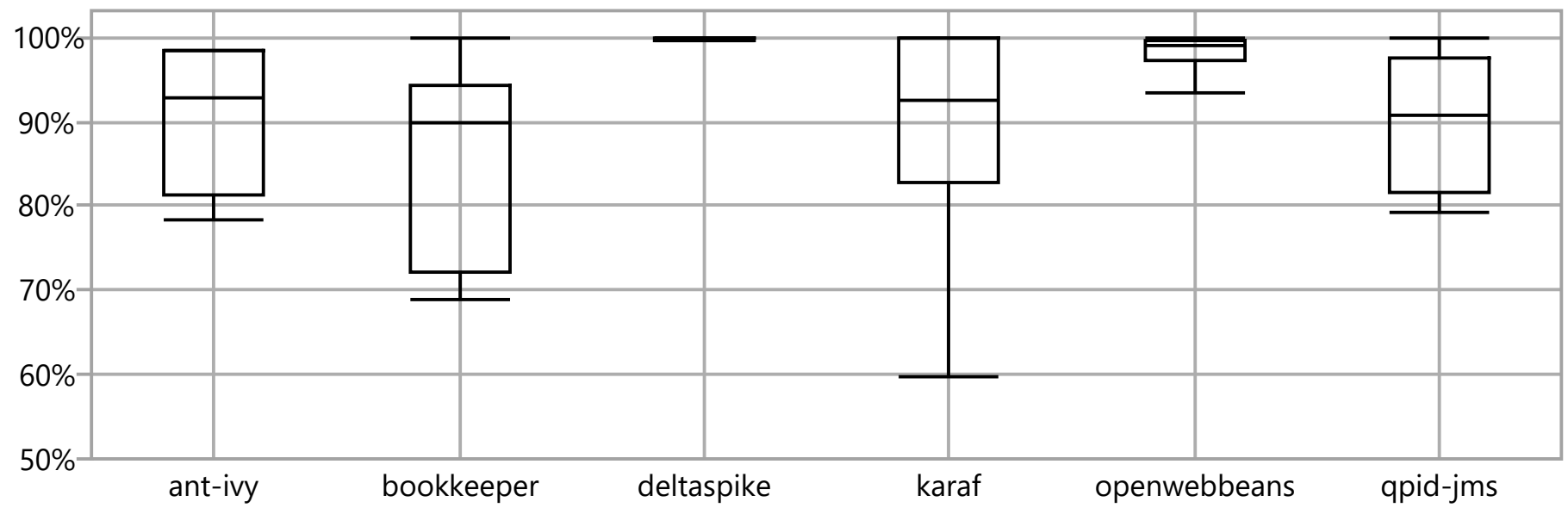

Figure 3.1.12: Proportion of untouched buggy methods over all buggy methods.

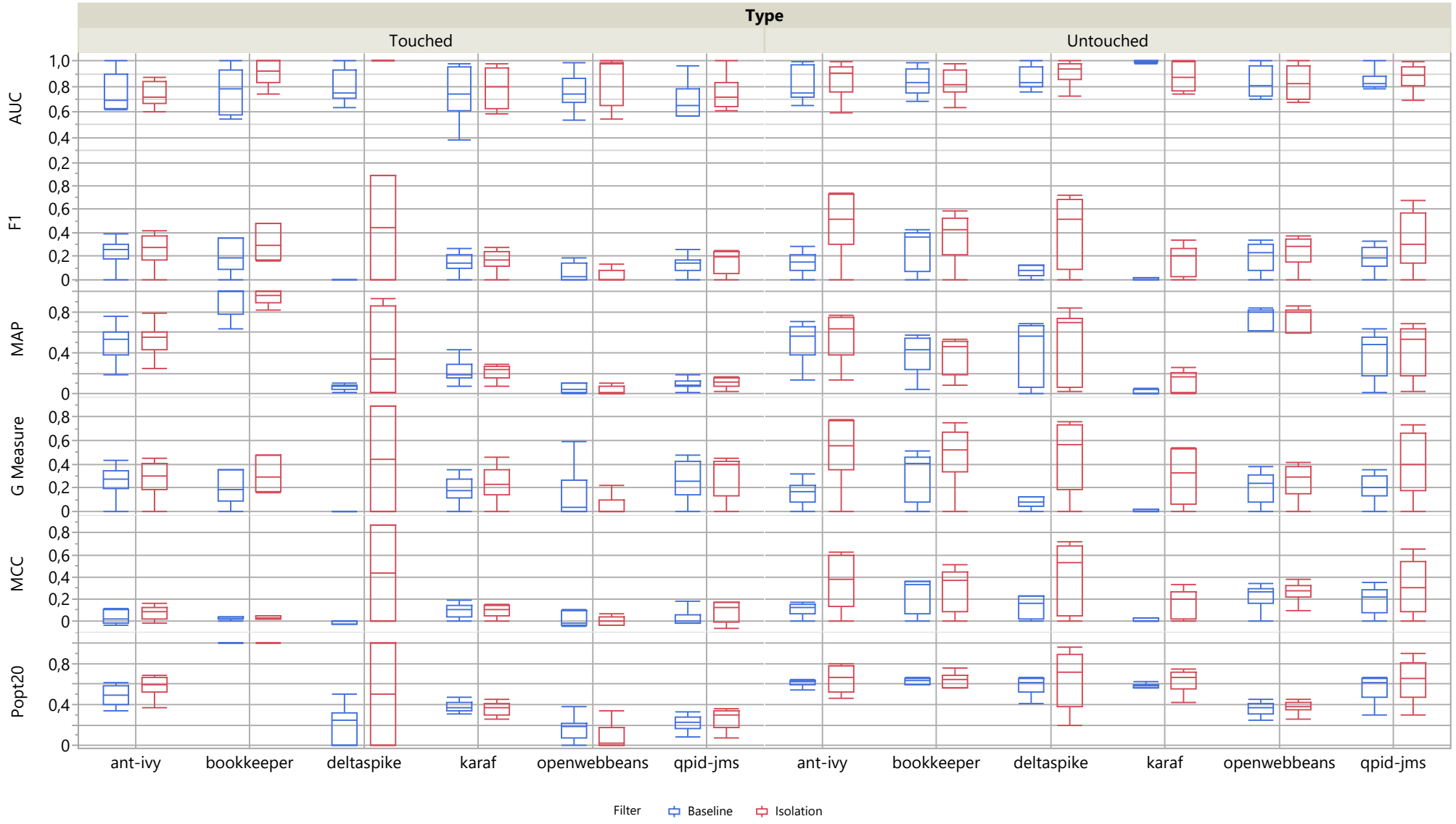

Figure 3.1.13: Accuracy metrics by project.

## RQ3: Does predicting in isolation increase accuracy?

Figure 3.1.13 reports the distribution of the value of AUC, F1, MAP, G-Measure, MCC and Popt20 metrics of TM and UM by project. Table 3.1.14 reports the median gain for each project in terms of AUC, F1, MAP, G-Measure, MCC and Popt20 metrics. Table 3.1.14 shows that isolation provides a weighted average gain of 10% in AUC and 8% in Popt20. Among the different metrics, MAP shows the highest average gain with a remarkable 2,469% across all projects. For UM, F1 and G-Measure have a positive gain in all projects, with a gain of 489% and 512%, respectively. Notably, ant-ivy and qpid-jms are the projects with positive gains in all metrics. These findings suggest that our approach of predicting the bugginess of methods in isolation benefits both UM and TM, resulting in significant improvements in performance metrics.

Table 3.1.15 presents the results of the Wilcoxon test comparing the accu-racy baseline versus the isolation prediction of buggy methods in six projects. The analysis shows that $H_{30}$ can be rejected for AUC in deltaspike, F1 in four projects (excluding openwebbeans and qpid-jms), MAP in deltaspike and karaf, G-Measure in five projects (excluding openwebbeans), and MCC in ant-ivy, deltaspike, and karaf. However, $H_{30}$ cannot be rejected for Popt20. These results indicate that the proposed approach of predicting buggy methods in isolation performs significantly better than the baseline approach for several evaluation metrics in various projects.

Table 3.1.15: Wilcoxon test on the difference in the accuracy metrics baseline versus isolation prediction of buggy methods.

| Project Name | AUC | F1 | MAP | G-Measure | MCC | Popt20 |
|---|---|---|---|---|---|---|
| **ant-ivy** | 0.5572 | 0.0074 | 0.1926 | 0.0105 | 0.0346 | 0.0687 |
| **bookkeeper** | 0.2799 | 0.0161 | 0.8418 | 0.0472 | 0.2451 | 0.4483 |
| **deltaspike** | 0.0043 | 0.0012 | 0.0286 | 0.0021 | 0.0010 | 0.0588 |
| **karaf** | 0.1623 | 0.0036 | 0.0247 | 0.0137 | 0.0093 | 0.4958 |
| **openwebbeans** | 0.3722 | 0.8763 | 0.8140 | 0.7474 | 0.5799 | 0.8599 |
| **qpid-jms** | 0.2961 | 0.0570 | 0.3980 | 0.0761 | 0.1693 | 0.3070 |

#### 3.1.2.3 Discussion

The results of our study reveal interesting findings related to the bugginess of methods in software systems and the effectiveness of predicting these bugs in isolation. Our study found that buggy methods are more likely to be UMs than TMs. This finding is somewhat surprising, as previous research suggested that only a small proportion of class bugs are introduced in previous releases.

Furthermore, we found that predicting in isolation increases the prediction accuracy of both types of methods. This is also surprising given that in some projects, UM is much more prevalent than TM. We suggest that process features such as lines of code touched are positively correlated with the bugginess of TM but always zero for UM. Thus, if process metrics are used in the training set to predict TM without removing UM, they lose prediction power for TM and may confound classifiers on the role of other product features. Similarly, predicting UM with process metrics that include TM data may decrease the prediction power for UM.

Furthermore, Figure 3.1.14 show the information gain ratio and the absolute value of Spearman's rho to measure the predictive power of each feature in isolation versus the baseline. Our results indicate that the predictive power of features in isolation is higher than the baseline, suggesting that isolating the training data is an effective approach to improve prediction accuracy.

Interestingly, we observed that the isolation approach worked better in some projects than others, and we have yet to determine why. However, our findings suggest that removing UM from the training set when predicting TM and, removing TM from the training set when predicting UM can improve prediction power.

#### 3.1.2.4 Threats to Validity

This section discusses the potential threats to the study's validity. For conclusion validity, the study's use of non-parametric tests increases the likelihood of type-2 errors, but since both hypotheses were rejected, the possibility of type-1 and type-2 errors is low. Internal validity is threatened by the lack of ground truth for method bugginess, which was mitigated by selecting projects with the most available affected versions. Construct validity is threatened by the missing method issue and the possibility of a study already discussing or providing a technique to predict bugginess, which was mitigated by using a large set of features and analysing commit histories. External validity is addressed using large datasets, but caution should be taken when generalising the findings to industrial projects. Finally, the study used various mechanisms to identify entities and mine versioning systems and other systems, which can pose threats to validity. These threats were mitigated by analysing commit messages, linking Jira and GitHub commits, and providing all datasets, results, and scripts in a replication package[9].

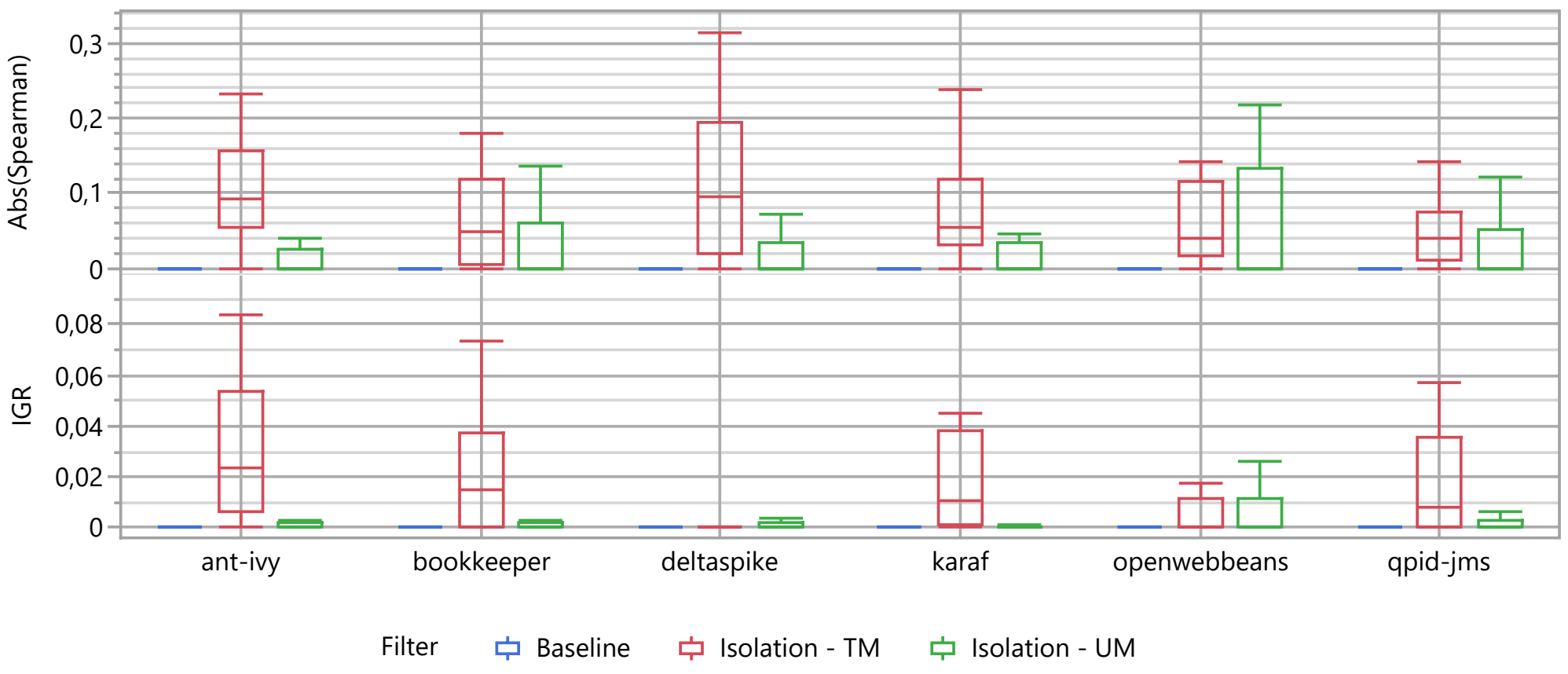

Figure 3.1.14: Predictive power of features in case of with versus without isolation.

### 3.1.2.5 Conclusions

We proposed a novel "isolation" methodology for predicting method bugginess, significantly increasing prediction accuracy across the selected project whilst decreasing data required for training. Our approach represents a departure from traditional code or process metrics and highlights the potential of alternative methodology for predicting method bugginess.

[9] `https://doi.org/10.5281/zenodo.8138497`

Future works in this area should consider the potential of predicting in isolation to improve past results and enhance the practical usage of our approach. Therefore, we plan to identify new project-wide metrics for dealing with untouched method data to improve our approach.

### 3.1.3 Enhancing the Defectiveness Prediction of Methods and Classes via JIT

[Context] Defect prediction can help at prioritizing testing tasks by, for instance, ranking a list of items (methods and classes) according to their likelihood to be defective. While many studies investigated how to predict the defectiveness of commits, methods, or classes separately, no study investigated how these predictions differ or benefit each other. Specifically, at the end of a release, before the code is shipped to production, testing can be aided by ranking methods or classes, and we do not know which of the two approaches is more accurate. Moreover, every commit touches one or more methods in one or more classes; hence, the likelihood of a method and a class being defective can be associated with the likelihood of the touching commits being defective. Thus, it is reasonable to assume that the accuracy of methods-defectiveness-predictions (MDP) and the class-defectiveness-predictions (CDP) are increased by leveraging commits-defectiveness-predictions (aka JIT). [Objective] The contribution of this paper is fourfold: (i) We compare methods and classes in terms of defectiveness and (ii) of accuracy in defectiveness prediction, (iii) we propose and evaluate a first and simple approach that leverages JIT to increase MDP accuracy and (iv) CDP accuracy. [Method] We analyse accuracy using two types of metrics (threshold-independent and effort-aware). We also use feature selection metrics, nine ma-

chine learning defect prediction classifiers, more than 2.000 defects related to 38 releases of nine open source projects from the Apache ecosystem. Our results are based on a ground truth with a total of 285,139 data points and 46 features among commits, methods and classes. [Results] Our results show that leveraging JIT by using a simple median approach increases the accuracy of MDP by an average of 17% AUC and 46% PofB10 while it increases the accuracy of CDP by an average of 31% AUC and 38% PofB20. [Conclusions] From a practitioner's perspective, it is better to predict and rank defective methods than defective classes. From a researcher's perspective, there is a high potential for leveraging statement-defectiveness-prediction (SDP) to aid MDP and CDP.

#### 3.1.3.1 Study Design

In this section, we explain the design of our study, which includes the procedures to choose our subject projects, our research questions (RQs) and our approaches to answer our RQs.

**Subject Projects**

We describe in the following how we chose the 10 datasets we used in this study.

1. We first retrieved the JIRA and Git URL of all existing Apache projects[10]. We focused on Apache projects instead of GitHub projects because Apache projects have a higher quality of defect annotation and are unlikely to be toy datasets [271].

2. We filtered out projects that are not tracked in JIRA or not versioned in Git.

[10] https://people.apache.org/phonebook.html

Table 3.1.16: Details of the datasets used in terms of commits, percent of defective commits, methods, percent of defective methods, classes, percent of defective classes, defects, linkage, releases, and manually validated commits.

| Dataset | CMT | D. CMT | Meth | D. Meth | CLS | D. CLS | Defects | Linkage | Releases | V. CMT |
|---|---|---|---|---|---|---|---|---|---|---|
| ARTEMIS | 589 | 34% | 83880 | 0.14% | 9425 | 0.41% | 133 | 82% | 1.0.0;1.0.1;1.2.0 | 21 |
| DIRSERVER | 367 | 6% | 19335 | 0.11% | 3367 | 1.85% | 14 | 83% | 1.01;1.02;1.03;1.04 | 10 |
| GROOVY | 701 | 2% | 14310 | 0.12% | 1496 | 1.54% | 20 | 77% | 1.0-1;1.0-2;1.0-3;1.0-4;1.0-5 | 12 |
| MNG | 1835 | 39% | 14519 | 2.33% | 2778 | 8.8% | 582 | 51% | 2.0a1;2.0a2;2.0a3;2.0-1;2.0-2 | 24 |
| NUTCH | 223 | 65% | 6934 | 4.86% | 1078 | 13.22% | 135 | 70% | 0.7;0.7.1;0.7.2 | 21 |
| OPENJPA | 538 | 54% | 44997 | 0.48% | 3119 | 2.65% | 263 | 92% | 0.9.0;0.9.6;0.9.7 | 23 |
| QPID | 1443 | 44% | 43049 | 3.42% | 5423 | 12.36% | 617 | 82% | M1;M2;M2.1;M3;M4 | 24 |
| TIKA | 246 | 29% | 1844 | 0.64% | 369 | 6.32% | 69 | 75% | 0.1-incubating;0.2;0.3 | 19 |
| ZOOKEEPER | 174 | 21% | 6249 | 0.49% | 838 | 7.01% | 27 | 76% | 3.0.0;3.0.1;3.1.0;3.1.1 | 13 |

3. We filtered out projects that are small and therefore not representative of a medium size industrial project. Specifically, we filtered out projects having: less than 20 releases, less than 20 commits per release, less than 20 linked and fixed defects per release, less than 50% of commits in the Java programming language. We define a commit in Java as a commit touching at least one Java file.

4. We filtered out two projects that although hosted on GitHub, their historical data did not cover a significant portion of their life time (which were actually covered in JIRA). These two projects were Openmeetings and Camel.

5. We selected the nine projects having the highest linkage rate, i.e., the highest proportion of defects with linked commits. A defect is considered to be linked if such a defect can be associated with at least one commit from the source code commit log. We selected nine projects due to the manual effort constraints required to perform the study.

Table 3.1.16 reports the details of the our 9 selected projects in terms of commits, percent of defective commits, methods, percent of defective methods, classes, percent of defective classes, defects, and releases.

In the reminder of this section we report the design of each research question.

#### RQ1: Do methods and classes vary in defectiveness?

Many studies evaluated techniques to perform MDP and CDP. However, there is a lack of studies that have compared the defectiveness across different granular levels (e.g., methods vs classes). Studying the defectiveness of methods and classes on the same releases can help us understand whether the defectiveness at a certain granularity is harder or easier to predict compared to other granularities. Specifically, despite we know that, by definition, defective classes are more frequent than defective methods, we do not know if defective classes are statistically more frequent than defective methods. In this study, we compare methods and classes in terms of defectiveness.

In this research question, we test the following null hypotheses:

•$H_{10}$: *the number of defective methods is equivalent to the number of defective classes.*

•$H_{20}$: *the proportion of defective methods is equivalent to the proportion of defective classes*

***Independent variables.*** The independent variable is the type of defective entity, either methods or classes.

***Dependent variables.*** The dependent variables are the number and the proportion of detective entities. The proportion of defective entities of a type is computed as the number of defective entities of that type divided by the total number of entities of that type.

***An extended RA-SZZ.*** In order to label commits, methods and classes as defective we performed the following steps, as detailed in Figure 3.1.15:

1. Defect identification: For each project, we find defect tickets on JIRA

where type == "defect" AND (status == "Closed" OR status =="Resolved") AND Resolution =="Fixed".

2. Labelling statements: We run RA-SZZ, which provides us with the defect-introducing statements.

3. Labelling commits: From the defect introducing statements we retrieved their defect-introducing commits and, hence, the defect-introducing releases.

4. Labelling methods: To identify which methods have been impacted by the defect-introducing statements (as retrieved by RA-SZZ), we developed a tool that uses the javaparser library[11] to parse the Java files implicated by the defect-introducing statements. Once the java files are parsed, the tool becomes aware of the *start* and *end* statements of each method within the Java files. Then, our tool identifies the methods that contain the defect-introducing statements that were previously retrieved. In doing so, we are able to identify the defective methods of a set of Java classes. Thus, in this study, we use an extended release of RA-SZZ, which provides a ground truth at the method granularity (in addition to the code statement and commit granularities).

5. Labelling classes: We labelled classes as defective if they contained at least one defective method.

Note that, in this study, an element at any granularity (i.e., a class, a method, or a statement) is defined as defective in all releases containing a defective statement.

[11] `https://javaparser.org/`

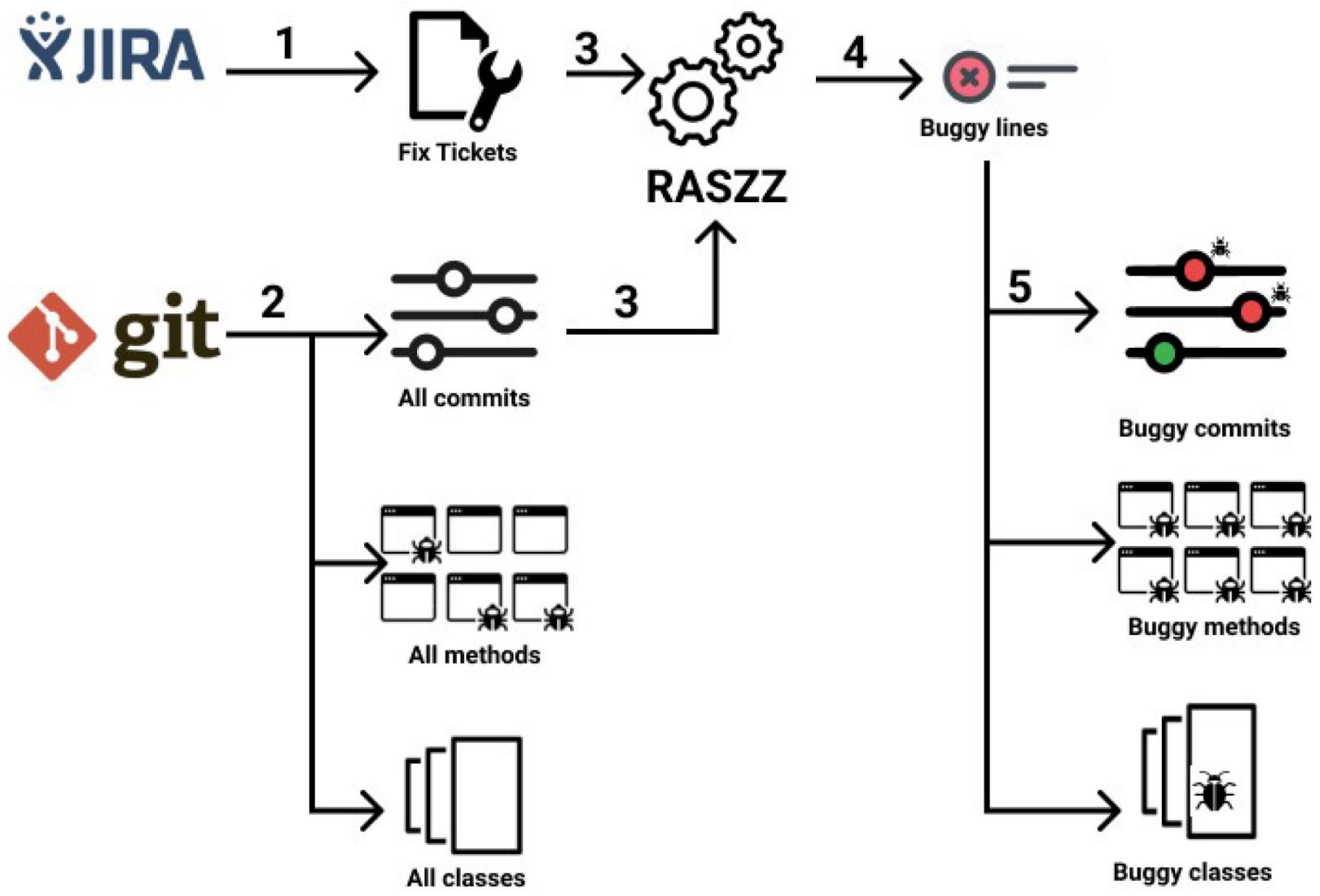

Figure 3.1.15: Steps RQ1.

***Manual validation.*** We have manually validated the defect-inducing commits used in this work. We used a 95% confidence level and a 20% margin of error; the resulting number of manually analyzed commits per project is reported in column 11 of Table 1. The third and fourth authors performed the manual validation of defect-inducing commits independently; a discussion resolved the disagreements between the two authors. The two authors resulted in agreement 95% of the cases (Kappa equals to 0.69) and the tool resulted in an agreement with the two authors 93% of the cases. The replication package reports the following information for each validated commit: bug ID, defect-fixing commit SHA, defect-inducing commit SHA, the label of the third author, the label of the fourth author, and the tool label.

***Hypothesis testing.*** To test our hypotheses, we used the paired Wilcoxon signed-rank test [327], which is a non-parametric test (i.e., there are no as-sumptions regarding the underlying distribution) to check whether two paired distributions are significantly different. We chose the non-parametric Kruskal–Wallis because our metrics (e.g., number and proportion of defective entities) do not follow a normal distribution (as we noted when performing Shapiro–Wilk tests [349]). Therefore, our approach is compliant to the suggestion to avoid using ScottKnottESD in case of non-normal distributions [160]. We use standard value of alpha $\alpha = 0.05$. To account for the chance of errors due to multiple comparisons, in all research questions we perform a Holm-Bonferroni correction of our $p - values$ [168]. We performed effect size analysis by using Cohen's d [160] which shall be interpreted as in Table 3.1.17.

Table 3.1.17: Intepretation of Cohen's d.

| **Effect size** | **d** |
|---|---|
| Very small | $<$0.01 |
| Small | $<$0.20 |
| Medium | $<$0.50 |
| Large | $<$0.80 |
| Very Large | $\geq$ 0.80 |

**RQ2: Does leveraging JIT information increase the accuracy of MDP?**

Every commit touches one or more methods; hence, the likelihood that a method is defective depends on the likelihood that the touched commits are defective. While many studies investigated techniques for MDP, no previous study investigated whether MDP may become more accurate if JIT information is used. In this paper, we propose and evaluate techniques to increase MDP accuracy by leveraging JIT.

In this research question, we test the following null hypothesis:

•$H_{30}$: *leveraging JIT does not improve the accuracy of MDP.*

***Independent variables.*** The independent variable is MDP with JIT information versus without JIT information. Specifically, we investigate the following MDP approaches:

- **Single**: It uses state of the art approach for MDP [146, 297]. Specifically, we used the following set of features as input to a machine-learning classifier:
  - size: LOC of a method.
  - methodHistories: number of times a method was changed.
  - authors: number of distinct authors that changed a method.
  - stmtAdded: sum of all source code statements added to a method body overall method histories.
  - maxStmtAdded: maximum number of source code statements added to a method body throughout the method's change history.
  - avgStmtAdded: average number of source code statements added to a method body per change to the method.
  - stmtDeleted: sum of all source code statements deleted from a method body over all method histories.
  - maxStmtDeleted: maximum number of source code statements deleted from a method body for all method histories.
  - avgStmtDeleted: Average number of source code statements deleted from a method body per method history.
  - churn: sum of stmtAdded plus stmtDeleted overall method histories.
  - maxChurn: maximum churn for all method histories.
  - avgChurn: average churn per method history

- cond: number of condition expression changes in a method body over all revisions.
  - elseAdded: number of added else-parts in a method body over all revisions.
  - elseDeleted: number of deleted else-parts from a method body over all revisions.

- **Combined:** It takes the **median** to combine the previously mentioned Single approach with two other scores:

  - SumC is the sum of defectiveness probabilities of the commits touching the method. The rationale is that the probability that a method is defective is related to the sum of probabilities of the commits touching the method.
  - MaxC is the max of defectiveness probabilities of the commits touching the method. The rationale is that the probability that a method is defective is related to the maximum probability of the commits touching the method.

The rationale of the Combined approach is that a defective commit incurs defective methods (i.e., those methods that are touched by the commit). We use the median as the combination mechanism because it is a simple way to combine several probabilities.

We use a standard JIT approach [191] to obtain probabilities of defectiveness of the commits that touch the methods. Specifically, we used the following set of features as input to a machine learning classifier:

  - Size: lines of code modified.

- Number of modified subsystems (NS): changes modifying many subsystem are more likely to be defect-prone.
- Number of modified directories (ND): changes that modify many directories are more likely to be defect-prone.
- Number of modified files (NF): changes touching many files are more likely to be defect-prone.
- Distribution of modified code across each file (Entropy): changes with high entropy are more likely to be defect-prone because a developer will have to recall and track higher numbers of scattered changes across each file.
- Lines of code added (LA): the more lines of code added the more likely a defect is introduced.
- Lines of code deleted (LD): the more lines of code deleted the higher the chance of a defect to occur.
- Lines of code in a file before the change (LT): the larger a file the more likely a change might introduce a defect.
- Whether or not the change is a defect fix (FIX): fixing a defect means that an error was made in an earlier implementation, therefore it may indicate an area where errors are more likely.
- Number of developers that changed the modified files (NDEV): the larger the NDEV, the more likely a defect is introduced because files revised by many developers often contain different thoughts and coding styles.
- Average time interval between the last and the current change (AGE): the lower the AGE, the more likely a defect will be introduced.

- Number of unique changes to the modified files (NUC): the larger the NUC, the more likely a defect is introduced because a developer will have to recall and track many previous changes.
  - Developer experience (EXP): more experienced developers are less likely to introduce a defect.
  - Recent developer experience (REXP): developers that have often modified the files in recent months are less likely to introduce a defect because they will be more familiar with the recent developments in the system.
  - Developer experience on a subsystem (SEXP): developers that are familiar with the subsystems modified by a change are less likely to introduce a defect.

***Dependent variables.*** The main dependent variable is the accuracy of MDP. As performance indicators of defect prediction we used the following metrics:

- AUC: Area Under the Receiving Operating Characteristic Curve [313] is the area under the curve of true positives rate versus false positive rate, which is defined by setting multiple thresholds. A positive instance is a defective entity, whereas a negative instance is a defect-free entity. AUC has the advantage of being threshold independent and, therefore, it is recommended for evaluating defect prediction techniques [216]. We decided to avoid metrics such as Precision, Recall and F1, since they are threshold-dependent.
- PofBX: as the effort-aware metric, we used PofBx [64, 399, 417, 385]. PofBx is defined as the proportion of defective entities identified by an-alyzing the first x% of the code base. For instance, a PofB10 of 30% signifies that 30% of defective entities have been found by analyzing 10% of the code base. We explored PofBx with an $x$ in the range of [10, 50]. While previous studies only focused on $x = 20$, we investigated a wider range to obtain more informative results. Note that PofB is different than Popt [189, 191, 258] in two aspects: normalization and range of $x$. Regarding normalization, while Popt normalizes the value according to a random approach, PofB does not perform such analysis, which aligns with our goals for two reasons:

1) AUC already provides the comparison against a random approach, since an AUC higher than 0.5 indicates that a classifier performed better than a random classifier,

2) In our study, we are interested in comparing classifiers that rank entities at different levels of granularity. Specifically, since methods and classes have a different defectiveness proportion (see Table 3.1.16), a random ranking would perform differently across methods and classes. Regarding the value of $x$, Popt represents an average of the complete spectrum of $x$, but we decided to neglect high values of $x$, as we believe that high values of $x$ would be unrealistic for practitioners when indicating which code should be inspected during testing. Specifically, the lower the amount of code tested, the higher the impact of the ranking approach; i.e., if 100% of the code needs to be inspected the ranking approach is effectively useless. Thus, we envisioned a metric that expresses the return of investing a specific amount of time in testing $x$% of the code as suggested by the ranking from a classifier. For all these reasons, PofB is a better match to our needs than Popt.

As an additional dependent variable, we measured the proportion of times features related to JIT are chosen to predict MDP. We performed this feature selection to complement the accuracy analysis since it is important to know whether our approach of using the median to leverage JIT information is beneficial. For example, if JIT information is selected as features for MDP but do not result in improving the accuracy of MDP, it indicates that using the median may not be the right approach to leverage JIT information.

***Measurement procedure.*** In this section, we describe the steps we performed to compute the accuracy metrics related to RQ2. As described in Figure 3.1.16, for each project, we:

1. Compute for each commit the above-mentioned features.
2. Label commits as defective or not by using RA-SZZ (as described in RQ1)
3. Create a commit dataset by combining features and defectiveness labels
4. Perform preprocessing:
    - Normalization: we normalize the data of all features with log10 since performed in many similar studies [179, 375],
    - Feature Selection: we filter the independent variables described above by using the correlation-based feature subset selection [156, 145, 207]. The approach evaluates the worth of a subset of features by considering the individual predictive capability of each feature, as well as the degree of redundancy between different features. The approach searches the space of feature subsets by a greedy hill-climbing augmented with a backtracking facility. The approach starts with an empty set of features and performs a forward search by considering all possible single-feature additions and deletions at a given point.

- Balancing: we apply SMOTE [60, 3] so that each dataset is perfectly balanced. As suggested [410], we apply feature selection and balancing to the training set only.

5. Create, Train, and Test commit datasets by splitting the preprocessed dataset into about 66% of releases as the training set and about 33% of releases as the testing set, while preserving the order of data [117]. We chose this split proportion since suggested in ML [410] and because it resemble the split proportion of the bootstrap approach [117]. Note that since the split is at the level of the release, and since different releases have a different number of commits, methods and classes, the specific proportion of spilt in training set and testing set vary across datasets and types of entities. Commits are assigned to the different releases given their timestamp. The status of methods and classes in a release takes into account the commits of that release.

6. Compute the predicted probability of defectiveness for each commit by using each of the 9 classifiers.

7. Compute the above mentioned features for each method.

8. Label methods as defective or not by using RA-SZZ (as described in RQ1).

9. Create a dataset of methods by combining features and defectiveness labels.

10. Perform preprocessing.

11. Create, Train, and Test method datasets by splitting the preprocessed dataset into 66% as the training set and 33% as the testing set, while preserving the order of data [117].

12. Compute *Direct* predicted the probability of defectiveness for each method by using each of the 9 classifiers.

13. Compute accuracy metrics of Direct.

14. Find commits related to each method. For the set of commits related to a method compute MaxC and SumC given the above commits predictions performed.

15. Perform feature selection, with the same technique described in item 4 (above), of the following features: Direct, MaxC and SumC.

16. Compute *Combined* by computing the median between Direct, MaxC and SumC.

17. Compute accuracy metrics of Combined.

In Figure 3.1.16, the dataset is split twice and some actions are repeated twice because one flow is for the data about commits and another is for data about the entity we want to predict (i.e., methods and classes).

In this paper, we used the following set of classifiers, since they have been successfully adopted in a previous study [117]:

- Random Forest: It generates several separate, randomized decision trees and provides as classification the mode of the classifications. It has proven to be highly accurate and robust against noise [48]. However, it can be computationally expensive as it requires the building of several trees.

- Logistic Regression: It estimates the probabilities of the different possible outcomes of a categorically distributed dependent variable, given a set of independent variables. The estimation is performed through the logistic distribution function [55].

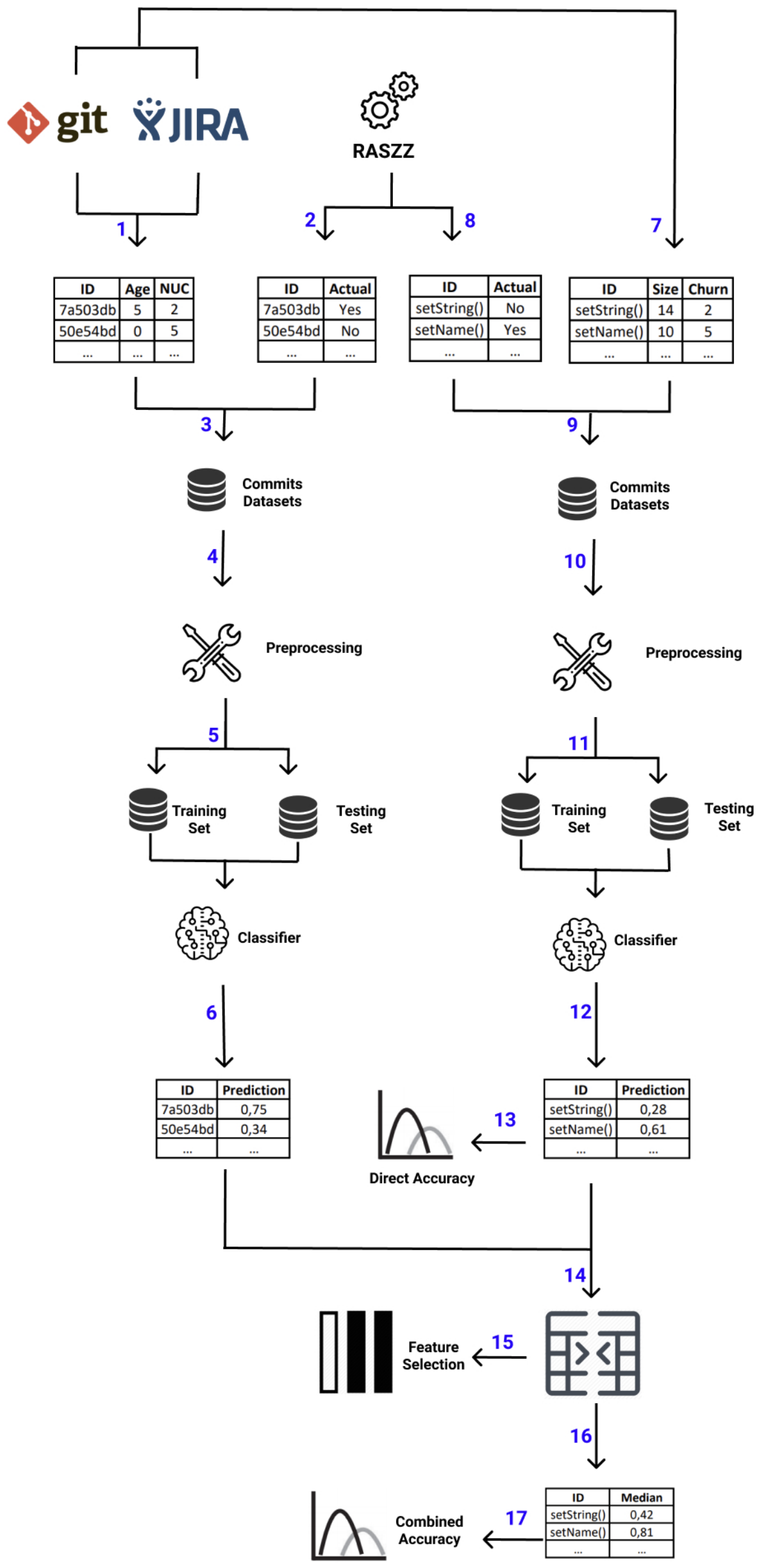

Figure 3.1.16: Measurement procedure of RQ2.

- Naïve Bayes: It uses the Bayes theorem, i.e., it assumes that the contribution of an individual feature towards deciding the probability of a particular class is independent of other features in that dataset instance [253].

- HyperPipes: It simply constructs a hyper-rectangle for each label that records the bounds for each numeric feature and what values occur for nominal features. During the classifier application, the label is chosen by whose hyper-rectangle most contains the instance (i.e., that which has the highest number of feature values of the test instance fall within the corresponding bounds of the hyper-rectangle) .

- IBK: Also known as the k-nearest neighbors' algorithm (k-NN) which is a non-parametric method. The classification is based on the majority vote of its neighbors, with the object being assigned to the class most common among its k nearest neighbors [18].

- IB1: It is a special case of IBK with K = 1, i.e., it uses the closest neighbor [18].

- J48: Builds decision trees from a set of training data. It extends the Iterative Dichotomiser 3 classifier [318] by accounting for missing values, decision trees pruning, continuous feature value ranges and the derivation of rules.

- VFI: Also known as voting feature intervals [92]. A set of feature intervals represents a concept on each feature dimension separately. Afterwards, each feature is used by distributing votes among classes. The predicted class is the class receiving the highest vote [92].

- Voted Perceptron: It uses a new perceptron every time an example is wrongly classified, initializing the weights vector with the final weights of the last perceptron. Each perceptron will also be given another weight corresponding to how many examples they correctly classify before wrongly classifying one, and at the end, the output will be a weighted vote on all perceptrons [137].

***Hypothesis testing.*** As in RQ1, we use the paired Wilcoxon signed-rank test [327] to test our hypothesis, $H_{30}$.

**RQ3: Does leveraging JIT information increase the accuracy of CDP?**

Similar to RQ2, as no previous work investigated whether CDP may become more accurate by using JIT information, we propose and evaluate a technique to increase CDP accuracy by leveraging JIT information.

In this research question, we test the following null hypothesis:

•$H_{40}$*: leveraging JIT does not improve the accuracy of CDP.*

***Independent variables.*** The independent variable is the use of JIT information in CDP. As in RQ2, we investigate the following MDP approaches:

- **Single**: It uses state of the art approach for CDP [115]. Specifically, we used the following set of features as input to a machine learning classifier:
  - Size (LOC): lines of code.
  - LOC Touched: sum over revisions of LOC added and deleted
  - NR: number of revisions.
  - Nfix: number of defect fixes.
  - Nauth: number of authors.

- LOC Added: sum over revisions of LOC added and deleted.
  - Max LOC Added: maximum over revisions of LOC added.
  - Average LOC Added: average LOC added per revision.
  - Churn: sum over revisions of added  deleted LOC.
  - Max Churn: maximum churn over revisions.
  - Average Churn: average churn over revisions.
  - Change Set Size: number of files committed together.
  - Max Change Set: maximum change set size over revisions.
  - Average Change Set: average change set size over revisions.
  - Age: age of release.
  - Weighted Age: age of release weighted by LOC touched.

- **Combined:** Similar to RQ2, it takes the median to combine the previously described Direct approach with JIT information.

***Dependent variables.*** As in RQ2, the main dependent variable is the accuracy of CDP. Again, as an additional dependent variable, we measured the proportion of times features related to JIT are chosen to predict CDP.

***Measurement procedure.*** In this section, we describe the steps we performed to compute the accuracy metrics related to RQ3. We performed the same exact set of steps of RQ2 with the only difference being that we used the above mentioned features related to classes rather than features related to methods. Therefore, in RQ3, *median* is computed among the direct probability of a class to be defective, MaxC and SumC.

***Hypothesis testing.*** As in RQ1 and RQ2, we use the paired Wilcoxon signed-rank test [327] to test our hypotheses, $H_{40}$.

### RQ4: Are we more accurate in MDP or CDP?

When using defect predictions, developers may choose to inspect methods or classes during testing (i.e., before code is shipped to production). However, no previous study has investigated which prediction granularity is more advantageous (methods or classes?) in terms of accuracy and effort. In this study, we compare the accuracy of MDP against CDP, using also effort aware metrics.

In this research question, we test the following null hypotheses:

• $H_{50}$: *MDP is as accurate as CDP.*

***Independent variables.*** The independent variable is the granularity of the entity that is subject to the defectiveness prediction. The independent variable can have the following values: methods or classes.

***Dependent variables.*** The dependent variable is the accuracy of MDP and CDP as measured by the same performance metrics we used in RQ2 and RQ3.

***Measurement procedure.*** We used the data already used in RQ2 and RQ3 related to the Combined approach.

***Hypothesis testing.*** As in our previous three research questions, we use the paired Wilcoxon signed-rank test [327] to test our hypotheses, $H_{50}$.

### Replication package

For the interested researchers, the present study can be replicated using the replication package available online[12]. The replication package provides the scripts used to measure the data and the data itself. Both scripts and data are organized by research questions.

[12] https://doi.org/10.5281/zenodo.7213412

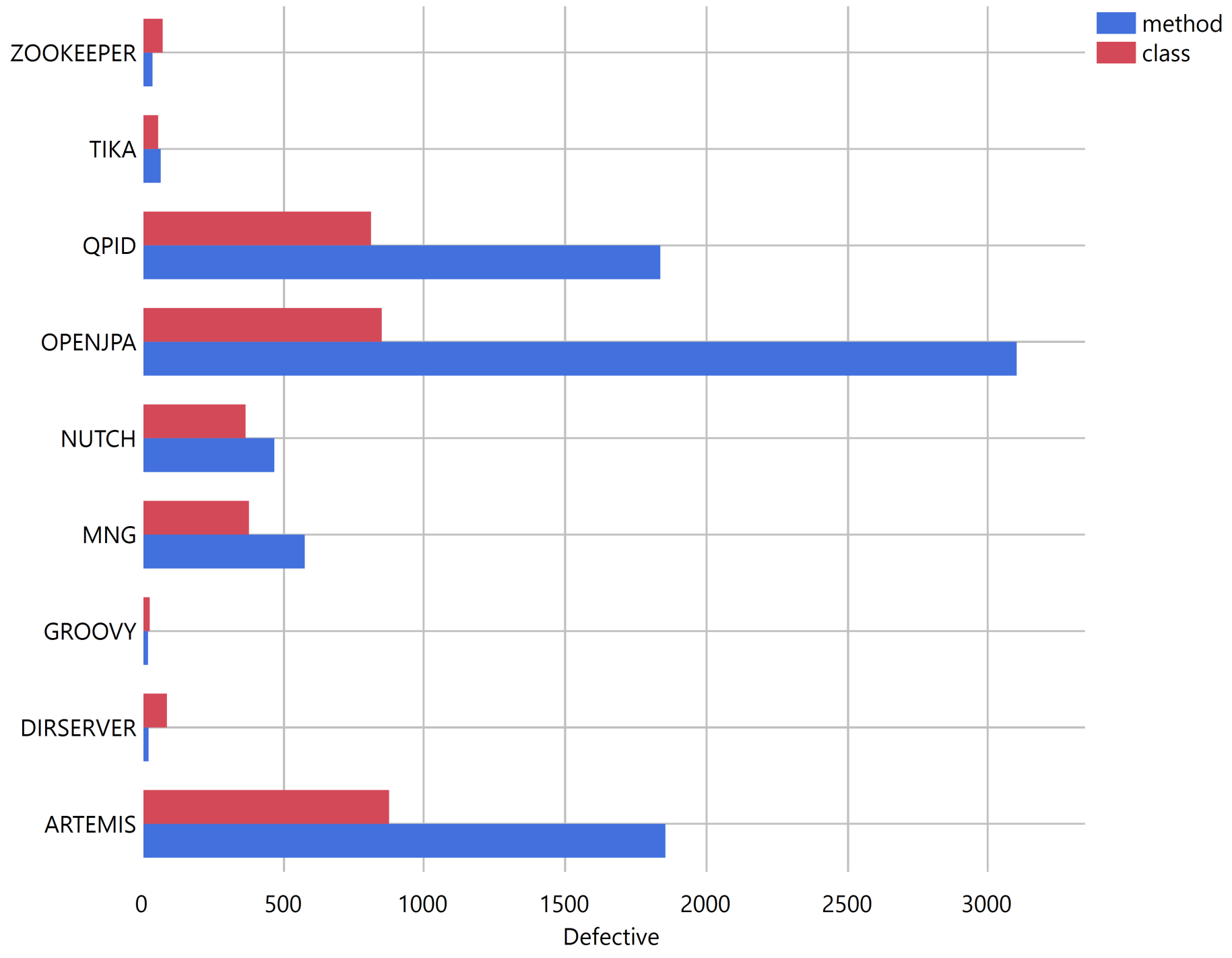

Figure 3.1.17: Number of defective entities (x-axis) in specific projects (y-axis) across different granularities of entities (color).

### 3.1.3.2 Study Results

**RQ1: Do methods and classes vary in defectiveness?**

Figure 3.1.17 reports the number of defective entities (x-axis) in specific projects (x-axis) across different granularities of entities (color). According to Figure, 3.1.17 there is no entity that is more defective than another in all projects. Moreover, the number of defective entities varies across projects.

Figure 3.1.18 reports the proportion of defective entities (x-axis) in specific datasets (x-axis) across different types of entities (color). According to Figure 3.1.18: **in all nine projects, the proportion of defective classes is higher than the proportion of defective methods**. Therefore, it is more likely to find by chance a defective class than a defective method. Moreover, the number of defective entities varies across datasets.

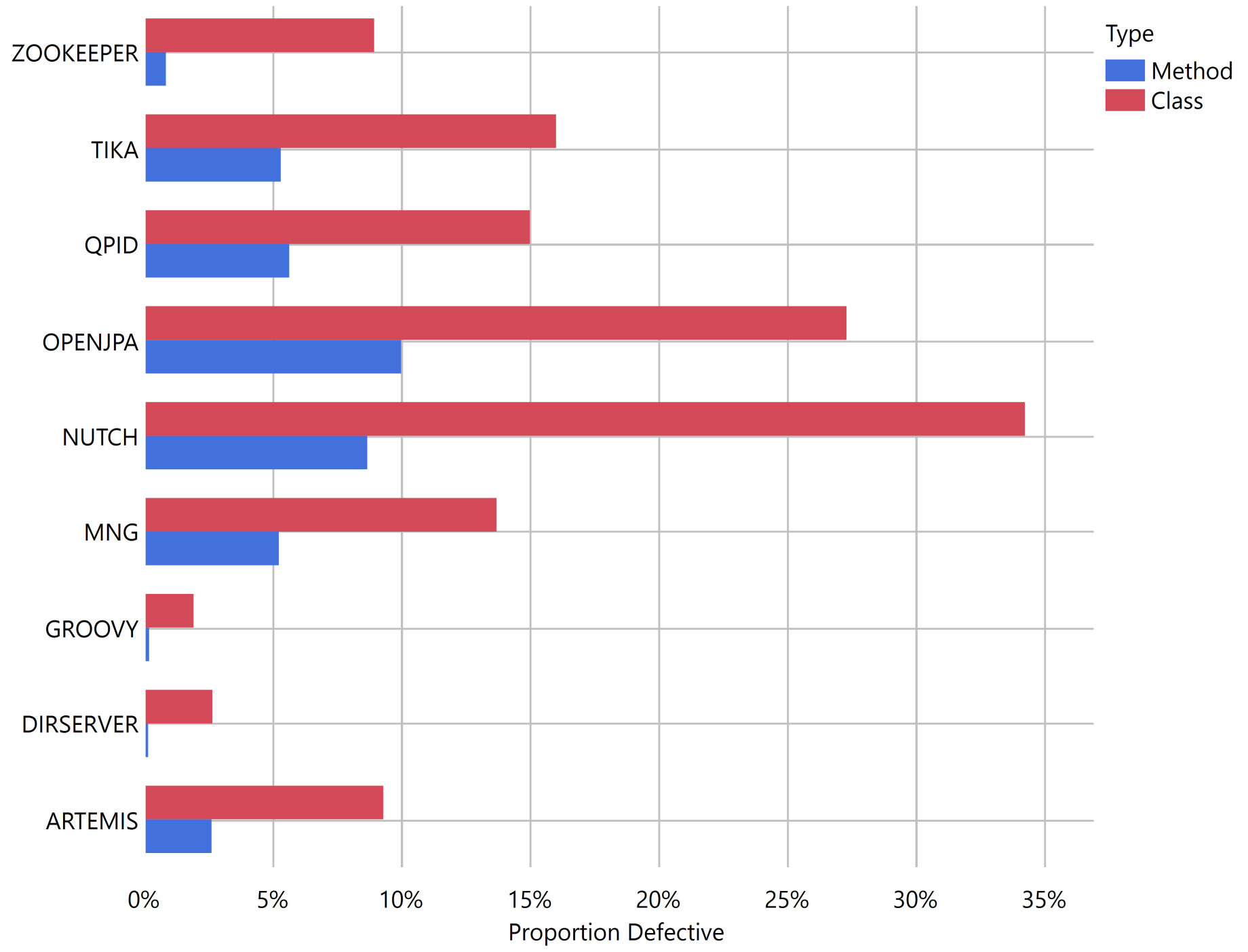

Figure 3.1.18: Proportion of defective entities (x-axis) in specific datasets (x-axis) across different types of entities (color).

Table 3.1.18 reports the statistical results (p-value) comparing the number of defective classes versus the number of defective methods. According to Table 3.1.18, there is a statistical difference between the number of defective methods and the number of defective classes; therefore, we can reject $H_{10}$. Moreover, the effect size is large.

Table 3.1.19 reports the statistical results (p-value) comparing the proportion of defective methods against the proportion of defective classes. According to Table 3.1.19, **methods are statistically less frequently defective than classes**. Therefore, we can reject $H_{20}$. Moreover, the effect size is medium.

Table 3.1.18: Statistical result (p-value) and Cohen's d effect size, comparing the number of defective methods and the number of defective classes in our projects.

| **Entity** | **Pvalue** | **Cohen's d** |
|---|---|---|
| Class Vs Method | 0.0488 | 0.598 |

Table 3.1.19: Statistical result (p-value) and Cohen's d effect size, comparing the proportion of defective methods against the proportion of defective classes.

| **Entity** | **Pvalue** | **Cohen's d** |
|---|---|---|
| Class Vs Method | 0.0488 | 0.359 |

**RQ2: Does leveraging JIT information increase the accuracy of MDP?**

***Accuracy.*** Figure 3.1.19 reports the distribution of AUC values across classifiers (y-axis) achieved by our proposed approaches (i.e., Combined and Direct) for MDP in the subject projects (x-axis). According to Figure 3.1.19, **the Combined approach is more accurate than Direct in all projects, except DRSERVER**. Moreover, the distribution of AUC values across classifiers of Combined is narrower than the distribution of Direct.

Figure 3.1.20 reports the distribution, across classifiers, of PofB values (x-axis) achieved by the Combined and Direct approaches (colors) for MDP in our subject projects (quadrant). According to Figure 3.1.20:

- Combined is better than Direct in all PofBs in Groovy, MNG, NUTCH, OPENJPA, QPID and TIKA.
- Combined is worse than Direct in all PofBs in DERSERVER.
- Similar to the previous results related to the AUC, the distribution of values from Combined is substantially more narrow than the distribution of values from Direct. This result indicates that the choice of classifiers is

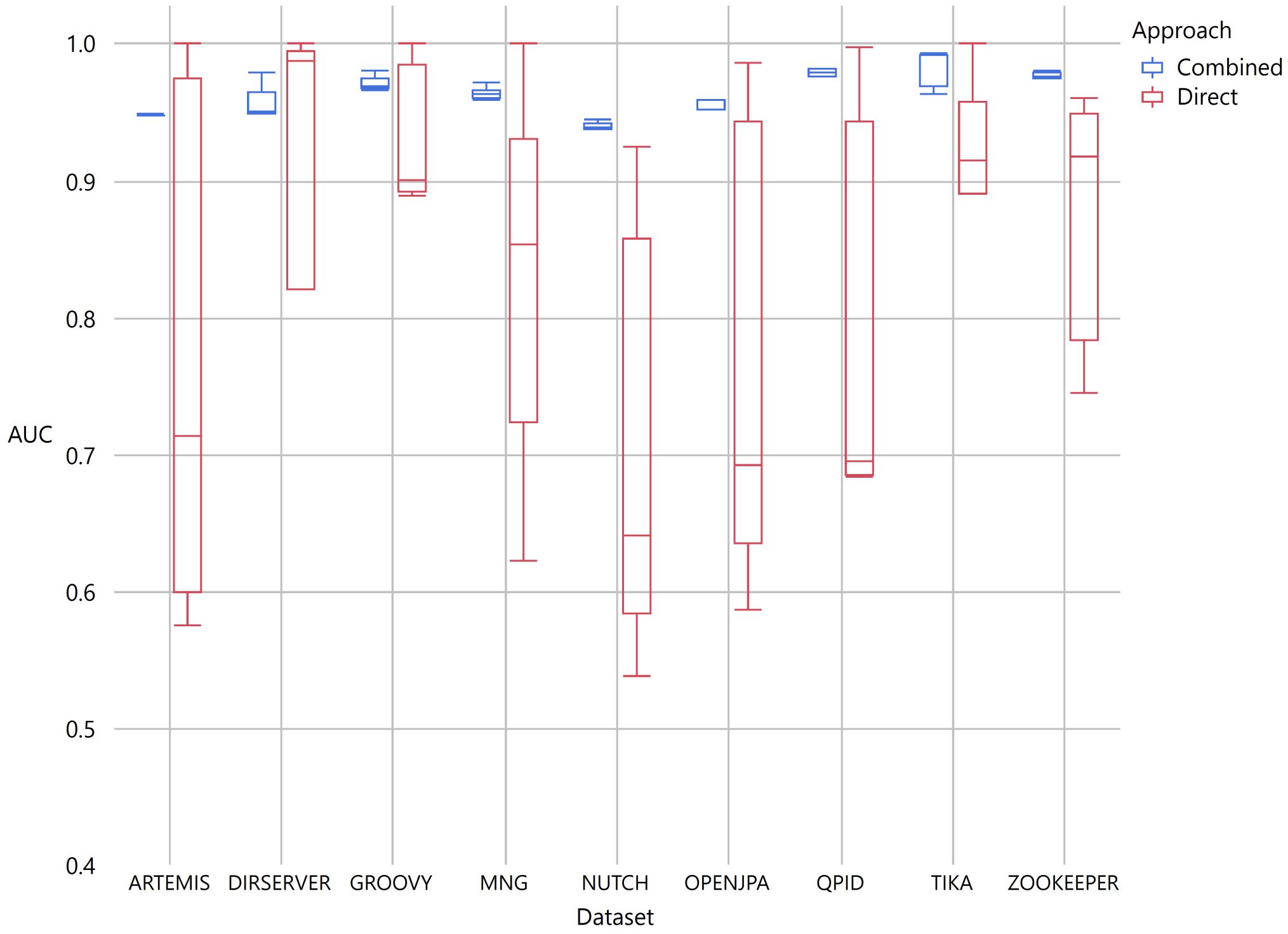

Figure 3.1.19: Distributions across classifiers of AUC values (y-axis) achieved by Combined and Direct (colors) for MDP in our subject projects (x-axis)

not as important when using Combined.

Figure 3.1.21 reports the mean of the relative gain in MDP by leveraging JIT across classifiers and projects. According to Figure 3.1.21: leveraging JIT increases the accuracy of MDP by an average of 17% in AUC and 46% in PofB10. It is interesting to note that the relative gain is inversely correlated with PofB; this is due to the fact that the margin of performance, and hence the relative gain, is reduced when considering a larger code base.

***Feature selection.***

Figure 3.1.22 reports the percent (x-axis) of times, across the nine classifiers, that a given feature (color) has been chosen in a given project (y-axis). According to Figure 3.1.22:

- **MaxC or SumC have been selected in all projects, except Zookeeper.**
- Direct has been selected in seven out of the nine projects.
- MaxC or SumC have been selected more than Direct in five out of the nine projects.

Table 3.1.20 reports the statistical results (p-value) comparing the MDP accuracy of the combined versus direct approach. An asterisk identifies cases where the pvalue is lower than alpha according to the Holm-Bonferroni correction and hence we can reject the null hypothesis. We note that MDP is statistically more accurate by leveraging JIT in AUC and in seven out of nine PofB. Therefore, we can reject $H_{30}$ and claim that leveraging JIT statistically and significantly improves the accuracy of MDP. Moreover, the effect size is at least medium in most of the metrics.

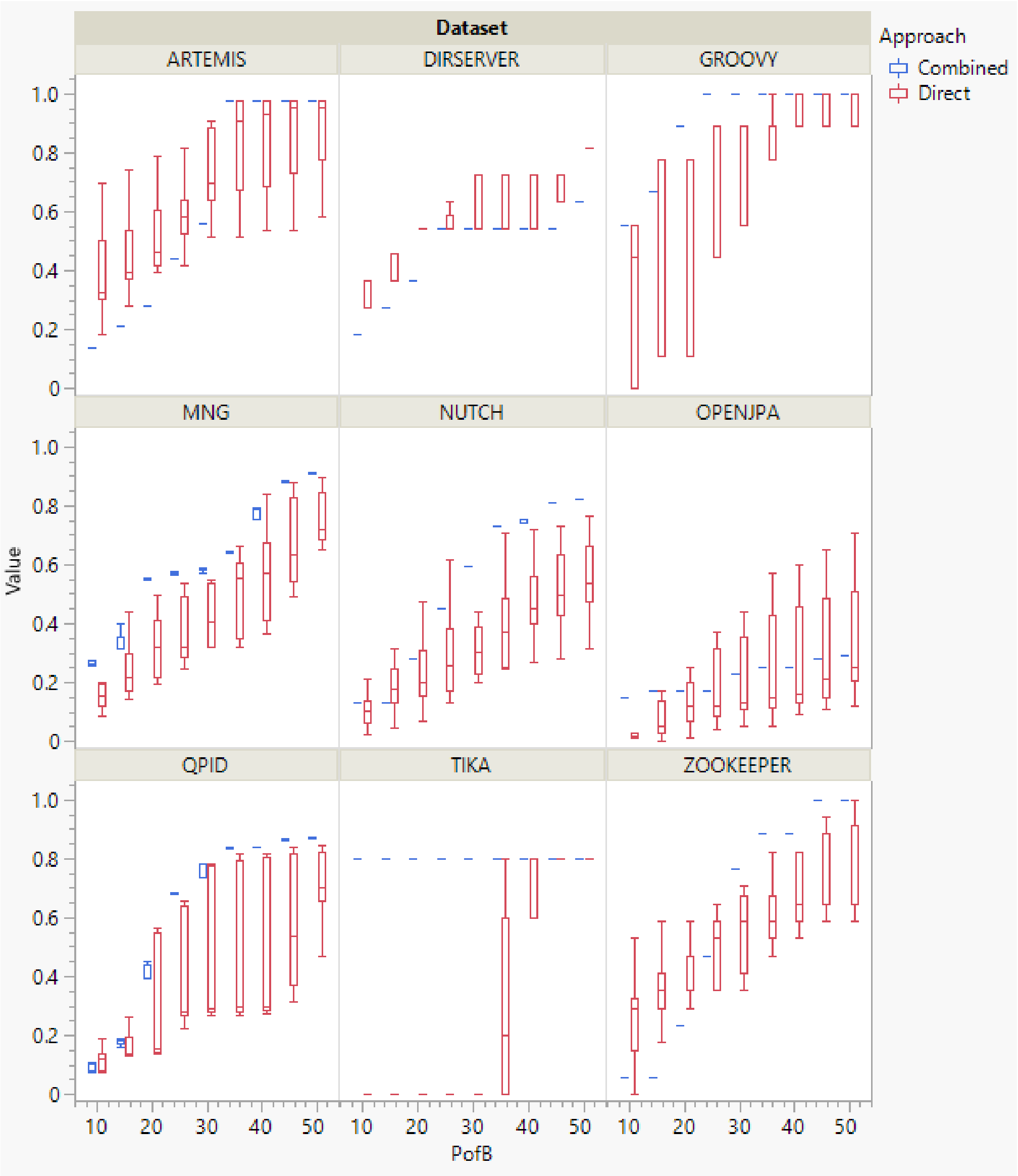

Figure 3.1.20: Distribution across classifiers of PofB values (x-axis) achieved by the Direct and Combined approaches (colors) for MDP in our studied projects (quadrant)

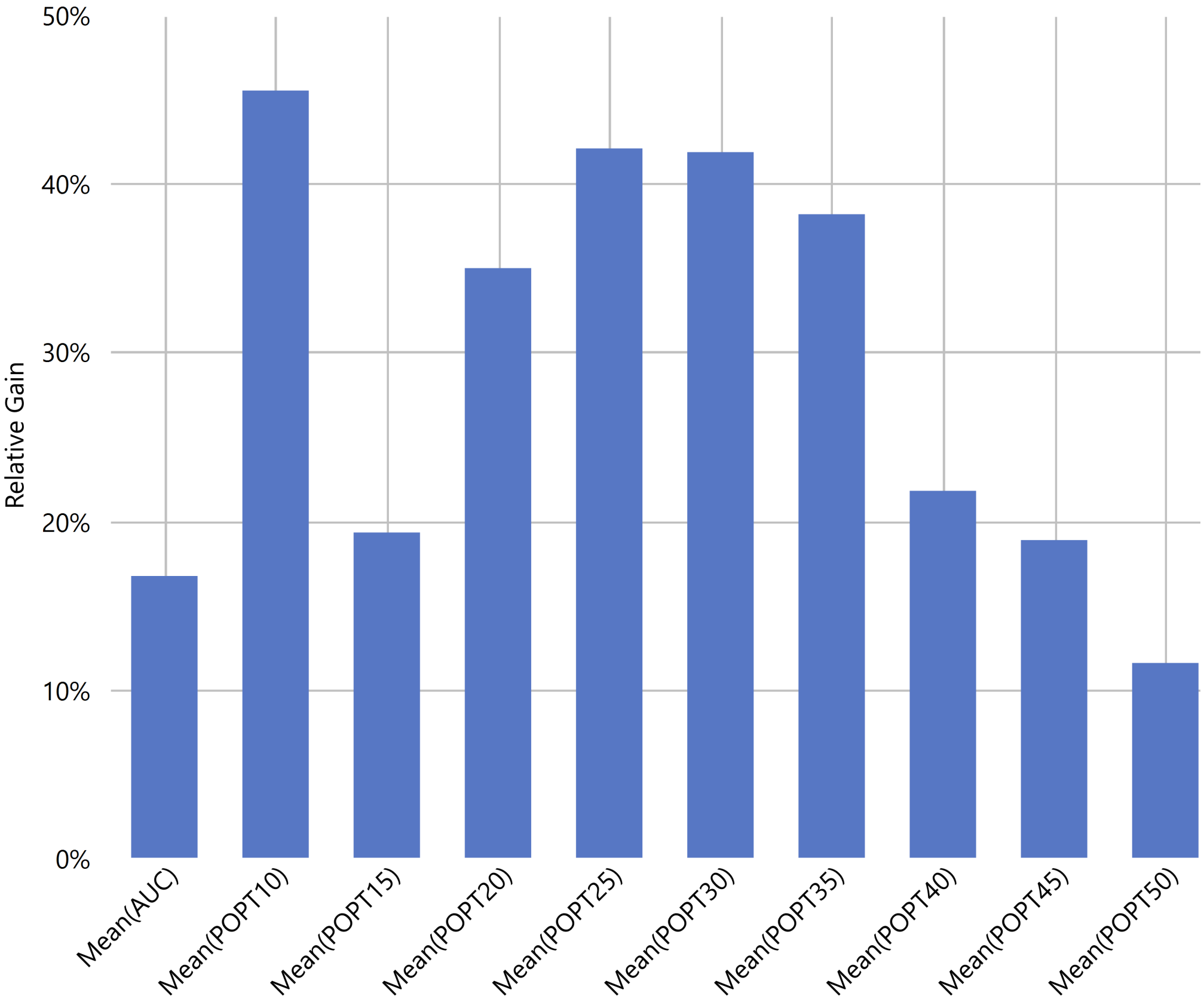

Figure 3.1.21: Mean relative gain in MDP by leveraging JIT across classifiers and datasets.

Table 3.1.20: Statistical result (p-value) and Cohen's d effect size, comparing the MDP accuracy of the combined versus direct approach.

| | AUC | Popt10 | Popt15 | Popt20 | Popt25 | Popt30 | Popt35 | Popt40 | Popt45 | Popt50 |
|---|---|---|---|---|---|---|---|---|---|---|
| Pvalue | 0.0001* | 0.1082 | 0.5813 | 0.0275* | 0.0001* | 0.0001* | 0.0001* | 0.0001* | 0.0001* | 0.0001* |
| Cohen's d | 1.1746 | 0.4052 | 0.2263 | 0.4900 | 0.7130 | 0.7970 | 0.8241 | 0.5896 | 0.5567 | 0.3949 |

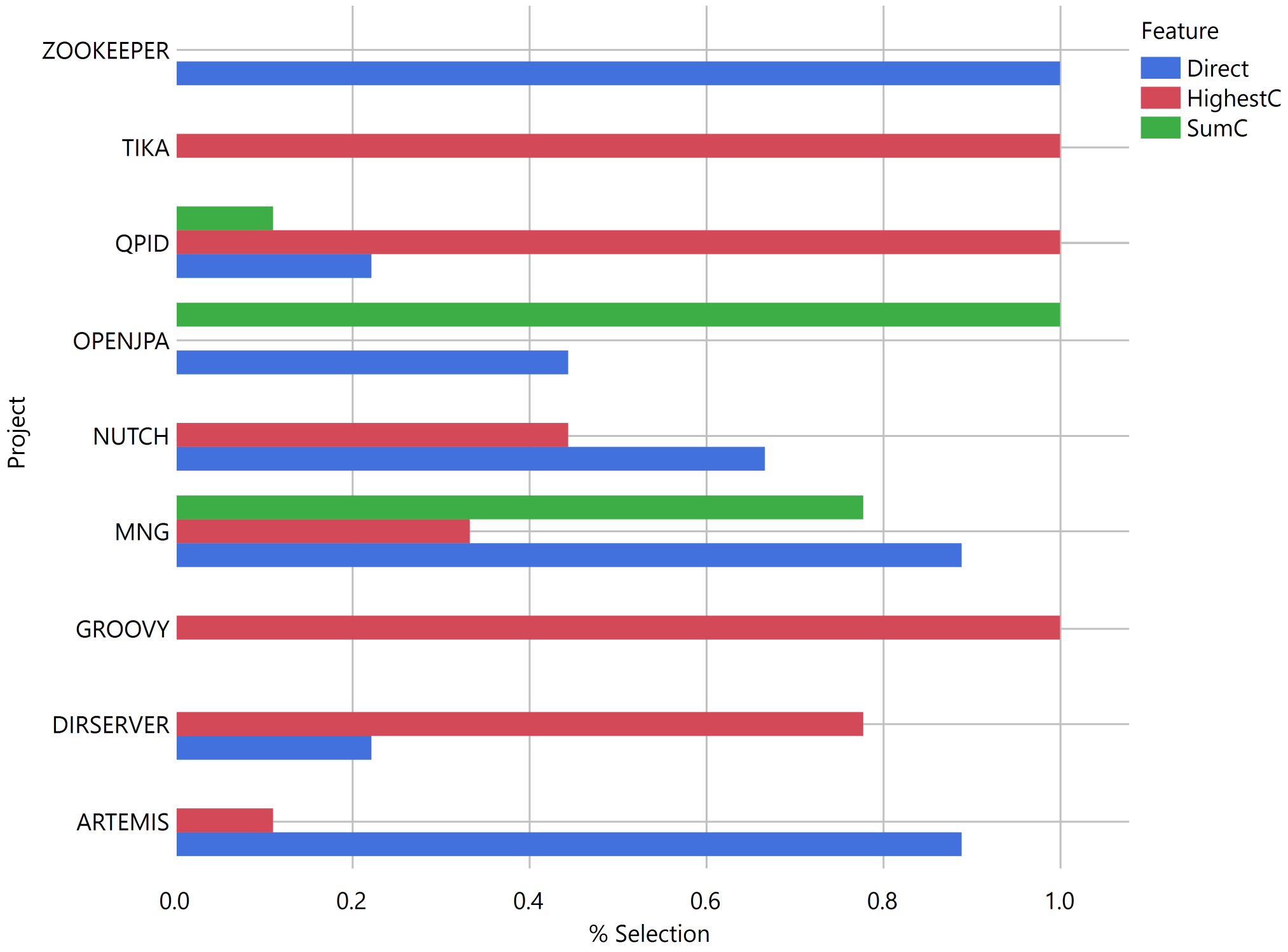

Figure 3.1.22: Percent (x-axis) of times, across the nine classifiers, that a given feature (color) has been chosen in a given project (y-axis), in MDP.

**RQ3: Does leveraging JIT information increase the accuracy of CDP?**

***Accuracy.*** Figure 3.1.23 reports the distribution of the AUC values across classifiers (y-axis) achieved by different our approaches (i.e., Combined and Direct) for CDP in our subject projects (x-axis). According to Figure 3.1.23:

- **The median of Combined is more accurate than Direct in all nine projects**.
- As in RQ2, the distribution of values across classifiers of Combined is extremely narrower than the distribution of Direct. Therefore, the choice of classifiers is not as important when using the Combined approach.

Figure 3.1.24 reports the distribution across classifiers of PofB values (x-axis) achieved by different approaches (Combined and Direct) for CDP in our subject projects (quadrant). According to Figure 3.1.24:

- Combined is better than Direct in all PofB values in seven out of nine projects: ARTEMIS, DIRSERVER, MNG, NUTCH, OPENJPA, QPID and TIKA.
- There is no dataset where Combined is worse than Direct in all PofB values.
- Similar to the results for AUC, the distribution of values across classifiers from Combined is extremely narrower than the distribution from Direct. Therefore, the choice of classifiers is not as important when using Combined.

Figure 3.1.25 reports the mean relative gain in CDP by leveraging JIT across classifiers and projects. According to Figure 3.1.25 **leveraging JIT increases**

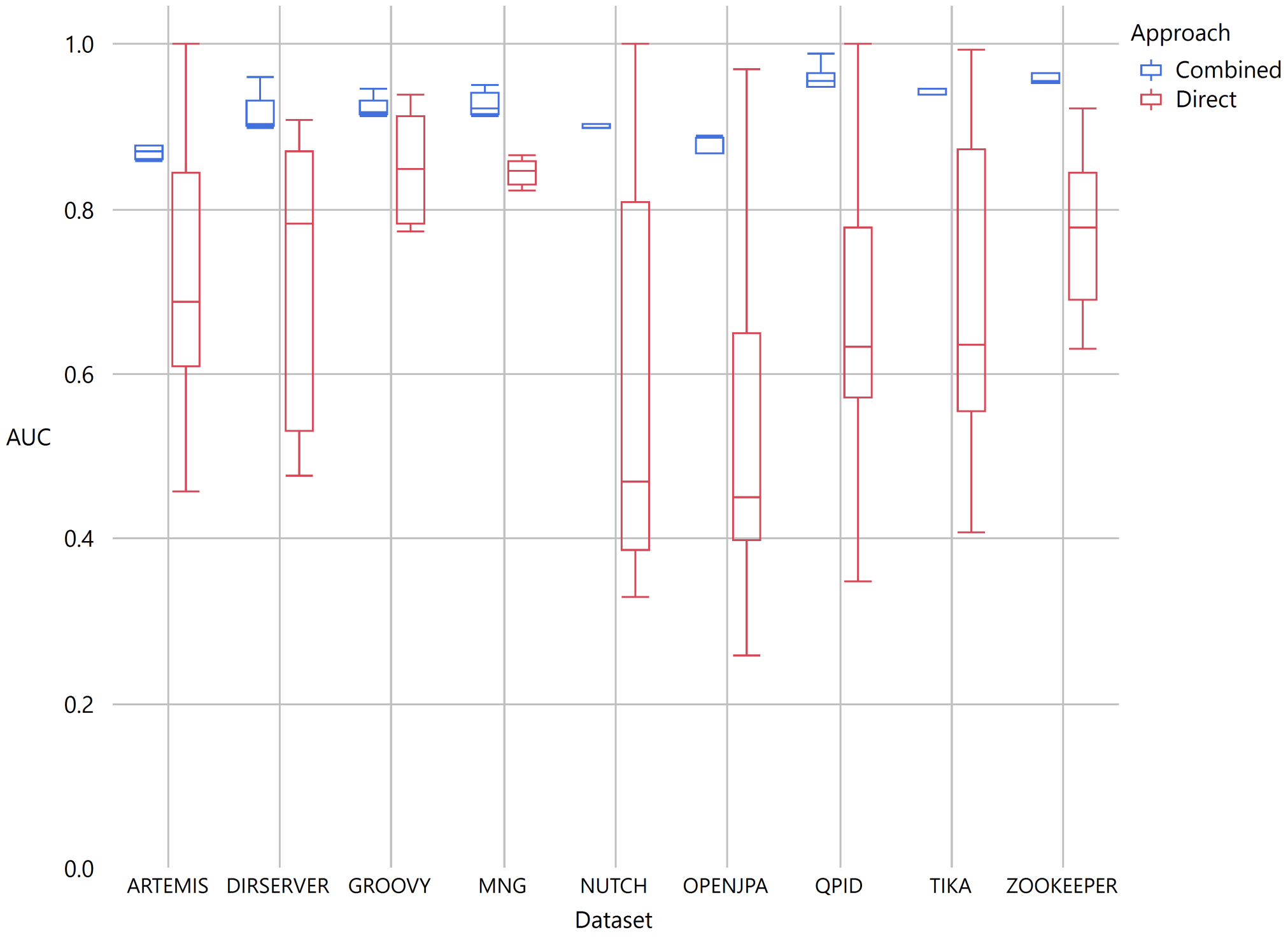

Figure 3.1.23: Distribution of the AUC values across classifiers (y-axis) achieved by different approaches (Combined and Direct) for CDP in our subject projects (x-axis)

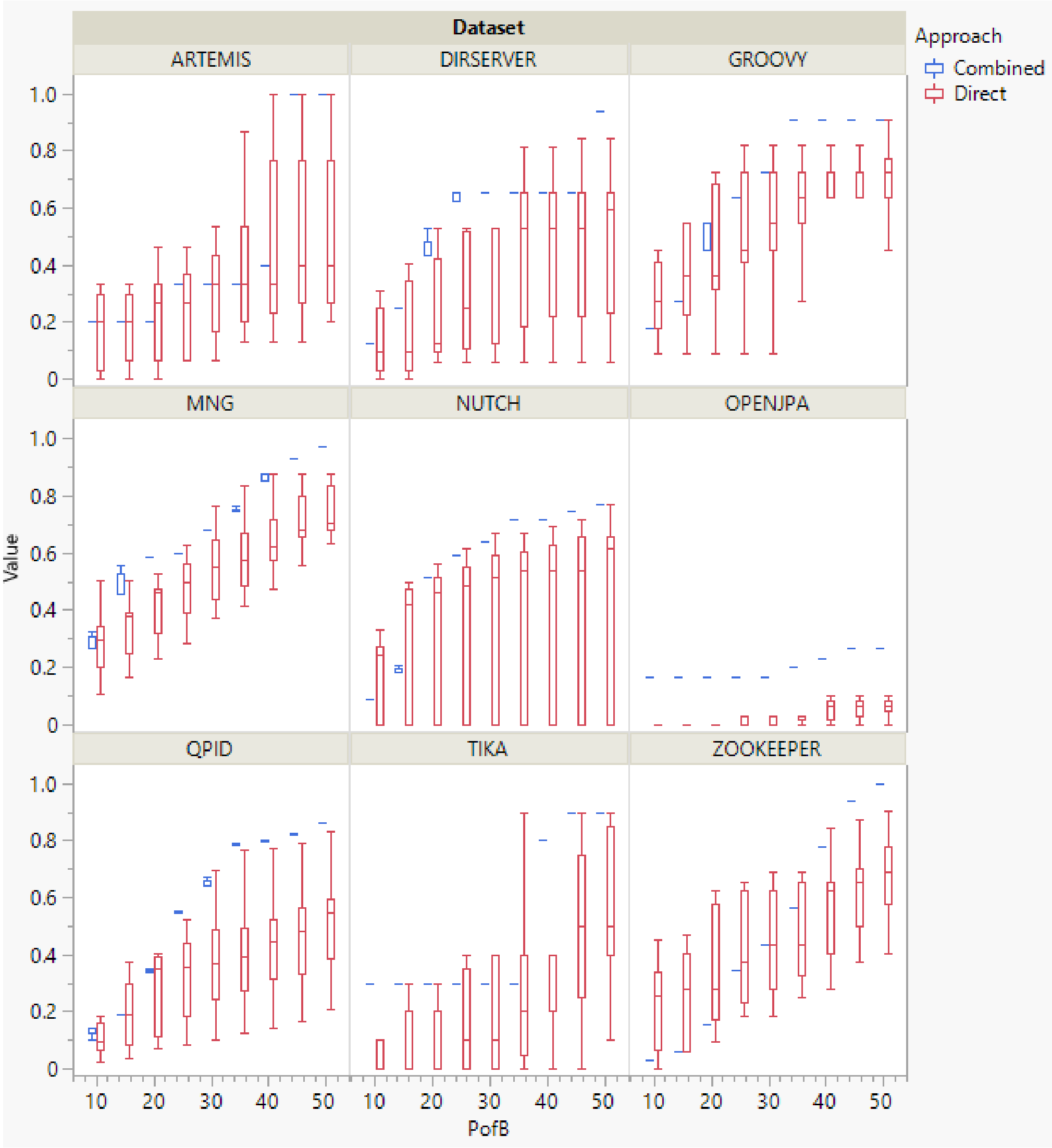

Figure 3.1.24: Distribution across classifiers of PofB values (x-axis) achieved by different approaches (Combined and Direct) for CDP in our subject projects (quadrant)

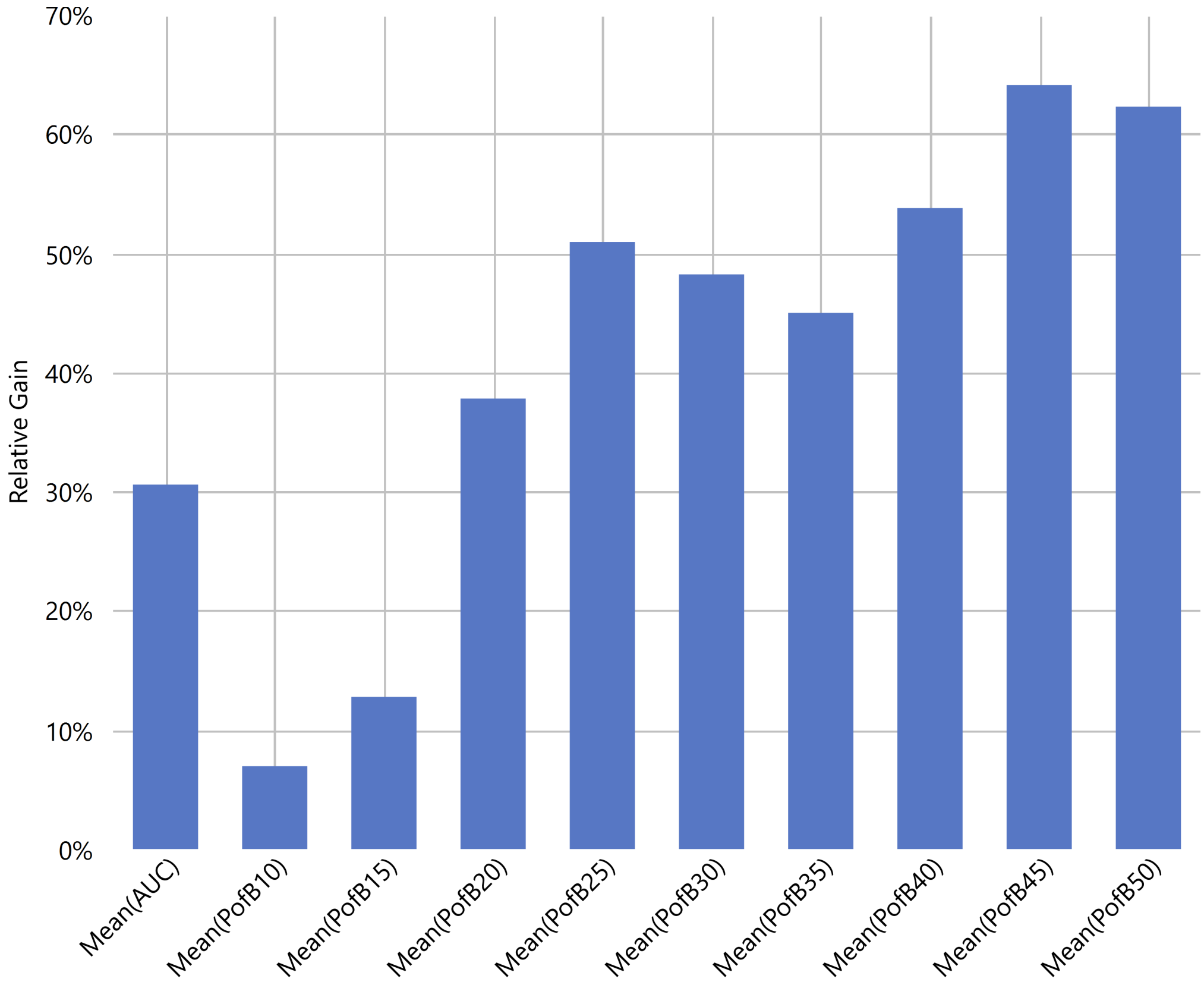

Figure 3.1.25: Mean relative gain in CDP by leveraging JIT across classifiers and datasets.

**the accuracy of CDP by an average of 31% in AUC and 38% in PofB20.** It is interesting to note that the relative gain is not inversely correlated with PofB (as observed in RQ2).

***Feature selection.***

Figure 3.1.26 reports the distribution across datasets of feature selection and reports which feature is selected. According to Figure 3.1.26:

- **MaxC or SumC have been selected in all nine projects.**
- Direct has not been selected in two out of the nine projects.

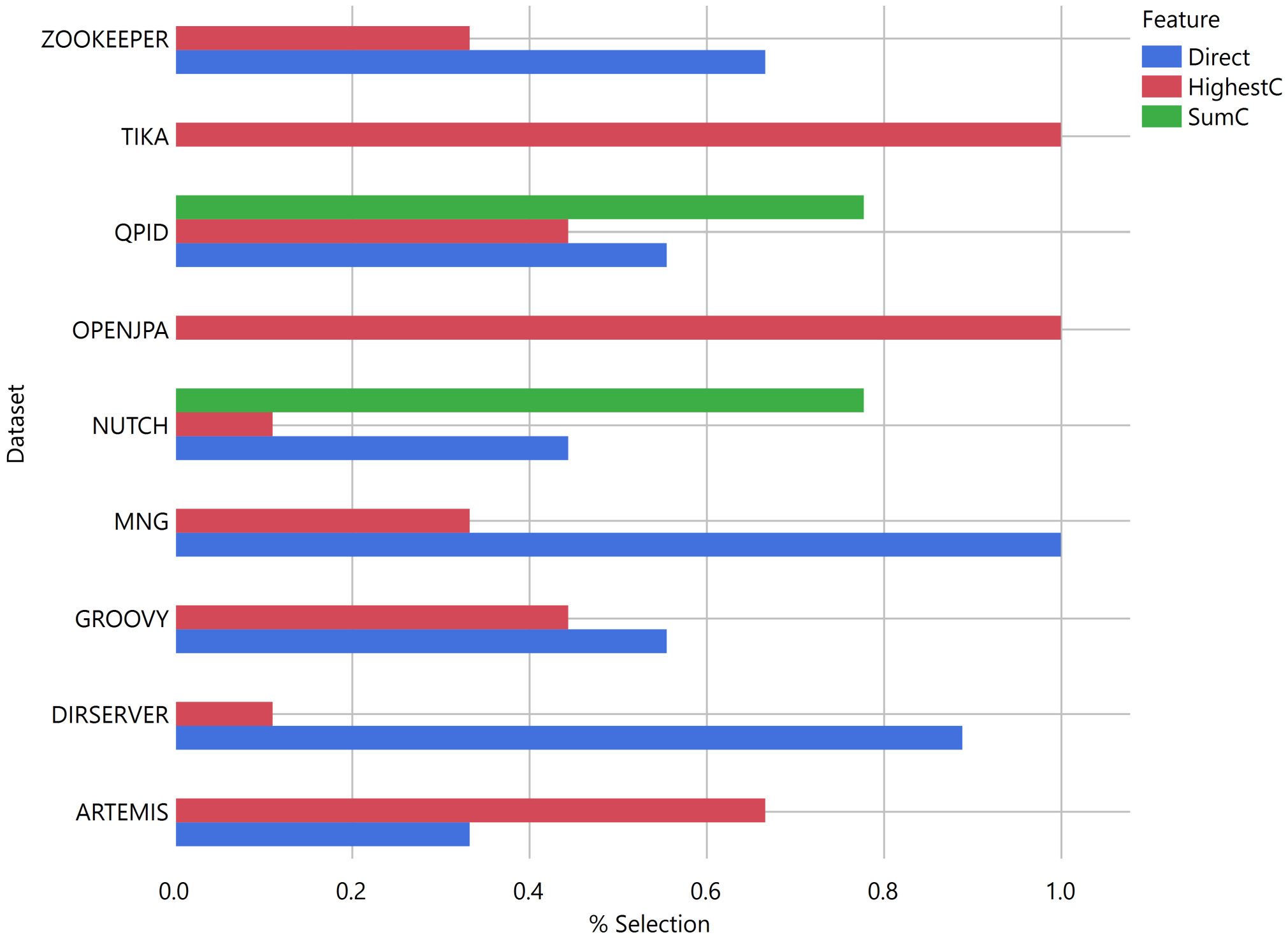

Figure 3.1.26: Distribution across datasets of feature selection for CDP

- MaxC or SumC were selected more than Direct in five out of the nine projects.

Table 3.1.21 reports the statistical results (p-value) comparing the CDP accuracy of the combined versus direct approach. An asterisk identifies cases where the pvalue is lower than alpha according to the Holm-Bonferroni correction and hence we can reject the null hypothesis. We note that MDP is statistically more accurate by leveraging JIT in AUC and in seven out of nine PofB. Therefore, we can reject $H_{40}$ and claim that leveraging JIT statistically and significantly improves the accuracy of CDP. Moreover, the effect size is at least medium in most of the metrics.

Table 3.1.21: Statistical result (p-value) and Cohen's d effect size, comparing the CDP accuracy of the combined versus direct approach.

| | AUC | Popt10 | Popt15 | Popt20 | Popt25 | Popt30 | Popt35 | Popt40 | Popt45 | Popt50 |
|---|---|---|---|---|---|---|---|---|---|---|
| Pvalue | 0.0001* | 0.2223 | 0.0913 | 0.0001* | 0.0001* | 0.0001* | 0.0001* | 0.0001* | 0.0001* | 0.0001* |
| Cohen's d | 1.5737 | 0.0943 | 0.1787 | 0.5269 | 0.7842 | 0.7590 | 0.7295 | 0.9890 | 1.2621 | 1.2941 |

**RQ4: Are we more accurate in MDP or CDP?**

Figure 3.1.27 reports the distribution across classifiers of AUC values (y-axis) achieved for MDP or CDP (colors) in our subject projects (x-axis). According to Figure 3.1.27:

- **MDP is more accurate than CDP in all nine projects.**
- The distribution of values across classifiers for MDP is extremely narrower than the distribution for CDP. Therefore, the choice of classifiers is less important in MDP than it is in CDP.

Since Figure 3.1.27 results are interesting considering that defective methods are harder to find by chance than defective classes, Figure 3.1.28 reports the distribution across projects of AUC values (y-axis) achieved for MDP or CDP (colors) by our classifiers (x-axis). According to Figure 3.1.28:

- **MDP is more accurate than CDP in all nine classifiers.**
- The distribution of values across projects for MDP is extremely narrower than the distribution for CDP. Thus, classifiers are more stable in MDP than CDP.

Figure 3.1.29 reports the mean across classifiers for a specific value of PofB (x-axis) achieved by MDP or CDP (colors) in our subject projects (quadrant). We reported the results in bar-charts instead of box-plots (as done in RQ2 and RQ3) because the distributions are very narrow, which hinders the visualization of results in this case. These distributions can also bee seen in previous figures (Figure 3.1.20 for MDP and Figure 3.1.24 for CDP). According to Figure 3.1.29:

- MDP is better than CDP in all PofB values in four projects.
- MDP is worse than CDP in all PofB values only in the Groovy project.

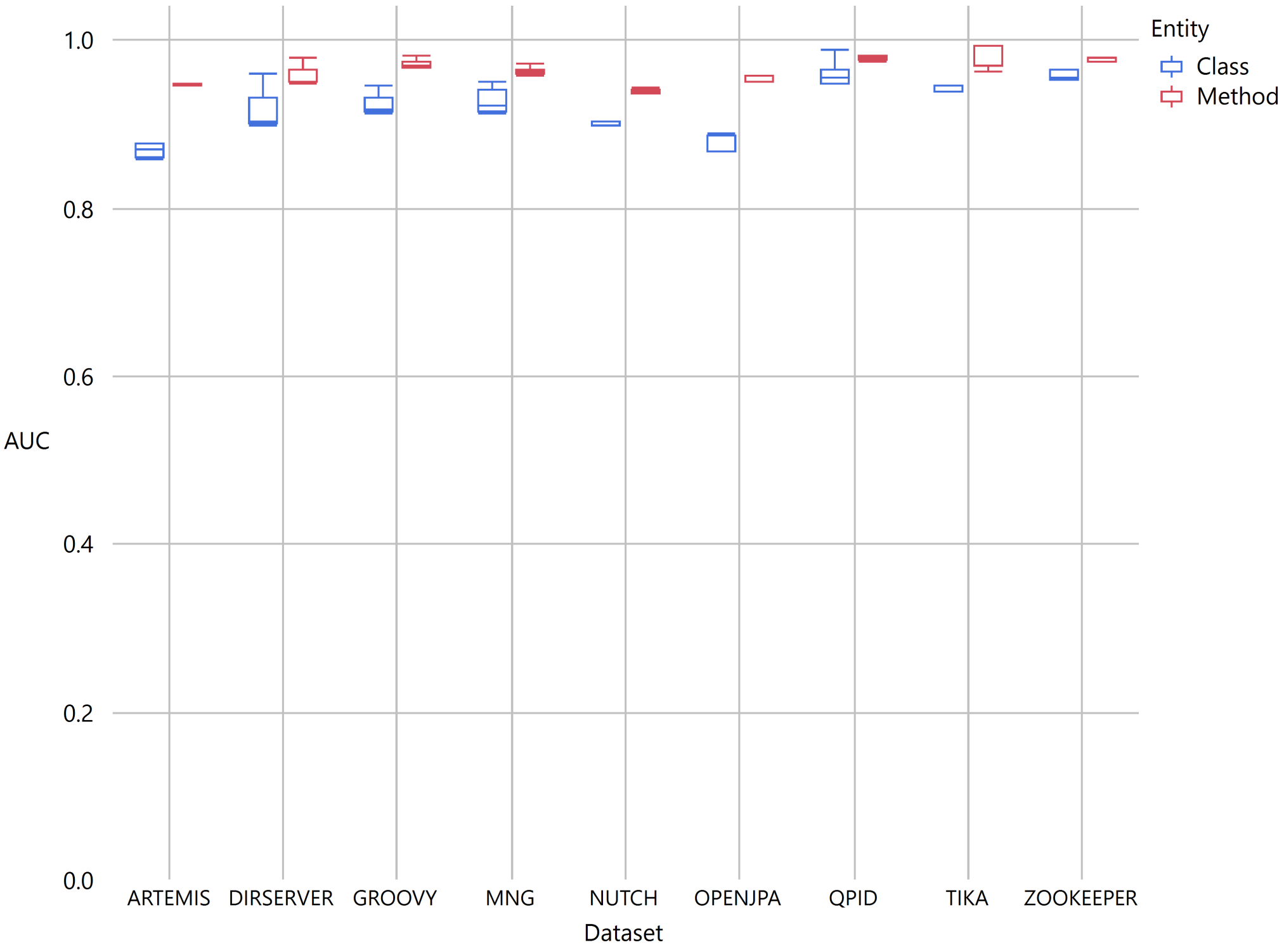

Figure 3.1.27: Distribution across classifiers of AUC values (y-axis) achieved for MDP or CDP (colors) in our subject projects (x-axis).

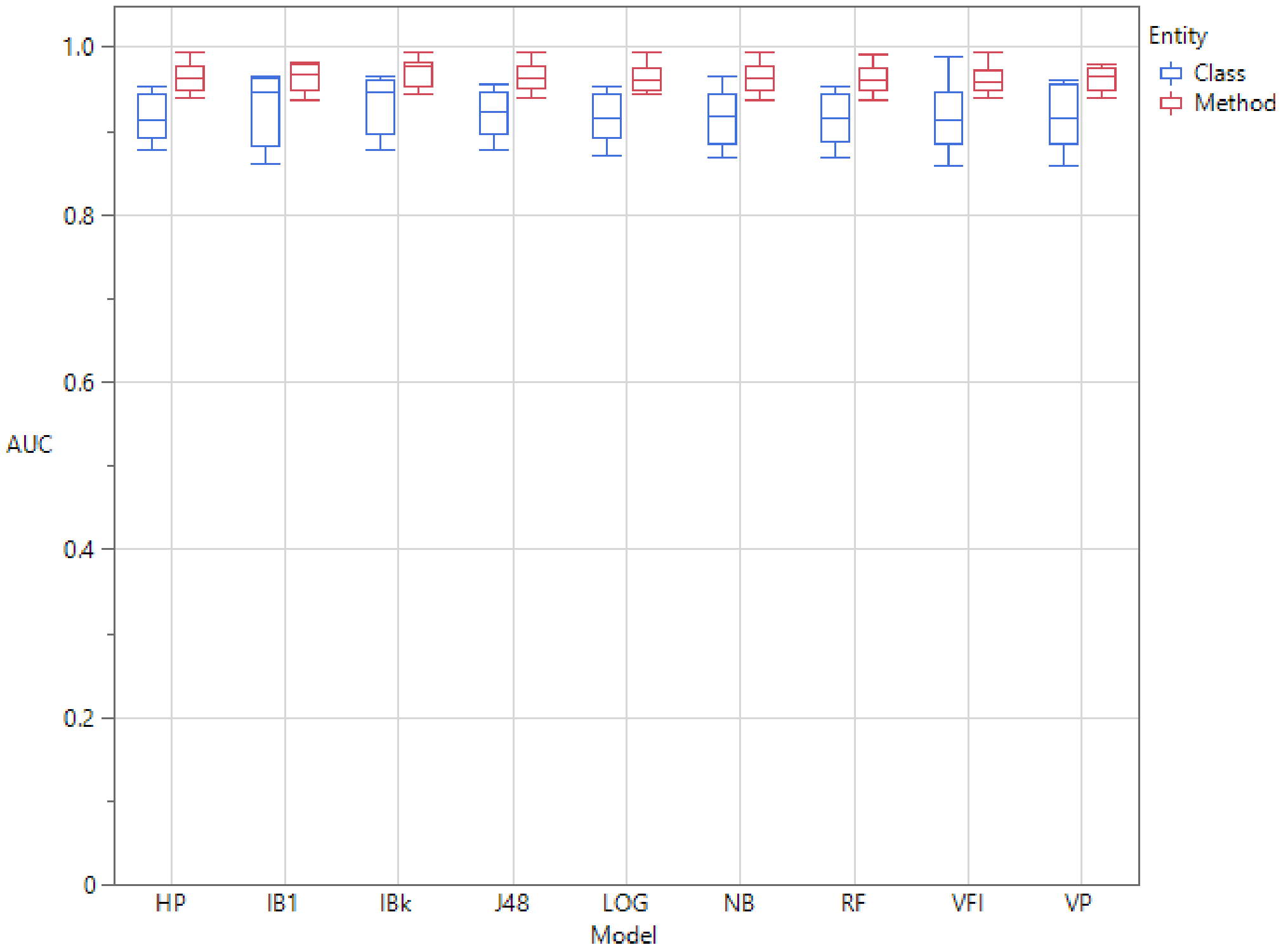

Figure 3.1.28: Distribution across projects of AUC values (y-axis) achieved for MDP or CDP (colors) by our classifiers (x-axis).

Table 3.1.22: Statistical result (p-value) and Cohen's d effect size, comparing the MDP versus CDP accuracy. An asterisk identifies a pvalue lower than alpha.

| | AUC | Popt10 | Popt15 | Popt20 | Popt25 | Popt30 | Popt35 | Popt40 | Popt45 | Popt50 |
|---|---|---|---|---|---|---|---|---|---|---|
| Pvalue | 0.0001* | 0.0001* | 0.0001* | 0.0002* | 0.0001* | 0.0001* | 0.0001* | 0.0003* | 0.1381 | 0.9942 |
| Cohen's d | 1.7579 | 0.5661 | 0.4334 | 0.3894 | 0.4737 | 0.5959 | 0.6106 | 0.2784 | 0.0367 | 0.1196 |

Figure 3.1.30 reports the mean of the relative gain in performing MDP over CDP across classifiers and projects. According to Figure 3.1.30 **MDP is more accurate than CDP by an average of 5% in AUC and 62% in PofB10.** It is worth noting that the relative gain is inversely correlated with PofB again.

Table 3.1.22 reports the statistical results (p-value) comparing the accuracy of MDP versus CDP. An asterisk identifies the eight cases out of ten where the pvalue is lower than alpha according to the Holm-Bonferroni correction and hence we can reject the null hypothesis. We note that MDP is statistically more accurate than CDP in AUC and in seven out of nine PofB. Therefore, we can reject $H_{50}$ and claim that MDP is more accurate than CDP. Moreover, the effect size is at least medium in four out of ten metrics.

#### 3.1.3.3 Discussion

This section discuss our main results, offering possible explanations for the results, implications, and guidelines for practitioners and researchers.

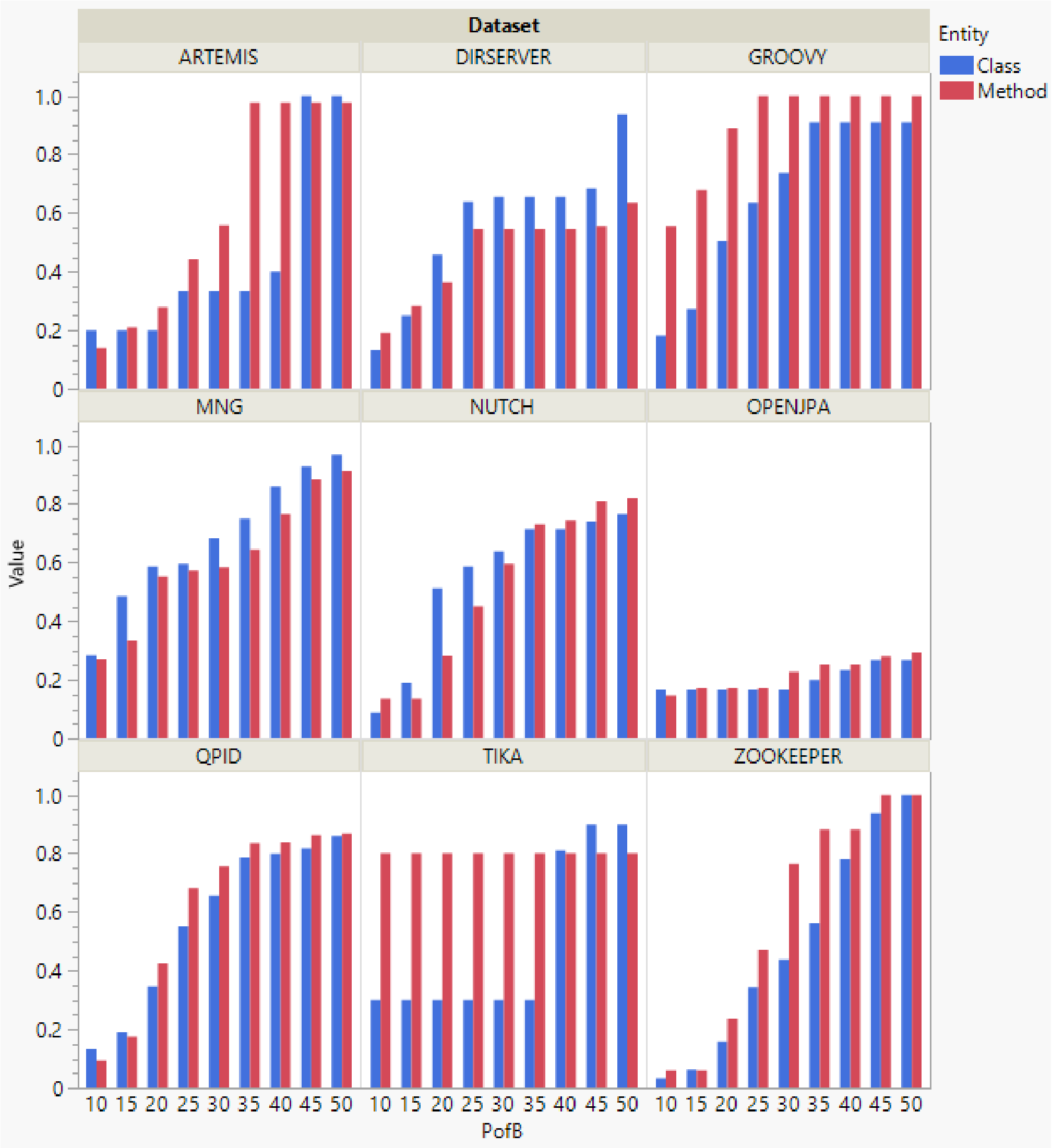

Figure 3.1.29: Mean across classifiers of PofB values (x-axis) achieved for MDP or CDP (colors) in our subject projects (quadrant).

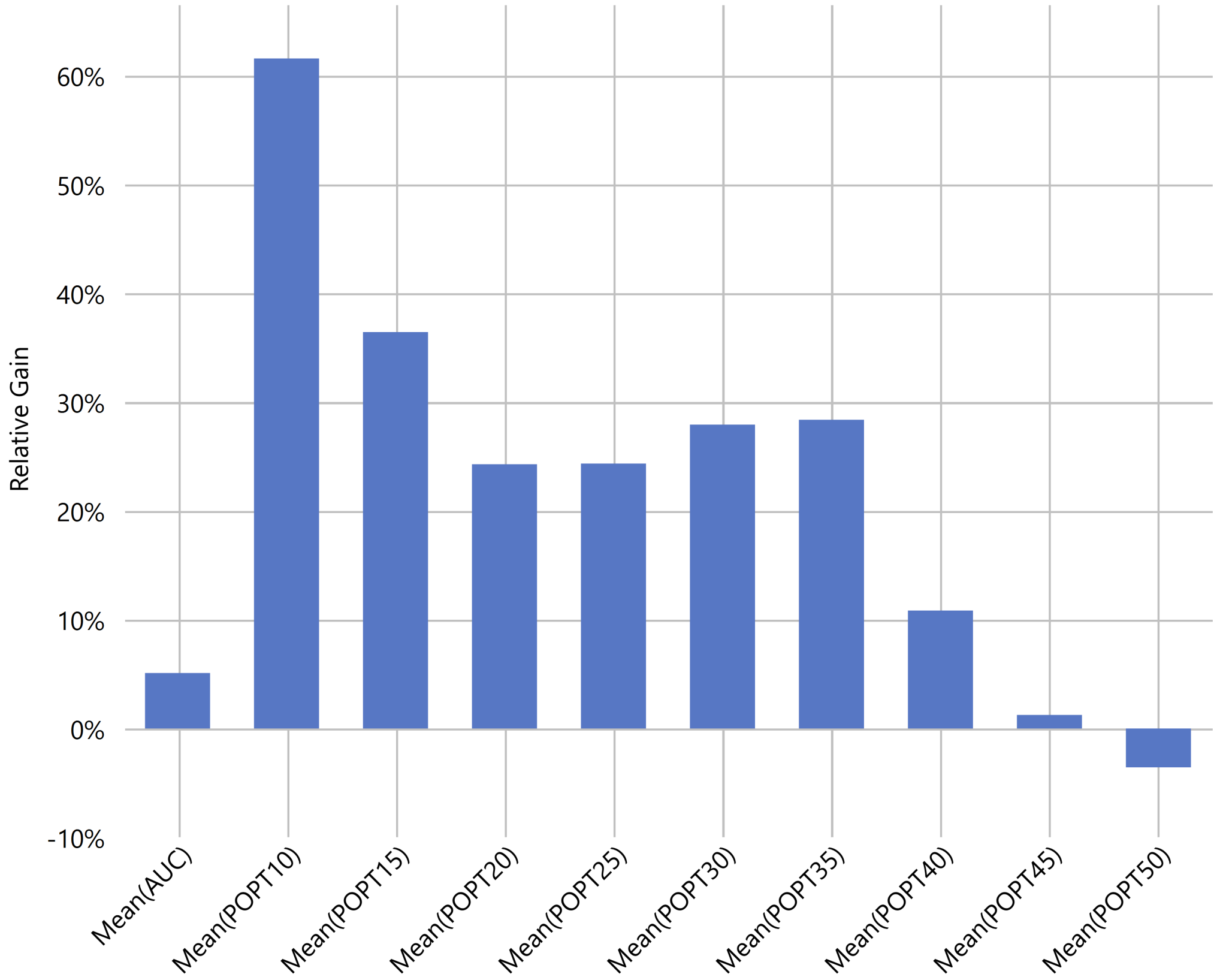

Figure 3.1.30: Mean of the relative gain in performing MDP over CDP across classifiers and projects

### Main results and possible explanations

The main result of RQ1 is that **defective methods are significantly less frequent than defective classes. This means that it is harder to find by chance defective entities if they are methods rather than classes**. This result confirms common wisdom for which MDP is more challenging than CDP because defective methods are rare. It was surprising at first to observe that in three out of nine projects the number of defective classes is higher than defective methods. We triple-checked the results, and we found no mistakes. Our investigation revealed that such a surprising result was due to defects pertaining only to attributes and that such defects are particularly numerous in those three projects. Let us take the DIRSERVER-1019 as an example defect, which affects the core/src/main/java/org/apache/directory/server/core/jndi/ServerDirContext.java class. By observing the content of the DIRSERVER-1019 fix commit, which has ID 09cc2c065fb36662ebf9f56486 af28d87ad09d4c, we realize that developers removed the static keyword from the attribute "static final FilterParserImpl filterParser = new FilterParserImpl();". Such an attribute, and hence the DIRSERVER-1019 defect, was inserted in the commit b392e8f69e2c6f30116459152db612e414f18724. The fact that defective classes can be more than defective methods supports the need for CDP as it can identify defects that MDP cannot.

The main results of RQ4 are that **MDP is substantially more accurate than CDP (a mean of +5% in AUC and +62% in PofB10)**. The higher accuracy of MDP in comparison with CDP is visible in all datasets in terms of AUC. This might be due to the coarser granularity of the classes that, by definition, are larger than methods and, hence, harder to rank as only partially defective. Another possible reason is that defective methods are lower in pro-portion than defective classes and, hence, a better ranking is more visible in effort-aware metrics like PofB.

The main result of RQ2 is that **leveraging JIT increases the accuracy of MDP by an average of 17% in AUC and 46% in PofB10**. Similarly, in RQ3, we observe that **leveraging JIT increases the accuracy of CDP by an average of 31% in AUC and 38% in PofB20**. Moreover, our statistical tests reveal that leveraging JIT increases the accuracy of MDP and CDP in AUC and in PofB (from PofB20 to PofB50). We note that leveraging JIT increases the accuracy of MDP and CDP in all datasets in terms of AUC.

Regarding PofB, **in some Combined datasets - PofBX-JIT decreases the accuracy of MDP and CDP**. For instance, in Zookeeper, JIT decreases the accuracy of CDP in PofB (from PofB10 to PofB25).

Regarding ZOOKEEPER, it was surprising at first to observe that according to Figure 3.1.19 Combined is better than Direct whereas for the same project in Figure 3.1.22 HighestC and SumC have never been selected. We triple-checked the results and we found no mistakes. If HighestC and SumC have not been selected, Combined cannot be better than Direct. If on the one side, this reasoning is correct, on the other side the datasets upon which the feature selection is applied differ between Figure 3.1.19 and Figure 3.1.22. To compute results in Figure 3.1.19, the feature selection is applied to the training set, as it aims at supporting the prediction on the testing set. To compute results in Figure 3.1.22, the feature selection is applied to the entire dataset, as it aims at providing results on a dataset. Thus, there is no inconsistency between results in Figure 3.1.19 and Figure 3.1.22.

***The role of partially defective commits.*** We note that leveraging JIT increases the accuracy of MDP or CDP under two conditions: 1) the JIT is accurate, 2)the entities touched by a defective commit are defective. This last point is important, since defective commits usually have only a small proportion of statements that are defect-inducing [296]. Thus, the methods and classes touched by the non-defect-inducing statements from a defective commit are actually not defective. In this study, when we leverage JIT, the defectiveness of a commit is cascaded over all the touching entities, therefore we cascade this defectiveness also over the entities touched by the non-defect-inducing statements of the commit. In other words, while our JIT prediction is performed at the commit level, our ground truth is computed at the defect-inducing-statements level. Thus, the positive impact of leveraging JIT to support MDP or CDP is correlated with the percent of defective entities (classes or methods) touched by defective commits.

To investigate whether the effect of partially defective commits is substantial in our results, we report in Figure 3.1.31 the mean percent, across commits, of defective entities touched by defective commits. A low value of this percentage indicates that defective commits were only partially defective and, hence, they had non-defect-inducing statements touching several entities.

Regarding RQ2, according to Figure 3.1.31, Artemis, Dirserver and Zookeeper have the lowest proportion of actually defective methods touched by a defective commit; this is in line with the Figure 3.1.20 and Figure 3.1.22 as, in Zookeeper, neither HighestC nor SumC have been selected in MDP. Moreover, according to Figure 3.1.31, Tika has the highest proportion of actually defective methods touched by a defective commit; this is in line with Figure 3.1.22 as, in Tika, Direct has never been selected in MDP.

Regarding RQ3, according to Figure 3.1.31, Zookeper has the lowest proportion of actually defective classes; this is in line with Figure 3.1.24 since in this project Combined has a lower PofB10 to PofB25 than Direct. Moreover, according to Figure 3.1.31, Nutch and Openjpa have the highest proportion of actually defective classed touched by a defective commit; this is in line with Figure 3.1.22 as, in Tika, Direct has never been selected in MDP.

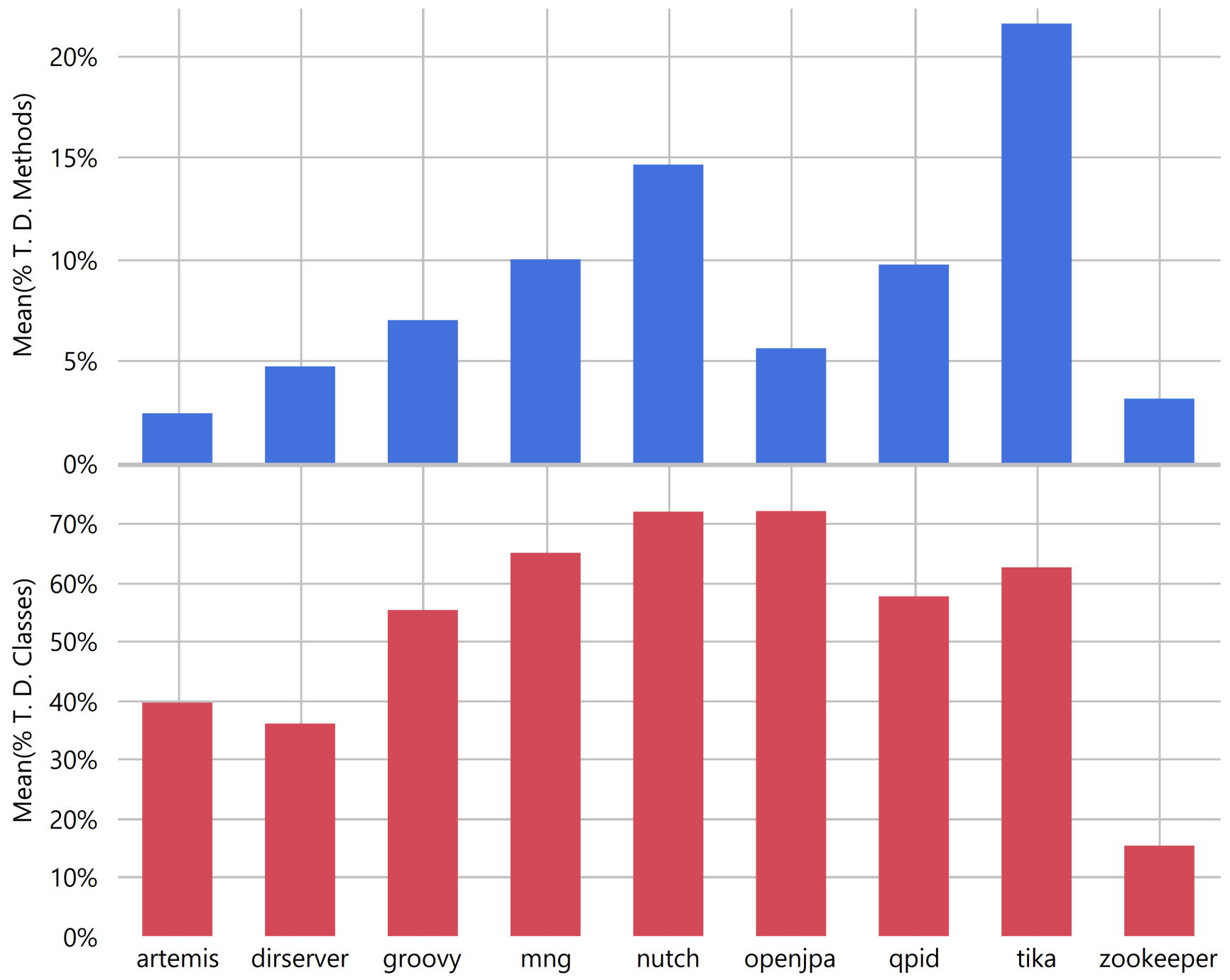

Figure 3.1.31: Mean percent across commits of defective entities that are touched by a defective commit.

Comparing RQ2 to RQ3, we observe that JIT helped more CDP than MDP. This can be explained by Figure 3.1.31, since the percent of entities that are actually defective when touched by a defective commit is about five times higher for classes than it is for methods. This means that leveraging a finer grained defect prediction than JIT, i.e. a statement-defectiveness-prediction [312], would likely benefit more MDP than CDP.

***The narrower distributions.*** Other important results from RQ2 and RQ3

are that the distributions of accuracy are extremely narrower in both MDP and CDP when leveraging JIT. This indicates that the choice of classifiers does not impact Accuracy as much and, hence, leveraging JIT does not only increases the accuracy of both MDP and CDP, but also makes them much more stable across a set of different classifiers. In order to analyze possible reasons as to why the distributions become narrower, Figure 3.1.32 and 3.1.33 report the STDV of Combined, Direct, MaxC and SumC in each specific dataset for MDP and CDP, respectively. According to Figure 3.1.32 and 3.1.33, Combined has a lower STDV than Direct in all datasets in both MDP and CDP. However, the difference in the STDV among Combined and Direct is not as big to explain the significant difference in the resulting accuracy metrics.

We do not see any correlation between defective commits ratio and the benefits of Combined for MDP or CDP. Specifically, Combined is better than Direct also in Groovy, see Figure 6 and Figure 7 for MDP and see Figure 10 and Figure 11 for CDP; i.e., leveraging JIT increases the accuracy of MDP and CDP even in a project with only 2% of defective commits. Moreover, we note that large projects might bias results across projects, e.g., Figure 8 and Figure 12.

#### Implications

The main implication of RQ1 is that **if practitioners do not use any support system, they should prefer testing instances that are classes instead of methods**. Conversely, RQ4 demonstrates that provided the same amount of effort, practitioners would find a much higher percentage of defective entities using a ranked list of methods instead of a list of classes. Specifically, analyzing results across datasets, the lowest accuracy achieved by MDP is higher than the highest accuracy achieved by CDP. Thus, the main implication of RQ4 to practitioners is that **it is better to predict and rank defective methods rather than defective classes**. The main implications for practitioners of RQ2 and RQ3 is that overall, **ranking and classification of both methods and classes shall be done by leveraging JIT information**.

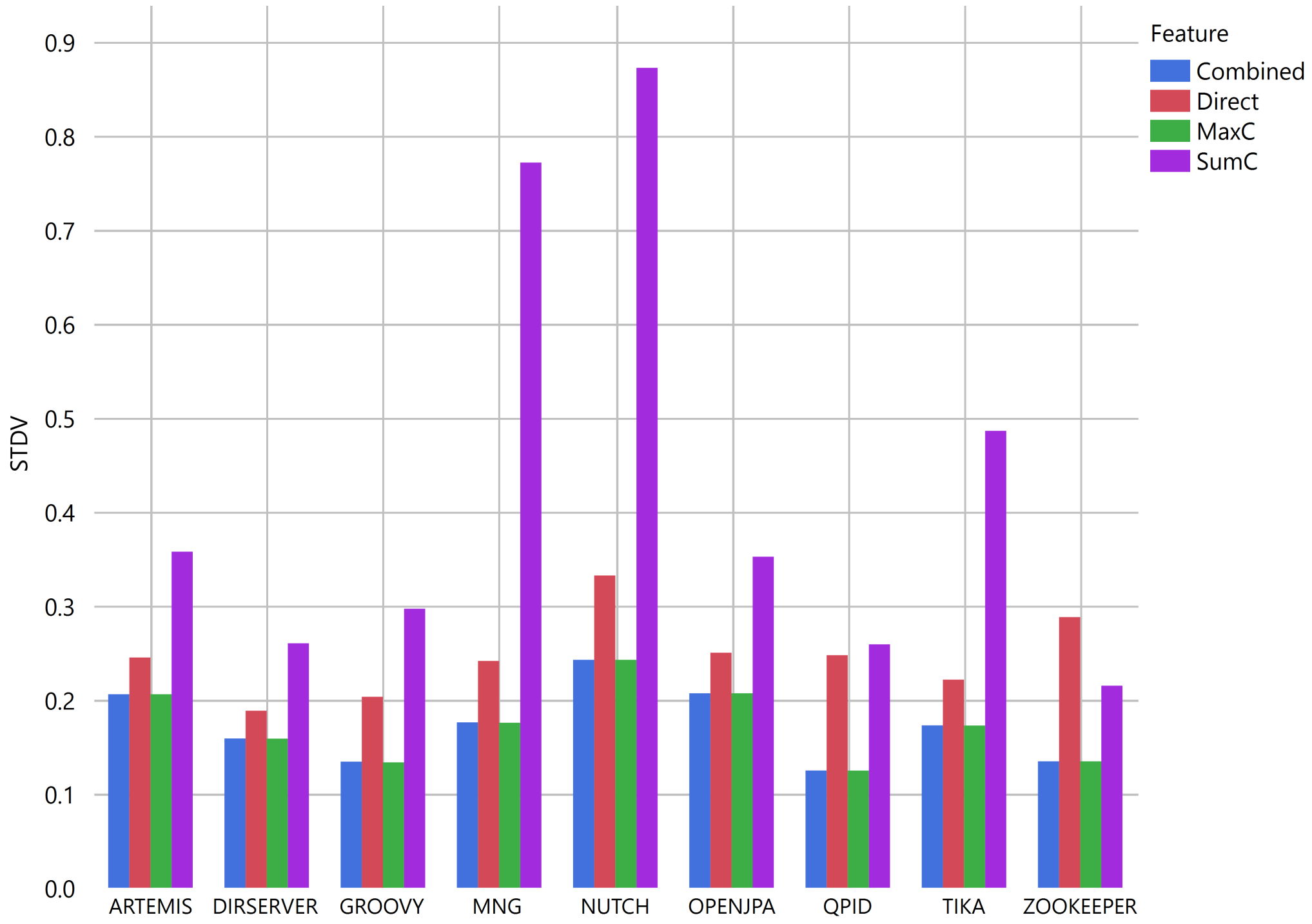

Figure 3.1.32: Distribution across classifiers of standard deviation achieved by different features (colors) for MDP in our projects

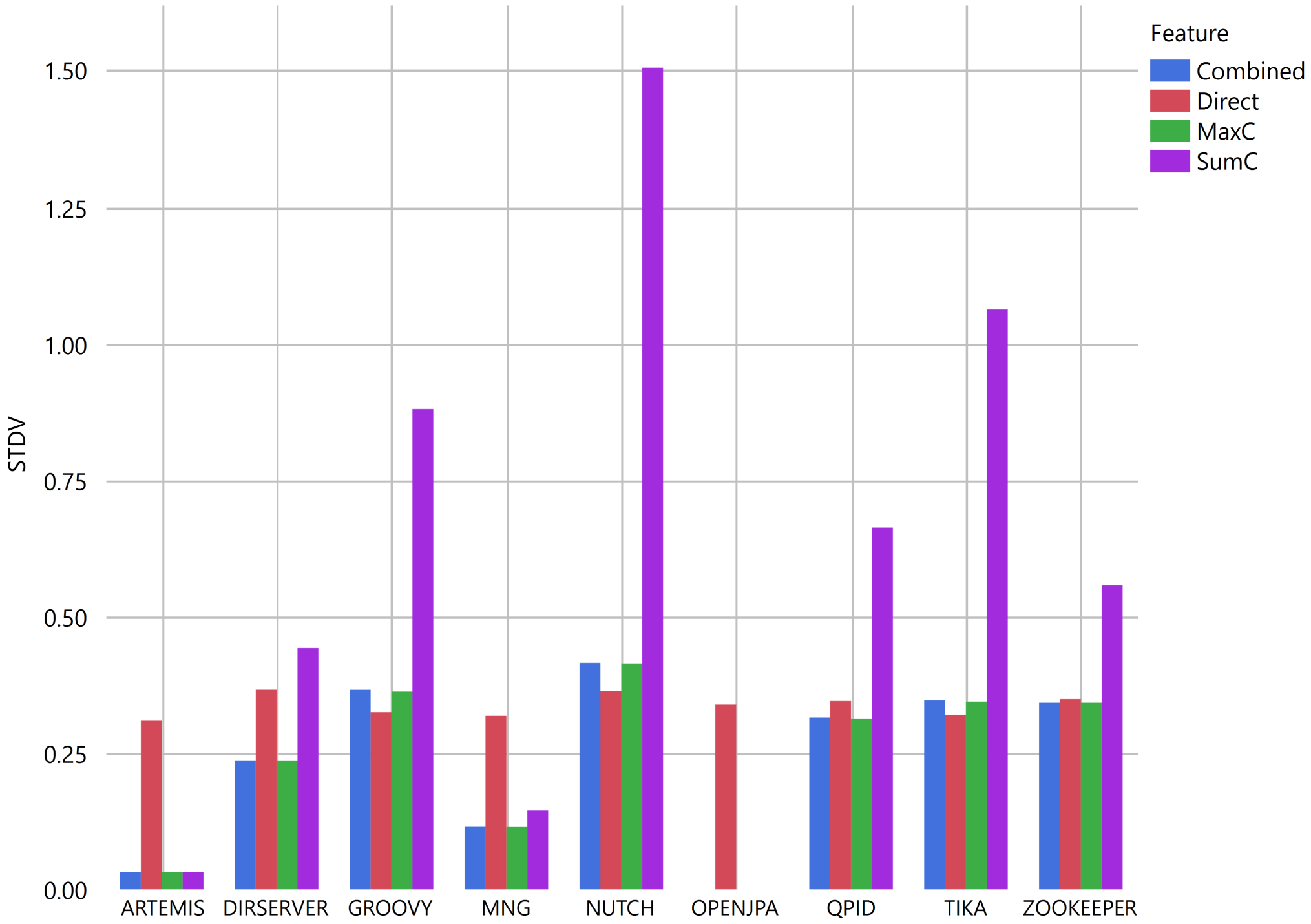

Figure 3.1.33: Distribution across classifiers of standard deviation achieved by different features (colors) for CDP in our projects

Regarding implications to researchers, we found a very limited number of related papers that compared MDP to CDP. For instance, in an ongoing systematic mapping study, we found that, in the last five years, no existing work investigated the accuracy of MDP in three of the major software engineering journals (IEEE Transactions of Software Engineering, ACM Transactions on Software engineering and Methodologies, and Empirical Software Engineering). Hence, there is a high potential for the research community in improving MDP instead of focusing on CDP. Moreover, given the observed low percent of defective methods touched by defective commits, i.e., the defective commits are only partially defective, there is a high potential in leveraging statement-level-defectiveness [312] for MDP.

#### 3.1.3.4 Threats To Validity

In this section, we report the threats to validity of our study. The section is organized by threat type, i.e., Conclusion, Internal, Construct, and External.

**Conclusion**

Conclusion validity concerns issues that affect the ability to draw accurate conclusions regarding the observed relationships between the independent and dependent variables [413].

We tested all hypotheses with non-parametric tests (e.g., Kruskal–Wallis) which are prone to type-2 error, i.e., not rejecting a false hypothesis. We have been able to reject the hypotheses in most of the cases; therefore, the likelihood of a type-2 error is low. Moreover, the alternative would have been using parametric tests (e.g., ANOVA) which are prone to type-1 error, i.e., rejecting a true hypothesis, which, in our context, is less desirable than type-2 error. Also, we acknowledge that our proposed method (i.e., median) to combine JIT with MDP and CDP is a simple and effective baseline to start with (as demonstrated by our results).

There is in the literature a gap between defect prediction accuracy and its actual value in software quality. To face this threat we adopted nine effort-aware metrics; these metrics relate to the effort savings provided by adopting defect prediction models.

Since AUC is sensitive to imbalanced data, and since our datasets are highly unbalanced, AUC results must be interpreted with care. The remaining nine performance metrics are insensitive to imbalanced data; thus, our overall results are not impacted by the imbalanced nature of our data."

**Internal**

Internal validity is concerned with influences that can affect the independent variables with respect to causality [413]. A threat to *internal validity* is the lack of ground truth for commits, methods and class defectiveness. In other words, the used RA-SZZ is not perfectly accurate. Nevertheless, we would argue that this is a common threat in most of empirical research in the area of software engineering [190]. Moreover, to face this threat, we have manually analyzed many commits.

Still regarding the lack of ground truth for commits, methods and class defectiveness, one relevant threat to validity is the possibility that non-linked tickets are related to defect fixes or injections. To face this threat, we tried our best to select projects with the highest linkage. Moreover, we reported the linkage proportion of each project in Table 1 to allow the reader to reason about the validity of the results to specific projects. Finally, we believe that the presence of non-linked commits could have inhibited the observed positive effects of using CDP to support MDP or CDP. Specifically, the inaccuracy in commits labeling, caused by the non-linked commits, reasonably inhibits the accuracy of the commits defectiveness prediction and hence its use.

The execution of a prediction study on defect prediction entails many, often subjective, design decisions such as validation technique, balancing, normalization, tuning, and many more, which might influence the prediction results. We do not expect that our design choices coincide with the choices of all readers, our intent is to use state-of-the-art techniques. We documented all our design choices in Section 2; moreover, we made our replication package available to researchers. Regarding tuning, many studies suggest the tuning of hyperparameters [139, 375]; however, in the present study, we use default hyperparameters due to resource constraints and due to the static time-ordering design of our evaluation. In the future, we plan to evaluate the interaction factor between hyperparameters tuning and the benefits of using JIT for MDT and CDP.

In this work, as in many other similar ones [375, 117, 23, 372, 140, 115, 391], we do not differentiate among the severity of defects. If, on one side, the severity of defects is important and practical, on the other side, to the best of our knowledge, there is no study suggesting that the severity of defects impacts defect prediction accuracy. Therefore, in the future, we plan to extend this work by analyzing the sensitivity of the current results to the severity or priority of the considered defects.

### Construct

Construct validity is concerned with the degree to which our measurements indeed reflect what we claim to measure [413].

In order to avoid that dormant defects would impact our ground-truth, we neglected the last 90% of the releases. This provides us the confidence that snoring is only about 1% in our datasets [115, 5].

Our results could also be impacted by our specific design choices including classifiers, features, and accuracy metrics. In order to face these threats, we based our choice on past studies.

One relevant threat to validity is the possibility that non-linked tickets are related to defect fixes or injection; this would bias our ground truth. To face this threat, we tried our best to select projects with the highest linkage. Moreover, we reported the linkage proportion of each project in Table 1 to allow the reader to reason about the validity of the results to specific projects. Finally, we believe that the presence of non-linked commits could have inhibited the observed positive effects of using CDP to support MDP or CDP. Specifically, the inaccuracy in commits labeling, caused by the non-linked commits, reasonably inhibits the accuracy of the commits defectiveness prediction and hence its use.

### External

External validity is concerned with the extent to which the research elements (subjects, artifacts, etc.) are representative of actual elements [413].

This study used a large set of datasets and, hence, could be deemed of high generalization compared to similar studies. Of course, our results cannot be generalized by projects that would significantly differ from the settings used in this present study. Moreover, since we focused on open-source projects, due to their high availability, we recommend care in generalizing these findings to industrial projects. Please note that considering only mature projects is a threat to external generalizability as results might not generalize to immature projects. One could argue that since only nine nontrivial Apache projects have issue linkage rate above 50%, then no other Apache project might reasonably benefit from the proposed method. We note that the linkage impacts the labeling mechanism, which might have no impact on our results. In practice, our independent variables are orthogonal to the labeling mechanism. In other words, if in the future we will be able to label the defectiveness of commits, classes, and methods according to a linkage-agnostic mechanism, then we believe that our results will still reasonably hold.

Finally, in order to promote reproducible research, all datasets, results, and scripts for this paper are available in our replication package [13].

#### 3.1.3.5 Conclusion

In this study, we: (i) compare methods and classes in terms of defectiveness; (ii) compare methods and classes in terms of accuracy in defectiveness prediction; (iii) propose and evaluate a first and simple approach that leverages JIT information to increase MDP and (iv) CDP accuracy.

Our analysis features two types of accuracy metrics (threshold-independent and effort-aware) and feature selection metrics, nine machine learning defect prediction classifiers, 1,860 defects related to 35 releases of 9 open source projects from the Apache ecosystem. Our results rely on a ground truth featuring a total of 269,004 data points and 46 features among commits, methods and classes.

Our results reveal that:

[13] 5

- MDP is significantly more accurate than CDP (+5% AUC and 62% PofB10). Thus, it is better to predict and rank defective methods instead of than defective classes from a practitioner's perspective. From a researcher's perspective, given the scarce number of MDP studies, there is a high potential for improving MDP accuracy.

- Leveraging JIT by using a simple median approach increases the accuracy of MDP by an average of 17% in AUC and 46% in PofB10 and increases the accuracy of CDP by an average of 28% in AUC and 31% in PofB20. However, in a few cases, leveraging JIT decreased the accuracy of MDP and CDP.

- Since many defective commits were only partially defective, only a small percent of methods touched by defective commits were actually defective. Therefore, we expect that leveraging statement-defectiveness-prediction [312] would better enhance MDP than JIT.

In conclusion, from a practitioner's perspective, it is better to predict and rank defective methods than defective classes. From a researcher's perspective, there is a high potential for leveraging statement-defectiveness-prediction (SDP) to aid MDP and CDP.

In the future we plan to:

- Propose and evaluate new approaches to improve MDP by leveraging JIT. Specifically, instead of using a static approach like median, we could use a machine learning approach to combine MDP with JIT information.

- Use smell information [134] to support MDP as suggested by previous works [198, 290, 369, 291, 369].

- Leverage statement level defect prediction [312] to augment MDP and CDP.

- Investigate whether dormant defects [5, 115] or other types of noise in the datasets [66, 373, 166, 321, 42] have more impact on MDP or CDP.

- Replicate the approach in the context of dependence [75], performance [65] or security [429] defects.

- Using multi-level features in a single prediction model. While in this work we evaluated the benefits of combining two predictions, e.g., commits with methods, in the future, we plan to investigate the benefits of performing a single prediction that uses features at different levels, i.e., features at commits and methods levels).

## 3.2 Tools and Techniques for Vulnerability Detection and Prediction

---

### 3.2.1 An Extensive Comparison of Static Application Security Testing Tools

**Context:** Static Application Security Testing Tools (SASTTs) identify software vulnerabilities to support the security and reliability of software applications. Interestingly, several studies have suggested that alternative solutions may be more effective than SASTTs due to their tendency to generate false alarms, commonly referred to as low Precision. **Aim:** We aim to comprehensively evaluate SASTTs, setting a reliable benchmark for assessing and finding gaps in vulnerability identification mechanisms based on SASTTs or alternatives. **Method:** Our SASTTs evaluation is based on a controlled, though synthetic, Java codebase, it involves an assessment of 1.5 million test executions, and it features innovative methodological features such as effort-aware accuracy metrics and method-level analysis. **Results:** Our findings reveal that SASTTs detect a tiny range of vulnerabilities. In contrast to prevailing wisdom, SASTTs exhibit high Precision while falling short in Recall. **Conclusions:** The paper suggests that enhancing Recall, alongside expanding the spectrum of detected vulnerability types, should be the primary focus for improving SASTTs or alternative approaches, such as machine learning-based vulnerability identification solutions.

#### 3.2.1.1 Methodology

The following section outlines our design decisions and provides the rationale behind them to address our research questions. We detail the tools employed, our measurement process, and the statistical tests conducted.

**RQ 1.** ***What is SASTTs coverage?***

Given the substantial number of available SASTTs and CWEs, it becomes crucial to determine which SASTT accurately detects particular CWEs and which ones go undetected.

We define coverage as the percentage of CWEs identified, at least once, by one or more SASTTs. To understand the expected coverage, we assess the SASTT's documentation claims regarding its ability to identify specific CWEs. To gauge the actual coverage, we evaluate the SASTT's performance in identifying specific CWEs. A SASTT may claim or actually identify a CWE. Therefore, a combination of SASTT and CWE can fall into one of the following statuses:

- *Not Expected*: The SASTT does not claim, in the documentation, to be able to identify the CWE.
- *Expected (**ECWE**)*: The SASTT claims, in the documentation, to identify the CWE. If expected, the combination of SASTT and CWE can be in two sub-statuses:
  - *Actual (**ACWE**)*: The SASTT identifies the CWE in at least one test case; i.e., there is at least one true positive.
  - *Not Actual*: The SASTT identifies the CWE in no test case; i.e., there is no true positive.

To analyze the distribution of ECWEs and ACWEs, we employ UpSet graphs [79]. UpSet is a graphical visualization technique describing the intersection of datasets [218]. The dimension of a rectangle represents the cardinality of the set. In our context, a set is the specific group of CWEs expected or actually identified by a specific group of SASTTs.

We also investigate how the number of unique ECWEs or ACWEs increases with the number of SASTTs. This helps us understand how many SASTTs are worth using if we want to cover a specific number of CWEs.

**RQ 2.** ***Are SASTTs accurate?***

Since multiple SASTTs can identify the same CWEs, it is important to understand the accuracy of SASTTs.

Our dependent variable is the accuracy of SASTTs measured by four accuracy metrics. Three of our accuracy metrics are standard in information retrieval:

- **Precision**: $\frac{TP}{TP+FP}$.
- **Recall**: $\frac{TP}{TP+FN}$.
- **F1-score**: $\frac{2*Precision*Recall}{Precision+Recall}$.

We adopt one effort-aware accuracy metric: NPofB20 [52]. NPofBX is the percentage of defective entities discovered by examining the first x% of the code base, ranked by their probabilities, to be vulnerable, divided by their size, i.e., LOC. For example, a method having an NPofB20 of 30% means that, following the ranking, we can detect 30% of vulnerable entities by examining only 20% of the codebase. Thus, NPofB20 is the percentage of defective entities discovered by examining the first 20% of the code base, ranked by their probabilities, to be vulnerable, divided by their size.

In this paper, we posit the following null hypotheses:

1. H01: There is no difference in the accuracy of SASTTs between ECWEs and ACWEs.

2. H02: There is no difference in the accuracy of SASTTs.

3. H03: The best SASTT across CWEs has the same accuracy as other SASTTs.

For each SASTT, we first compare the accuracy between ECWEs and ACWEs. This is done to understand how reliable past comparisons of SASTTs focusing on ECWEs rather than ACWEs are. Afterwards, to compare the accuracy of SASTTs, we consider ACWEs only.

Since different SASTTs observe different CWEs, a standard comparison of SASTTs based on the accuracy of their ACWEs can be biased by the characteristics of their ACWEs, such as the number of test cases and intrinsic difficulty level. To mitigate the impact of the cardinality of test cases on our SASTTs comparison, we analyze the accuracy per CWE and the weighted average accuracy across CWEs.

We test H01 and H02 via the Wilcoxon signed-rank test [408]. The Wilcoxon signed-rank test (**WT**) is a non-parametric statistical test that compares two related samples or paired data. WT uses the absolute difference between the two observations to classify and then compare the sum of the positive and negative differences. The test statistic is the lowest of both. We select WT to test H01 and H02 because the four SASTTs' accuracy metrics resulted as not normally distributed; hence, we use the Wilcoxon test instead of the paired t-test, which assumes a normal data distribution.

We set our alpha to 0.01. We decided to reduce alpha from the standard value of 0.05 due to the many statistical tests we perform and hence to reach a good balance between type 1 and type 2 error [199].

To test H03, we use Dunn's all-pairs test [99]. The Dunn All-Pairs Test, also known as the Dunn Test (**DT**) or Dunn-Bonferroni Test, is a post hoc test used for statistical analysis. It is a non-parametric test that compares the differences between all group pairs in a dataset. DT compare the rank sums of all possible group pairs and obtains a p-value set for each paired comparison. These p values are adjusted with Bonferroni corrections to mitigate type I errors. The adjusted p-values are compared to a pre-defined significance level to determine the statistically significant comparison pairs. We select DT to test H03 because the four SASTTs' accuracy metrics are not normally distributed, and unlike WT, DT is a pair-wise test.

#### The Dataset

To evaluate SASTTs, we use the Java version of the JTS V. 1.3, its purpose is to evaluate the capabilities of SASTTs specifically. The version used in the present paper includes 112 CWEs. Each CWE has a different number of test cases; each test case is a method in Java that is defined to contain or not contain a specific CWE. Specifically, each CWE has a set of Java classes; each Java class has multiple methods implementing the same logic. Each test case is named "goodXY" if it implements logic Y without CWE X or "badXY" if it implements logic Y with CWE X. The total number of test cases in JTS is 176,451. To illustrate the concepts above, Listing 3.2.2 presents a simple example of a "bad" method for CWE 209, and Listing 3.2.1 presents a simple example of the "good" method. CWE-209, i.e., Generation of Error Message Containing Sensitive Information, refers to a security weakness where an application reveals sensitive information through error messages displayed to users or logged in a file. When an application displays an error message, it should provide only the necessary information to help users understand the cause of the error and how to resolve it. However, suppose an error message contains sensitive information, such as passwords, private keys, stack traces or other confidential data. In that case, attackers can exploit it to grasp the info of source files.

Listing 3.2.1: Example of a method affected by CWE 209.

```
public void bad() throws Throwable {
    try
    {
        throw new UnsupportedOperationException();
    }
    catch (UnsupportedOperationException exception)
    {
        // FLAW: Print stack trace to console on error
        exception.printStackTrace();
    }
}
```

Listing 3.2.2: Example of a method not affected by CWE 209.

```
private void good1() throws Throwable{
    try
    {
        throw new UnsupportedOperationException();
```

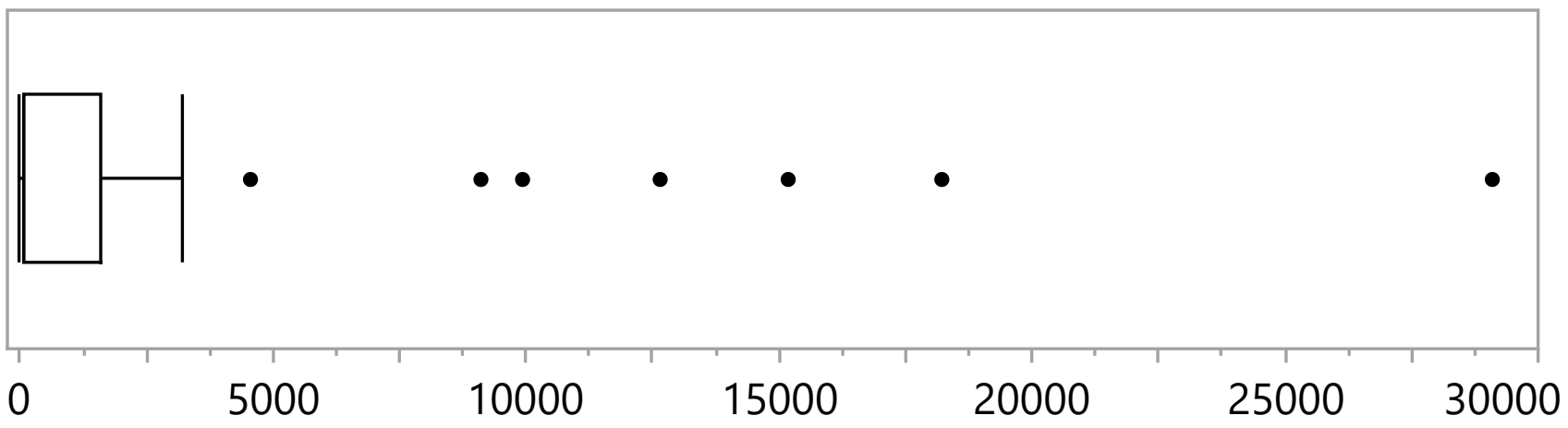

Figure 3.2.1: Distribution of number of test cases across CWEs.

```
    }
    catch (UnsupportedOperationException exception)
    {
        // FIX: print a generic message
        IO.writeLine("Unsupported operation error");
    }
}
```

Figure 3.2.1 reports the distribution of the number of test cases across CWEs in JTS. According to Figure 3.2.1, the number of test cases for CWE ranges within 1 (CWE 111) and 29,095 (CWE 190), with an average value of 1.250 across CWEs.

#### Static Application Security Testing Tools

In this paper we analyze eight SASTTs: SpotBugs 4.6, FindSecBugs 1.12, VCG 2.2, PMD 6.44, SonarQube 9.5, Infer 1.1, Jlint 3.1.2, and Snyk 1.1073. For further details about the SASTTs, please see Section 2.3.

#### Measurement Procedure

Figure 3.2.2 presents our measurement procedure. We set up the environment requested by each SASTT. We then performed the measurement process in five steps: Since SASTTs report violations without explicitly referring to specific CWEs, in *Step 1* we manually map each SASTT reported violation to its specific CWE developing the Tool Mapping. In *Step 2* we scan the JTS with each SASTT which produces the Juliet Test Suite Results. Afterwards, in *Step 3* we develop the Parsed Output by the Juliet Test Suite Results with the Tool Mapping. Thereafter, in *Step 4* we measure the accuracy of the Parsed Output by comparing the results against the JTS documentation [342] obtaining the *Juliet accuracy report.* Subsequently, in *Step 5* we interpret the SASTTs results according to Algorithm 1 to get the *confusion matrix.* Specifically, we computed the confusion matrix elements as follows:

- *True Positive*: the test case is defined as affected by CWE and the SASTT identifies it.
- *True Negative*: the test case is defined as not affected by CWE and the SASTT doesn't identify it.
- *False Positive*: the test case is defined as not affected by CWE and the SASTT identifies it.
- *False Negative*: the test case is defined as having a CWE and the SASTT doesn't identify it.

Afterwards, we identify the ACWEs for a SASTT (see Section 3.1) as the ECWEs having at least a true positive for the SASTT.

#### 3.2.1.2 Results

This section aims to answer the research questions and test the hypotheses posed in the design section.

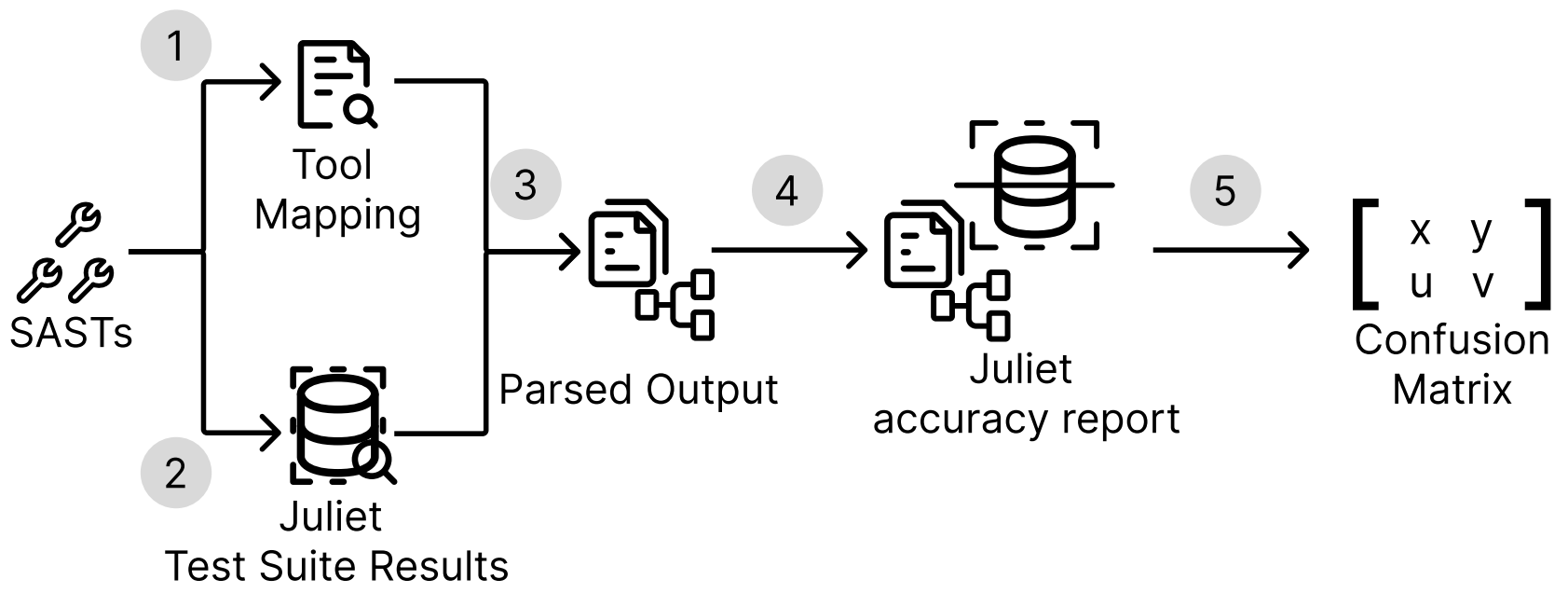

Figure 3.2.2: Workflow of the measurement procedure.

**Algorithm 1** Confusion matrix value.

```
if FileName$MethodName contains 'bad' then
    if cwe_predicted == null || cwe_predicted ≠ cwe_id then
        predicted ← no
        output False Negative
    if cwe_predicted == cwe_id then
        predicted ← yes
        output True Positive
else
    if cwe_predicted == null || cwe_predicted ≠ cwe_id then
        predicted ← no
        output True Negative
    if cwe_predicted == cwe_id) then
        predicted ← yes
        output False Positive
if cwe_predicted == (?) then
    predicted ← ?
    output ?
```

Table 3.2.1: Number of ECWEs and ACWEs per SASTT.

| | **T1** | **T2** | **T3** | **T4** | **T5** | **T6** | **T7** | **T8** |
|---|---|---|---|---|---|---|---|---|
| **Expected** | 58 | 19 | 7 | 6 | 26 | 18 | 8 | 4 |
| **Actual** | 9 | 11 | 7 | 6 | 12 | 12 | 6 | 3 |

**RQ 1.** ***What is SASTTs coverage?***

We investigate the SASTTs coverage in terms of ECWEs and ACWEs. Table 3.2.1 presents the distribution of ECWEs over our eight SASTTs. According to Table 3.2.1, the number of ECWEs and ACWES varies between four (T8) and 58 (T1). Moreover, we note that the total number of non-unique ECWEs across our eight SASTTs is 146. Table 3.2.1 presents the number of ACWEs and ECWEs per SASTT. According to Table 3.2.1, the ACWEs are less than half of the ECWEs in total and for most of SASTTs. Moreover, two SASTTs identify all ECWEs, i.e., ACWEs == ECWEs.

Figure 3.2.3 presents the UpSet intersection graph of ACWEs across SASTTs. There is no CWE that is identified by more than four SASTTs. Moreover, six out of eight SASTTs have at least one exclusive ACWE. According to Figure 3.2.3, T1 has only two exclusive ACWEs, i.e., only 10% of its exclusive ECWEs. Furthermore, T5 has the highest number of exclusive ACWEs, i.e., 9. Finally, it is interesting that most ACWEs relate to only a single SASTTs.

Table 3.2.2 presents how the number of ECWEs and ACWEs increases by increasing the number of SASTTs. Table 3.2.2 shows that the actual and expected coverage does not increase over six SASTS. Specifically, our set of eight SASTTs expects to identify 70% of CWEs in JTS but, according to our results, they actually identify only 34% (see Table 3.2.2, column 6). Thus, most of the CWEs are expected but not actually identified and two-thirds of CWEs are not identified by any SASTT. Finally, we note that a single SASTT covers a maximum of 12 ACWEs, i.e., 11% of JTS, and 58 ECWEs (i.e., 52% of JTS).

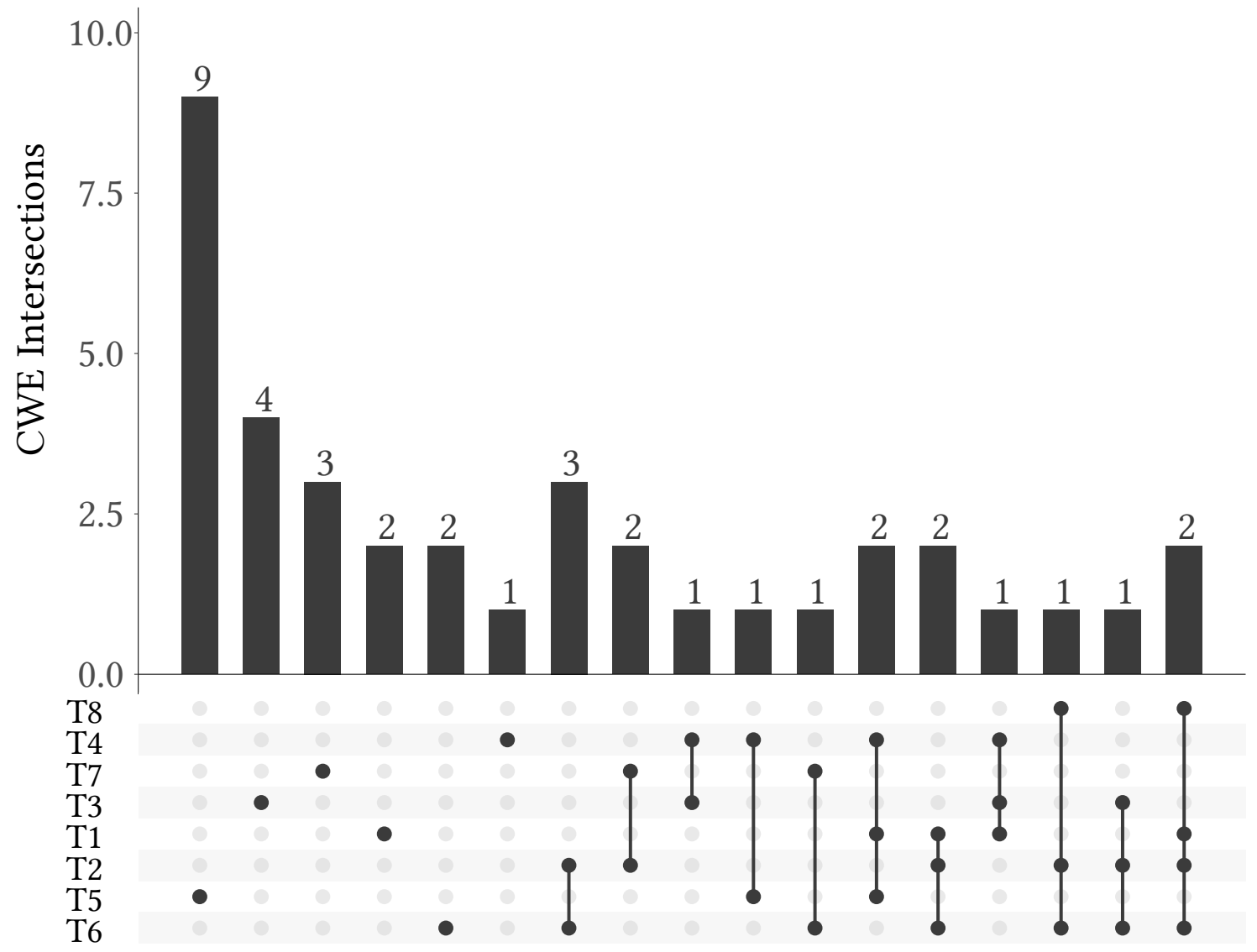

Figure 3.2.3: UpSet graph on the ACWEs per SASTT.

Table 3.2.2: Max unique ACWEs and CCWEs by SASTT and JTS Coverage (%).

| # SASTTs | 1 | 2 | 3 | 4 | 5 | 6 | 7 | 8 |
|---|---|---|---|---|---|---|---|---|
| ECWEs | 58 (52%) | 68 (60%) | 72 (64%) | 76 (67%) | 78 (69%) | 79 (70%) | 79 (70%) | 79 (70%) |
| ACWEs | 12 (11%) | 24 (21%) | 30 (27%) | 35 (31%) | 37 (33%) | 38 (34%) | 38 (34%) | 38 (34%) |

**RQ 2.** ***Are SASTTs accurate?***

Regarding the accuracy of ECWEs versus ACWEs, Figure 3.2.4 provides the distributions across ECWEs versus ACWEs, of each SASTTs accuracy metric. According to Figure 3.2.4, the difference between ECWEs and ACWEs is more evident in Recall than Precision. Specifically, the median Recall of ACWEs is higher than ECWEs in all SASTTs. Interestingly, T1 shows zero Recall for ECWEs and a positive Recall for ACWEs. In six out of eight SASTTs, ACWEs exhibit higher F1 scores than ECWEs. There is no observable difference in NPofB20 between ACWEs and ECWEs.

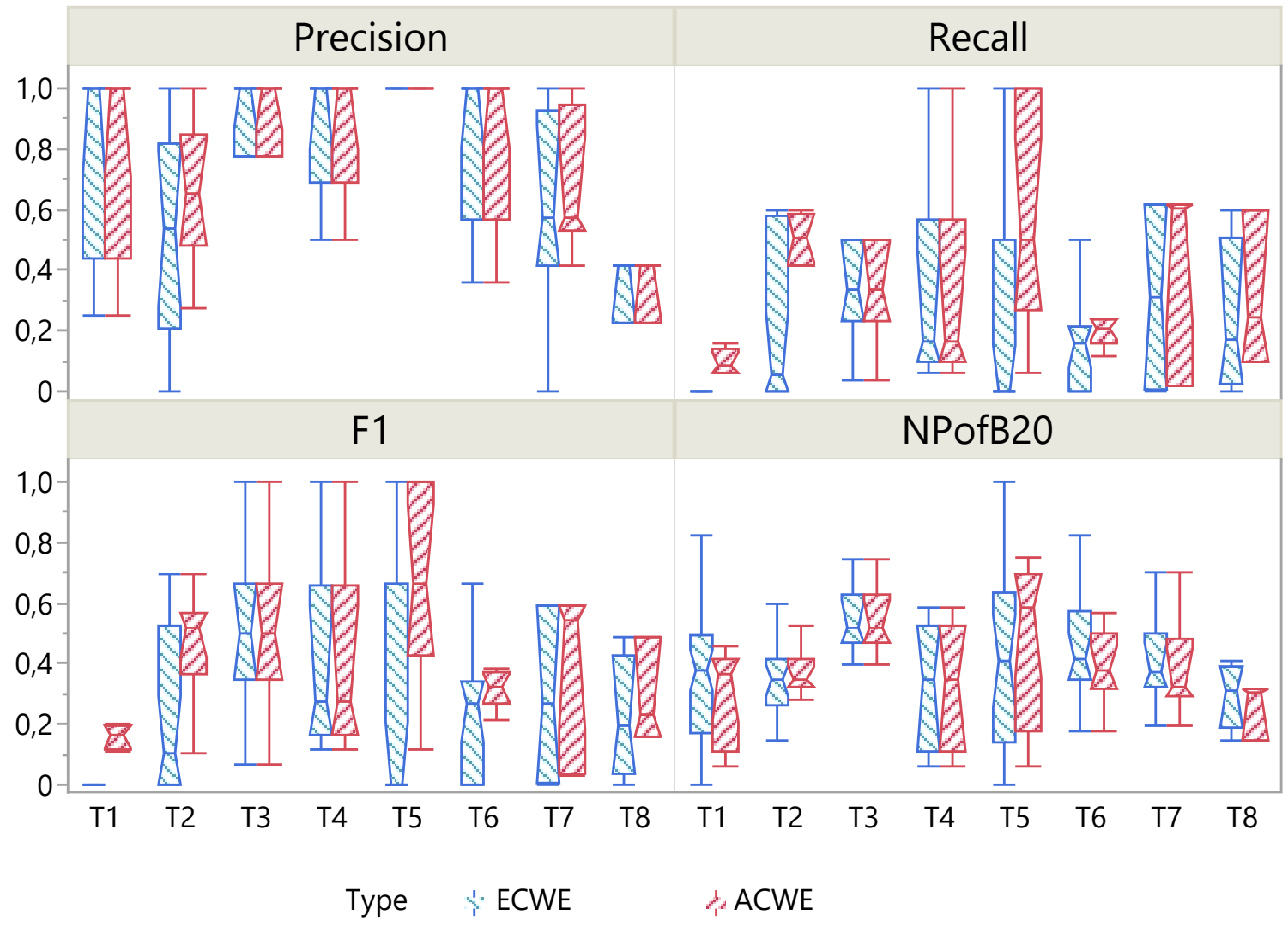

Figure 3.2.4: Distribution of SASTTs accuracy between ECWEs and ACWEs.

Table 3.2.3 reports the statistical test results about the differences, and the gain, in our four accuracy metrics, between ECWEs and ACWEs of SASTTs. We report the gain as statistically significant (i.e., p-value $< 0.0001$) with an asterisk (*). According to Table 3.2.3, we can reject H01 in 15 out of 32 cases. Moreover, we can reject H01 for all T7 four accuracy metrics. Considering only the ACWEs rather than all ECWEs significantly increases the Recall of most SASTTs; in particular, the Recall increases by more than 150% for three SASTTs (T1, T5, and T7). Interestingly, in four out of eight SASTTS, NPofB20 exhibits a negative and statistically significant variation. Consequently, ACWEs are ordered in a manner unfavorable to NPofB20.

Regarding the accuracy of SASTTs, Figure 3.2.5 reports the accuracy met-

Table 3.2.3: Accuracy gain of SASTTs on ACWEs over ECWEs, i.e., H01 test results. $*$ = p-value $< 0.001$

| | **T1** | **T2** | **T3** | **T4** | **T5** | **T6** | **T7** | **T8** |
|---|---|---|---|---|---|---|---|---|
| **Precision** | 0% | 0% | 0% | 0% | 0% | 0% | 162% $*$ | 0% |
| **Recall** | 156% $*$ | 12% $*$ | 0% | 0% | 237% $*$ | 4% $*$ | 501% $*$ | 0% |
| **F1** | 156% $*$ | 12% $*$ | 0% | 0% | 237% $*$ | 4% $*$ | 501% $*$ | 0% |
| **NPofB20** | -14% $*$ | -3% $*$ | 0% | 0% | 20% $*$ | -1% | -24% $*$ | 0% |

rics per SASTTs on all the ACWEs. According to Figure 3.2.5, in all eight SASTTs, the median Precision across ACWEs, is higher than the median Recall. Moreover, T3, T4, T5, and T6, show a perfect median Precision; thus, most of the SASTTs provide no false positive. We note that five out of eight SASTTs show a median Recall below 0.5; moreover, 11 out of our 38 ACWEs, i.e., 29%, have a maximum Recall lower than 0.5. In about a third of ACWEs, all SASTTs provide more false negatives than true positives, i.e., the number of undetected vulnerabilities is higher than the detected ones. Finally, regarding NPofB20, by ordering the test cases and inspecting only the first 20% we would discover, on average, at least 30% of positive cases.

Table 3.2.4 presents the weighted average accuracy, across test cases. T5 shows the maximum average across tests and the highest Recall across ACWEs. However, according to Figure 3.2.5, the Recall, F1, and NPofB20 distributions of T5 are very wide, ranging from 0.08 to 1.0. Thus, even the most accurate SASTT in Recall misses many test cases.

Regarding comparing the accuracy across SASTTs, p-values resulted lower than $\alpha$ in all four accuracy metrics, and hence we can reject H02 and claim that SASTTs differ in accuracy.

Regarding how the best SASTT compares to other SASTTs, according to Table 3.2.4, our best SASTT is T5. According to the DT results comparing differences in the accuracy between T5 and other SASTTs, we can reject H03 in 26 out of 28 cases. Thus, in all six metrics, we can claim that T5 is statistically more accurate than all SASTTs, other than T4. More specifically, T5 is not significantly better than T4 in F1.

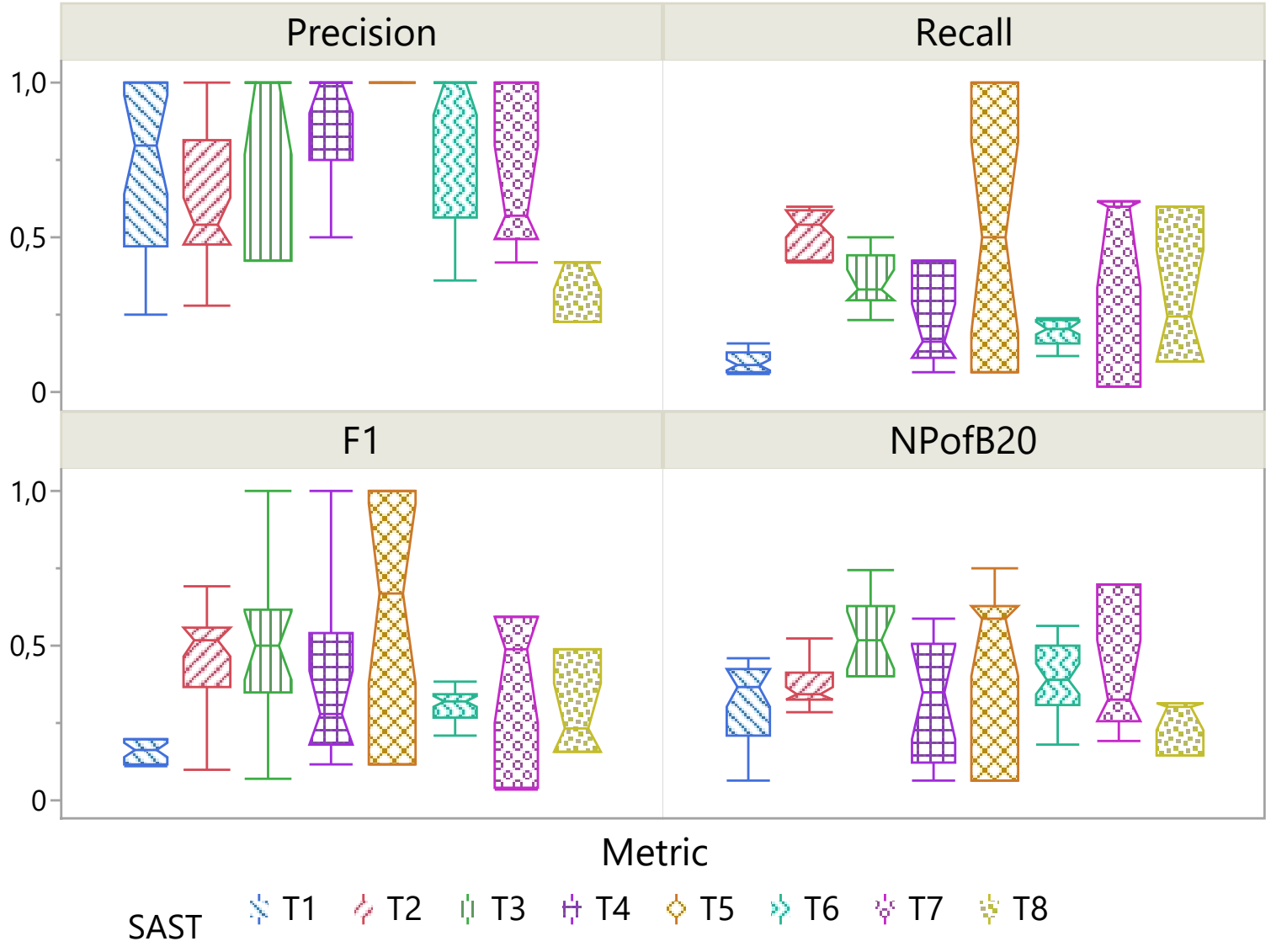

Figure 3.2.5: Distribution of SASTT accuracy on ACWEs.

We also analyzed how often a SASTT is the most accurate across the different combinations of specific ACWEs and accuracy metrics. Table 3.2.5 shows that T5 is the best choice in 54 cases out of 228. Furthermore, T2 is the best option 41 times. Conversely, T8 is the best SASTT, the lowest number of times, i.e., twice. Hence, some SASTTs are the best more times than others; however, all SASTTs are the best at least once.

### 3.2.1.3 Discussion

This section discusses our results by focusing on their impacts on practitioners and researchers.

Table 3.2.4: Distribution of SASTT weighted average accuracy on ACWEs.

| | **T1** | **T2** | **T3** | **T4** | **T5** | **T6** | **T7** | **T8** |
|---|---|---|---|---|---|---|---|---|
| **Precision** | 0,516 | 0,493 | 0,713 | 0,769 | 1,000 | 0,933 | 0,636 | 0,267 |
| **Recall** | 0,130 | 0,518 | 0,124 | 0,421 | 0,579 | 0,174 | 0,440 | 0,256 |
| **F1** | 0,197 | 0,475 | 0,172 | 0,532 | 0,699 | 0,282 | 0,404 | 0,248 |
| **NPofB20** | 0,388 | 0,361 | 0,510 | 0,482 | 0,565 | 0,457 | 0,340 | 0,183 |

Table 3.2.5: Frequency of best SASTT for a specific combination of ACWE and accuracy metric.

| SAST | T1 | T2 | T3 | T4 | T5 | T6 | T7 | T8 |
|---|---|---|---|---|---|---|---|---|
| Frequency | 23 | 41 | 36 | 18 | 54 | 32 | 22 | 2 |

#### Dataset

In Section 3.2.1.1 we show that different CWEs have a different number of test cases and in Section 3.2.1.2, we show that different SASTTs identify different CWEs. Past studies [288, 220, 59, 152, 28, 16, 277], have overlooked the issue of under/over-representation of CWEs in JTS. Thus, we must exercise caution when interpreting or aggregating results from such studies as they used the simple average. In this paper, we provide **recommendation 1: we shall use the weighted average rather than the simple average when evaluating SASTTs over JTS.**

#### RQ 1. *What is SASTTs coverage?*

One of the main results of RQ1 is that SASTTs do not identify most of the CWEs they claim to identify, i.e., ACWEs $<$ 50% of ECWEs. Hence, practitioners should lower their expectations of the documented capabilities of SASTTs. In the same vein, Li et al. [220] highlighted the differences between the advertised capabilities of SASTTs and their actual performance. We provide **recommendation 2: we shall trust SASTTs performances in empirical results** **rather than in SASTTs documentation.**

Another significant result of RQ1 is that most CWEs are identified by no more than one SASTT, so for most of the CWEs, there is no question of which SASTT to use if we want to identify it. However, our results show that a single SASTTs can identify only 11% of CWEs. This result aligns with Oyetoyan et al. [288] as they emphasize the significance of avoiding reliance on a single SASTT. Oyetoyan et al. [288] highlight the limitations of using a single SASTT, as it cannot comprehensively cover the vast array of CWEs or encompass all potential patterns associated with a particular CWE. Thus, we provide **recommendation 3: we shall use multiple SASTTs to identify many vulnerability types.**

Our set of eight SASTTs can identify only 34% of CWEs in JTS. Similarly, Oyetoyan et al. [288] conducted a qualitative study wherein they interviewed eight developers from a security research team that specifically focused on the analysis of SASTTs inclusion within agile processes. They highlighted the complementary role of techniques such as code inspection, which can provide a broader perspective on the overall state of the codebase. Thus, we provide **recommendation 4: we shall complement the use of SASTTs with other vulnerability identification techniques such as code inspection**. Moreover, **recommendation 5: when improving SASTTs, or providing alternative techniques for vulnerability identification, such as the ones based on ML, we shall focus on increasing the number of identified vulnerability types.**

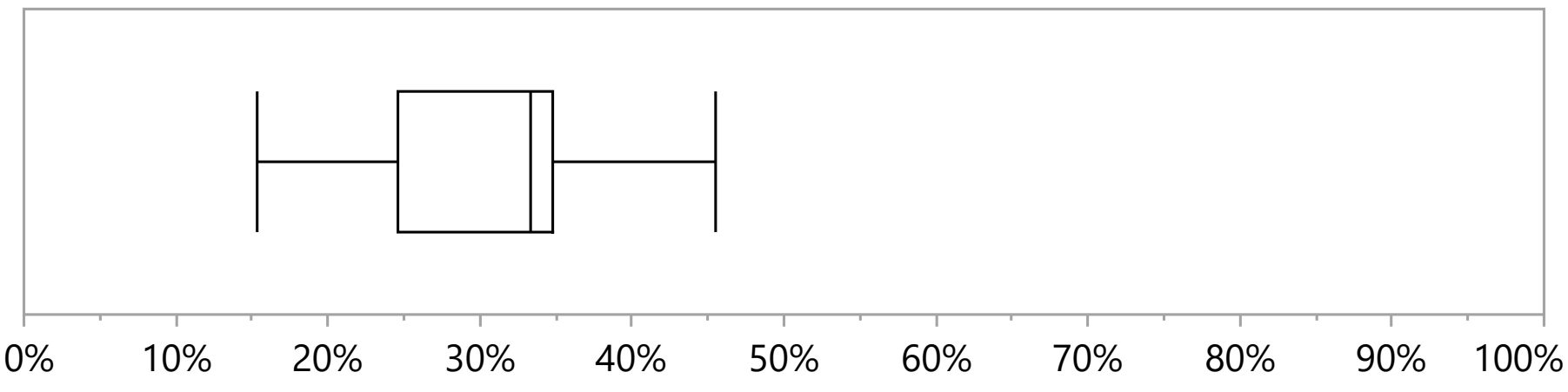

Figure 3.2.6: Distribution of the proportion of positive test cases across CWEs.

**RQ 2.** ***Are SASTTs accurate?***

Regarding which SASTTs to use when there are several options, our results show that T5 is a dominant SASTT in accuracy and in ACWEs i.e., 12. Still, about which SASTTs to use, our results show that all eight SASTTs were best in at least one CWE-accuracy metric combination. Hence, we can infer that all SASTTs have a reason to be used, though some SASTTs have more reasons than others. Thus, we provide **recommendation 6: we shall choose the SASTTs to use depending on the CWEs and accuracy aspect we care about.**

One of the main results of RQ2 is that the accuracy of SASTTs computed on the ACWEs is much higher than on the ECWEs, especially for three SASTTs where Recall increases by over 150%. Thus, we provide **recommendation 7: we shall compute SASTTs accuracy on ACWEs rather than ECWEs or CWEs.** Moreover, **recommendation 8: we shall exercise caution when generalising the results of previous work since they rely on ECWEs or CWEs.**

A further significant result of RQ2 is that SASTTs show a pretty high Precision and a low Recall. Specifically, half of the SASTTs have perfect Precision, i.e., they provide no false positives, aka no false alarms, and most SASTTs have a median Recall below 0.5, i.e., they provide more false negatives than true positives. Hence, our study reveals that the Achille's heel of SASTTs is false negatives rather than false positives. We debunked the common myth of high false positive rate of SASTTs promoted in many past studies [288, 220, 59, 152, 28, 16, 277, 245]. Moreover, Li et al. [220] shares a common focus with numerous prior researchers [28, 16]: the examination of false positives provided by SASTTs; we reveal that this is not the most important challenge in using SASTTs. In conclusion, we provide **recommendation 9: when improving SASTTs or providing alternative techniques for vulnerability identification, such as the ones based on ML, we shall focus on increasing Recall over Precision of actual SASTTs.** Moreover, **recommendation 10: when deciding what to test with multiple SASTTs, we shall focus on code fragments identified as not vulnerable since the ones identified as vulnerable are likely so.**

#### 3.2.1.4 Threats to Validity

In this section, we report the threats to the validity of our study.

All experiments took about a week on one McAffee sever model BG5500 running Windows Server 2022. The server has two Intel Xeon (R) CPU X5660, a base clock speed of 2.79 GHz and 72.0 GB of RAM with a clock speed of 1067 MHz. To support replicability, we report our datasets and scripts availability in Section 3.2.1.6. Our replication package presents the steps to replicate our study. If a SASTTs is changed or added, the steps should be performed precisely in the same order as described in the replication package. The scripts allow researchers to re-generate data at each specific step.

One possible threat lies in the representativeness of CWEs. Specifically, JTS, although a widely used and respected suite, may have its own biases. It might have been designed with specific vulnerability types or coding practices in mind, potentially leading to overrepresentation or underrepresentation of specific vulnerabilities [136, 233, 404]. Since JTS targets common vulnerabilities in software applications, it may not capture vulnerabilities specific to particular domains or niche applications. Moreover, we may face language-specific biases, i.e., JTS may inadequately cover vulnerabilities specific to programming languages different from Java. Therefore, our results apply to Java code only. Furthermore, JTS is affected by an evolution bias, which might cause it not to keep pace with the evolving vulnerability landscape. New vulnerabilities and exploitation techniques constantly emerge, and if the test suite is not regularly updated, the SASTT may miss detecting these novel vulnerabilities, leading to false negatives.

Since JTS is a synthetic code base, another threat is the representativeness of the test cases in JTS for specific CWEs. Although CWEs are commonly represented by hundreds or thousands of test cases in JTS, such test cases might be trivial or, anyhow, not representative of current development. In general, we note that no experimental artefact or subject is perfect [118], there is always a tradeoff among competing quality attributes like, in our case, representativeness and control, rather than a perfect solution [73]. If, on the one side, the test cases in JTS provide a high level of control, as we are confident about the presence or absence of specific CWE, on the other side, JTS provides a questionable level of realism, as we cannot know how much the test cases are representative of current development. However, realism is a hard quality attribute to fulfil since we note that a code base of a domain or a company cannot be deemed realistic to other domains or companies and would provide a questionable level of control. Thus, we recommend caution in generalizing our results; we plan to complement this study with future investigations on other types of software code.

Unfortunately, the output of SASTT is not standardised, and each SASTT has its format to report vulnerabilities. Moreover, SASTTs do not always report the CWE identifier of a violation. Therefore, as Step 2 of Fig 2 discussed, we manually created scripts to map each SASTT output to CWE identifiers. Misclassification errors may occur while creating this mapping and scripts to process SASTT outputs. The first two authors tested the scripts and mapping multiple times to mitigate this threat. Moreover, we note that even a tool providing a 99% Precision may be useless in practice if the number of analyzed elements are billions.

The selected SASTTs represent a subset of the countless available. To mitigate this threat, we used all SASTTs used in previous studies [288, 220, 59, 152, 28, 16, 277]. A possible construct validity threat regards using a single metric to draw conclusions. We mitigate this problem by evaluating individual SATTs considering many metrics, i.e., Precision, Recall, NPofB20, and F1. According to Algorithm 1, if SASTT finds CWE_X in a test case defined as without CWE_Y, with X different than Y, then this is measured as a true negative. The rationale is that the test case does not have CWE_Y, and the SASTT did not find it. Similarly, if SASTT finds CWE_X in a test case defined as with CWE_Y, with X different than Y, this is measured as a false negative. The rationale is that the test case has CWE_Y and the SASTT did not find it. Despite not being completely intuitive, this aligns with the JTS definition [186].

Given the novelty and impact of this high Precision - low Recall result, we looked into design choices that could have caused this result. We note that using ECWEs is a conservative measure as it somehow removes CWEs when Recall is zero; hence, using ECWEs can increase Recall but cannot impact

Precision. A further possible explanation for the high number of false negatives could have been a high proportion of positives in the test case distribution of JTS. Therefore, we analysed the distribution of positive test cases over ACWEs. Finally, given our peculiar experimental design, our results cannot be directly compared to those of other benchmarks such as OWASP.

According to Figure 3.2.6, all ACWEs have a proportion of positive test cases below 45%, and the average proportion of positive test cases across CWEs is 34%. Thus, a random classification approach that classifies half of JTS as positive and the other as negative would achieve many more false positives than false negatives. Therefore, in light of this information, the low Recall - high Precision result is even more surprising as the characteristic of the adopted dataset biases results in the opposite direction, as does focusing on ECWEs only.

#### 3.2.1.5 Conclusion

We conducted an extensive and in-depth analysis of SASTTs' ability to detect Java code vulnerabilities. This thorough evaluation, based on 1.5 million test executions on a controlled, though synthetic, codebase, sets a reliable benchmark for assessing and enhancing the effectiveness of SASTTs or other alternatives, such as machine-learning-based vulnerability identification solutions.

To summarize our findings, we provide ten key recommendations and highlight numerous SASTTs limitations. These include the overpromising of SASTTs capabilities in their documentation and incomplete coverage of vulnerability types across all SASTTs. This result suggests that multiple SASTTs and other vulnerability identification techniques, such as code inspection or machine-learning-based vulnerability identification solutions, should complement a single SASTT.

The significant limitations of SASTTs identified in this study strongly encourage further research in this area. One important SASSTs limitation is the observed low Recall, which resulted even lower than Precision. In other words, false negatives outnumber false positives. In light of the assumption that a false negative, i.e., the consequences of a not discovered vulnerability in a software application, is more severe than a false positive, i.e., the human effort for inspecting a software application without a vulnerability, we recommend that future advances on vulnerability identification prioritize reducing false negatives rather than false positives.

#### 3.2.1.6 Data Availability

We provide both the raw data. raw accuracy metrics and the scripts required to replicate our research findings on Zenodo (`https://doi.org/10.5281/zenodo.10524245`).

### 3.2.2 VALIDATE: A Deep Dive into Vulnerability Prediction Datasets

**CONTEXT**: Vulnerabilities are an essential issue today, as they cause economic damage to the industry and endanger our daily life by threatening critical national security infrastructures. Vulnerability prediction supports software engineers in preventing the use of vulnerabilities by malicious attackers, thus improving the security and reliability of software. Datasets are vital to vulnerability prediction studies, as machine learning models require a dataset. Dataset creation is time-consuming, error-prone, and difficult to validate.

**OBJECTIVES**: This study aims to characterise the datasets of prediction studies in terms of availability and features. Moreover, to support researchers in finding and sharing datasets, we provide the first VulnerAbiLty predIction DatAseT rEpository (*VALIDATE*).

**METHODS**: We perform a systematic literature review of the datasets of vulnerability prediction studies.

**RESULTS**: Our results show that out of 50 primary studies, only 22 studies (i.e., 38%) provide a reachable dataset. Of these 22 studies, only one study provides a dataset in a stable repository.

**CONCLUSIONS**: Our repository of 31 datasets, 22 reachable plus nine datasets provided by authors via email, supports researchers in finding datasets of interest, hence avoiding reinventing the wheel; this translates into less effort, more reliability, and more reproducibility in dataset creation and use.

#### 3.2.2.1 Systematic Dataset Review

In the present work, we perform the first SDR in the field of VPS [437, 203]. We build our methodology on findings and observations of Croft et al. [83] and Nong et al. [280].

##### Goal and Research Questions

We formalised the goal of this study according to the Goal Question Metric (GQM) approach [34] as follows:

| | |
|---|---|
| *Investigate* | datasets, |
| *for the purpose of* | characterisation, |
| *with respect to* | nine dimensions, |
| *from the point of view of* | researchers, |
| *in the context of* | VPS. |

Based on the aforementioned goal, we defined our main research question, which serves as the primary focus of our investigation: *What are the defining characteristics of state-of-the-art datasets in VPS?*"

### Research Methodology

The SDR consist of two phases. Phase 1 gathers studies using PICO [159]. Phase 2 focuses on filtering the studies according to inclusion and exclusion criteria.

Phase 1 involves the collection of 2802 studies spread across three different sources. Specifically:

- ACM Digital Library: 69 studies, 67 of which coincide with Croft et al. [83],
- IEEEXplorer: 2360 studies, 1118 of which coincide with Croft et al. [83], and
- Scopus: 373 studies, 357 of which coincide with Croft et al. [83].

We use an SLR procedure based on Zhang et al. [437], Kitchenham [203], and Lewowski and Madeyski [217]. Moreover, our primary study selection is inspired by Croft et al. [83]. Specifically, Croft et al. [83] performed a study selection process in February 2021 and obtained 61 studies. Table 3.2.6 describes our PICO enquiry. Croft et al. [83] provides the base of our query to which we add to the population Category the word "repository" and "dataset" to expand and accommodate the different scopes of the literature review. Croft et al. [83] limited their research to studies published until February 2021. Our query extends the selection of studies of Croft et al. [83] to those published up to January 2023.

We derive our groups and dimensions from studies highlighting aspects of ML for VPS. For instance, Falessi et al. [117] point out the importance of preserving the mined data time order. Similarly, several studies approached VPS with different grains [431, 12, 370, 275, 104, 329, 119]; thus, we decided to define a dimension on the predicted entity granularity. Zhang et al. [441] combined software metrics and text features for VPS. We acknowledge this idea of combination, and therefore we decide to focus our attention on different feature sets [398, 231, 309, 221] by defining a dimension that analyses the available feature sets of the datasets. Russo et al. [338] point out the relevance of CVE in VPS [68, 255, 125, 265]; therefore, we focus our attention on the availability of CVE info in the datasets.

Figure 3.2.7 presents our steps in the selection process of studies, i.e., Phase 2. We divide phase 2 into three specific steps:

1. Duplicate Removal: we eliminate duplicates in the three distinct data sources. This resulted in the discarding of 501 studies.

2. Criteria Application: we apply inclusion and exclusion criteria to the remaining 2301 studies, eliminating an additional 2251 studies. Croft et al. [83] proposed a set of criteria in their systematic literature review (SLR). We find Croft et al. [83] inclusion and exclusion criteria to match our aim, so we decided to avoid reinventing the wheel. Table 3.2.7 presents our inclusion and exclusion criteria inspired from [83].

3. Snowball Sampling: we use the snowball sampling technique [129, 412] to incorporate potentially related studies that the query may not have captured. This results in the identification of 8 additional studies.

4. Quality Assessment: Table 3 presents the quality checklist inspired by

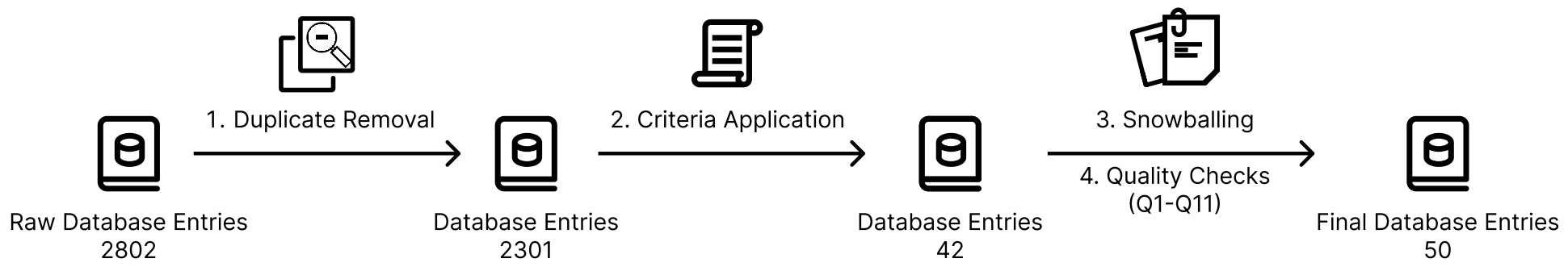

Figure 3.2.7: Workflow for study selection and inclusion.

the established guidelines by Kitchenham [203]. According to Table 3, we assessed whether the original studies' authors clearly stated the design's aims correctly aligned with them. Furthermore, we evaluated if the metrics used in the measurement procedures of the original studies were appropriate for answering the research questions and if the sample represented the specific vulnerability type and granularity. We check whether the authors justified smaller sample sizes and if they thoroughly described the specific tools used in the study. At the end of the fourth step in Figure 1, all the papers selected satisfied questions one to 12. Finally, Figure 3.2.8 graphically presents the last quality checks, i.e., Q12 and Q13 (see Section 3.2.2.1).

The final selection of the studies in this paper consists of 50 studies, including 171 unique projects and seven programming languages.

### Coding

This study systematically characterises datasets by categorising them into 9 dimensions through thematic synthesis [86]. Each of the abovementioned dimensions represents a specific characteristic of the dataset.

Table 3.2.6: PICO Query String

| Category | Subject | Search Terms |
|---|---|---|
| **Population** | Software | “software” OR “code” OR "repository" OR "dataset" |
| **Intervention** | Machine Learning Static Application | “learn” OR “neuralnetwork” OR “artificial intelligence” OR “AI-based” OR “predict” NOT(“fuzz” OR “test” OR “attack” OR “adversarial” OR “malware” OR “description”) |
| **Comparison** | - | - |
| **Outcomes** | Software Vulnerability Prediction | “vulnerability” AND (“predict” OR “detect” OR “classify” O “identify” OR “discover” OR “uncover” OR“locate”) |

Table 3.2.7: Inclusion and exclusion criterias

| | **Inclusion Criteria** |
|---|---|
| I1. | The study relates to the field of VPS, and informs the practice of Software Engineering |
| I2. | The study presents a unique VPS process or evaluation. |
| I3. | The study is a full paper longer than six pages. |

| | **Exclusion Criteria** |
|---|---|
| E1. | Solely a literature review or survey article. |
| E2. | Non peer-reviewed academic literature. |
| E3. | Academic articles other than conference or journal papers, such as book chapters or dissertations. |
| E4. | Studies not written in English. |
| E5. | Studies whose full-text is unavailable. |
| E6. | Studies published to a venue unrelated to the discipline of Computer Science. |
| E7. | Studies published to a journal or conference with a CORE ranking of less than A and H-index less than 40, and that have a citation count of less than 20. |

Table 3.2.8: Quality Checklist

| | |
|---|---|
| **Design** | |
| Q1 | Are the aims clearly stated? |
| Q2 | Was the study designed with these questions in mind? |
| Q3 | Do the study measures allow the questions to be answered? |
| Q4 | Is the sample representative of the population to which the results will generalise? |
| Q5 | Was the sample size justified? |
| Q6 | If the study involves technology assessment, is the technology clearly defined? |
| Q7 | Are the measures used in the study fully defined? |
| Q8 | Are the measures used in the study the most relevant for answering the research questions? |
| Q9 | Is the study's scope (size and length) sufficient to identify changes in the outcomes of interest? |
| **Analysis** | |
| Q10 | Were the basic data adequately described? |
| **Conclusions** | |
| Q11 | Are all study questions answered? |
| **Availability of Datasets** | |
| Q12 | Was the replication package provided in the paper? |
| Q13 | Is the replication package available? |

**RQ 1.** ***What is SASTTs coverage?*** Given the increasing importance of reproducibility and replicability in SE studies, [151, 235, 98, 192, 292], and since the dataset is vital to replicate VPS, in this dimension, we characterise whether VPS provides a dataset. In addition, we want to understand the stability of the storage location used for the dataset. We used Cruzes and Dybå [86] recommendations for deriving these dimensions and their values. Our dimensions aim to support researchers in finding a dataset of interest. The existence of possible overlaps or relations across dimensions does not impact the aim of the dimensions. To assess the availability of the datasets, we read the studies and manually checked for any link or reference to the dataset. Eventually, once we found a reference to a dataset in the form of an URL, we manually followed the reference and downloaded the referenced dataset. We note that the type of storage location of datasets is an important aspect. We classify the location as *Stable Repository* if the authors provide an external reference to the dataset, the dataset can be downloaded, and the dataset is in a remote stable repository, e.g. Zenodo. We classify the location as *Unstable Repository* if the authors provide an external reference to the dataset, the dataset can be downloaded, and the dataset is in a remote unstable repository, for example, GitHub or personal websites. It is important to discriminate between stable and unstable locations because datasets in unstable locations can be deleted or permanently moved. Hence, a study might not be replicable if its dataset is stored in an unstable repository. If the dataset was unavailable or unavailable, we emailed all authors requesting such an unavailable dataset. We waited for the answer for three months and thanked all authors who could provide the dataset. All dimensions are based on datasets available in their studies or retrieved via emails; we call these datasets the gathered datasets (**GD**).

The two authors of the paper acted as coders. We used Microsoft Forms to input the data and then outputted a spreadsheet; this supported smooth individual coding. We used the outputted spreadsheet to compare the answers provided by each coder. In case of disagreement, we discussed the matter thoroughly to understand the reasons and then reached a consensus. We analyse our agreements via Cohen's $\kappa$ [76]. When assigning items to different categories, the kappa statistic is commonly used to evaluate the agreement between raters or classifiers. In the context of VPS datasets, it can be used to measure the level of agreement between two independent authors who categorised the datasets according to different characteristics. When comparing the observed level of agreement to the expected level of agreement, Cohen's $\kappa$ provides a metric for the reliability and consistency of the categorisations. Both authors characterised the dimensions individually. In the event of disagreements, we reached a consensus through discussion. Table 3.2.9 presents the interpretation $\kappa$ as suggested by Cohen [76] and Sim and Wright [359].

Table 3.2.9: Interpretation of $\kappa$ values for measuring agreement

| Value of $\kappa$ | Interpretation |
|---|---|
| $\kappa < 0$ | No agreement |
| $0 \leq \kappa < 0.4$ | Poor agreement |
| $0.4 \leq \kappa < 0.6$ | Discrete agreement |
| $0.6 \leq \kappa < 0.8$ | Good agreement |
| $0.8 \leq \kappa < 1$ | Excellent agreement |

In the following subsections, we present the 9 dimensions along which we characterise the datasets. We note that all of them are relevant to vulnerability prediction, whereas some are only for VPS, e.g., CVE/CWE information. VALIDATE supports researchers in finding the dataset of interest, and it is independent of the specific decision-making approach (see Section 2.4.3).

**Dimension 1:** ***Granularity of the labeled entity*** Different VPS can focus on entities of different granularity such as classes and files [230], commits [211], methods [297], code fragments[225], or machine code [237]. Understanding the granularity is important since Morrison et al. [269], analysing Microsoft products, reports that the granularity of the predicted entity impacts the accuracy and actionability of the prediction model. Le et al. [211] and Chakraborty et al. [56] focus on comparing the granularity of the predicted entity. This dimension aims to characterise the trends in the granularity of predicted entities. Our methodology consists in downloading the dataset and manually inspecting it. We also checked the granularity is defined in the study.

**Dimension 2:** ***Nature of the labelled entity*** The data in the dataset can be collected from open-source projects hosted online through VCS like GitHub or the data can be synthetically created by an authority like NIST [78] with the SARD [43]. Understanding the domain of the data is important since Chakraborty et al. [56] discuss the domain of the labelled entities and their impact on the model's performance. Specifically, synthetic entities might not fully grasp the complexity of real-world projects, thus harming the prediction model's ability to generalise. This dimension aims to characterise the trends in data domain; i.e., synthetic versus open-source. Our methodology consists of finding details of the source of the collected data.

**Dimension 3:** ***Type of labelling*** Multiclass VPS aim to determine which code is affected by a vulnerability and the type of predicted vulnerability. For instance, Zou et al. [452] proposed and evaluated the first deep learning-based system for multiclass vulnerability prediction. This dimension aims to characterise the trends in the type of prediction: binary or multiclass. Our methodology consists in looking for details about the class type of the label.

**Dimension 4:** ***Feature sets availability*** The features are the input to an ML model [135, 336]. Different features aim to achieve different goals [256]. In SE, we can measure many aspects of a software entity, such as software or a development process. Chidamber and Kemerer [71] introduce object-oriented software metrics. Moser et al. [270] introduce software development metrics also known as change metrics that aim to measure the process. Recent studies focused on the interpretation and use of specific metrics for VPS[2] and their possible subsets [397]. This dimension characterises the set features in the dataset. Our methodology consists of analysing the name of the features in the dataset and their definition in the correlated study.

**Dimension 5:** ***Availability of either a CVE or a CWE information*** Mann and Christey [244] introduced the concept of Common Vulnerability Enumeration (CVE), also known as Common Vulnerabilities and Exposures. Martin et al. [246] introduced the concept of Common Weakness Enumeration (CWE), a dictionary of publicly known vulnerabilities and security hotspots maintained by The MITRE Corporation [262]. The CWE aims to classify a specific vulnerability that belongs to the release of a project maintained by a specific vendor with a unique identification number. CVE is fundamental for new vulnerability studies [422]. Therefore, the research community, following this new trend, made available large-scale datasets of CVE Infos [326, 416, 122, 309] and tools to automatically retrieve CVE [41]. This dimension characterises the trend in the availability of CVE information. Our methodology involves looking for details about the availability of CVE info in the collected data.

**Dimension 6:** ***Labelling Process*** Labelling the data is crucial for VPS [83]. Due to automatism or not considering real-world scenarios, incorrect labelling can lead to model problems [182]. For example, Tantithamthavorn et al. [373], Eilers et al. [101], and Falessi et al. [115] underlined the impact of mislabeled data on the performance of ML model. Therefore, the researchers developed tools and techniques [306, 343] to cure software repositories and made manually curated data repositories available for SVP[309, 306, 343, 406]. This dimension aims to investigate whether the authors manually curated the dataset. A dataset is manually curated if researchers checked at least one aspect during the labelling process; the next dimension focuses on the specific curated aspect. Our methodology consists in looking for how the data have been labelled.

**Dimension 7:** ***Manually Curated*** The labelling process may contain automatic steps. For example, a researcher might decide to check if a bug ticket is a bug or a feature [101] or can check which commit actually fixed the ticket or which line of code induced a vulnerability [311]; after this check, a set of automated steps leads to a dataset. This dimension aims to understand what has been manually curated in a manually curated dataset. Our methodology seeks details about the labelling phase of the collected data.

**Dimension 8:** ***Ground Truth Source*** Ground truth (**GT**) source dimension characterise where authors gained their GT. The GT is an essential component of VPS , as it provides the basis for evaluating the effectiveness of different techniques. GT can suffer from issues related to data availability and quality [345]. Researchers often gain vulnerability data from public databases such as NVD [368, 221, 123, 310, 279]. These databases may not be comprehensive, as not all vulnerabilities are reported or discovered. The data quality can also vary, as the information may not be standardised or consistent [89]. Our methodology seeks details about the GT source of the collected data.

**Dimension 9:** ***Negatives Assessed*** As already said, mislabeling can impact the performance of prediction models [373, 101, 115]. There are two possible types of mislabeling: false positive or false negative. In the false positive case, a non-vulnerable code is labelled as vulnerable; in the false negative case, a vulnerable code is labelled as non-vulnerable. In the absence of evidence, we must assume that both mislabeling types impact prediction models' performance. This dimension aims to investigate whether researchers manually assessed both vulnerable and nonvulnerable code or if researchers assessed only the vulnerable code and labelled the remaining code as non-vulnerable. Our methodology consists of carefully checking if the study's authors manually checked if the code labelled as negative has no vulnerability.

#### 3.2.2.2 Results

In this section, we discuss the results of the characterisation of our groups. However, before analysing the results, we analyse the level of agreement of the authors in characterising the dataset over the 9 dimensions. Regarding the author's agreement on each dimension: While we fully agreed (i.e., 100%) on over 80% of the dimensions, it's important to note instances of divergence. Notably, we found a variance in opinions, with an 80% agreement on Dimension 1 (Granularity of the labeled entity) and a 93% agreement on Dimension 4 (Feature sets availability).

**RQ** ***Dataset availability***

In this group, we assess whether primary studies provide a means of reproducing the results. The values of the current dimension are:

- **Provided**: the authors provide an external reference to the dataset.
- **Not Provided**: the authors do not provide an external reference to the dataset.
- **Reachable**: the authors provide an external reference to the dataset, and the dataset can be downloaded.
- **Not Reachable**: the authors provide an external reference to the dataset, and the dataset cannot be downloaded, i.e., the provided reference is broken.

- **Retrieved**: the authors do not provide an external reference to the dataset, and the authors provide the dataset to us after an e-mail request.
- **Not Retrieved**: (the authors do not provide the dataset to us after an email request) AND (the authors do not provide an external reference to the dataset) OR [ (the dataset was provided) AND (it was unreacheable)].
- **Stable Repository**: the authors provide an external reference to the dataset, the dataset can be downloaded and the dataset is in a remote stable repository, e.g., Zenodo [113].
- **Unstable Repository**: the authors provide an external reference to the dataset, the dataset can be downloaded, and the dataset is in a remote unstable repository, e.g., GitHub [148] without DOI commits [360].

Figure 3.2.8, reports the availability of the datasets in our 50 primary studies according to the options mentioned above. According to Figure 3.2.8 only 25 of the 50 studies provide their dataset but three of these are currently not reachable. Thus, only 44% of the studies provide a reachable dataset. Finally, 21 studies are currently hosted in an unstable repository; thus, only one dataset is hosted in a stable repository (i.e. Zenodo).

To obtain datasets that were not publicly available, we contacted the corresponding author of each respective dataset. If no corresponding author was specified, we contacted all authors. Three months were given to the authors to provide us with the datasets. After the three months had elapsed, we expressed our gratitude to the authors who had provided us with their datasets. We also sent a follow-up email to the authors who had not replied to the initial email, giving them an additional week to respond. After a total period of three months and one week, we began the process of classifying the datasets.

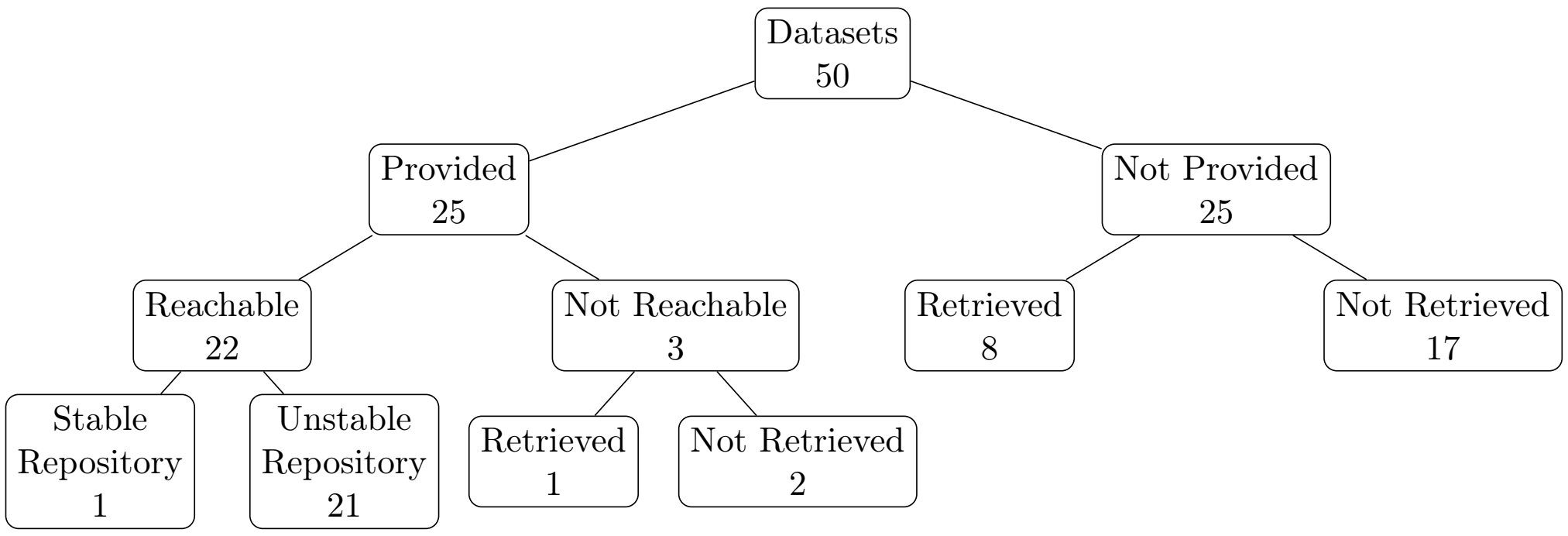

Figure 3.2.8: Availability of datasets.

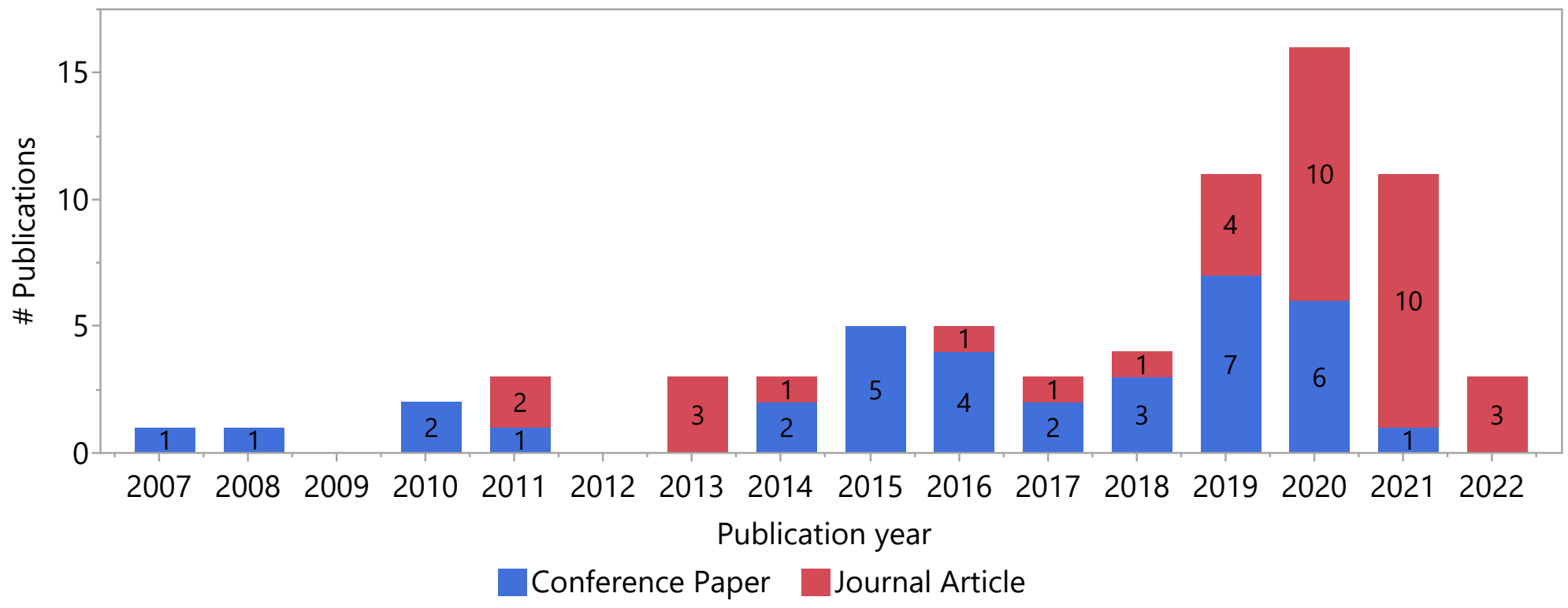

Figure 3.2.9: Publication and venue distribution over the years.

Figure 3.2.9 presents the year of publication distribution of the 50 GDs published between 2007 and 2022 and their publication venue distribution. Notably, there is a spike in publications during 2020, with 16 out of the 50 datasets published during that year. Moreover we note a gap in 2009 and 2012. It is noteworthy that, despite conferences being the most preferred venue in the early 2010s, there has been a shift in the community's interest towards publishing in journals in the following years. In particular, in 2022, all of the studies that met our criteria for inclusion came from journal venues.

In conclusion, in this work, according to Table 3.2.8 we gathered a total of 31 datasets (GDs): 22 Reachable, 1 Not Reachable Retrieved, and 8 Not Provided Retrieved, which resulted in 21 unique datasets.

#### Dimensions

In this subsection, we characterize the 31 GDs along the 9 dimensions.

**Dimension 1:** ***Granularity of the labeled entity*** In this dimension, we characterise the types of entities involved in the prediction. Studies can predict the following types of entity, ordered from the coarsest to the finest-grained.

- **Class or File**: in Object Oriented Programming (OOP) [367] a class is an extensible code template for object creation. Provides initial values for its state and implementations of its behaviour [142]. Usually, a class corresponds to a single file; nevertheless, in OOP it is possible to have multiple classes inside a single file. In non-OOP, a file contains only functions [395].

- **Commit**: the study predicts commits on the Version Control System (**VCS**)[310].

- **Fragment**: the study predicts only a fragment of the code (i.e., a specific line of a method/function)[431].

- **Machine Code** : instruction and data expressed in a form directly recognisable by the CPU [452]. The study predicts the representation of the source code by machine code[237].

- **Method**: the study predicts a method (e.g., Java) or a function (e.g., C)[69]

Figure 3.2.10a reports the distribution of the granularity of the predicted entities. We note that:

- the most predicted entity is Class or File (17),

- methods are predicted less than Class or File (10) but more than the other types, and

- the least predicted entity is the commit, with only one study predicting it.

**Dimension 2:** ***Nature of the labelled entity*** In this dimension, we characterise the nature of the predicted entities. The predicted entities can be of the following types:

- **Artificial**: The predicted entities are created manually for testing and evaluation purposes; examples include the Juliet Test Case from NIST/SARD. The main disadvantage of this type of entity is that it might not represent the industrial code [221].

- **Real**: The predicted entities are derived from open-source projects and online VCS. The main disadvantage of this nature of entities is that it may not be related to all types of vulnerability [238].
- **Mixed**: a combination of the above options [227].

Figure 3.2.11d reports the nature of the entities. We can observe that:

- the most chosen nature is Real with 23 datasets,
- artificial data is a minority, with only six datasets, and
- only two datasets use a mixed data source.

**Dimension 3:** ***Type of labelling*** In this dimension, we characterise the class of the label. The class of the label of the predicted entities can be of the following types:

- **Binary**: the label can only be true or false, i.e., the presence or absence of vulnerabilities [183].
- **Multiclass**: the label can have more than two values and represents the absence of a specific vulnerability, e.g., CWE-23: Relative Path Traversal, CWE-79: Improper Neutralisation of Input During Web Page Generation [452].

Figure 3.2.11a reports the class of the label. We note that:

- 87% of the gathered dataset uses binary labelling, and
- 13% of the gathered datasets uses a multi-class label, using MITRE classification to label the entity.

**Dimension 4:** ***Feature sets availability*** In this dimension, we analyse the feature sets in GDs. The possible feature sets can be of the following types:

- **String**: the feature set consists of strings derived from string operations in the source code or in the compiled machine code [276].

- **P/C Metric**: [252, 71, 289]: The feature set consists of process and complexity metrics [68].

- **CVE-Commit**: the feature set consists of characteristics of a CVE and its related commit [310].

Figure 3.2.10b reports the feature sets available in GDs. We note that:

- the majority of GDs has string features, and

- the minority of GDs has CVE-commit.

**Dimension 5:** ***Availability of either a CVE or a CWE information*** In this dimension, we analyse the information in the GD related to the vulnerable code. The values of the current dimension are:

- **CVE\CWE Info available**: the dataset reports either a CVE or a CWE information for each positive entity [398].

- **No CVE\CWE Info available**: the dataset reports neither a CVE nor a CWE information for each positive entity [221].

Figure 3.2.11b reports how many GDs provide a CVE or a CWE information for each positive entity. We note that 26 GDs (84%) provide a CVE or a CWE.

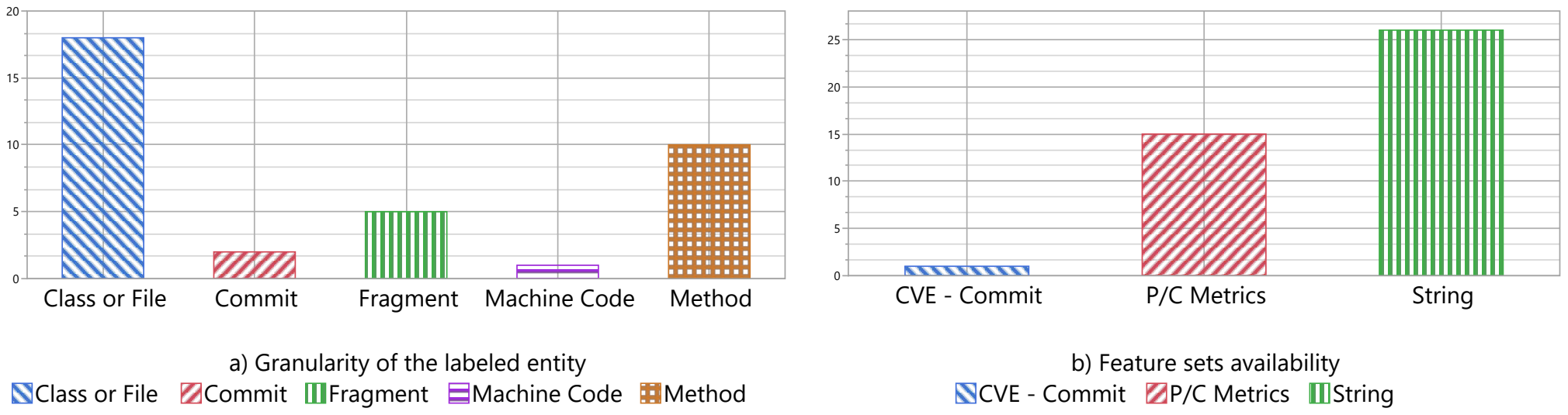

Figure 3.2.10: Granularity of the labeled entity and Feature sets availability

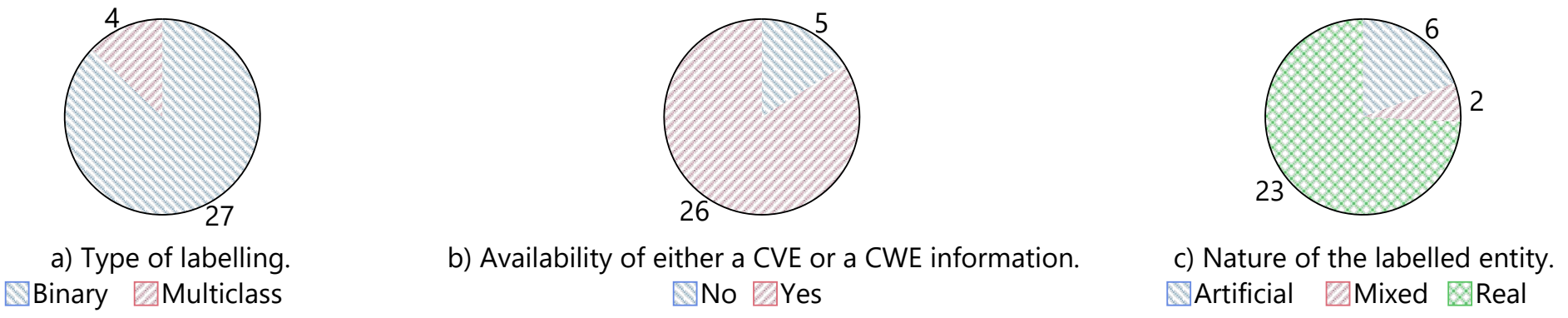

Figure 3.2.11: Type of labelling, Nature of the labelled entity, and Availability of either a CVE or a CWE information.

**Dimension 6:** ***Labelling Process*** In this dimension, we characterise the process of labelling entities. The values of the current dimension are:

- **Automated**: the authors generated the dataset via a script [123].
- **Given**: the authors used the dataset as provided by some sort of official entity (e.g., the Juliet test case provided by NIST) [341].
- **Manual**: the authors manually curated at least one aspect of the dataset (e.g., SAP Dataset [309].
- **Reused**: the authors reused the dataset from another study [20] reused their own datasets in Alves et al. [19].

Figure 3.2.12c reports the type of labelling process. According to Figure 3.2.12c:

- 15 out of the 31 GDs (48%) have at least one entity that is manually curated in the process, and

- only one out of the 31 GDs (3%) are reused from a previous study.

**Dimension 7:** ***Manually Curated*** This dimension investigates what is manually curated during the labelling process. We note that manual curation may only be done on a sample rather than the entire dataset, and a mixed dataset may only have a partial amount of entries containing CVE information. Researchers should analyse the datasets to understand whether the amount and type of curation fit their needs. The values of the current dimension are:

- **Code**: the author manually reviews the project's source code (code) or a fragment (slice) of it [345].
- **Commit**: the commit was manually reviewed [149].
- **Ticket**: the Jira ticket / NVD's CVE was manually reviewed [368].
- **Nothing**: nothing was manually curated [56].

Figure 3.2.12a reports what is manually curated during labelling. According to Figure 3.2.12a:

- most of the gathered datasets (i.e., 15 out of 31 GDs) have nothing manually curated, and
- the least curated entity is the commit with only 5 out of the 31 GDs.

**Dimension 8:** ***Ground Truth Source*** In this dimension, we characterise the source of the GT. The authors used the following sources to obtain the GT:

- **Bugzilla**: GT from Bugzilla [355],
- **Fortify SCA**: GT from Fortify SCA tool [345],

- **Human Expert**: GT from human expert knowledge [430],
- **NVD**: GT from CVE information extracted from NVD [395],
- **SARD**: GT from ad-hoc test suites [226], and
- **SQLI-LABS**: GT from SQLI-LABS published data [125].

Figure 3.2.12b presents the distribution of sources for the GT in VPS. We note that the NVD is the most prevalent source of the ground truth, i.e., 77% GDs used NVD. Additionally, human experts are the second most effective source of the ground truth. Notably, in the single instance where a tool was employed, Fortify SCA was a static application security testing tool, no one used a dynamic security testing tool.

**Dimension 9:** ***Negatives Assessed*** In this dimension, we characterise the nature of negative entities. The values of the current dimension are:

- **Negatives Assessed**: what is labelled as not vulnerable results from an expert decision assessing the absence of vulnerability [255].
- **No Negatives Assessed**: what is labelled as not vulnerable is the complement of what experts assessed as the presence of vulnerability [355].

According to our findings:

- 71% of GDs do not assess negatives, and
- 29% of GDs do assess negatives.

We note that the 29% of GDs assessing negatives consider only one CWE at a time. In other words, a fragment of code deemed negative means that that fragment does not have a specific CWE, rather than that that fragment does not have any CWE. Therefore, no study assessed the absence of vulnerabilities in a code fragment.

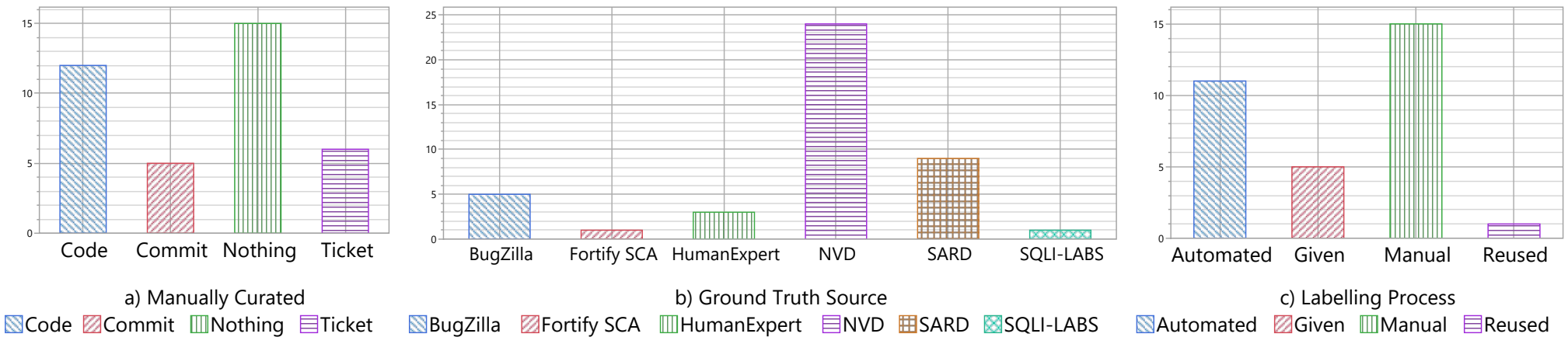

Figure 3.2.12: Manually Curated, Negatives Assessed and, Labelling Process.

#### 3.2.2.3 Replication Study

In this section, we show how VALIDATE can be used for conducting a non-exact replica study, thus replicating the use of VALIDATE by researchers.

**Dataset selection**

Among the possible datasets in VALIDATE, we were interested in finding an ”ML-ready” dataset with software product and process metrics as features. As an ML toolbox, we chose WEKA due to our successful research experience [115, 117, 120, 391]. Thus, we selected the work of Tang et al. [370]; they evaluated three open-source projects, namely Moodle [96], PhpMyAdmin [302], Drupal [77]. Combining the information from the security advisors and the NVD, they created a dataset containing 223 vulnerabilities. They used such a dataset to investigate whether text mining prediction produces more accurate results than software features. Their results show that the text mining features delivered better performance than the software metrics.

### Design

In this section, we explain the key design concept of our non-exact replica. Table 3.2.10 reports the design elements of the original study and our non-exact replica. Our non-exact replica shares many design elements with the original study. We use the same six datasets of the original study based on the authors' open-source projects (i.e., Drupal, Moodle and PhpMyAdmin). As an ML tool, we use Weka [411] as the original paper. As per the independent variable (IV), we choose the same Metrics / Token. As accuracy metrics, we chose to use the same ones as in the original study: Inspection Ratio [370] ($\mathcal{I} = \frac{TP+FN}{TP+FP+TN+FN}$) and Recall [425, 17, 127, 314] ($\mathcal{R} = \frac{TP}{TP+FN}$).

Our non-exact replica improves the original study on many design elements [181, 322, 443, 371, 139, 375]. As a validation procedure, we use the same three-fold cross-validation of the original study. To avoid randomness in the split procedure that affects our results, we repeated the three-fold cross-validation 100 times instead of the three times in the original study. In VPS, feature selection is crucial as it allows researchers to identify and prioritise the most influential factors contributing to system vulnerabilities. Researchers can streamline their analysis and improve the accuracy of ML models [443]. Hence, as a feature selection, we use symmetric uncertainty (**SU**) as suggested by Zhao et al. [443]. Moreover, defect prediction and VPS datasets often exhibit imbalance [104]. We employ class balancing techniques to mitigate biases favouring majority classes in ML models. Notably, SMOTE [60] proves advantageous compared to other methods like oversampling and undersampling. SMOTE effectively tackles class imbalance by creating synthetic instances for the minority class, offering a robust solution without sacrificing data from the majority class to achieve dataset balance [371]. Finally, tuning ML models is essential in VPS because it allows for the optimisation of hyperparameters, enhancing the models performance and predictive accuracy [379]. Manual tuning involves adjusting parameters based on domain knowledge and experimentation to find the best configuration for a given dataset. In the context of VPS, where the nature of vulnerabilities and their manifestations in code can vary widely, manual tuning is particularly tedious and challenging. Auto-tuning models, such as AutoWeka [379], automate the hyperparameter tuning process, systematically exploring the parameter space to find the optimal configuration. The output of Auto-WEKA is the best classifier and the related best parameters given the provided dataset. Table 3.2.10 reports the resulting classifier and parameters for each project.

Table 3.2.10: Study comparison

| | **Original** | **Replica** |
|---|---|---|
| **Datasets** | Drupal, PhpMyAdmin, Moodle | |
| **Machine Learning Tool** | Weka | |
| **Independent Variable** | Metrics / Token | |
| **Accuracy Metrics** | Inspection \| Recall | |
| **Validation Techniques** | 1 time 3-fold cv | 100 times 3-fold cv |
| **Feature Selections** | none | Symmetrical Uncertainty |
| **Class Balancing** | none | SMOTE |
| **Hyperparameters Tuning** | none | AutoWEKA |
| **Statistical Test** | none | Wilcoxon signed-rank |
| **Classifiers** | Random Forest | **Drupal** |
| | | Metrics: RandomSubSpace + Random Forest |
| | | Tokens: SGD |
| | | **Moodle** |
| | | Metrics: AdaBoostM1 + DecisionTable |
| | | Tokens: SMO + NormalizedPolyKernel |
| | | **PhpMyAdmin** |
| | | Metrics: RandomForest |
| | | Tokens: SMO + NormalizedPolyKernel |

Our analysis procedure includes a statistical test to reject the null hypothesis of no difference between $\mathcal{I}$ and $\mathcal{R}$ in defect prediction provided by using Metrics or Tokens. Since our data strongly deviate from normality, we use the Wilcoxon signed-rank test [408] to test our null hypothesis.

**Results**

Figure 3.2.13 reports the accuracy of text-mining-based models and software-metrics-based models. According to Figure 3.2.13 the use of tokens provides higher $\mathcal{I}$ and $\mathcal{R}$ than the metrics. The results of the statistical test show a p-value lower than 0.0001 in the three datasets and two accuracy metrics. Therefore, our non-exact replica confirms the results of the original study even with a more sophisticated empirical procedure: text mining models have higher accuracy than software metrics-based models [370].

All experiments took about a week on one McAffee server model BG5500 running Windows Server 2022. The server has two Intel Xeon (R) CPU X5660, a base clock speed of 2.79 GHz and 72.0 GB of RAM with a clock speed of 1067 MHz.

#### 3.2.2.4 Discussion

Our systematic review of the VPS datasets shows that 50% of the studies provide a dataset. This result is in line with Liu et al. [235], so it is necessary to enforce a replicability package for ML studies. Liem and Panichella [229] discuss the replicability and randomness in using ML in empirical SE studies. Specifically, they note a high variability in using ML techniques and tools with small to non-existent discussions on the randomicity of the specific ML technique or tool. Figure 3.2.8 shows that out of the 22 studies that provide a reachable dataset (44% of the total), only one dataset is currently hosted in a stable repository. The main issue with unstable repositories lies in the volatility of the archive, which is due to their intrinsic non-trackable nature. For example, a GitHub repository can be archived without prior notice from the author; similarly, the central administration can update a personal website of an academic website.

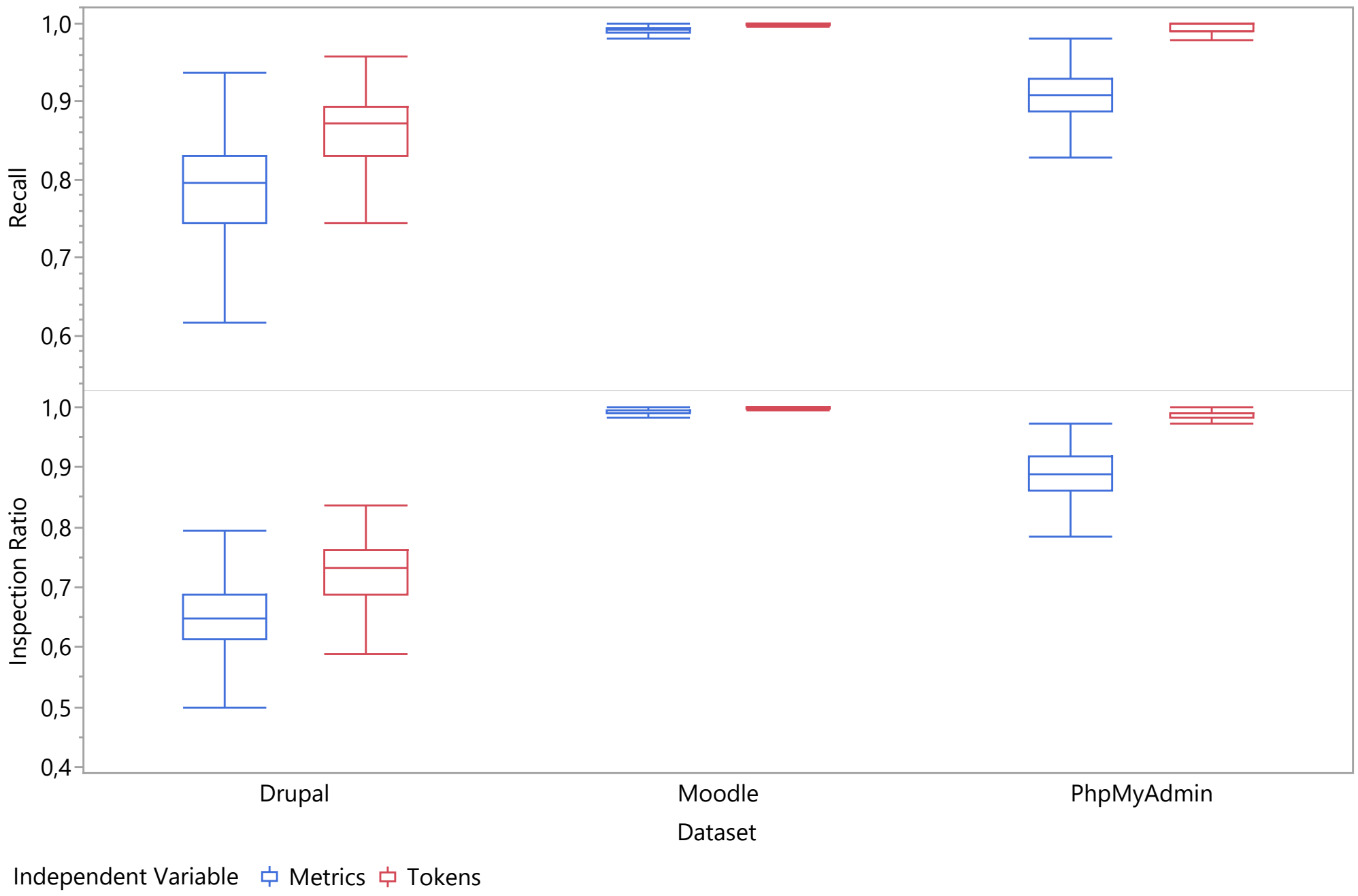

Figure 3.2.13: Accuracy of text-mining-based models and software-metrics-based models.

The issue of historical and locational traceability of digital objects leads GitHub to add academic features to its repositories [360]. Similarly to Mozilla Science Labs [133], Figshare [347], and Zenodo [113], GitHub allows users to generate DOI commits and integrate their repository with external archiving platforms. We encourage researchers to provide a reachable replication package and host it in a stable repository.

Regarding Dimension 1, Figure 3.2.10 shows a majority of coarse-grained entities, that is, class or file; this result is in line with Arisholm and Briand [26] and Morrison et al. [269].

Regarding Dimension 2, Figure 3.2.11b shows that the majority (70% of the 31 GD) use real data (i.e., open-source projects) that are easier to analyse through their coarser grain than to build the project and analyse the byte-code, thus inducing the focus toward class or files. It should be noted that 85% of the GD uses a binary classification; therefore, many studies can use a large body of knowledge on classifiers. Although only 15% of the studies use multi-class classification, recently, the VPS community seems to be focussing on this kind of classification [195, 452]. Regarding Dimension 6, Figure 3.2.12c shows that the most used labelling process involves manually curating at least one aspect of the labelling process; nonetheless, it represents only 42% of the GD. We note that curating at least one aspect of the labelling process provides a more refined dataset; in the same vein, Ponta et al. [309] propose a five-year manually-reviewed dataset.

Our SDR shows that only 33% of the 31 GD assess the entity labelled negative (see Dimension 8, Figure 3.2.12b). Furthermore, no study manually checked the absence of any vulnerability; i.e., what is labelled as negative is the absence of a particular vulnerability rather than all vulnerabilities. For instance, in the Juliet Test Case, "CWE321_Hard_Coded_Cryptographic_Key__ basic_81_goodG2B" is labelled as negative for CWE321 rather than for any CWE. Therefore, we can argue that what has been labelled as negative for a vulnerability might be positive for another vulnerability. Thus, past ML studies might have overestimated the false positive rate [62]. In conclusion, the labelling process is not yet standardised [83], and it is essential to assess the negative labels better.

Finally, on the possibility of reusing datasets according to their content, according to Dimension 4, Figure 3.2.10b, the scientific community is driving its attention on the chance given by the NLP-alike technique for feature extraction directly from the source code or the machine code we call "string feature". As discussed in Section 3.2.2.1, the presence of a standardised vulnerability nomenclature, that is, CVE / CWE information, helps researchers focus on a specific family of vulnerabilities or gain more knowledge of their existence and interaction. Therefore, according to Figure 3.2.11c, the scientific community appears to be aware of the issue since most GD, i.e., 71%, provide the CVE/CWE info.

Regarding Dimension 8, the findings from Figure 3.2.12b have several implications for VPS. The first implication is that the prevalence of NVD as the most prolific source of ground truth suggests that it is the most accessible. Moreover, Figure 3.2.12b suggests that utilising multiple sources to establish the ground truth or a combination of different sources, such as automated tools and human expertise, can provide a more comprehensive understanding of the potential vulnerability landscape, which is crucial for developing effective security measures.

According to Dimension 9, only 9% of the GD provides datasets compliant with this demand.

To understand coverage across multiple dimensions, we use a Burt table

[194] which is a contingency table that displays the frequency of cases for two categorical variables in a suitable format for analysis. The Burt table helps identify the association patterns between two variables and provides an easy visual overview of the relationship between categorical variables. To interpret a Burt table, we should look for cells with high or low frequencies relative to the total number of cases. These indicate solid or weak associations between the two variables. Table 3.2.11 reports the Burt table values for each combination of categories in the MCA analysis.

To better understand Table 3.2.11, and specifically to facilitate the identification of unavailable datasets, we counted in Table 3.2.11, for each dimension value, the number of unavailable datasets in combination with all other dimensions values. For instance, if the number of missing combinations of a cell is zero, there is at least a dataset with that dimension value and all values of the other dimensions. A dimension with all zeros indicates the presence of at least a dataset with all values of that dimension and all the values of the other dimensions. A cell with many missing combinations indicates a low availability of datasets with that dimension value and all values of the other dimensions. Table 3.2.12 reports the number of missing combinations. According to Table 3.2.12:

- "CVE Info" and "Binary" are the two values with only one missing combinations. This means that datasets with "CVE Info" and "Cross Project" are available with all but one values of other dimensions.

- The value “Reused” of the dimension Labelling Process show the highest number of missing combinations, i.e., 25. This means there is a scarce availability of datasets with those values and values of other dimensions.

- The value “Human Expert” of the dimension Ground Truth Source show 22 missing combinations. Thus, too few studies involves human expert to curate or validate the ground truth

The above result implies that researchers can leverage existing datasets when in need of "CVE Info" and "Binary" features. In contrast, they will likely build their own when needing other features.

#### 3.2.2.5 Threats to Validity

This section discusses different threats to the validity [73] of the present study. We organise the discussion on two topics: SDR and non-exact replica.

Regarding SDR, our classification of datasets is driven by our experience in supporting SE through ML [181, 322, 443, 371, 139, 375, 37, 116, 6, 331]. We did not use card sorting in our workflow as done in previous studies [449], because we felt that spreadsheets and comparisons proved more suited to our specific context as characterised by a relatively small number of dimensions, values, and coders. If on the one side, we are confident that our classification, and hence the filters in VALIDATE, can help researchers find useful datasets, on the other side there is a chance that researchers would like to search the datasets in VALIDATE according to unavailable filters. Researchers might require filters that differ in topic or granularity from the current ones. Therefore, we plan to accommodate future requests in changing the datasets classifications according to researchers' needs. The dataset provided by the researchers via email or through the repository referenced in the paper is assumed to be the data set used in the study. To mitigate this issue, we checked that the dataset was in accordance with the description in the paper of the data.

The authors of this paper manually performed the classification; therefore,

there is the possibility that the classification is not reliable. We mitigated this threat to validity by:

- removing intrinsically subjective categories such as the ease of use or the ease of retrieving the datasets, and

- classifying each dataset in each category independently across researchers. Subsequently, we calculated Cohen [76] Kappa and checked a high level of agreement (see Section 3.2.2.2).

The exclusion criteria E7, i.e., citation count of less than 20, can hinder the validity of the work by excluding valuable studies that have not yet gained significant recognition in the academic community. For instance, the low number of studies in 2022 could be due to the exclusion criteria E7; i.e., recent relevant papers need more time to gain 20 citations. However, in light of the replicability principle, we have used the exclusion criteria in Croft et al. [83].

Regarding the non-exact replica, we use classifiers suggested by Auto-Weka. Auto-Weka has different parameters, including the running time and the optimised accuracy metric. We set the running time to two hours per dataset; we selected the proportion of accurate classifications as an optimised accuracy metric. There is a chance that with different parameters, auto-weka would find a better classifier that would lead to a different result from the current ones. Unfortunately, this is a common issue with tuning. To mitigate this issue, we tried different classifiers such as Random Forest [48] and IBK [4], and the current results still hold.

The ongoing evolution of vulnerability datasets, driven by the inclusion of undisclosed vulnerabilities, can potentially change our ground truth. Moreover, interpretability is essential in VPS as it facilitates model improvement and enables effective decision-making by clearly understanding how and why vulnerabilities are identified, thus promoting overall cybersecurity resilience. For instance, authors can revise a published dataset by changing a few values. We aim to keep VALIDATE up to date with researchers' changes.

We decided not to include size as a dimension to differentiate datasets because our datasets are highly heterogeneous in type. For instance, no single metric can characterize the size of datasets containing code and datasets containing project characteristics.

#### 3.2.2.6 Conclusion

In this study, we analysed the availability and characteristics of the prediction studies' datasets. We performed an SDR of VPS datasets. Our results show that of the 50 primary studies, only 22 studies provide a reachable dataset. Of these 22 studies, only one study provides a dataset in a stable repository. Therefore, researchers should focus on where to publish their datasets.

We provide the first repository for VPS datasets to support researchers in finding and sharing datasets (VALIDATE). Our repository of datasets supports researchers in finding datasets of interest, hence avoiding reinventing the wheel; this translates into less effort, more reliability, and more reproducibility in dataset creation and use. Although there are many datasets, no dataset might meet the needs of a researcher. Therefore, our repository provides evidence of the need to create a new dataset. An exciting result of our SDR is that no study manually verified the absence of any vulnerability, i.e., negative labels were not adequately evaluated. Therefore, previous studies may have overestimated false negatives.

We show how VALIDATE can be used by performing a non-exact replica fea-turing a more sophisticated design. Our non-exact replica confirms the original study's results: text-mining-based models have higher accuracy than software-metrics-based models. Therefore, the availability of ready-to-use datasets helps in replication.

We provide a replication package that contains all the data used to generate graphs, statistical analysis, VALIDATE user guide, and all the references to the original studies [1].

In the future, we plan to improve the knowledge of our field in the following key areas:

- **Security by design**: Since many security vulnerabilities are at the design level [325], we plan to investigate how to codify architectural styles and patterns into datasets ready to use for vulnerability prediction.
- **Analyse study replicability**. Recent studies [334, 235, 151, 90, 53, 187, 85, 21, 328, 97] focused on the replicability of empirical SE experiments, while other studies, such as the one conducted by Neto [274] focused on strategies to help researchers create replicability packages. We envision a model to automatically analyse the replicability of studies and to support researchers in facilitating the replicability of future studies.
- **Automated characterisation of the studies and their datasets.** In this study, we manually curated the content of VALIDATE. In the future, we plan to automate the categorisation process by creating an ML/DL model to categorise new datasets automatically.

Following, we provide both the unprocessed and summarized versions of the Burt Table referenced in Section 3.2.2.4.

[1] https://doi.org/10.5281/zenodo.10135002

Table 3.2.11: Burt Table

| Dimension | | Labelling Process | | | | Granularity | | | | |
|---|---|---|---|---|---|---|---|---|---|---|
| | Values | Automated | Given | Manual | Reused | Class or File | Commit | Fragment | Machine Code | Method |
| Labelling Process | Automated | 28 | 0 | 0 | 0 | 15 | 0 | 2 | 0 | 11 |
| | Given | 0 | 10 | 0 | 0 | 5 | 0 | 5 | 0 | 0 |
| | Manual | 0 | 0 | 38 | 0 | 24 | 2 | 2 | 1 | 9 |
| | Reused | 0 | 0 | 0 | 1 | 0 | 0 | 1 | 0 | 0 |
| Granularity | Class or File | 15 | 5 | 24 | 0 | 44 | 0 | 0 | 0 | 0 |
| | Commit | 0 | 0 | 2 | 0 | 0 | 2 | 0 | 0 | 0 |
| | Fragment | 2 | 5 | 2 | 1 | 0 | 0 | 10 | 0 | 0 |
| | Machine Code | 0 | 0 | 1 | 0 | 0 | 0 | 0 | 1 | 0 |
| | Method | 11 | 0 | 9 | 0 | 0 | 0 | 0 | 0 | 20 |
| Manually Curated | Code | 2 | 2 | 18 | 0 | 10 | 0 | 6 | 1 | 5 |
| | Commit | 0 | 0 | 9 | 0 | 6 | 1 | 0 | 0 | 2 |
| | Nothing | 26 | 8 | 0 | 1 | 20 | 0 | 4 | 0 | 11 |
| | Ticket | 0 | 0 | 11 | 0 | 8 | 1 | 0 | 0 | 2 |
| Assessed Negatives | | 0 | 8 | 11 | 0 | 7 | 0 | 5 | 0 | 7 |
| CVE Info | | 23 | 10 | 36 | 0 | 40 | 2 | 9 | 1 | 17 |
| Source Type | Artificial | 2 | 10 | 0 | 1 | 5 | 0 | 8 | 0 | 0 |
| | Mixed | 0 | 0 | 3 | 0 | 0 | 0 | 2 | 0 | 1 |
| | Real | 26 | 0 | 35 | 0 | 39 | 2 | 0 | 1 | 19 |
| Label Type | Binary | 24 | 6 | 36 | 1 | 39 | 2 | 5 | 1 | 20 |
| | Multiclass | 4 | 4 | 2 | 0 | 5 | 0 | 5 | 0 | 0 |
| Feature Set | String | 0 | 0 | 3 | 0 | 3 | 0 | 0 | 0 | 0 |
| | CVE - Commit | 0 | 0 | 2 | 0 | 0 | 2 | 0 | 0 | 0 |
| | P/C Metrics | 13 | 0 | 15 | 1 | 22 | 0 | 1 | 0 | 6 |
| | String | 15 | 10 | 18 | 0 | 19 | 0 | 9 | 1 | 14 |
| Ground Truth Source | BugZilla | 7 | 0 | 3 | 0 | 6 | 0 | 0 | 0 | 4 |
| | Fortify SCA | 0 | 0 | 6 | 0 | 6 | 0 | 0 | 0 | 0 |
| | HumanExpert | 0 | 0 | 4 | 0 | 4 | 0 | 0 | 0 | 0 |
| | NVD | 17 | 3 | 21 | 1 | 22 | 2 | 6 | 1 | 11 |
| | SARD | 4 | 6 | 4 | 0 | 5 | 0 | 4 | 0 | 5 |
| | SQLI-LABS | 0 | 1 | 0 | 0 | 1 | 0 | 0 | 0 | 0 |

**Continued:** Burt Table

| Dimension | | Manually Curated | | | | Assessed Negatives | CVE Info | Source Type | | | Label Type | |
|---|---|---|---|---|---|---|---|---|---|---|---|---|
| | Values | Code | Commit | Nothing | Ticket | | | Artificial | Mixed | Real | Binary | Multiclass |
| Labelling Process | Automated | 2 | 0 | 26 | 0 | 0 | 23 | 2 | 0 | 26 | 24 | 4 |
| | Given | 2 | 0 | 8 | 0 | 8 | 10 | 10 | 0 | 0 | 6 | 4 |
| | Manual | 18 | 9 | 0 | 11 | 11 | 36 | 0 | 3 | 35 | 36 | 2 |
| | Reused | 0 | 0 | 1 | 0 | 0 | 0 | 1 | 0 | 0 | 1 | 0 |
| Granularity | Class or File | 10 | 6 | 20 | 8 | 7 | 40 | 5 | 0 | 39 | 39 | 5 |
| | Commit | 0 | 1 | 0 | 1 | 0 | 2 | 0 | 0 | 2 | 2 | 0 |
| | Fragment | 6 | 0 | 4 | 0 | 5 | 9 | 8 | 2 | 0 | 5 | 5 |
| | Machine Code | 1 | 0 | 0 | 0 | 0 | 1 | 0 | 0 | 1 | 1 | 0 |
| | Method | 5 | 2 | 11 | 2 | 7 | 17 | 0 | 1 | 19 | 20 | 0 |
| Manually Curated | Code | 22 | 0 | 0 | 0 | 7 | 22 | 4 | 3 | 15 | 16 | 6 |
| | Commit | 0 | 9 | 0 | 0 | 2 | 9 | 0 | 0 | 9 | 9 | 0 |
| | Nothing | 0 | 0 | 35 | 0 | 8 | 29 | 9 | 0 | 26 | 31 | 4 |
| | Ticket | 0 | 0 | 0 | 11 | 2 | 9 | 0 | 0 | 11 | 11 | 0 |
| Assessed Negatives | | 7 | 2 | 8 | 2 | 19 | 19 | 8 | 2 | 9 | 15 | 4 |
| CVE Info | | 22 | 9 | 29 | 9 | 19 | 69 | 12 | 3 | 54 | 59 | 10 |
| Source Type | Artificial | 4 | 0 | 9 | 0 | 8 | 12 | 13 | 0 | 0 | 7 | 6 |
| | Mixed | 3 | 0 | 0 | 0 | 2 | 3 | 0 | 3 | 0 | 3 | 0 |
| | Real | 15 | 9 | 26 | 11 | 9 | 54 | 0 | 0 | 61 | 57 | 4 |
| Label Type | Binary | 16 | 9 | 31 | 11 | 15 | 59 | 7 | 3 | 57 | 67 | 0 |
| | Multiclass | 6 | 0 | 4 | 0 | 4 | 10 | 6 | 0 | 4 | 0 | 10 |
| Feature Set | String | 1 | 1 | 0 | 1 | 0 | 3 | 0 | 0 | 3 | 3 | 0 |
| | CVE - Commit | 0 | 1 | 0 | 1 | 0 | 2 | 0 | 0 | 2 | 2 | 0 |
| | P/C Metrics | 6 | 3 | 14 | 6 | 2 | 25 | 1 | 0 | 28 | 27 | 2 |
| | String | 15 | 4 | 21 | 3 | 17 | 39 | 12 | 3 | 28 | 35 | 8 |
| Ground Truth Source | BugZilla | 1 | 0 | 7 | 2 | 0 | 8 | 0 | 0 | 10 | 10 | 0 |
| | Fortify SCA | 2 | 2 | 0 | 2 | 0 | 6 | 0 | 0 | 6 | 6 | 0 |
| | HumanExpert | 2 | 0 | 0 | 2 | 0 | 3 | 0 | 0 | 4 | 4 | 0 |
| | NVD | 14 | 6 | 18 | 4 | 10 | 37 | 6 | 1 | 35 | 35 | 7 |
| | SARD | 3 | 1 | 9 | 1 | 8 | 14 | 6 | 2 | 6 | 11 | 3 |
| | SQLI-LABS | 0 | 0 | 1 | 0 | 1 | 1 | 1 | 0 | 0 | 1 | 0 |

**Continued:** Burt Table

| Dimension | | Feature Set | | | | Ground Truth Source | | | | | |
|---|---|---|---|---|---|---|---|---|---|---|---|
| | Values | String | CVE - Commit | P/C Metrics | String | BugZilla | Fortify SCA | HumanExpert | NVD | SARD | SQLI-LABS |
| Labelling Process | Automated | 0 | 0 | 13 | 15 | 7 | 0 | 0 | 17 | 4 | 0 |
| | Given | 0 | 0 | 0 | 10 | 0 | 0 | 0 | 3 | 6 | 1 |
| | Manual | 3 | 2 | 15 | 18 | 3 | 6 | 4 | 21 | 4 | 0 |
| | Reused | 0 | 0 | 1 | 0 | 0 | 0 | 0 | 1 | 0 | 0 |
| Granularity | Class or File | 3 | 0 | 22 | 19 | 6 | 6 | 4 | 22 | 5 | 1 |
| | Commit | 0 | 2 | 0 | 0 | 0 | 0 | 0 | 2 | 0 | 0 |
| | Fragment | 0 | 0 | 1 | 9 | 0 | 0 | 0 | 6 | 4 | 0 |
| | Machine Cod | 0 | 0 | 0 | 1 | 0 | 0 | 0 | 1 | 0 | 0 |
| | Method | 0 | 0 | 6 | 14 | 4 | 0 | 0 | 11 | 5 | 0 |
| Manually Curated | Code | 1 | 0 | 6 | 15 | 1 | 2 | 2 | 14 | 3 | 0 |
| | Commit | 1 | 1 | 3 | 4 | 0 | 2 | 0 | 6 | 1 | 0 |
| | Nothing | 0 | 0 | 14 | 21 | 7 | 0 | 0 | 18 | 9 | 1 |
| | Ticket | 1 | 1 | 6 | 3 | 2 | 2 | 2 | 4 | 1 | 0 |
| Assessed Negatives | | 0 | 0 | 2 | 17 | 0 | 0 | 0 | 10 | 8 | 1 |
| CVE Info | | 3 | 2 | 25 | 39 | 8 | 6 | 3 | 37 | 14 | 1 |
| Source Type | Artificial | 0 | 0 | 1 | 12 | 0 | 0 | 0 | 6 | 6 | 1 |
| | Mixed | 0 | 0 | 0 | 3 | 0 | 0 | 0 | 1 | 2 | 0 |
| | Real | 3 | 2 | 28 | 28 | 10 | 6 | 4 | 35 | 6 | 0 |
| Label Type | Binary | 3 | 2 | 27 | 35 | 10 | 6 | 4 | 35 | 11 | 1 |
| | Multiclass | 0 | 0 | 2 | 8 | 0 | 0 | 0 | 7 | 3 | 0 |
| Feature Set | String | 3 | 0 | 0 | 0 | 0 | 0 | 0 | 3 | 0 | 0 |
| | CVE - Commit | 0 | 2 | 0 | 0 | 0 | 0 | 0 | 2 | 0 | 0 |
| | P/C Metrics | 0 | 0 | 29 | 0 | 7 | 3 | 3 | 14 | 2 | 0 |
| | String | 0 | 0 | 0 | 43 | 3 | 3 | 1 | 23 | 12 | 1 |
| Ground Truth Source | BugZilla | 0 | 0 | 7 | 3 | 10 | 0 | 0 | 0 | 0 | 0 |
| | Fortify SCA | 0 | 0 | 3 | 3 | 0 | 6 | 0 | 0 | 0 | 0 |
| | HumanExpert | 0 | 0 | 3 | 1 | 0 | 0 | 4 | 0 | 0 | 0 |
| | NVD | 3 | 2 | 14 | 23 | 0 | 0 | 0 | 42 | 0 | 0 |
| | SARD | 0 | 0 | 2 | 12 | 0 | 0 | 0 | 0 | 14 | 0 |
| | SQLI-LABS | 0 | 0 | 0 | 1 | 0 | 0 | 0 | 0 | 0 | 1 |

Table 3.2.12: Burt Table Summary

| Dimension | Values | No. of missing combinations |
|---|---|---|
| Labelling Process | Automated | 14 |
| | Given | 17 |
| | Manual | 6 |
| | Reused | 25 |
| Granularity | Class or File | 7 |
| | Commit | 22 |
| | Fragment | 14 |
| | Machine Code | 24 |
| | Method | 14 |
| Manually Curated | Code | 9 |
| | Commit | 14 |
| | Nothing | 11 |
| | Ticket | 13 |
| Assessed Negatives | | 10 |
| CVE Info | | 1 |
| Source Type | Artificial | 14 |
| | Mixed | 21 |
| | Real | 6 |
| Label Type | Binary | 1 |
| | Multiclass | 15 |
| Feature Set | String | 22 |
| | CVE - Commit | 22 |
| | P/C Metrics | 8 |
| | String | 5 |
| Ground Truth Source | BugZilla | 18 |
| | Fortify SCA | 21 |
| | HumanExpert | 22 |
| | NVD | 5 |
| | SARD | 11 |
| | SQLI-LABS | 23 |

### 3.2.3 Can We Trust the Default Vulnerabilities Severity?

As software systems become increasingly complex and interconnected, the risk of security debt has risen significantly, increasing cyber-attacks and data breaches. Vulnerability prioritization is a critical activity in software engineering as it helps identify and address security vulnerabilities in software systems promptly and effectively. With the increasing complexity of software systems and the growing number of potential threats, it is essential to have a systematic approach to vulnerability prioritization to ensure that the most critical vulnerabilities are addressed first. The present study aims to investigate the agreement between the default and the National Vulnerability Database (NVD) severity levels. We analyzed 1626 vulnerabilities encompassing 12 unique types of vulnerabilities associated with 125 Common Platform Enumeration identifiers belonging to 105 Apache projects. Our results show a scarce correlation between the default and NVD severity levels. Thus, the default severity of vulnerabilities is not trustworthy. Moreover, we discovered that, surprisingly, the same type of vulnerability has several NVD severity; therefore, no default prioritization can be accurate based only on the type of vulnerability. Future studies are needed to accurately estimate the priority of vulnerabilities by considering several aspects of vulnerabilities rather than only the type.

#### 3.2.3.1 The Empirical Study Design

Our empirical study was designed as a case study following established guidelines [337]. In the upcoming sections, we detail the specific research questions and goals that drive our study and the procedures we used for data collection and analysis.

### Goal and Research Questions

We formalized the goal of this study according to the Goal Question Metric (GQM) approach [34] as follows:

*Investigate* SQ Vulnerabilities and CVE, *for the purpose of* evaluation, *with respect to* their relation, *from the point of view of* developers, *in the context of* OSS.

Based on the aforementioned goal, we defined two main Research Questions ($RQ_s$) which serve as the primary focus of our investigation.

> $RQ_1$. *How many rules relate to a single CVE?*

The SQ platform provides predefined rules that detect various types of security vulnerabilities in code. However, multiple CWE categories can impact a single CVE. On the other hand, multiple SQ rules can identify a single CWE. The multiplicity involved in this relationship can lead to ambiguity in understanding the root cause of the vulnerability and its remediation. Moreover, prioritizing which vulnerabilities to address can become tedious and challenging.

Therefore, investigating the relationship multiplicity can help us gather more accurate information and insights into the causes, remediation, and prioritization of vulnerabilities in software. Using CWEs as *trait d'union* between CVE and Sonar rule, we can grasp the specific SQ rules associated with a given CVE. Thus developers can focus their efforts on the most critical vulnerabilities and address them promptly and effectively. After establishing the relationship between CVEs and SQ rules and mitigating the ambiguity caused by the many-to-many multiplicity, we can focus on prioritising the vulnerabilities. Hence we

can ask:

*$RQ_2$. Do default and NVD severity differ?*

Vulnerability prioritization is a critical aspect of vulnerability management [57], as it enables developers to prioritize their efforts and address the most severe vulnerabilities first, reducing the risk of security breaches and potential harm to users. Analyzing the correlation between the default severity and the NVD base score (i.e., NVD severity) can help practitioners tailor vulnerability prioritization tasks to their specific needs and enhance our understanding of the vulnerability life-cycle. By understanding how the default severity and the NVD base score relate, practitioners can better evaluate the severity of vulnerabilities and prioritize their efforts accordingly. This approach allows practitioners to focus on critical vulnerabilities that pose a higher risk to their system's security rather than wasting time and resources on less significant vulnerabilities. Consequently, analyzing the correlation between these two factors can lead to a more effective vulnerability management process, ultimately enhancing the overall security posture of the system.

### Context of the Study

The context of our study is the relationship between the default and NVD severity. We aim to identify the most common default severity associated with the NVD severity and their correlation. CVEs provide valuable information regarding multiple vulnerability metrics, including severity. More specifically, the NVD severity refers to the Common Vulnerability Scoring System (**CVSS**) [257] base score, a standard for assessing software systems' severity of security vulnerabilities. We focused on the CVSSv3, which provides a numerical and an ordinal score representing the severity of a vulnerability. Organizations can use those scores to prioritize their remediations to different security issues.

#### Project Selection

The project selection focused on the ASF ecosystem for two main factors: the first one is that we have access to a large number of CVEs associated with the ASF projects, enabling us to analyze a large amount of data. Furthermore, being open-source projects, source code is readily available to give us a broader view of the vulnerability impacting the specific project. Hence, we explored the relationship between NVD and default severity on a dataset containing 1626 vulnerabilities with 12 unique types of vulnerabilities associated with 125 common platform enumeration (**CPE**) identifiers belonging to 105 Apache projects.

#### Study Setup and Data Collection

This section outlines our data collection methodology. Our approach entailed accessing the CVE feed from the NVD encompassing vulnerability data spanning from 2004 to 2023.

To achieve our research goal, we filter out CVE items with null CVSSv3 fields ($RQ_1$) and without "apache" in the vendor part of the CPE. CPE is a standardized method for identifying IT systems and software products. the identifier consists of three components: a vendor name, a product name, and a version number, it is maintained by NVD and is an integral part of a CVE.

We merged the NVD base score with the matching default severity to conclude our analysis. This enabled us to investigate the correlation between vulnerability severity and default severity. It is noteworthy that we chose to conduct a correlation analysis [365] rather than an agreement analysis [250, 76], given that the variables under consideration are ordinal rather than nominal. To an-alyze the relationships between CVEs, CWE categories, and SQ rules ($RQ_1$), we extracted the relevant CVEs from the NVD feed and computed the multiplicity of the CVE-CWE relationships. We also extracted all SQ rules and computed the relationship between the rules and the CWE categories using a manually curated mapping. We decided to filter out all the CWEs with a one-to-many relationship with SQ rules and all CVEs with the same relationship to CWEs. Finally, we linked CVEs with their matching SQ rules via CWE (i.e., CVE→CWE→SQ Rules), obtaining the final dataset in which a single CVE is mapped to a single SQ rule. Hence, we can describe the relationship multiplicity among CVE, CWEs, and SQ rules in the ASF ecosystem.

#### Data Analysis

This section presents the data analysis procedure we employed in addressing our research questions.

To answer $RQ_1$, we linked CVEs, trough their CWE, to the matching SQ rules and obtain the final datasets that map one CVE to one SQ rule. Therefore, we compute the cardinality of CVE, CWE, and SQ rules relationships.

Finally, for $RQ_2$, we investigate the correlation between the default and the NVD base score severity computing Spearman's $\rho$. Hence, we gain crucial insights into the effectiveness of default severity in vulnerability management.

#### Replicability

We prepared a replication package[2] that includes all the scripts used in our analysis and the raw data. In addition, we have provided CSV files that contain the results for each research question. This enables other researchers to verify

[2] `https://doi.org/10.5281/zenodo.8139908`

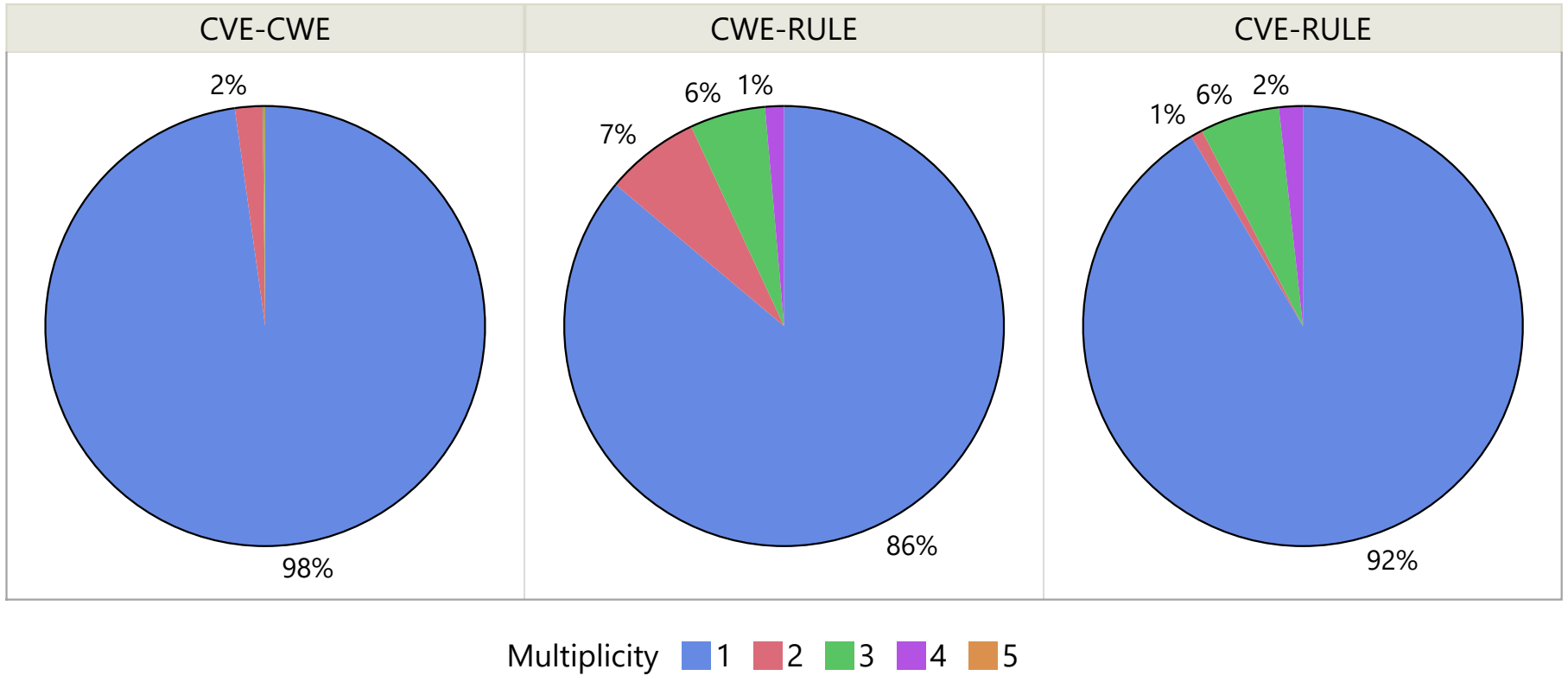

Figure 3.2.14: Proportion of multiplicities among CVE, CWE and rules.

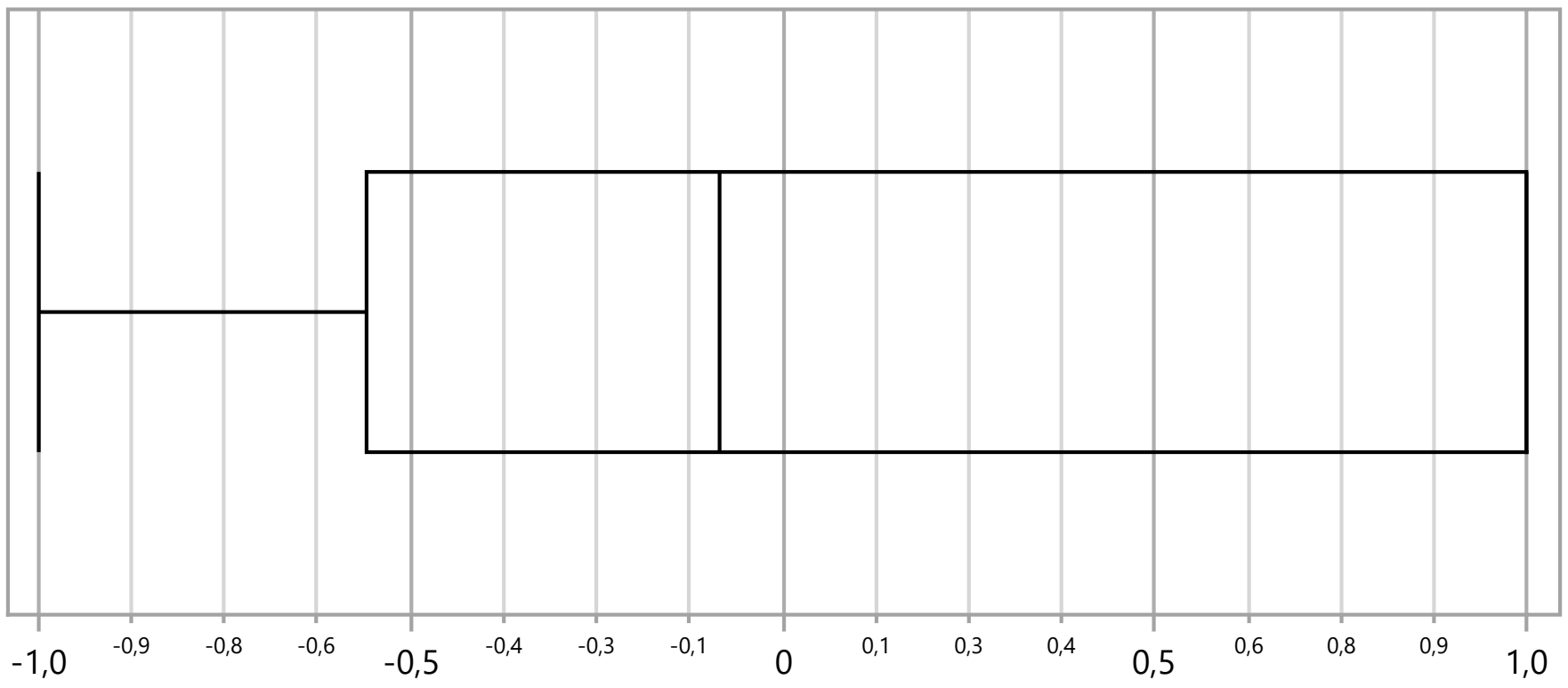

Figure 3.2.15: Distribution of Spearman's $\rho$ across projects.

our findings and build upon our work.

#### 3.2.3.2 Results

**RQ$_1$. How many rules relate to a single CVE?**

Figure 3.2.14a presents how many CWEs are reported in a CVE; this table provides insight into the relationships between CVE and CWE entries and cardinality distribution across these entries. According to Figure 3.2.14a, 98% of 1,626 CVE entries (i.e., 1590 entries) report only one CWE. One CVE reports

five CWEs. Figure 3.2.14b presents how many rules relate to how many CWEs. According to Figure 3.2.14b, 86% of 72 CWE entries (i.e., 62 entries) relate to only one rule. Furthermore, five CWE entries relate to two rules, 4 CWE entries relate to three rules and one CWE entry relates to four rules.

Figure 3.2.14c presents how many CVEs relate to how many rules. According to Figure 3.2.14c, 92% of 342 CVE entries (i.e., 313 entries) relate to only one rule. Furthermore, three CVE entries relate to two rules, twenty CVE entries relate to three rules, and six CVE entries relate to four rules.

**$RQ_2$. Do default and NVD severity differ?**

Figure 3.2.15 presents Spearman's $\rho$ correlation between default and NVD severity regarding project distribution. The mean $\rho$ across projects resulted as 0.07; thus, the correlation is in average null. According to Figure 3.2.15, the distribution of $\rho$ widely varies across projects. Specifically, most of the projects have negligible negative correlation and there is even a project where the correlation is fully negative. **This result suggests that we cannot trust the default severity level**.

One possible reason for the observed null correlation between default and NVD severity is that the same rule has several priorities in different CVEs. Figure 3.2.16 presents the proportions of NVD severity levels across 12 rules, i.e., all with at least two CVE entries. According to 3.2.16 the same rule has different NVD priorities in 11 out of 12 cases (i.e., 92%). **This result suggests that a default severity is not possible as it even changes for the same rules**.

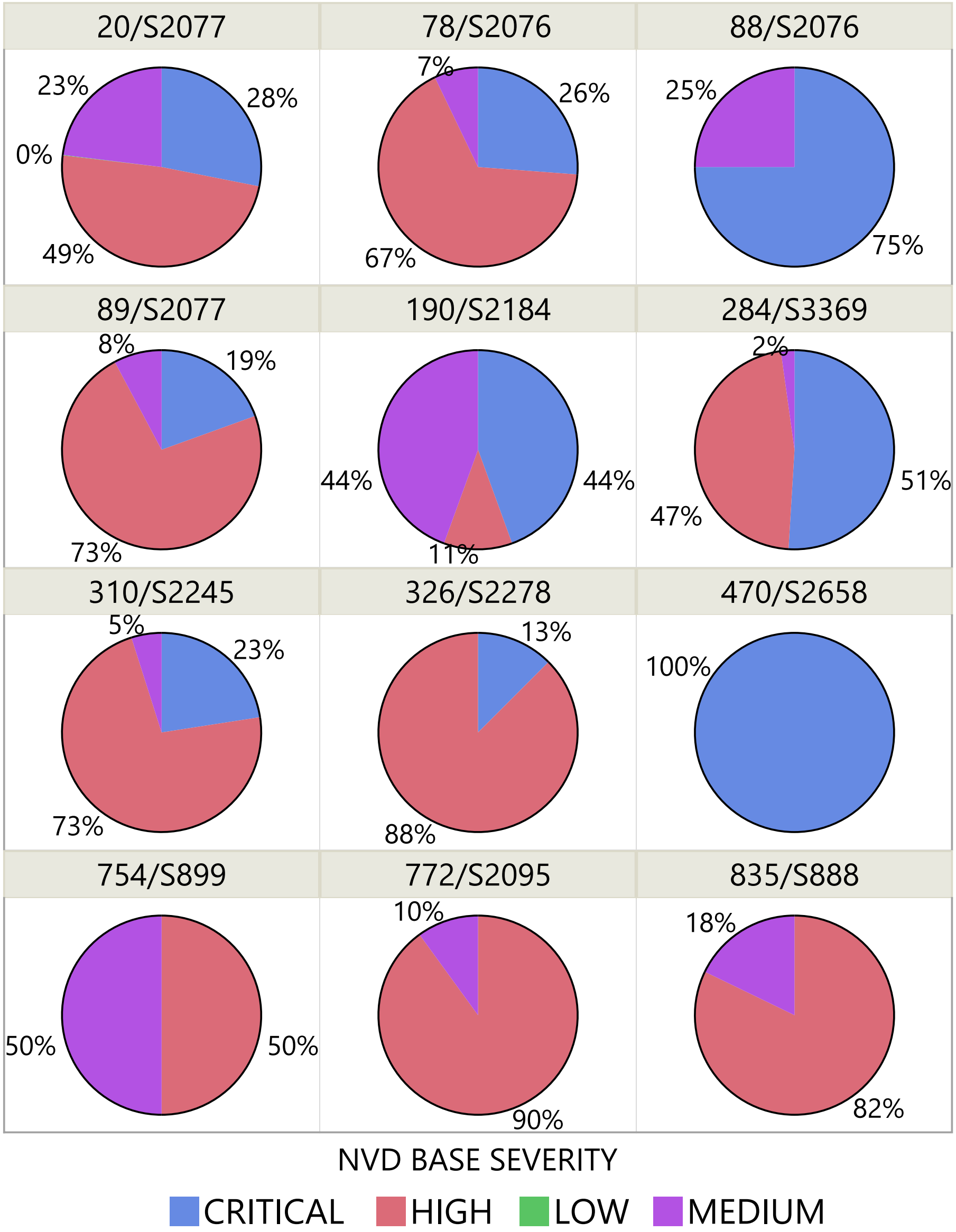

Figure 3.2.16: Distribution of NVD Base Severity across CWEs/SQ rules.

Table 3.2.13: SQ rule Type distribution

| **SQ rule Type** | **Count** | **Percentage** |
|---|---|---|
| Bug | 30 | 33% |
| Code Smell | 20 | 22% |
| Vulnerability | 42 | 45% |

#### 3.2.3.3 Discussion

Regarding $RQ_1$, Figure 3.2.14c reveals that most CVE entries are tied to a single rule. Thus, the vulnerabilities identified in the ASF ecosystem are usually associated with a single rule. This finding suggests **that vulnerabilities in the studied ecosystem have a distinctive structure and can be detected using a single rule. Hence, identifying those vulnerabilities does not pose significant challenges**.

Regarding $RQ_2$, the low correlation presented in Figure 3.2.15 could suggest that a default severity is inaccurate. Therefore that is necessary to customize the priority in each industrial context. However, very interestingly, the fact that most of the rules (i.e., 92%) have several priorities, see Figure 3.2.15, suggests that context-based customization of rule priorities would be inaccurate. Specifically, Figure 3.2.15 suggests that **the priority of a rule violation must be established according to many factors other than the rule ID only**.

Finally, Table 3.2.13 presents the number of SQ rules with a particular type despite having "cwe" in their tags. This finding supports the conclusions of a previous study [215], which had identified inconsistencies between the type and the severity that SQ had assigned to Sonar-Issue affected classes.

#### 3.2.3.4 Threats to Validity

This section presents the threats to the validity of our study. We have categorized the threats by type, including Conclusion, Internal, Construct, and External validity [73].

**Conclusion Validity**

Spearman's $\rho$ is a non parametric measure of the strength and direction of a correlation between two variables, but due to multiple factors, this approach may hinder the conclusion's validity. For instance, the correlation coefficient is unreliable with a small sample size and cannot accurately represent the relationship between variables. Moreover, Spearman assumes a linear monotonic correlation between the variables; if the assumption were to prove invalid or not applicable, Spearman's correlation might not accurately measure the strength and direction of the association. We have mitigated this issue by ensuring a representative sample and checking the data distribution. We could mitigate this issue by using other non-parametric measures of correlation. Still, our analysis found that Spearman's $\rho$ was more suitable than other correlation measures, such as Kendall's $\tau$ because it can effectively handle tied observations in the data. On the other hand, Kendall's $\tau$ relies on concordant and discordant pairs of observations.

**Internal Validity**

The internal validity of this study may be threatened by the lack of an official SQ rule - CWE mapping API or dataset. Nonetheless, we have taken measures to address this concern by conducting a comprehensive examination of the SQ documentation and meticulously linking each SQ rule to the corresponding CWE, as specified on Sonar's official website[3]. It is worth mentioning that the rule dump, containing the squid identifier (legacy name of SQ rules), pertains to SQ 7.5. Subsequent releases may have introduced additional rules and refined the existing ones. Therefore, future research should expand our rule set to mitigate the potential threats to internal validity.

#### Construct Validity

Construct validity concerns how our measurements reflect what we claim to measure [73]. Our specific design choices, including our measurement process and data filtering, may impact our results. To address this threat, we based our choice on well-established guidelines in designing our methodology [337, 34].

#### External Validity

External validity concerns how the research elements (subjects, artifacts, etc.) represent actual elements [73].

The present study utilized a substantial dataset obtained through scripted scraping of public records from the NVD concerning open-source ASF projects. Consequently, the study may be considered open-source focused. To address this limitation, future research should broaden the scope of the analysis to include industry projects. It is important to note that the NVD dataset was scraped during the initial week of June 2023, thus limiting the scope of replication efforts to that specific time frame. To facilitate the replication of our study, we have included a comprehensive replication package in Section 3.2.3.1

[3] `https://rules.sonarsource.com/java/`

#### 3.2.3.5 Conclusion

In this section, we briefly draw our conclusions. Our study examined the relationship between the NVD severity and the severity assigned to violations of SQ rules. Our key finding was that relying on default severities for rule violations may lead to inaccuracies, as the same SQ rule may be associated with different NVD severities. This highlights the challenge of creating a universal severity rating for each vulnerability, as the vulnerability context and development team can have a significant influence. Therefore, customization of priorities in each industrial context is necessary [121]. In addition, we identified inconsistencies between the type and the severity that SQ had assigned to Sonar-Issue affected classes, which raises concerns about the reliability of SQ's classification system. Future studies are needed to accurately estimate the priority of vulnerabilities by considering several aspects of vulnerabilities rather than only the type. Specifically, the community should investigate how SQ rule severity can predict the likelihood of exploiting a vulnerability in the wild. Moreover, research should examine the impact of SQ on the vulnerability management process and mitigation. Finally, this research can benefit from, for instance, OSS project analysis to measure the impact of vulnerability management on real-world applications. Future works should focus on gathering OSS project analysis data, on a large scale, to further improve our impact in the field.

### 3.2.4 Leveraging Large Language Models for Preliminary Security Risk Analysis: A Mission-Critical Case Study

Preliminary security risk analysis (**PSRA**) provides a quick approach to identify, evaluate and propose remeditation to potential risks in specific scenarios. The extensive expertise required for an effective PSRA and the substantial ammount of textual-related tasks hinder quick assessments in mission-critical contexts, where timely and prompt actions are essential. The speed and accuracy of human experts in PSRA significantly impact response time. A large language model can quickly summarise information in less time than a human. To our knowledge, no prior study has explored the capabilities of fine-tuned models (**FTM**) in PSRA. Our case study investigates the proficiency of FTM to assist practitioners in PSRA. We manually curated 141 representative samples from over 50 mission-critical analyses archived by the industrial context team in the last five years. We compared the proficiency of the FTM versus seven human experts. Within the industrial context, our approach has proven successful in reducing errors in PSRA, hastening security risk detection, and minimizing false positives and negatives. This translates to cost savings for the company by averting unnecessary expenses associated with implementing unwarranted countermeasures. Therefore, experts can focus on more comprehensive risk analysis, leveraging LLMs for an effective preliminary assessment within a condensed timeframe.

#### 3.2.4.1 Methodology

Our empirical study was designed as a case study following established guidelines [337]. In the upcoming sections, we detail the specific research questions and goals that drive our study and the procedures we used for data collection and analysis.

**Goal and Research Questions**

We use the Goal Question Metric (GQM) approach [34] to formalise our goal as follows:

*Investigate* a fine-tuned model, *for the purpose of* evaluation, *with respect to* its proficiency in PSRA, *from the point of view of* practitioners, *in the context of* mission critical IT security.

Based on our goal, we defined two Research Questions ($RQ_s$).

*$RQ_1$ Can an LLM perform PSRA?*

General purpose large language models (**GPLLMs**) assist humans in everyday tasks, from creative writing to data analysis [444, 58]. As GPLLMs, we selected OpenAI's *gpt-3.5-turbo-1106* model. Despite the broad capabilities of GP models, fine-tuning becomes essential for tasks requiring specific domain expertise [171]. For instance, Yang et al. [423] achieved success by fine-tuning LLAMA on a manually curated financial dataset, enhancing the effectiveness of financial professionals. Moreover, Tufano et al. [386] utilized fine-tuning in the development of AthenaTest, an automated approach for generating unit test cases. Nevertheless, to the best of our knowledge, no previous study has explored the potential of FTM for preliminary security risk analysis. Therefore, investigating the proficiency of an FTM in this context enables practitioners to

Table 3.2.14: Expert Profiles

| ID | YE | Role | Level | AR |
|---|---|---|---|---|
| 1 | 35 | Security Secretariat | Manager | 51-60 |
| 2 | 5 | Cipher Operator | Senior | 31-40 |
| 3 | 45 | Director of Security | Director | 81-90 |
| 4 | 26 | Coordinator of Interventions | Manager | 51-60 |
| 5 | 5 | Cipher Operator | Senior | 41-50 |
| 6 | 3 | System Administrator | Junior | 21-30 |
| 7 | 15 | Security Officer | CEO | 41-50 |

YE: Years of Experience. AR: Age Range

promptly identify security threats within the tested environment, particularly in mission-critical scenarios. After establishing the effectiveness of the model, we ask:

*$RQ_2$ Can an LLM outperform human expert?*

The level of human expertise and years spent in the field directly impact PSRA quality and proficiency. Table 3.2.14 provides an overview of the experts' profiles. We can measure the time required for a new team member to attain the proficiency level of a senior team member in years or even decades. In contrast, LLMs can rapidly assimilate years of training within mere minutes. Therefore, it is essential to investigate the FTM proficiency in terms of accuracy and time, comparing it to human experts. We provided 40 samples to the FTM and seven human experts. We measured the proficiency of both human experts and the FTM by measuring their timing and computing metrics such as Accuracy, Precision, Recall, and F1-score [58].

### Industrial context of the study

The context of our case study is an Italian company that has been operating in the civil and military security sector for over 30 years. It is dedicated to re-searching and developing new technologies for information security and provides products and services aimed at safeguarding data, both at rest and in motion. Additionally, it conducts design, verification, implementation, and certification interventions for security, including "What-if" Analysis.

#### Sample Selection

The sample selection derives from previously finalised risk analyses, thus enabling us to obtain the ground truth necessary for fine-tuning and evaluating the model. We engaged with the company Risk Analysis and Management Team (**RAMT**) and randomly selected excerpts from the existing finalised documents. To ensure the replicability of our findings, we excluded all sensitive information and samples that could compromise data anonymity. The final dataset comprises 141 samples, encompassing over 50 mission-critical analyses conducted over the last five years. The authors and the RAMT reviewed each sample and agreed unanimously on each classification.

#### Study setup and data collection

This section delineates our data collection methodology. Our study entailed analysing excerpts from PSRA interviews. The RAMT provided 200 samples from previously finalised PSRAs. We excluded 59 samples due to sensitive data, representativity concerns, and the need for data anonymity. Regarding representativity, our goal was to fine-tune the model with diverse scenarios, avoiding anchoring towards specific keywords or scenario descriptions. The final dataset comprised 141 samples, with 100 samples allocated for training and validation and the remaining 41 samples designated for testing.

OpenAI API's documentation states that fine-tuning requires providing con-versation samples in a JSONL-formatted document[4]. When fine-tuning a model, three roles are available: the *system* role guides the model on how to behave or respond. In our case, we directed the model to reply in a JSON-like output to facilitate automated analysis. We asked the model to respond briefly with only three possible answers in the result field: **yes** if it identified a potential security threat, **no** otherwise, and **more** in case the user-provided insufficient information for a proper PSRA. Moreover, we allowed the model to provide a more open reply in the message field of the custom-defined JSON for future works. The *user* role simulates the end user asking a question. Similarly to how we label instances in machine learning, the *assistant* role simulates the model to provide examples of correct answers. We used 70 samples as a training set and 30 as a validation set.

Finally, we tasked the model and the seven human experts with analyzing the last 41 samples (**testing samples**). We instructed the experts to classify the samples following the model's three possible answers, i.e., yes, no, and more.

### Data Analysis

This section presents the data analysis procedure we employed in addressing our research questions. To answer $RQ_1$, we fine-tuned the selected GPLLM, acting as the baseline, obtained our FTM, and evaluated both on the testing samples. For $RQ_2$, we conducted an evaluation involving seven human experts on the same testing samples and compared their performance against the FTM. The assessment covered multiclass classification carried out by the models and the human experts, employing widely used metrics: Accuracy, Precision, Recall, and F1-Score. In RA not all misclassifications are deemed equal [384], and their significance can vary from one context to another. In our context, a "yes" instead of a "no" is less severe than a "no" instead of a "yes" or a "no" instead of a "more". To account for the varying severity of errors, we weighted the metrics. We utilized ScikitLearn (**SL**)[5] to calculate accuracy metrics. Specifically, we used "weighted" as the *average* parameter for weighted evaluation and "micro" for unweighted evaluation. When specifying "micro", SL computes metrics globally, counting total true positives, false negatives, and false positives across all classes. It then calculates accuracy metrics using these global counts, thus giving equal weight to each instance. It's important to note that when we mention the "*proficiency*" of a model or a human expert, we consider accuracy metrics, the number of errors, and the time taken to evaluate the samples.

[4] `https://platform.openai.com/docs/guides/fine-tuning`

#### Replicability

Our replication package includes a Python notebook importable into Google Colab with fine-tuning data, questionnaire answers, error type weights, and raw model responses. The package is available on Zeondo: `https://zenodo.org/doi/10.5281/zenodo.10501335`.

### 3.2.4.2 Results

#### $RQ_1$ Can an LLM perform PSRA?

Table 3.2.15 presents the proficiency comparison between human experts and LLMs with the 'micro' and 'weighted' average types. According to Table 3.2.15, FTM consistently outperforms the baseline, i.e., GPLLM, in each accuracy metric. Moreover, FTM exhibits high precision in both average types, suggesting low rates of false positives. Furthermore, FTM achieves a weighted recall of 0.8814, suggesting it can effectively discover preliminary security risks with a low rate of false negatives. **This result suggests that FTM can effectively perform PSRA**.

[5] `https://scikit-learn.org/stable/modules/generated/sklearn.metrics.precision_recall_fscore_support.html`

Table 3.2.15: Proficiency comparison between human experts and LLMs.

| Metric | Average Type | E1 | E2 | E3 | E4 | E5 | E6 | E7 | FTM | GPLLM |
|---|---|---|---|---|---|---|---|---|---|---|
| **Accuracy** | micro | 0.8 | 0.8 | 1 | 0.65 | 0.6 | 0.8 | 0.75 | 0.9 | 0.525 |
| | wheighted | 0.8 | 0.8 | 1 | 0.65 | 0.6 | 0.8 | 0.75 | 0.9 | 0.525 |
| **Precision** | micro | 0.8 | 0.8 | 1 | 0.65 | 0.6 | 0.8 | 0.75 | 0.9 | 0.525 |
| | wheighted | 0.751 | 0.751 | 1 | 0.473 | 0.225 | 0.752 | 0.680 | 0.881 | 0.463 |
| **Recall** | micro | 0.8 | 0.8 | 1 | 0.65 | 0.6 | 0.8 | 0.75 | 0.9 | 0.525 |
| | wheighted | 0.593 | 0.615 | 1 | 0.433 | 0.369 | 0.593 | 0.517 | 0.818 | 0.269 |
| **F1 Score** | micro | 0.8 | 0.8 | 1 | 0.65 | 0.6 | 0.8 | 0.75 | 0.9 | 0.525 |
| | wheighted | 0.574 | 0.614 | 1 | 0.380 | 0.278 | 0.561 | 0.456 | 0.813 | 0.196 |
| **Time (# Errors)** | | 8m 12s (8) | 9m 24s (8) | 12m 51s (0) | 15m 25s (14) | 14m 4s (9) | 11m 35s (8) | 6m 52s (10) | 18s (4) | 1m (19) |

### $RQ_2$ Can an LLM outperform human expert?

According to Table 3.2.15, human experts outperform the GPLLM in each accuracy metric and the overall number of incorrectly classified samples (i.e., Errors #). Nonetheless, FTM outperforms six of seven human experts in all accuracy metrics, in the number of errors, and in analysis time. **This result suggests that FTM can outperform human experts in PSRA.**

Furthermore, despite Expert 4 (E4) reporting the highest number of incorrectly classified samples (i.e., 14), the weighted precision and recall scores are higher than those of, for instance, E5, highlighting that errors from E5 were more severe than those from E4. **These results suggest that weighting error types in PSRA is essential for result interpretation**.

### 3.2.4.3 Discussions

Regarding $RQ_1$, Table 3.2.15 shows that although the GPLLM performed poorly, it was still able to correctly identify 47% of the samples in less than a minute. This outcome aligns with the inherent advantages LLMs have over human speed in analyzing data. Therefore, we highlight that **fine-tuning on a small dataset can significantly enhance proficiency** [282, 446], resulting in improvements for the weighted metrics of up to 204% in Recall and 314% in F1.

Regarding $RQ_2$, more experienced experts lead to less severe errors. E3 exemplifies that years of experience make a difference in the human world. On the other hand, in 18 seconds, the FTM challenged the accuracy of 45 years of experience. Nevertheless, our case study aimed not to replace human experts but to investigate a preliminary tool for quickly scanning a context and detecting preliminary security risks. **Experts should leverage the FTM to enhance their analysis capabilities, and to focus to the more extensive RA**.

### 3.2.4.4 Threats to Validity

In this section, we discuss the threats to the validity of our case study. We categorized the threats in Construct, Internal, External, and Conclusion validity following established guidelines [73]. **Construct validity** concerns how our measurements reflect what we claim to measure [73]. Our specific design choices may impact our results, including our measurement process and data filtering. To address this threat, we based our choice on past studies and used well-established guidelines in designing our methodology [337, 34].

**Internal Validity** is the extent to which an experimental design accurately identifies a cause-and-effect relationship between variables [73]. Our study relies on 141 samples, which can potentially be biased from the sample selection and the MCC. We addressed this issue by sampling over 50 mission-critical analyses conducted by the industrial context team on different fields, from national security to health and education.

**External validity** concerns how the research elements (subjects, artefacts) represent actual elements [73]. Our case study focused on an Italian company operating in the civil and military security field. The use of the Italian language and the specific characteristics of this company may limit the generalizability of the findings to other organisations or fields. We addressed this concern similar to the internal validity threats by sampling from over 50 mission-critical risk analyses across various fields. Moreover, we chose a GPLLM with no specific Italian language advantages or restrictions. Given the inherent language capabilities of the model and the context-less nature of RA, having the samples in Italian should not pose any generalizability issues [282, 72].

**Conclusion Validity** focuses on how we draw conclusions based on the design of the case study, methodology, and observed results [73]. Our conclusions rely on the specific accuracy metrics chosen, and there may be other aspects or dimensions of performance that we did not consider. To address this potential limitation, we selected metrics from recent related studies that have faced the challenge of validating FTM proficiencies in specific tasks [58].

#### 3.2.4.5 Conclusions

In this section, we briefly draw our conclusions. Our study delved into the proficiency of LLMs in PSRA. The findings emphasized that GPLLMs are suitable for PSRA, but fine-tuning is essential to narrow down the model's focus toward specific tasks. Notably, our FTM outperformed six out of seven human experts in both accuracy metrics, the number of errors, and evaluation time. Moreover,

FTM misclassifications were less severe than those made by their human counterparts, as highlighted by the weighted metrics. Hence, suggesting that FTM can be used as a 'copilot' in the decision-making process of PSRA.

In the industrial context, our approach has effectively reduced errors in PSRA, accelerated risk detection, and minimized both false positives and, most importantly, false negatives, which are critical for the reliability of operations. Consequently, our approach prevented unnecessary costs for the company, avoiding expenses incurred in implementing unnecessary countermeasures (i.e., false positives). As a result, the RAMT can now concentrate on the more extensive risk analysis, obtaining an effective preliminary analysis in a short period by leveraging our FTM. Therefore, the key benefit of the FTM is to support the decision-makers in focusing, quickly, on the vulnerabilities that need to be addressed without concerns about incorrect indications on the items to be worked on. Our current research efforts are focusing on leveraging retrieval augmented generation to further enhance the model adding more risk analysis frameworks, to allow a cross-evaluation between different reference standards, specific threat detection, personnel cost, and remediation estimation for a more comprehensive security risk analysis.

# Chapter 4

# Other Areas of Research

*This Chapter introduces the other areas of research, i.e., quantum software engineering. We discuss the state of the art and two main contributions to the field.*

# 4.1 Quantum Software Engineering

This chapter presents our contribution to the emerging field of QSE. We present a brief section with the required background and motivating scenarios that inspired our latest two contributions to the field.

## 4.1.1 Background and Motivating Scenarios

This chapter introduces the background and motivating scenarios of our two recent contributions to QSE.

The word quantum in "quantum computing" indicates the quantum mechanics that a system utilize to run computational intensive operations [442, 278]. In Physics, quantum is the smallest unit of any physical entity and generally refers to atomic or subatomic particles, e.g., electrons, neutrons, and photons. QC leverages the principles of quantum mechanics to process information and perform specific computational tasks much faster than conventional computer systems. It also allows practitioners and researchers to probe many possibilities simultaneously. QC is well suited for optimization, analysis, simulation, cybersecurity, cryptography, and molecular modeling [442, 251, 46].

The critical difference between the classical and quantum computer is the qubit [442, 202]. Differently than classical computer-exclusive bits (0,1), qubits can be in a state that is a superposition of 0 and 1. Superposition allows for the representation of qubits as a linear combination of two basis states |0 and |1. Consequently, any qubit can be represented as $x|0\rangle + y|1\rangle$, where x and y are complex numbers that adhere to the normalization condition $|x|^2 + |y|^2 = 1$. This condition ensures that the probability of finding the qubit in either state upon measurement is one.

The distinct feature of quantum mechanics is the superposition [442][95]. Superposition means that a quantum system at a time could be at all possible states rather than one specific state. In quantum computers, a quantum register exists in all possible 0s and 1s superposition states, unlike classical, where the register has only one value at a time. Therefore, two qubits combine as a whole in a superposition of four quantum states, i.e. |00|01|10 and |11, where n qubits can be described as the superposition state of qubits. It is the significant advantage of quantum computers over classical ones where n bits have a fixed single state [442].

Designing quantum applications is particularly challenging due to the inherent characteristics of quantum mechanics like superposition and entanglement [169]. Therefore, a novel scientific field, quantum software engineering (**QSE**) [442, 304, 305], has emerged and focus on fostering the application of traditional software engineering methods to quantum software development. A key point of QSE is integrating classical and quantum systems, which build the so-called hybrid systems [305, 442]. The requirement for coexistence and collaboration between classical and quantum infrastructures has led to the emergence of hybrid systems. [323]. In hybridization, specific and complex tasks are assigned to quantum devices, while the less demanding components to classical computing systems, which minimize the risk of unstable and unreliable applications [403, 402]. However, this hybrid concept of quantum and classical computing poses various challenges to application developers, who are required to not only effectively exploit the potentials of quantum hardware and algorithms but also ensure the development of industry-strengths applications [8, 401, 241]. It is reasonable to assume that most classical developers and designers might lack the essential skills and expertise to perform low-level plumbing activities in quantum computing infrastructures. They need to go through an intensive learning process to understand pulses, circuits, hamiltonian parameters, and complex algorithms. Approaches focused on workflows, such as those mentioned by Weder et al. [403], are crucial for effectively orchestrating classical applications with quantum circuits. However, these approaches have limitations in their capacity to systematically reuse modular and parameterized knowledge and implement a structured lifecycle model. Moreover, they typically do not consider the specific context of applications and often overlook factors of quality and resource constraints. What is required are the methodologies and tools that enable the design, development, deployment, monitoring, and management of applications which can be executed on quantum infrastructure without an in-depth understanding of the underlying principles of quantum mechanics. At the same time, the applications should be dynamically adaptable and reliable, capable of handling (unexpected) changes in the underlying quantum hardware infrastructure. We can address the hybridization challenges by leveraging service-oriented computing/architectures[8, 7, 293, 294].

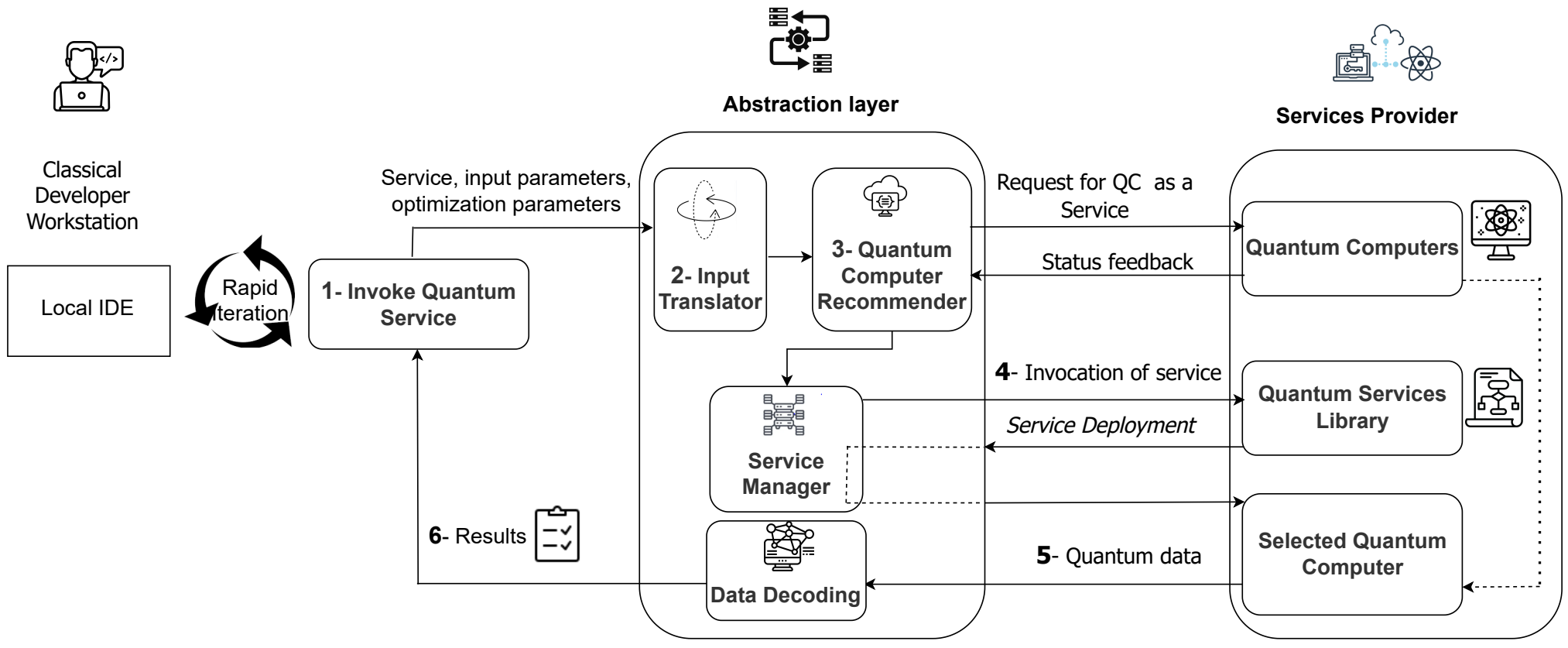

Figure 4.1.1: Proposed QCSHQD Framework

Service-oriented architecture (SOA) is an architectural style in which business and IT systems are designed in terms of services available and interrelated with the outcomes of these services. A service is a logical representation of a set of activities with specified outcomes, is self-contained and may be composed of other services. Still, consumers of the service need not be aware of any internal structure [176].

Following the SOA principles, we introduce a framework aimed at exploiting the potential of SOA within the QC domain. QCSHQD enables users, i.e., classical developers, to locate and employ various QC capabilities, which consist of computing resources and applications provided by service providers. It offers a strategy for developing quantum applications by exploiting a set of reusable functional units, which have well-defined interfaces and are implemented using classical and quantum hybrid computing methodologies.

We now hypothesize a motivating scenario to elaborate on the challenges classical developers face to leverage QC and highlight our scientific endeavors' importance. This scenario also aims to show the impacts and benefits of QCSHQD.

**Motivating Scenario**: Drug discovery is one of the most significant industries in health. Various pharmaceutical companies are trying to develop life-changing medicines for financial purposes or to improve human life experiences. To this end, they need to simulate molecular interactions and solve optimization problems, which consume a lot of time and energy with classical computers due to the complex nature of chemical processes. While the engineers working in these companies know that by using quantum computers, these simulations can be done with higher accuracy and speed, they will face two significant challenges:

- Complexity of Quantum Computing: The team lacks expertise in quantum

mechanics and finds the quantum computing landscape more complex than classical computing. This gap in knowledge makes it difficult for them to leverage quantum computing for their project.

- Accessibility of Quantum Resources: Even if they had the necessary knowledge, gaining access to quantum computing resources is another hurdle. Quantum computers are not widely accessible, and interfacing with them requires specific tools, methods, and processes.

We introduce QCSHQD to address such challenges. Our framework offers a solution to make QC resources available as a service to classical developers, eliminating the need for them to have expertise in quantum mechanics. Our framework strives to make QC practical and easily accessible within software development scenarios.

**QC current challenges.** Classical Computing (**CC**) represents information with bits that can take 0 or 1 as a state; on the contrary, QCs employ quantum bits or qubits, qubits leverage the principles of quantum superposition and entanglement, allowing them to exist in a combination of both states simultaneously, represented by a quantum state vector. This property enables qubits to perform parallel computations and explore multiple possibilities simultaneously, vastly increasing computational power. Additionally, qubits are highly sensitive to their environment, leading to phenomena like decoherence, which poses significant challenges for maintaining the integrity of quantum information over time. Furthermore, qubits require specialized quantum gates and operations for manipulation, distinct from classical logic gates used with bits. Overall, qubits fundamentally extend computation capabilities beyond classical bits, offering the potential for exponentially faster processing and solving problems intractable for classical computers. This unique capability enables

QC to outperform CC in specific computational tasks [303]. De Stefano et al. [91], investigated the state of quantum programming, recognizing its evolution from a scientific interest to an industrially available technology that challenges the limits of CC. The findings revealed limited current adoption of quantum programming.

***Classi*$|Q\rangle$ enabling factors.** Researchers in non-computer science fields, such as physicists, mathematicians, data scientists, biologists, and many others, cannot write cumbersome and complex code such as QASM for leveraging QC powers.

ASTs and OpenQASM 3.0 are the key technologies that make $Classi|Q\rangle$ development possible. ASTs are a hierarchical tree structure representing the syntactic structure of source code in a programming language. It is an abstraction of the code's syntax that disregards specific details such as formatting and focuses on the essential structure. ASTs are a valuable tool in compiler design and programming language processing. They capture the relationships between code elements, such as expressions, statements, and declarations. By abstracting the essential components of a programming language, an AST enables tools and compilers to analyse and manipulate code in a language-agnostic manner. Therefore, **our framework can potentially convert every language to QASM**.

Similarly, OpenQASM is a specialised language crafted for quantum computer programming. It acts as a user-friendly interface, enabling the expression of quantum circuits and algorithms in a format executable by quantum processors. OpenQASM offers a structured and easily understandable method for specifying quantum gates, operations, and measurements, facilitating the creation of quantum programs for diverse quantum devices. The recent release of OpenQASM's third major version marked a significant milestone by introducing classical control flow, instructions, and data types. As the OpenQASM specification outlines, this addition allows for defining circuits involving real-time computations on classical data, **paving the way for genuinely hybrid solutions and serving as a driving force of $Classi|Q\rangle$, motivating and empowering it**.

#### 4.1.1.1 Hypothetical Motivating Scenarios

Before presenting $Classi|Q\rangle$ design, we have hypothesized a motivating scenario. Anchoring our research endeavors in real-world problems helps set the context for $Classi|Q\rangle$, highlighting the importance of our study, engaging readers, and showing our work's impact and potential benefits.

Meet Alex, an experienced software engineering researcher leading a team at the intersection of CC and QC. Alex's team brings together quantum physicists, chemists, biologists, and data scientists for a multidisciplinary study. Each subteam favors different programming languages with varying proficiency levels. While physicists and mathematicians prefer Fortran and PyC, computer scientists are fluent in PyC but emphasize the need for QASM to run quantum code. Chemists and biologists have basic Python knowledge.

$Classi|Q\rangle$ serves as the pivotal tool for their collaboration. This framework seamlessly translates PyC-based algorithms into QASM, enabling quantum-hybrid computation. The translator accelerates research and innovation by overcoming language barriers, allowing Alex's team to explore novel applications while accommodating subteams' preferred languages.

Felderer et al. [128] highlighted a significant challenge in the disparity of backgrounds and adopted terminologies among individuals working in diverse fields when referring to quantum technologies. For instance, developers with a physics background in QC often lack expertise in software engineering, particularly in software processes and quality practices. Furthermore, a challenge arises from the intricate nature of quantum algorithms and programming, which tends to be inaccessible to a broader spectrum of developers due to its inherent complexity. In response to these challenges, the authors emphasise the need to address the interdisciplinary gap by bridging the knowledge divide between developers with different backgrounds. The identified issues underscore the importance of technical proficiency, interdisciplinary collaboration, and knowledge transfer in QC and SE. Our work stems from these challenges, presenting a possible bridge for R&P to harness QC power more efficiently. The programming languages and resource power limitations of state-of-the-art QC providers have led to a preference for hybrid approaches surpassing classical computational power [9]. Despite prevailing research trends, $Classi|Q\rangle$ seeks to bridge the gap between proficient classical practitioners and the QC potentials, presenting a seamless transition path.

$Classi|Q\rangle$ is tailored to a diverse team; to our knowledge, such a framework was never developed. Leveraging Python's simplicity, the code is easily understood by researchers with low coding proficiency yet robust enough to optimise recurrent computational patterns when translating to QASM.

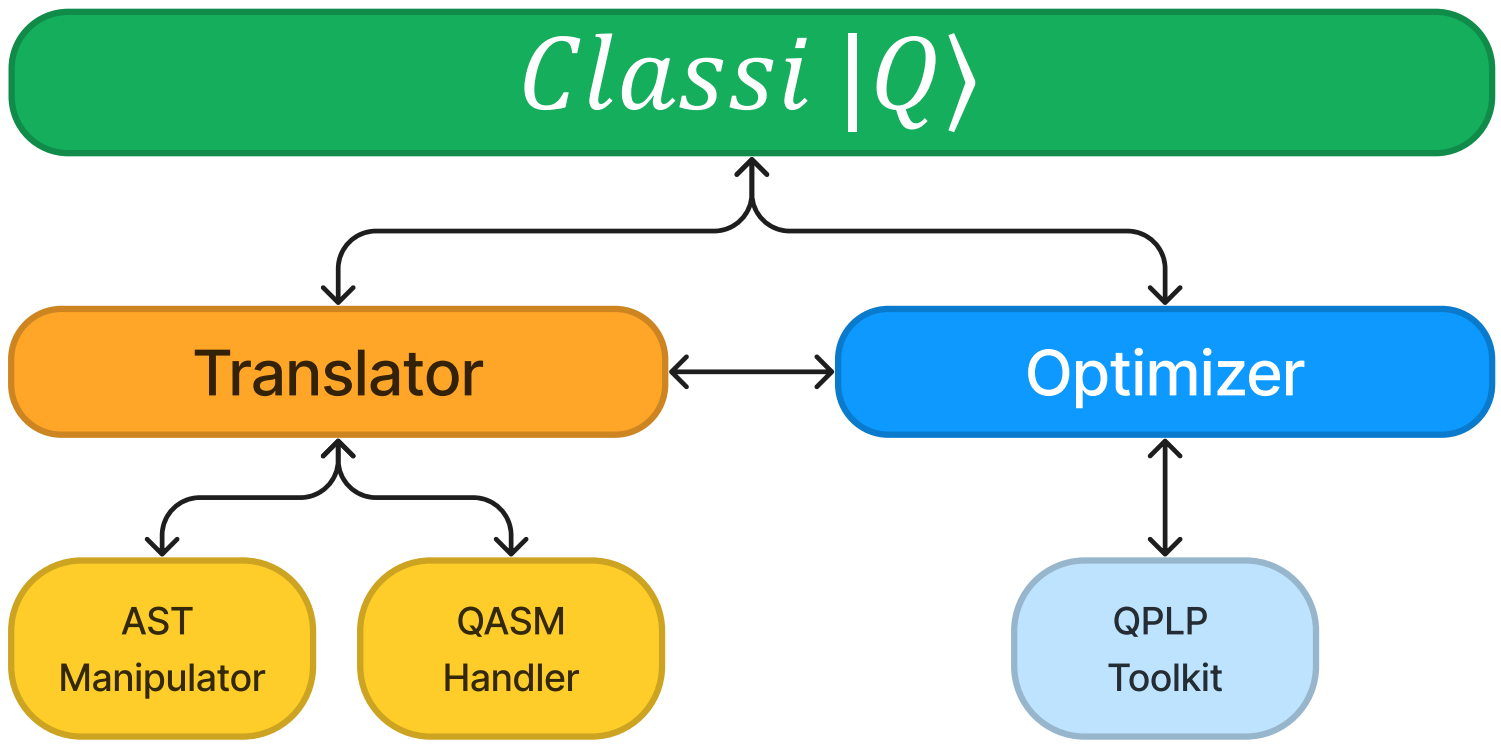

Figure 4.1.2: Overview of the $Classi|Q\rangle$ Framework

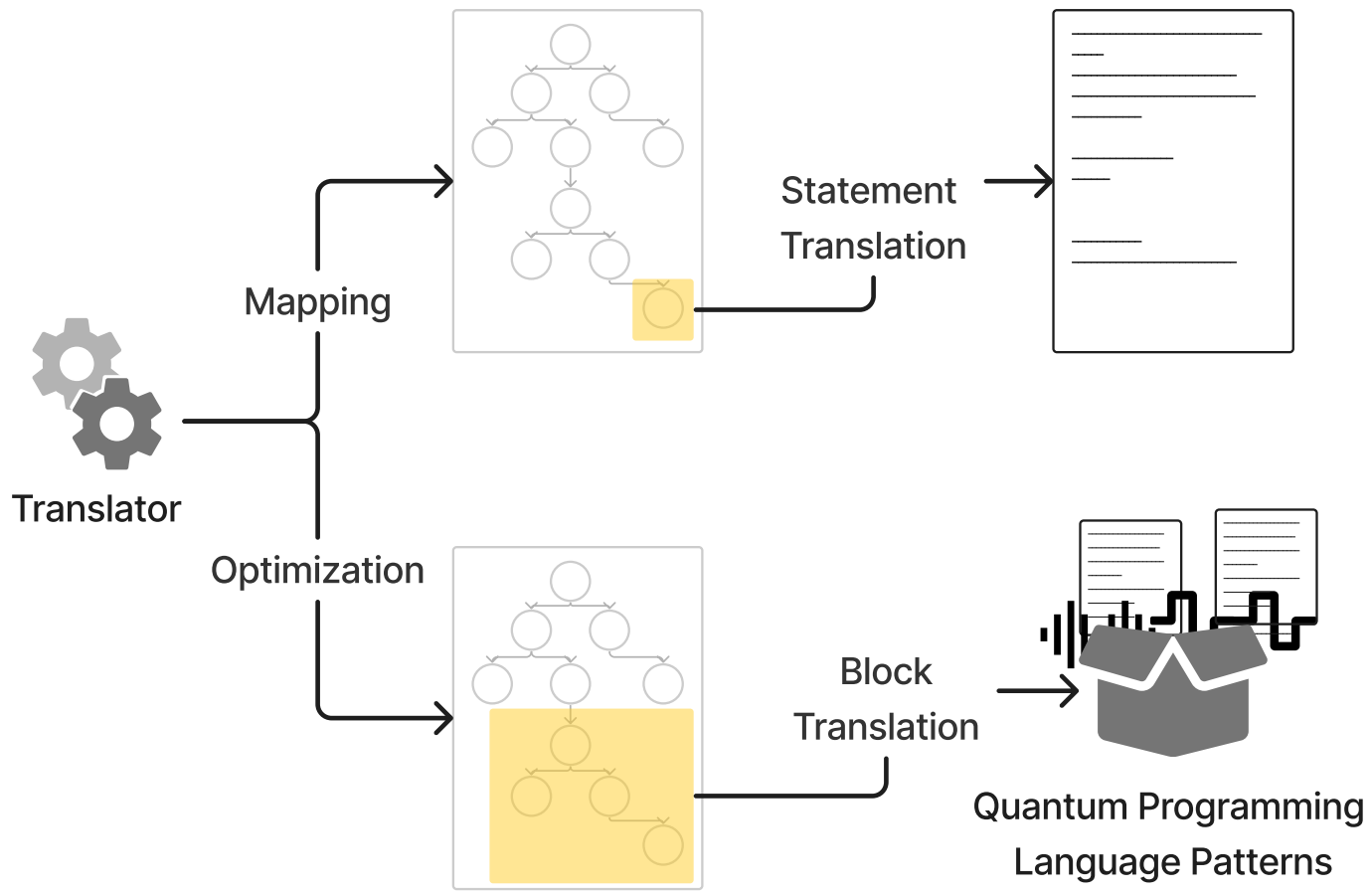

Figure 4.1.3: Workflow Overview

### 4.1.2 QCSHQD: Quantum computing as a service for Hybrid classical-quantum software development: A Vision

Quantum Computing (**QC**) is transitioning from theoretical frameworks to an indispensable powerhouse of computational capability, resulting in extensive adoption across both industrial and academic domains. QC presents exceptional advantages, including unparalleled processing speed and the potential to solve complex problems beyond the capabilities of classical computers. Nevertheless, academic researchers and industry practitioners encounter various challenges in harnessing the benefits of this technology. The limited accessibility of QC resources for classical developers, and a general lack of domain knowledge and expertise, represent insurmountable barrier, hence to address these challenges, we introduce a framework- Quantum Computing as a Service for Hybrid Classical-Quantum Software Development (QCSHQD), which leverages service-oriented strategies. Our framework comprises three principal components: an Integrated Development Environment (IDE) for user interaction, an abstraction layer dedicated to orchestrating quantum services, and a service provider responsible for executing services on quantum computer. This study presents a blueprint for QCSHQD, designed to democratize access to QC resources for classical developers who want to seamless harness QC power. The vision of QCSHQD paves the way for groundbreaking innovations by addressing key challenges of hybridization between classical and quantum computers.

#### 4.1.2.1 The Vision

We introduce QCSHQD, a middleware allowing developers to use QC services in their current classical software development setup. QCSHQD aims to simplify the process of connecting to and using quantum services, so developers can delegate complex computational tasks to QC.

QCSHQD consists of three key components: the Local Integrated Development Environment (IDE) - a dedicated space for classical developers to work; the abstraction layer- responsible for managing quantum services; and the service provider- primarily tasked with the execution of the invoked services. Figure 4.1.1 presents QCSHQD workflow, structured in six steps, to streamline access to QC services for users. According to our design, the workflow starts when the developers need the available QC services, which they iteratively invoke using the local IDE interface (1) by defining the input parameters and the optimization parameters of the service (See Figure 4.1.1). Next based on the specified parameters, the abstraction layers takes these information and translate using input translator (2) into a form that quantum computer can process. Based on the translatesd parameters, the quantum computer recommender (3) identifies the quantum computer that is available and optimal to execute a given request [389]. Once the quantum computer is selected, the service manager requests the required information to deploy the service to the selected quantum computer (4). After the execution of the service, the abstraction layer receives the response (quantum data) from the quantum computer (5) and decode/translate the quantum results back into a classical format (6), as shown in Figure 4.1.1, which the developers can access using the IDE interface.

For the interaction across the above components, the abstraction layer functions as an intelligent intermediary/middleware and simplifies the complexity of hybridization between classical and quantum computers (See Figure 4.1.1).

#### 4.1.2.2 Implementation and Execution Roadmap

This section details the tools and methodologies for implementing QCSHQD. We plan to extend the PyDev plugin[1] for Eclipse to develop the IDE. Furthermore, for version control and research data management, the project will integrate Git [2], which can be effectively used within the Eclipse environment through existing Git plugins.

The abstraction layer design is based on the work of Valencia et al. [389] and Weder et al. [402] through OpenAPI[3] and TOSCA-based orchestration [234]. The role of a TOSCA-based orchestration mechanism, [402, 234] will be automating the process of translating classical input to quantum-compatible formats, selecting appropriate quantum computers, handling the deployment of quantum computing services, interpreting the quantum data and translating them back into classical representation of the results. The OpenAPI offers a standardized specification for developing and generating APIs to ensure seamless task execution and data integrity. The importance of OpenAPI not only can be defined in users' understanding and exploring services' capabilities, but it will also facilitate services development. Its specification needs to be modified to support quantum services [335], therefore, we plan to utilize OpenAPI with an extension, considering custom properties tailored specifically for defining quantum applications. These custom properties serve as a means to incorporate supplementary information into the API contract definition, expanding the scope of the specification [335].

[1] `https://www.pydev.org/`
[2] `https://git-scm.com/`
[3] `https://www.openapis.org/`

Our primary deployment targets are QC services available on the cloud as well as real-world QC infrastructure, such as VTT quantum computer HELMI connected with CSC LUMI[4] for QC provider, which depends on the evolving situation of QC.

Future efforts will be put into defining and evaluating the existing process for developing quantum software based on the QCSHQD. To anticipate this task, we benefit from continuous software engineering practices for team collaboration, short development iterations, and continuous delivery [131].

### Potential Limitations

QC is still a new domain, and technologies are quickly evolving. Therefore, in this work, we foresee a set of possible scenarios that might impact the proposed road-map.

- The **integration** of quantum computing services with classical systems may encounter compatibility issues due to the quantum infrastructure readiness.
- **Allocation of quantum resources** in a cloud-based service model could be complex, in case of high demand for quantum computing power.
- **Scalability:** As the demand for quantum computing resources grows, the framework might face scalability issues.
- **Evolution of Quantum Technologies:** The rapidly evolving nature of quantum computing technologies may outpace the development of the proposed framework, necessitating continuous updates and adaptations.

[4]https://www.csc.fi/lumi

#### 4.1.2.3 Conclusion

We have introduced QCSHQD, aiming to broaden access to QC resources for classical developers. Through adopting service-oriented computing strategies, QCSHQD streamlines the integration of QC capabilities and reduces the complexity associated with quantum programming. The proposed roadmap for QCSHQD implementation employs modern tools, positioning QCSHQD as a bridge that enables classical developers, regardless of their QC expertise, to exploit the extensive capabilities of QC resources. Future efforts will focus on implementing QCSHQD, followed by conducting a thorough evaluation of its efficiency and practical applicability. Future efforts will also be put into investigating the impact of our QCSHQD framework on the existing quantum software development process.

### 4.1.3 $Classi|Q\rangle$: Towards A Translation Framework To Bridge The Classical-Quantum Programming Gap

In the rapidly advancing field of quantum computing (**QC**), seamless leverage of quantum architecture's power is paramount. Moreover, the diverse needs of researchers, practitioners, data scientists, and educators lead to different solutions involving diverse programming skills and quantum technology. In this study, we introduce $Classi|Q\rangle$, a translation framework idea to bridge Classical and Quantum Computing by translating high-level programming languages, e.g., Python or C++, into a low-level language, e.g. Quantum Assembly (**QASM**). Our framework addresses needs such as accessibility, enabling individuals from various backgrounds to apply their existing knowledge of classical program-ming languages to quantum computing tasks; efficiency, allowing researchers and practitioners to make use of QC power with minimal prior knowledge; interdisciplinary collaboration, facilitating teamwork among experts from different fields by removing programming language barriers in quantum projects; and education, enabling the teaching and learning of quantum concepts using familiar classical programming languages, thereby lowering the entry barrier for students. Our idea paper serves as a blueprint for ongoing efforts in quantum software engineering, offering a roadmap for further $Classi|Q\rangle$ development to meet the diverse needs of researchers and practitioners. $Classi|Q\rangle$ is designed to empower researchers and practitioners with no prior quantum experience to harness the potential of hybrid quantum computation. We also discuss future enhancements to $Classi|Q\rangle$, including support for additional quantum languages, improved optimization strategies, and integration with emerging quantum computing platforms.

#### 4.1.3.1 Design

This section introduces the concept of QPLPs and the design of $Classi|Q\rangle$ .

**Quantum Programming Language Patterns**

Recent studies, introduced patterns in QC realm [219, 49, 143, 174, 299]. For instance, Leymann [219] propose patterns for handling essential operations within the QC realm, such as state preparation, entanglement and unentanglement, phase shift, and many others. In the same vein, Georg et al. [143] presents execution patterns such as Standalone Circuit Execution, Ad-hoc Hybrid Code Execution, Pre-deployed Execution, Prioritized Executionl, and Orchestrated Execution while Bühler et al. [49] showcased development-pattern such as: Quantum

Module, Quantum Module Template, Hybrid Module, and Classical-Quantum Interface and one new execution pattern idea of Quantum Circuit Translator, a pattern for translating a circuit among different QC vendors. R&P can find the previous patterns in the quantum computing patterns online library[5].

Gamma et al. [141] pioneered the concept of patterns in software engineering. The motivation behind their book was to leverage the power of object-oriented programming and provide reusable elements that would make software more flexible, modular, and reusable. We want to define the QPLPs as *elements of reusable QC algorithm.* We envision collecting or developing patterns that allow R&P to replace entire blocks of classical computation, such as the computation of the discrete algorithm [102] or a mean [47], the search for an element in an array [154], with black-quantum boxes that leverage the full power of the underlying quantum hardware. The optimization module will then leverage our collection of QPLPs and replace CC blocks with QC-powered blocks.

To our knowledge, none of the existing patterns consider programming language patterns. We can trace this research gap to two root causes: 1) each QC vendor proposes its implementation and design of a language that leads to 2) scarce interoperability among vendors. In this scenario, the most promising QC language is QASM [219, 84], and it can have the impact that Java had when first released.

#### *Classi*$|Q\rangle$

Figure 4.1.2 shows the building block of *Classi*$|Q\rangle$ . The framework comprises two main blocks: the **Translator** and the **Optimizer**. The translator module (**TM**) handles the translation of CP via its AST representation. Within TM, the AST manipulator sub-module handles AST operations like tree traversal and source code statement interpretation. The QASM Handler sub-module leverages a custom grammar, i.e., mapping file, to interpret and translate the abstract representation of the code into the corresponding OpenQASM equivalent.

[5] `https://quantumcomputingpatterns.org/`

The optimizer module (**OM**) optimizes entire code blocks. We design OM to analyze multiple statements, i.e., blocks, of code and leverage QPLPs to replace the entire block with a predefined quantum subroutine.

More specifically, Figure 4.1.3 presents the overview of the translation workflow. $Classi|Q\rangle$ will enable R&P to translate CP with our without optimizations. In the "mapping translation" scenario, $Classi|Q\rangle$ transverses the AST, analyzes its content, and translates each statement in the corresponding OpenQASM classical representation. On the other hand, block translation aims to translate entire source code blocks, replacing their content with an improved quantum algorithm. This block replacement stems from our definition of QPLPs. Hence, it will read the source code, understand what kind of computation the R&P envisioned, and provide one or more possible alternatives to compute the same output with the improved QC algorithm version.

#### 4.1.3.2 Roadmap

This section briefly presents the roadmap we will pursue to bring $Classi|Q\rangle$ to fruition. We are developing the TM and the OM functionalities in parallel. Specifically, we target Python as the first proof of concept (**POC**). We believe to be able to release the first beta of the TM by the end of spring 2024. **The translation module** is the core of $Classi|Q\rangle$ basic functionalities; hence, it is our main focus. We finished the development of the AST transversal for Python and are currently implementing the statement translation leveraging the grammar provided by OpenQASM.

We are currently exploring the existing literature and experimental implementations in detail to identify QPLPs suitable for classical sub-problems within the quantum algorithm. The goal is to understand the nuances of these techniques and their potential applicability to map a specific solution to a broader type of similar computation instances. Once promising patterns are identified, the focus shifts to seamlessly incorporating them into the existing translation workflow in the QPLP toolkit sub-module.

#### 4.1.3.3 Limitations & Future Directions

This section acknowledges limitations to our idea and possible future directions.

*Classi*$|Q\rangle$ is designed to comprehensively support PyC grammars and syntaxes. However, it is essential to note that object translation remains unavailable at present. This limitation stems from the intrinsic nature of QASM as a "low-level" language, lacking support for high-level constructs such as objects. Nevertheless, we will support all built-in functions, enabling the translation of complex calculations by composing programs exclusively with built-in functions. Future works should address this limitation through smart recursive translation or expanding the QASM capabilities, essential for enhancing the hybrid approach.

While the current focus of translation is on Python and C++, it is noteworthy that *Classi*$|Q\rangle$ will ultimately evolve into a source-language-agnostic tool. Consequently, with customised grammar, it can translate other source languages, such as Q# or Java, to QASM.

#### 4.1.3.4 Conlusions

In this work, we have introduced $Classi|Q\rangle$, the design of the ideal PyC to QASM translator, bridging the classical and quantum realms. To illustrate the practical relevance of $Classi|Q\rangle$, we outline three hypothetical scenarios grounded in real-world contexts. Moreover, we detail the design, key decisions, and the introduction of QPLPs. The ultimate goal of $Classi|Q\rangle$ is to empower researchers and practitioners to harness the capabilities of quantum computing without requiring specific training. We will achieve this by leveraging state-of-the-art techniques and utilizing the features provided by QASM 3.0 to democratizing the access to quantum-hybrid solutions, intelligently managing the interaction between classical and quantum realms.

# Chapter 5

# Discussions and Impact

*This Chapter comprehensively discusses the main contributions and their scientific and industrial impact.*

Table 5.1: Impact of Contributions on Industry and Scientific Community

| Contribution | Industry | Scientific Community |
|---|---|---|
| Effort-aware defect prediction [52] | Improved prediction accuracy; Reproducibility tool | Improved reproducibility, generalizability, and scientific rigour |
| Hidden risks in untouched code [104] | Proactive risk management; Risk awareness | New Research area; Improved predictive modeling |
| JIT defect prediction [119] | Informed decision-making; Accuracy insights | MDP/CDP improvements |
| SASTTs for Java [103] | Informed tool selection; Security improvements | Extensive SASTTs analysis |
| VPS datasets [105] | Dataset accessibility; Improved vulnerability management | Facilitated reproducibility; standardized dataset use |
| Vulnerability prioritization [108] | Severity clarity; Targeted vulnerability management | Addressed classification ambiguities; Experts insights |
| Risk analysis in mission-critical IT systems [109] | Scalable AI-based risk assessment | AI-driven PSRA methodology |
| QSE [340, 112] | Democratized QC access; Faster QC onboarding | QSE advancement; broader participation |

Table 5.1presents the impact of the thesis contributions on industry and the scientific community. We sought to address critical challenges in SE through AI-driven methodologies and developing tools and frameworks for defect and vulnerability prediction; we sought to improve decision-making, enhance security, and expand accessibility to innovative technologies. Our work targeted critical areas in SE, such as reproducibility, method validation, proactive risk management, and QSE, each with a distinct impact on industry and academia.

Our contributions in SE practices strive towards higher accuracy, accessibility, and applicability of AI-driven methodology. Industrial tools and models that enable us to proactively manage organizations' vulnerabilities for optimizing defect prediction facilitate increased software quality and security standards. For instance, our effort-aware defect prediction framework mitigates the lack of reproducibility in defect studies with a robust tool to perform effective measurement and comparison of defect predictions by developers. With the incorporation of JIT information in defect prediction, industries will also be in a better position to make decisions on which predictive models best suit their needs. This could help reduce maintenance costs with improved code quality.

Our work on unseen risks of unmodified code introduces a new perspective to defect prediction, making both industry and academia aware of the prevalence of such risks while providing predictive models that could be used to reduce said risks. This proactive approach is carried over to vulnerability prioritization with our work on severity metrics and classification systems, offering new clarity that allows security teams to target the most critical threats, reducing security debt.

The scientific merit of the VALIDATE tool consists of how it tries to assist in overcoming the reproducibility crisis with high-quality datasets in machine learning-based vulnerability prediction studies. This tool aids in searching for an appropriate dataset by a researcher and curating the dataset, thus standardizing VPS research and allowing comparability across different studies. This focused review of SASTTs and CWEs for Java programming advances the frontier by underlining the gaps in existing tools. It motivates further refinement of static analysis, thus acting as a guideline for improvement by researchers.

Our contributions range from classical SE to adopting quantum computing and AI for critical applications. $Classi|Q\rangle$ and QCSHQD democratize the chain in QSE by enabling classical developers to tap into quantum computing resources and by pioneering the use of LLMs in the risk analysis of mission-critical systems, offering a practical, scalable alternative to human-driven assessment and bringing AI-driven insight to contexts where rapid, reliable risk assessment is critical.

Finally, our contributions point toward a proactive, scientifically sound approach to AI4SE. Our contributions aimed to fill gaps between theory and practice, contributing to a more approachable, resilient, and scientifically sound approach toward SE challenges.

# Chapter 6

# Threats To Validity

*This Chapter comprehensively discusses the main threats to validity of the thesis.*

**Conclusion Validity**. Conclusion validity concerns issues that affect the ability to draw accurate conclusions regarding the observed relationships between the independent and dependent variables [413]. We tested all hypotheses with non-parametric tests, such as the Kruskal-Wallis and Wilcoxon Signed Rank tests, which, although effective, are prone to type-2 errors (i.e., failing to reject a false hypothesis). However, we rejected the hypotheses in all cases, minimizing the likelihood of type-2 error. We opted for non-parametric tests to reduce the risk of type-1 errors (i.e., rejecting a true hypothesis), which we deemed less desirable in this context.

Spearman's $\rho$ further allowed us to explore correlations between variables in a way suited for tied data, although it assumes linear monotonic relationships that may not always hold. We verified data distributions and used Spearman's $\rho$ over alternatives like Kendall's $\tau$ due to its effectiveness with our dataset's tied observations. We acknowledge, however, that different measures could yield slightly different interpretations, which may affect the overall strength of the conclusions.

**Internal Validity**. Internal validity addresses factors that may affect the causality of our findings. A primary internal threat arises from the absence of an official SQ rule-to-CWE mapping API or dataset, which we addressed by conducting a manual mapping based on SonarQube documentation[1]. We carefully cross-referenced SonarQube rules to CWE identifiers to ensure alignment despite minor misclassification risks. Moreover, we mitigated the lack of ground truth for defectiveness in methods and classes by selecting widely used projects in prior studies and cross-validating the classification with multiple researchers.

For instance, our industrial study [109] study relies on multiple samples from mission-critical analyses in various fields, including national security and education. We made design choices, such as using a substantial sample (141 instances) and employing state-of-the-art techniques like Auto-Weka. However, tuning options, such as running time (set at two hours per dataset) and using the proportion of accurate classifications, could vary slightly, potentially impacting findings. We confirmed that results were robust across different classifiers, including Random Forest and IBK, reducing this threat.

[1] `https://rules.sonarsource.com/java/`

**Construct Validity**. Construct validity relates to how our measurements truly reflect what we claim to measure [413]. We addressed construct validity by using the walk-forward technique and selecting metrics in alignment with well-established studies. Design choices like classifiers, features, and metrics could influence outcomes, but we based these selections on widely accepted practices in defect prediction research.

A threat lies in the representativeness of datasets, as we excluded subjective measures (e.g., ease of retrieval) and classified datasets consistently across researchers to maintain reliability. We validated the classification process with a high inter-rater agreement measured by Cohen's Kappa, strengthening our confidence in the data's alignment with our constructs. The evolving nature of vulnerability datasets and the need for interpretability, especially in Vulnerability Prediction Systems (VPS), are also acknowledged limitations.

**External Validity**. External validity reflects the extent to which our findings generalize beyond this study. Our dataset comprises open-source projects, limiting its direct applicability to industrial settings. However, to increase generalizability, we selected various datasets, including Java-based and other commonly used project types. The National Vulnerability Database (NVD) dataset reflects data as of June 2023, and our focus on CWEs limits the generalizability to other programming languages or domains.

The ongoing evolution of vulnerability data poses further limitations, as future versions of datasets may impact reproducibility. To promote transparency and support replication, we provide a comprehensive replication package that includes all datasets, scripts, and results[2]. To ensure long-term applicability, we plan to incorporate additional dataset filters and address user feedback in future studies, adapting our classifications to emerging research needs.

Similarly, generalizability in terms of the technological transfer, Wieringa and Daneva [407] propose a lab-to-field strategy that is particularly appealing, as it talks of scaling the findings from controlled environments, like open-source datasets, to real-world industrial settings where conditions can hardly be predicted. Wieringa and Daneva [407] are also in line with the challenges of our study since adapting results obtained based on datasets like NVD to larger industrial domains requires careful validation and iterative refinement. Their case-based and sample-based generalization strategies also provide techniques to reduce project variability by shifting the focus toward recurring patterns and smaller, more stable components in the data. These strategies could improve the transferability of our findings, such that they continue to be relevant for both open-source and industrial contexts.

Moreover, Ghaisas et al. [144] complement Wieringa and Daneva [407] by underlining generalization based on similarity, obtained from recurring lessons learned in industrial case studies. Their approach will be particularly relevant in transferring knowledge gained from open-source environments to similar industrial environments by leveraging domains, types of projects, or risk profiles. With the thesis, one can achieve similarity-based generalization that will help him find commonalities between open-source projects and their industrial counterparts and allow him to transfer insights into practical applications. This perspective also underlines the importance of embedding our findings in real-life scenarios, as illustrated by the industrial focus in their work. To address both issues, we also engaged with industry surveying practitioners and collaborated with the industrial research team to assess the validity of our findings in an industrial scenario [109, 110].

[2]`https://doi.org/10.5281/zenodo.8138497`

# Chapter 7

# Future Works

*This chapter briefly mentions works published after compiling the present thesis and possible future works.*

The key areas that this thesis has highlighted in the context of software development and risk analysis have to do with future research directions and enhancement studies. First, in the context of untouched code, improvement regarding algorithms for estimating defects for these segments might come from deep learning or ensemble methods. With an understanding of the root causes of risks for untouched code, new mitigation strategies might be constructed, especially when longitudinal studies track how these risks evolve and evaluate proactive management strategies over time.

In the future, research in JIT defect prediction could focus on further efficiency in JIT information across different software development methodologies. Integrating JIT data with automated testing or continuous integration allows real-time defect detection and prevention, assessing scalability on large-scale projects and distributed systems.

For SASTTs, such an extension of analysis to more programming languages and frameworks would give a wider perspective. Industry collaborations would, therefore, enable the validation of refinements based on real-world scenarios. At the same time, studies on the impact of SASTT recommendations on developer productivity and software quality would provide important insights.

In vulnerability prioritization studies, extending the review methods of datasets for domains other than vulnerability prediction may be helpful to gain further insights. Standardization of benchmarks in evaluating systems of vulnerability prediction would contribute to stronger assessments, and synthetic vulnerability datasets could be developed to address the biases in improving prediction accuracy.

Second, vulnerability prioritization could be improved by researching temporal factors affecting severity. Again, development by automated tool integration with data from CVE, CWE, and SQ rules should help streamline this; further, such industry partnerships would help review the effectiveness in real-world security response.

Case studies across various industries will help validate Large Language Model-based risk analysis in PSRA for mission-critical IT systems. Integrating these techniques with the currently existing frameworks for their large-scale performance evaluation will form the basis for better decision-making.

Finally, on QSE, creating training materials for classical developers and asking for feedback regarding the usability of the QCSHQD platform should increase the number of contributors to quantum software developments. Collaboration with quantum hardware manufacturers would ensure that tools are compatible with various quantum architectures, allowing broader adoption.

My current research focuses on understanding and mitigating technical debt, improving risk analysis in AI-driven systems, and enhancing security practices in software engineering. One of my recent studies explores the correlation between architectural smells and static analysis warnings, revealing insights into how these elements impact code quality and maintainability [111]. This contributes to a broader understanding of refactoring practices and the potential for guiding developers toward a more resilient codebase [329].

Much of my work addresses the role of large language models (LLMs), specifically their utility in risk analysis for mission-critical systems. In this context, I assess the actionability of generative AI (GenAI) within security frameworks to understand how these models can improve vulnerability identification and risk mitigation strategies [110]. This aligns with my ongoing efforts to leverage GenAI for advancing evidence-based software engineering [107].

Moreover, my research includes contributions to secure software solutions within interconnected environments. For instance, in vehicle-to-vehicle communication, I have examined cybersecurity challenges and proposed solutions to fortify secure connections between vehicles [388]. Furthermore, I am involved in developing software infrastructure for the edge-to-cloud continuum, an area critical for the evolution of 6G technology, addressing scalability and security across distributed networks [10].

Another focus of my research is understanding developers' perspectives on code complexity. I have conducted studies on early-career developers' perceptions, particularly regarding complexity metrics, highlighting the need for clearer, more understandable coding standards to improve developer productivity and code comprehension [106]. This work dovetails with our efforts to identify metrics that support developer-based refactoring recommendations, further facilitating maintainability [330].

Similarly, we are investigating the application of temporal network analysis to study and mitigate Microservice Architectural Degradation by leveraging static analysis methods to model and understand the temporal evolution of service dependencies [31].

Finally, I am examining the implications of technical debt in the broader software lifecycle. With my current colleagues, we are recently analyzing on both the benefits and risks associated with technical debt, particularly as perceived in developer discussions [222].

# Chapter 8

# Conclusions

*This chapter comprehensively concludes our thesis and wraps up our contributions.*

Our thesis has taken important steps toward overcoming the challenges and seizing the opportunity that AI presents to SSE. Therefore, our contribution is valuable to academia and industry as it sets the ground for new research venues and delivers practitioner-oriented implications to improve industrial practices.

Our work represents only one step toward such a future, in which educated and responsible AI use empowers developers and researchers to have confidence and accuracy while braving the intricacies of a modern software system.

The future beckons with the promise of further innovation and collaboration, giving ample opportunities for findings, tools, and techniques developed here to grow with the ever-evolving software engineering.

# Chapter 9

# Appendix

# A CRediT Author Statement

Table A.1: CRediT Authorship Contribution Percentages Across Papers

| **Paper Reference** | **[52]** | **[119]** | **[108]** | **[104]** | **[103]** | **[105]** | **[109]** | **[340]** | **[112]** |
|---|---|---|---|---|---|---|---|---|---|
| Conceptualization | 40% | 35% | 70% | 85% | 75% | 85% | 30% | 75% | 85% |
| Methodology | 30% | 25% | 80% | 75% | 80% | 70% | 75% | 45% | 90% |
| Software | N/A | N/A | N/A | N/A | N/A | 100% | N/A | N/A | N/A |
| Validation | 20% | 20% | 75% | 80% | 70% | 75% | 90% | 85% | 80% |
| Formal Analysis | 30% | 25% | 90% | 75% | 85% | 85% | 70% | 70% | 85% |
| Investigation | 30% | 25% | 85% | 90% | 80% | 95% | 85% | 60% | 90% |
| Writing - Original Draft | 50% | 45% | 95% | 90% | 75% | 90% | 65% | 80% | 85% |
| Writing - Review & Editing | 30% | 35% | 80% | 85% | 80% | 80% | 80% | 90% | 75% |
| Visualization | 20% | 20% | 70% | 70% | 70% | 80% | 85% | 75% | 80% |

The Contributor Roles Taxonomy (CRediT) was developed to acknowledge the distinct contributions of each author, address authorship disputes, and encourage collaborative research [14]. Originating from a 2012 workshop organized by Harvard University and the Wellcome Trust, CRediT was shaped through contributions from researchers, the International Committee of Medical Journal Editors (ICMJE), and publishing stakeholders like Elsevier, represented by Cell Press [81]. This taxonomy provides a structured way for authors to disclose their specific roles in a publication, enhancing transparency in author contributions [81, 14]. Table A.1 presents the thesis author's contributions to each published work according to the applicable CRediT categories.

# Chapter 10